\documentclass[
tightenlines,
nofootinbib,
 amsmath,amssymb,
 aps,
]{revtex4-1}

\usepackage{color,graphicx}
\usepackage{tcolorbox}
\usepackage{dcolumn}
\usepackage{bm}
\usepackage{slashed}
\usepackage{cancel}
\usepackage{mathtools}
\usepackage{soul}
\newcommand{\slas}[1]{\not\!{#1}}

\newcounter{comment}

\newcommand{\myprod}[2]{(#1\,#2)}
\newcommand{\realpol}{\Lambda_\gamma'}
\newcommand{\realeps}[2]{\varepsilon^{#1}_{\realpol \, #2}}
\newcommand{\Lcal}[2]{
\mathcal{L}_{h \Lambda_\gamma'}^{\nu \rho\,(#1,#2)}}
\newcommand{\sigmaunpol}[2]{\hat{\sigma}_{BH,\realpol}^{(#1,#2)}}
\newcommand{\sigmapol}[2]{\hat{\sigma}_{BH, \, h, \, \realpol}^{pol\,(#1,#2)}}

\begin{document}


\title{Extraction of Generalized Parton Distribution Observables from Deeply Virtual Electron Proton Scattering Experiments}

\author{Brandon Kriesten}%
\email{btk8bh@virginia.edu}
\affiliation{%
 Department of Physics, University of Virginia, Charlottesville, VA 22904, USA. 
}%

\author{Andrew Meyer}%
\email{ajm5an@virginia.edu}
\affiliation{%
 Department of Physics, University of Virginia, Charlottesville, VA 22904, USA. 
}%

\author{Simonetta Liuti}%
\altaffiliation[Also at ]{Laboratori Nazionali di Frascati, INFN, Frascati, Italy}
\email{sl4y@virginia.edu}
\affiliation{%
 Department of Physics, University of Virginia, Charlottesville, VA 22904, USA. 
}%

\author{Liliet Calero Diaz}%
\altaffiliation[Also at ]{Laboratori Nazionali di Frascati, INFN, Frascati, Italy}
\email{lc2fc@virginia.edu}
\affiliation{%
 Department of Physics, University of Virginia, Charlottesville, VA 22904, USA. 
}%
\author{Gary R. Goldstein}
  \email{gary.goldstein@tufts.edu}
\affiliation{%
Department of Physics and Astronomy, Tufts University, Medford, MA 02155 USA.}

\author{J. Osvaldo Gonzalez-Hernandez}
  \email{joghdr@gmail.com}
\affiliation{%
INFN, Torino}

\author{Dustin Keller}%
\email{dustin@jlab.org}
\affiliation{%
 Department of Physics, University of Virginia, Charlottesville, VA 22904, USA. 
}%


\date{\today}

\allowdisplaybreaks

\begin{abstract}
We provide the general expression of the cross section for exclusive deeply virtual photon electroproduction from a spin 1/2 target using current parameterizations of the off-forward correlation function  in  a  nucleon  for  different  beam  and  target  polarization  configurations  up  to  twist  three accuracy.  All contributions to the cross section including deeply virtual Compton scattering, the Bethe-Heitler process, and their interference, are described within a helicity amplitude based frame-work which is also relativistically covariant and readily applicable to both the laboratory frame and in a collider kinematic setting.  
Our formalism renders a clear physical interpretation of the various components of the cross section by making a connection with the known characteristic structure of the electron scattering coincidence reactions.
In particular, we focus on the total angular momentum, $J_z$, and on the orbital angular momentum, $L_z$. On one side, we uncover an avenue to a precise extraction of $J_z$, given by the  combination of generalized parton distributions, $H+E$,
through a generalization of the Rosenbluth separation method used in elastic electron proton scattering. On the other, we single out for the first time, the twist three angular modulations of the cross section that are sensitive to $L_z$. The proposed generalized Rosenbluth technique adds constraints and can be extended to additional observables relevant to the mapping of the 3D structure of the nucleon.
\end{abstract}

\pacs{Valid PACS appear here}
\maketitle

\section{Introduction}
\label{sec:Intro}
Current experimental programs of Jefferson Lab and COMPASS at CERN, as well as the planned future Electron Ion Collider (EIC) \cite{Accardi:2012qut,Geesaman:2015fha} are providing new avenues for concretely accessing the 3D quark and gluon structure of the nucleon and of the atomic nucleus. Knowledge of both the momentum and spatial distributions of quarks and gluons inside the nucleon will be conducive to understanding, within quantum chromodynamics (QCD), the mechanical properties of all strongly interacting matter. This includes the mass, energy density, angular momentum, pressure and shear force distributions in both momentum and coordinate space.	
%
The key to unlocking direct experimental access to spatial distributions of partons inside the proton was provided by Ji in Ref.\cite{Ji:1996ek}, where he suggested Deeply Virtual Compton Scattering (DVCS), $ e p \rightarrow e'p' \gamma$, as a fundamental probe where the high virtuality of the exchanged photon
makes it possible to gain insight on the partonic structure of the proton.
Simultaneously, by measuring the four-momentum transfer between the initial and final proton, similarly to elastic scattering experiments, one can obtain information on the location of the partons inside the proton by Fourier transformation. 
%

A challenging question since its inception has been to provide the formalism and theoretical framework for deeply virtual exclusive-type experiments including DVCS,  \cite{Kroll:1995pv,Guichon:1998xv,Belitsky:2001ns,Belitsky:2008bz,Belitsky:2010jw,Belitsky:2012ch,Diehl:2005pc,Braun:2014sta}, Deeply Virtual Meson Production, $ep \rightarrow e'p' M$ (DVMP), and Timelike Compton Scattering (TCS), $\gamma p \rightarrow {\it l}^+ {\it l}^- p'$, where a large invariant mass lepton pair is produced \cite{Belitsky:2002tf,Moutarde:2013qs,Boer:2015cwa,Boer:2015fwa}.  Separately measuring all of the helicity amplitudes which contribute to the hadronic current can allow us to constrain the underlying theoretical picture in terms of Generalized Parton Distributions (GPDs) \cite{Radyushkin:1997ki,Ji:1996ek} (see reviews in Refs.\cite{Diehl:2003ny,Belitsky:2005qn,Kumericki:2016ehc}). This stringent constraint on theoretical hypotheses will only be possible if the polarizations of the initial and final particles are measured.   

In this paper we derive a formulation of the deeply virtual photon electroproduction cross section in terms of helicity amplitudes. We calculate all configurations where either the beam and/or the target polarizations are measured for DVCS, for the Bethe-Heitler (BH) process, and for the interference term between the two. Extensions to include recoil polarization measurements and TCS will be provided in future publications. 
While many dedicated previous publications on this subject \cite{Kroll:1995pv,Guichon:1998xv,Belitsky:2001ns,Belitsky:2005qn,Belitsky:2008bz,Belitsky:2010jw,Belitsky:2012ch,Diehl:2005pc,Braun:2014sta} have been useful
to guide an initial set of experiments (reviewed in \cite{Kumericki:2016ehc}), we are now entering a more quantitative and
accurate experimental era that will extend from the modern Jefferson Lab program into the future EIC kinematic range. For a reliable extraction and interpretation of physics observables from experiment it is, therefore, timely to introduce the formalism for all deeply virtual exclusive processes according to the following set of benchmarks: 

\vspace{0.5cm}

\noindent\textit{i)} Be general, covariant, and exactly calculable\\
\textit{ii)} Provide kinematic phase separation\\
\textit{iii)} Provide clear information extraction\\
 
To clarify benchmark \textit{i)}, the formalism should be general so as to consistently describe and compare observables from all of the deeply virtual exclusive processes.  All steps from the construction of the lepton and hadron matrix elements to the final observable should be clearly displayed and directly calculated, including any instance of kinematic approximations.  The formalism should be present in a covariant description which can be used to interpret experimental results in any reference frame. 

Benchmark {\it ii)} implies that a clear pathway to data analysis should be provided where, for any independent polarization configuration, one has control over both the dynamic $Q^2$ dependence (twist expansion) and the kinematic dependence, including ${\cal O}(1/Q^2)$ sub-leading terms. In particular, each polarization correlation in the DVCS cross section can be written as the sum of terms of different twist, each one of these terms in turn appearing with a characteristic dependence on the azimuthal angle, $\phi$, the virtual photon polarization vector's phase. 
Both the $\phi$ and $Q^2$ dependence of the BH cross section are, instead, of pure kinematic origin resulting from the components of the four vector products in the transverse plane. The interference term contains $\phi$ dependence originating from both sources which has to be carefully disentangled. 

The ultimate goal of benchmark {\it iii)} is to bring out the physical interpretation of the different contributions to the cross section. The standard treatment of all exclusive lepto-production processes has been to organize the cross section in a generalized Rosenbluth form \cite{Rosenbluth:1950yq} (see {\it e.g.} \cite{Frullani:1984nn,Donnelly:1985ry,Raskin:1988kc,Sofiatti:2011yi,Milner:2018mmz}).  The same formalism is extended here to $ep\rightarrow e'p' \gamma$.
For example, this opens the way to uniquely determine the direct contribution of angular momentum as parametrized in \cite{Ji:1996ek} by the sum of GPDs, $H+E$ by Rosenbluth separation. The contribution of other GPDs can be disentangled within the same approach. The extraction of observables by Rosenbluth separation grants us a much needed extraction tool as well as a model independent methodology. 

\vspace{0.5cm}

The structure of the Virtual Compton Scattering and BH cross sections was previously studied in several papers, starting from the pioneering work of Ref.\cite{Kroll:1995pv} to the more recent helicity based formulations of Refs. \cite{Arens:1996xw,Diehl:2005pc,Goldstein:2010gu,Belitsky:2012ch,Braun:2014sta}. While some of the benchmarks were met in previous works, this is the first time, to our knowledge, that all criteria are satisfied within a unified description.
Specifically, helicity based formulations were outlined but not fully worked out by Diehl and collaborators in Refs.\cite{Diehl:1997bu,Diehl:2005pc,Diehl:2003ny}. 
Detailed derivations were subsequently given in Refs.\cite{Belitsky:2005qn,Belitsky:2008bz,Belitsky:2010jw,Belitsky:2012ch}. However, in an attempt to organize systematically the various kinematic dependencies, 
the contributions of the various polarization configurations were expanded into a Fourier series in $\phi$. This step provided a convenient, although approximate scheme to organize an otherwise rather complicated kinematic structure into harmonics. The most evident drawback of the ``Fourier harmonics" approach is that it disallowed a straightforward physical interpretation. 
Contributions that are vital to extract, for instance, the angular momentum terms have been either disregarded or deemed as subleading. A confusing situation has arisen on the role of various terms contributing at twist two and twist three as well as on the kinematic power corrections (see talk in \cite{Defurne_INT}) to which we provide a remedy.

We present the general structure of the cross section in terms of its BH, DVCS, and BH-DVCS interference terms in Section \ref{sec:2}.
The DVCS contribution to the cross section is  written 
in terms of structure functions for the various beam and target polarization configurations in Section \ref{sec:3}. 
The DVCS cross section displays the characteristic azimuthal angular dependence of coincidence scattering processes  that stems from the phase dependence of the $\gamma^* p \rightarrow \gamma'p'$, helicity amplitudes with the virtual photon, $\gamma^*$, aligned along the $z$ axis \cite{Frullani:1984nn,Donnelly:1985ry,Raskin:1988kc,Arens:1996xw,Diehl:2005pc,Sofiatti:2011yi,Milner:2018mmz}. 
In Sec. \ref{sec:3} we also provide an interpretation of the various polarization structures in terms of twist two and twist three GPDs.

The BH contribution is described in Section \ref{sec:BH}. 
For each parity conserving polarization configuration the BH cross section is written in a Rosenbluth-type form, displaying two quadratic nucleon form factor combinations multiplied by coefficients functions. In the unpolarized case, for instance, the two form factors correspond to the nucleon electric and magnetic form factors. The coefficient functions are  given by non trivial expressions in $\phi$.
The complicated structure of the $\phi$ dependence of these coefficients, in comparison to the DVCS one shown in Sec.\ref{sec:3}, is due to the fact that: {\it i)}  the lepton part of the cross section also contains an outgoing photon compared to the simpler $e \rightarrow \gamma^* e'$ vertex in DVCS; {\it ii)}  the BH virtual photon momentum, $\Delta$, is also offset from the $z$ axis by the polar angle, $\theta$.

The complication introduced in the BH kinematics also affects the BH-DVCS interference term. 
In Section \ref{sec:BHDVCS} we present a formulation that keeps the kinematic $\phi$ dependence stemming from four-vector products of the various momenta distinct from the helicity amplitudes (dynamic) phase dependence inherent to the polarization vectors. The Conclusions and Outlook are presented in Section \ref{sec:conc}. The Appendices contain many details of the calculation that are useful for both a direct verification of the helicity amplitudes formalism results and possible extensions to other configurations.

The main advantage of adopting our newly proposed formalism is that it brings out the inherent Rosenbluth-type structure of the deeply virtual exclusive scattering processes. Similarly to the BH contribution, the interference term can be written in an extended Rosenbluth-type form where, instead of two quadratic nucleon form factor combinations, we now have three combinations containing products of form factors and GPD dependent terms. For illustration, we show the BH contribution to the unpolarized cross section in Eq.(\ref{eq:BHintro}), and the leading order BH-DVCS interference term in Eq.(\ref{eq:BHDVIntro}),
\begin{eqnarray}
\label{eq:BHintro}
\frac{d^5\sigma^{BH}_{unpol}}{d x_{Bj} d Q^2 d|t| d\phi d\phi_S } &=&
\frac{\Gamma}{t^2} \left[ A_{BH} \left(F_1^2 + \tau F_2^2 \right)+ B_{BH} \tau G_M^2(t)  \right] \\
\label{eq:BHDVIntro}
\frac{d^5\sigma^{{\cal I}}_{unpol}}{d x_{Bj} d Q^2 d|t| d\phi d\phi_S } &=&   \frac{\Gamma}{Q^2 (- t ) } \,  \Big[ 
A_{\cal I}   \left(F_1 \Re e \mathcal{H} + \tau F_2  \Re e \mathcal{E} \right)  
+ B_\mathcal{I} \,  G_M  \, \Re e (\mathcal{H+E})
+ C_\mathcal{I} \,   
G_M \, \Re e \widetilde{\mathcal{H}} \Big] .
\end{eqnarray}
The detailed equations are derived and discussed in the following sections. 
This is not an exhaustive listing, and, for illustration purposes, only the unpolarized case  for BH (Eq.\eqref{eq:BHintro}) and the unpolarized leading order  for the BH-DVCS interference term (labeled ${\cal I}$ in Eq.\eqref{eq:BHDVIntro}) are quoted. 
\begin{figure}
\includegraphics[width=8cm]{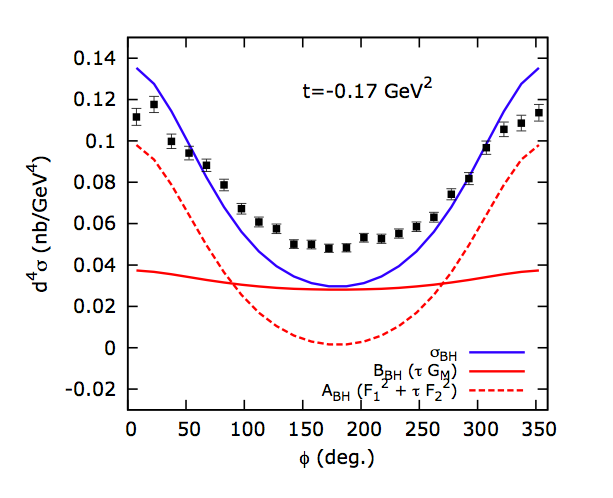}
\includegraphics[width=8.3cm]{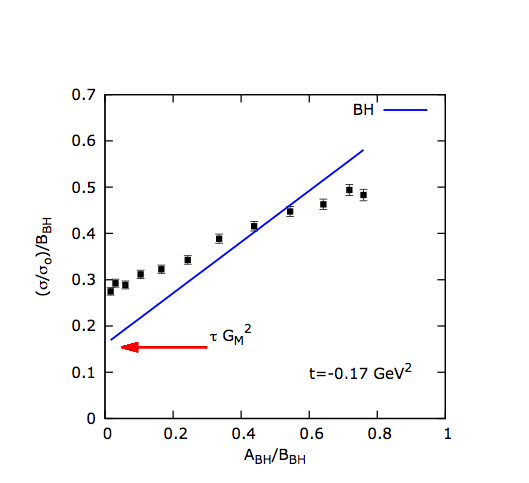}
\caption{(color online) {\em Left panel}: A sample of the new precise $ep\rightarrow e'p' \gamma$ data from Ref.\cite{Defurne:2015kxq} for unpolarized lepton proton scattering in the kinematic setting: $E_e=5.75$ GeV, $Q^2 = 1.82$ GeV$^2$, $x_{Bj}=0.34$, $t=-0.172$ GeV$^2$. The cross section is plotted vs. the azimuthal angle, $\phi$. The curves represent the BH contribution: (red dashed line) $\tau G_M^2$ term; (red full line) $(F_1^2 + \tau F_2^2)$ term; (blue) sum of the two; {\em Right panel}: Reduced cross section obtained from the same set of data  plotted vs. the kinematic variable $A_{BH}/B_{BH}$ from the Rosenbluth-type formula defined in Sec.\ref{sec:BH}. The straight line represents the BH calculation  intercepting the y-axis at $\tau G_M^2$. The difference between the data points and the curve reflects the contribution from the DVCS process.The formulation of the BH cross section is given in detail in Section \ref{sec:BH}}
\label{fig:defurne1}
\end{figure}

In both equations, $F_1$, $F_2$ are the Dirac and Pauli form factors , $G_M=F_1+F_2$ is the magnetic form factor $G_E = F_1 - \tau F_2$; $t$ is the momentum transfer squared,  ($\tau=-t/4M^2$); in Eq.(\ref{eq:BHDVIntro}) $\mathcal{H},  \mathcal{E}, \widetilde{\cal H}$  are  Compton form factors containing the GPDs that integrate to $F_1$, $F_2$ and $G_A$, respectively \cite{Diehl:2003ny}. 
$A_{BH}, B_{BH}$ are kinematic coefficients which are exactly calculable and rendered in covariant form in the following sections; $A_\mathcal{I}, B_\mathcal{I}, C_\mathcal{I}$ are also covariant kinematic coefficients which, however, contain an extra dependence on the phase $\phi$ as we also explain in what follows. 
The new formalism allows us to emphasize the physics content of the cross section: Eqs. \eqref{eq:BHintro} and \eqref{eq:BHDVIntro}  show a similar form where in both cases we can identify the first term in the equation with the electric form factor type contribution, and the second term with the magnetic form factor contribution. For the BH-DVCS interference we also have an extra function which includes the axial GPD (interestingly, a similar term would also be present in BH but parity violating). Similar structures are found for other polarization configurations. 

To be clear, we replace the ``harmonics-based" formalism adopted in most DVCS analyses with a Rosenbluth-based formulation which emphasizes the physics content of the various contributions, {\it e.g.} by making a clear parallel with coincidence scattering experiments, even if this implies introducing more complex $\phi$ dependent kinematic coefficients. Instead of following a harmonics based  prescription which, as shown in many instances, is fraught with ambiguities, we organize the cross section by both its phase dependence, disentangling the twist two and transversity gluons from higher twist contributions, and by its form factor content. The price of evaluating more complex $\phi$ structures is paid off not only by having a much clearer physics-based formulation, but also by the fact that the coefficients are exactly calculable: no approximation enters the calculation within the Born approximation adopted here. The numerical dependence on the various kinematic variables will be discussed in an upcoming publication.

\begin{figure}
\begin{center}
\includegraphics[width=7.2cm]{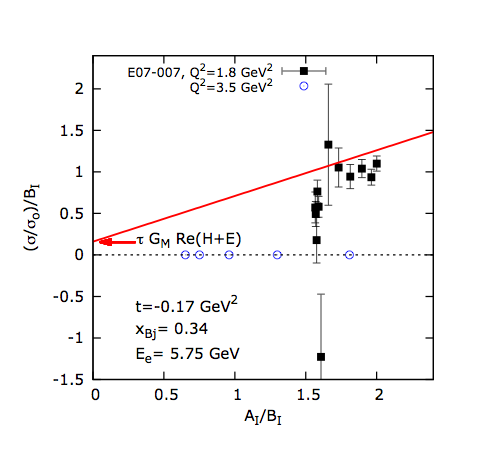}
\caption{
 Rosenbluth separation applied to the same data set as in Fig.\ref{fig:defurne1} \cite{Defurne:2015kxq}. The reduced cross section is plotted vs. the kinematic variable $A_{{\cal I}}/B_{{\cal I}}$ defined in Eq.(\ref{eq:BHDVIntro}) and Sec.\ref{sec:BHDVCS}. The line represents a model calculation and it intercepts the y-axis at $\tau G_M \Re e ({\cal H}+ {\cal E})$. The data points are the BH-DVCS contribution extracted by subtracting the calculated BH term from the unpolarized absolute cross section data of Ref.\cite{Defurne:2015kxq}. The line is a model evaluation obtained using GPDs from Ref.\cite{Goldstein:2010gu}.  The graphs also shows kinematic projections at higher $Q^2$, or by varying the angle $\theta_{e e'}$ between the initial and final lepton.}
\label{fig:defurne2}
\end{center}
\end{figure}

In Figures \ref{fig:defurne1},\ref{fig:defurne2} we illustrate the working of the Rosenbluth separation for typical kinematic settings from the  Jefferson Lab experiment E00-110 \cite{Defurne:2015kxq}. 
In Fig.\ref{fig:defurne1} we show, on the {\em lhs}, the $e p \rightarrow e'p'\gamma$ unpolarized cross section data plotted vs. $\phi$;  
on the {\em rhs} we plot the reduced cross section for the same set of data vs. the kinematic variable $A_{BH}/B_{BH}$ (the detailed definition of this quantity is given in Section \ref{sec:BH}). The BH cross section appears as a linear function of the variable $A_{BH}/B_{BH}$, with intercept given by $\tau G_M^2$ and slope given by $F_1^2 + \tau F_2^2$. The difference between the data and the BH line reflects the contribution from the DVCS process.

A generalized Rosenbluth separation can be performed for the BH-DVCS interference case, Eq.\eqref{eq:BHDVIntro}, by defining an analogous kinematic variable, $A_{\cal I}/B_{\cal I}$ similar to $A_{BH}/B_{BH}$ defined for BH (Fig.\ref{fig:defurne2}).
 The coefficient $C_{\mathcal I}$ is negligible compared to the other two.  The intercept with the y-axis is given by $\tau  G_M \Re e ({\cal H}+ {\cal E})$.
 \footnote{
 In the kinematic regime considered the DVCS contribution is expected to be dominated by the BH-DVCS interference term; the correction from the pure DVCS contribution is estimated to be $\approx 10 \%$.}
 Therefore, by exploiting the generalized Rosenbluth form of the BH-DVCS cross section one can directly extract the  Compton form factor combination describing angular momentum \cite{Ji:1996ek}. This term was deemed of higher order in all of the previous analyses because, similarly to what happens in elastic scattering and in the $\tau G_M^2$ term in BH, it is kinematically suppressed; however, it can be extracted if one disentangles it according to our proposed generalized Rosenbluth formulation. We reiterate that our example is for illustration purpose only. To obtain a precise value of both $\Im m (\cal{H+E})$ and $\Re e (\cal{H+E})$, a systematic analysis is in preparation. 
 
\section{General Framework}
\label{sec:2}
DVCS is measured in leptoproduction of
a real photon in the region of large momentum transfer between the initial and final lepton, where also an interference with the Bethe-Heitler radiation occurs, according to the reaction,
\begin{equation}
l(k, h)+p(p, \Lambda) \rightarrow l^\prime(k^\prime, h) + 
\gamma(q^\prime, \Lambda_\gamma^\prime) + p^\prime(p^\prime, \Lambda^\prime)
\label{process}
\end{equation}
with indicated momenta and helicities (Figures.~\ref{fig:BHDVCS},\ref{fig:kinematics}).
In this paper we present the formalism for a spin $1/2$ (nucleon) target.

The cross section is differential
in the four-momentum squared
of the virtual photon, $Q^2$, the four-momentum transfer squared between the initial and final protons,  
$t$, Bjorken $x_{Bj}=Q^2/2M\nu$, $\nu$ being the energy of the virtual photon, and two azimuthal angles measured relatively to the lepton
scattering plane, the angle $\phi$ to the photon-target scattering plane and
the angle $\phi_S$ to the transverse component of the target polarization
vector, as displayed in Fig.~\ref{fig:kinematics}. In what follows we give a detailed definition of both the general cross section and the various observables for deeply virtual photon production off a spin $1/2$ target. 

\subsection{Cross Section}
The cross section describing process (\ref{process}) is given by,
\footnote{Note that the dimensions of the cross section are given in nb/GeV$^4$. Eq.(\ref{eq:xs5foldgeneral}) is consistent with the definition of $\mid T \mid^2$ having dimension 1/(energy) squared while the helicity amplitudes, defined below, are dimensionless (see Appendix \ref{appa}).}

\begin{equation}
\frac{d^5\sigma}{d x_{Bj} d Q^2 d|t| d\phi d\phi_S } =
\Gamma \,  
\big|T\big|^2 \;.
\label{eq:xs5foldgeneral}
\end{equation}
$T$ is a
coherent superposition of the DVCS and Bethe-Heitler amplitudes,
\begin{equation}
T(k, p, k^\prime, q^\prime, p^\prime) = T_{DVCS}(k, p, k^\prime, q^\prime, p^\prime) + T_{BH}(k, p, k^\prime, q^\prime, p^\prime),
\end{equation}
%
%
yielding,
\begin{equation}
|T|^2 = |T_{\rm BH} + T_{\rm DVCS}|^2
=|T_{\rm BH}|^2 + |T_{\rm DVCS}|^2 + \mathcal{I}\;.
\label{eq:coherent}
\end{equation}
\begin{eqnarray}
\label{interf}
\mathcal{I} & = & T_{BH}^{*} T_{DVCS}
+ T_{DVCS}^{*} T_{BH} .
\end{eqnarray}
\begin{figure}
\begin{center}
\includegraphics[width=14cm]{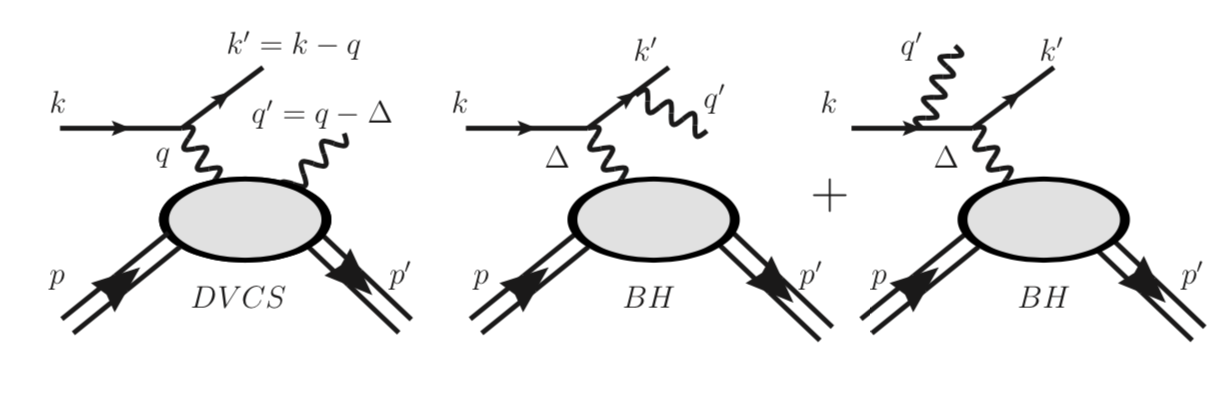}
\caption{Exclusive electroproduction of a photon through the DVCS and BH processes.}
\label{fig:BHDVCS}
\end{center}
\end{figure}

In the one photon exchange approximation the leptonic parts for the DVCS and BH are (Fig.\ref{fig:BHDVCS}),  
\begin{subequations}
\begin{eqnarray}
{\rm (DVCS)} \;\;\;  l(k) \rightarrow l^\prime(k^\prime) + {{\gamma^*}}(q) 
\label{DVCS1} \\
{\rm (BH)} \;\;\;   l(k)\rightarrow l^\prime(k^\prime) + {{\gamma^*}}(\Delta)+ \gamma'(q^\prime), 
\label{BH1}
\end{eqnarray}
\end{subequations}
while the DVCS and BH hadronic processes are given by,  
\begin{subequations}
\begin{eqnarray}
{\rm (DVCS)} \;\;\;  {\gamma^*}(q) + p \rightarrow \gamma^\prime(q') + p^\prime, \\
{\rm (BH)} \;\;\;  {\gamma^*}(q) + p \rightarrow p^\prime 
\end{eqnarray}
\end{subequations}

We define,
\begin{equation}
\label{eq:Gamma}
\Gamma = \frac{\alpha^3}{16\pi^2 (s-M^2)^2 \sqrt{1+\gamma^2}\, x_{Bj} },
\end{equation}
where $\alpha$ is the electromagnetic fine structure constant. $\Gamma$ has dimensions GeV$^{-4}$; the modulus squared of the matrix elements have therefore dimension GeV$^{-2}$, consistently with the cross section definition \eqref{eq:xs5foldgeneral}. We define,
\begin{equation}
Q^2 =- q^2= -(k-k')^2, \quad x_{Bj} = \frac{Q^2}{2 (pq)} = \frac{Q^2}{2 M \nu}, \quad  \nu = \frac{(pq)}{M},  \quad   \label{eq:invariants1}
\end{equation}
with $q  =  k-k'$, and $Q^2=-q^2$, $M$ being the proton mass.
Other kinematic variables are,

\footnote{ {We use the light cone kinematics notation  $v^\pm = (v^0 \pm v^3)/\sqrt{2}$, and metric $g_{{\rm oo}} = 1, g_{11}=g_{22}=g_{33}=-1$. }   }
\begin{subequations}
\label{eq:t}
\begin{eqnarray}
\Delta & \equiv & p^\prime - p =- q^\prime + (k - k^\prime) ;  \quad   P  =  \frac{p+ p'}{2}
\\
t & = & \Delta^2 \quad\quad  \xi  =  - \frac{(\Delta q)}{2(P q)} = \frac{(q+q')^2}{2[(Pq)+(Pq')]} =  x_{Bj} \frac{1+\displaystyle\frac{t}{2Q^2}}{2-x_{Bj}+ \displaystyle\frac{x_{Bj} t}{Q^2}} 
\\
\gamma &=& \frac{2x_{Bj} M}{Q} =\frac{Q}{\nu}, \quad s=(k+p)^2, \quad y =\frac{(pq)}{(pk)},
\quad 
\end{eqnarray}
\end{subequations}
where  {$\xi$ represents the skewness parameter, or the difference in the ``+" momenta of the incoming and outgoing quarks,  $- \Delta^+/(2P^+)$, in the large $Q^2$ limit   \cite{Meissner:2009ww}.} 

Notice that the virtual photons in the BH and DVCS amplitudes are different: in DVCS the virtual photon has four momentum $q$, setting the hard scale of the scattering process, $Q^2$, while in BH it has four momentum $\Delta$. 

The DVCS amplitude is written as,
\begin{eqnarray}
\label{eq:TDVCScov}
T_{DVCS} & = &  \left[\overline{u}(k',h)\gamma^\nu u(k,h) \right] {\cal W}_{\mu \nu}(p,p') \left(\varepsilon^{\Lambda_{\gamma'} \mu}(q') \right)^* \, ,
\end{eqnarray}
where the quantity in square brackets denotes the leptonic process. $W_{\mu \nu}$ is the DVCS hadronic tensor to be described in Section \ref{sec:3}; $\varepsilon^{\Lambda_{\gamma'} \mu}(q')$ is the polarization vector of the outgoing photon, $\gamma'$.  

For BH one has,
\begin{eqnarray}
T_{BH} & = & \left[ \left(\varepsilon^{\Lambda_{\gamma'} \mu}(q')\right)^* L_{\mu \nu}^h(k,k',q') \right] \overline{U}(p',\Lambda')\Gamma^\nu U(p,\Lambda), 
\end{eqnarray}
where one factors out the quantity in the square bracket denoting the lepton part, and
the nucleon current.
We denote the electron helicity as $h$, the initial (final) proton helicities as $\Lambda (\Lambda')$, the final photon helicity as $\Lambda_\gamma'$, and the exchanged photon helicity for DVCS as $\Lambda_{\gamma^*}$.
The helicity dependence of the two types of amplitudes can be made explicit by expressing them as,
\footnote{The formalism considered throughout this paper is valid at order {$\alpha_{EM}$.}}
\begin{eqnarray}
T_{DVCS, \Lambda \Lambda^\prime}^{h \Lambda^\prime_\gamma}  &=&  \sum_{\Lambda_{{\gamma^*}}} A_h^{\Lambda_{\gamma^*}}(k, k^\prime, q) f_{\Lambda \Lambda^\prime}^{{\Lambda_{{\gamma^*}} \Lambda_\gamma^\prime}} (q, p, q^\prime, p^\prime) 
\label{eq:DVCS1b}
\\
T_{BH, \Lambda,\Lambda^\prime}^{h \Lambda_\gamma^\prime}  &=&  \left[ B_{h \Lambda_\gamma'} (k, k^\prime, q^\prime,\Delta) \right]_\nu \left[J_{\Lambda\Lambda^\prime}(\Delta, p, p^\prime)\right]^\nu .
\label{eq:BH1b}
\end{eqnarray}
$A_h^{\Lambda_{\gamma^*}}$ corresponds to the lepton-photon  interaction in Eq.(\ref{DVCS1}) and Fig.\ref{fig:BHDVCS} (left),
$\left[B_{h \Lambda_\gamma^\prime}\right]_\nu $, corresponds to the lepton process in Eq.(\ref{BH1}), Fig.\ref{fig:BHDVCS} (right),
\begin{eqnarray}
\label{eq:A_intro}
A_h^{\Lambda_{\gamma^*}} & = & \frac{1}{Q^2} \overline{u}(k',h)\gamma^\mu u(k,h) \left(\varepsilon^{\Lambda_{\gamma^*}}_\mu(q)\right)^*  \\
\label{eq:B_intro}
B^{h \Lambda_\gamma^\prime}_\nu  & = &  \frac{1}{\Delta^2} \left(\varepsilon^{\Lambda_{\gamma'} \mu}(q')\right)^* L_{\mu \nu}^h(k,k',q') 
\end{eqnarray}
The helicity amplitudes for the $\gamma^* p \rightarrow \gamma' p'$ scattering process in DVCS, and the nucleon current in BH are respectively, defined as,
\footnote{We adopt here the formalism for the helicity amplitudes as in Jacob and Wick \cite{Jacob:1959at} for states with momenta at angles $\theta, \phi$ (see \cite{Leader} for a detailed description of this formalism). 
}
\begin{eqnarray}
 f_{\Lambda \Lambda^\prime}^{{\Lambda_{\gamma^*} \Lambda_\gamma^\prime}} (q, p, q^\prime, p^\prime) & = & \left[\varepsilon^{\Lambda_{\gamma^*}}(q)\right]^\mu {\cal W}_{\mu\nu} [\varepsilon^{\Lambda_\gamma'}(q')]^{\nu *} \\
\label{eq:famp}
J_{\Lambda \Lambda^\prime}^\nu(\Delta, p, p')  &= &\overline{U}(p',\Lambda')\left[ \left(F_1+ F_2\right)\gamma_\nu 
  -\frac{(p+p')_\nu}{2M} F_2 \right] U(p,\Lambda) ,
\end{eqnarray}
where $F_1$ and $F_2$ are the proton Dirac and Pauli form factors. $W_{\mu\nu}$ is parameterized in terms of GPDs Compton Form Factors (CFFs), which are complex amplitudes. In this paper we adopt the parameterization of Ref.\cite{Meissner:2009ww} including twist two and twist three GPDs.  
The explicit expressions for
the DVCS lepton, $A_h^{\Lambda_{\gamma^*}}$, and hadron $f_{\Lambda \Lambda^\prime}^{{\Lambda_{\gamma^*} \Lambda_\gamma^\prime}}$, helicity amplitudes, are given in Section \ref{sec:3}; the BH lepton tensor, $L_{\mu \nu}^h$ (see also \cite{Guichon:1998xv}), and hadronic current $J^\mu_{\Lambda \Lambda'}$, are given in Section \ref{sec:BH}. 

In the expressions above we introduced the polarization vector for the virtual photon in the DVCS process, $\varepsilon^{\Lambda_{\gamma^*}}(q)$. 
While in the BH term the helicity of the exchanged photon with momentum $\Delta$ is summed over,  in DVCS the virtual photon helicity is singled out to separate the contributions of different twist. In particular, similarly to DIS, the twist two term corresponds to transversely polarized photons, the twist three term contains one longitudinally polarized photon, and the twist four term contains two longitudinally polarized photons. We will see in Section \ref{sec:3} that DVCS allows for transversely polarized photons helicity flip, $1 \rightarrow -1$, described by the transversity gluons GPD terms.  

\subsection{Kinematics}
\begin{figure}
\begin{center}
\includegraphics[width=8cm]{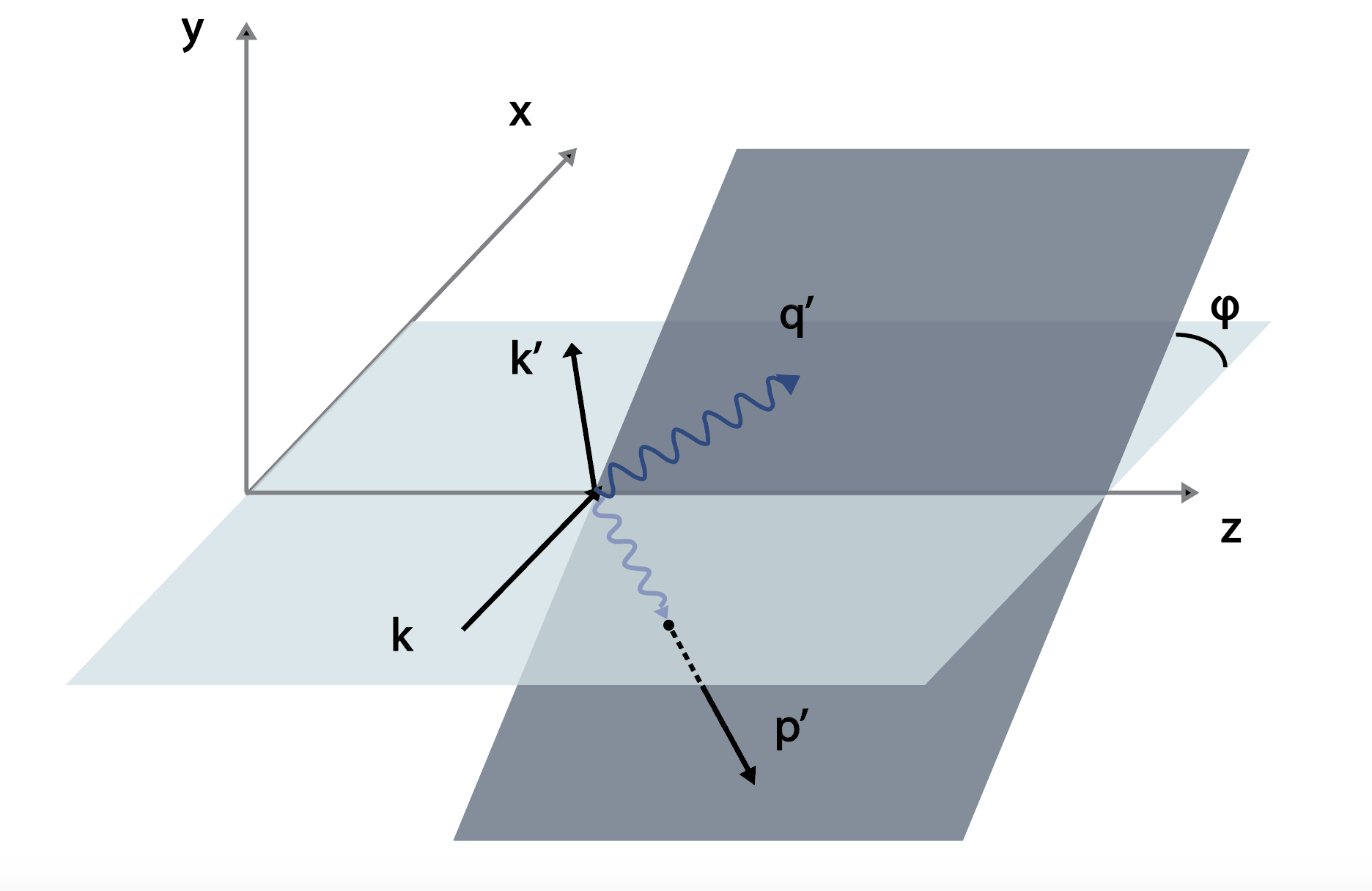}
\includegraphics[width=8cm]{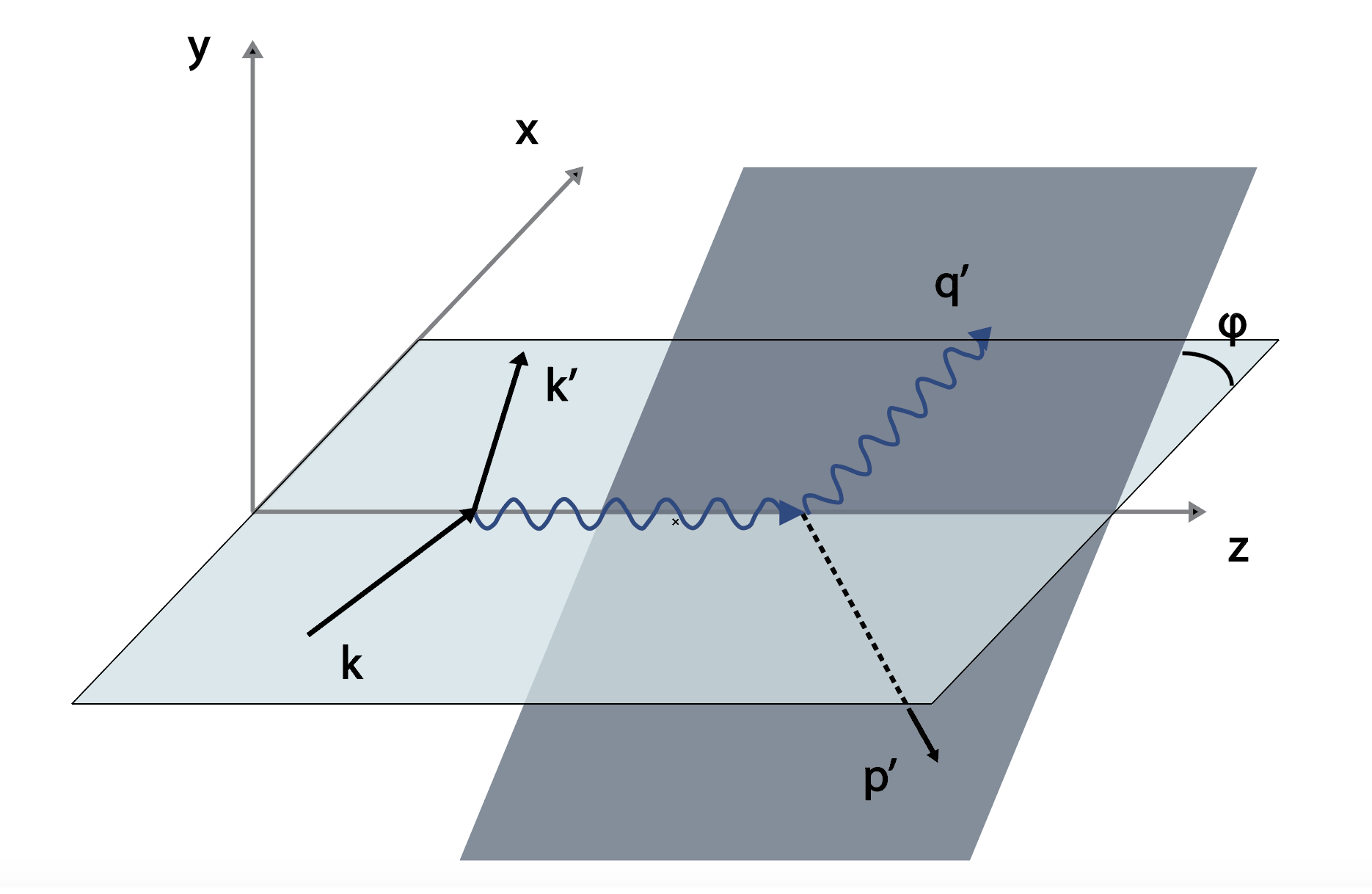}
\caption{Kinematics setting for the DVCS (right panel) and BH (left panel) processes.}
\label{fig:kinematics}
\end{center}
\end{figure}

We begin choosing the kinematics in the target rest frame, {\it i.e.} as in Fig.\ref{fig:kinematics}. 
Notice, however, that
the formalism developed in Sections \ref{sec:3}, \ref{sec:BH} and \ref{sec:BHDVCS} is fully covariant and it can be therefore extended to collider kinematics with either collinear or crossed beams. 
The incoming and outgoing electrons define the lepton plane, which 
is chosen here to be the $x$-$z$ plane; the hadron plane is fixed by the outgoing photon 
and the outgoing proton momentum at an azimuthal angle $\phi$ from the $x$-axis.
%
%
In this frame the four-momenta for the overall process, with ${\vec q}={\vec k} - {\vec k}^\prime$ along the negative $z$ axis read,
\label{eq:momenta}
\begin{align}
k & \equiv  \mid \vec{k} \mid(1;\sin \theta_l , 0 , \cos \theta_l) \nonumber\\
k^\prime & \equiv  \mid \vec{k}^\prime \mid(1; \sin \theta_l^\prime , 0 , \cos \theta_l^\prime) \nonumber \\
q & \equiv  (\nu; 0, 0, -\nu \sqrt{1+\gamma^2} ), \nonumber \\
q^\prime & \equiv    \mid \vec{q}^\prime \mid (1;   \sin\theta \cos\phi, \sin\theta \sin\phi,  \cos\theta), \nonumber\\
p & \equiv  (M;0,0,0) \nonumber \\
p^\prime & \equiv  (p_0^\prime ; \mid \vec{p}^\prime \mid \sin\theta^\prime \cos\phi, \mid \vec{p}^\prime \mid \sin\theta^\prime \sin\phi, \mid \vec{p}^\prime \mid \cos\theta^\prime)  \nonumber\\
\Delta & =  p'-p  \equiv  (p_0^\prime-M ; \mid \vec{p}^\prime \mid \sin\theta^\prime \cos\phi, \mid \vec{p}^\prime \mid \sin\theta^\prime \sin\phi, \mid \vec{p}^\prime \mid \cos\theta^\prime) ,
\end{align}
where we have taken the leptons to be massless.
Note that the exchanged photon in BH has momentum $\Delta = p^\prime-p$,
while in DVCS 
it has momentum $q \neq p'-p$.

All kinematic variables defined in Eq.\eqref{eq:momenta} are written below in terms of invariants.
The angular dependence of the various momenta can be written in terms of the invariants $x_{Bj}$, $y$, $\gamma^2$, $t$, and $M$, Eqs.(\ref{eq:invariants1},\ref{eq:t}). For the lepton angles one has,
\begin{subequations}
\begin{eqnarray}
\cos\theta_l &=& -\frac{1}{\sqrt{1+\gamma^2}}\left(1 + \frac{y \gamma^2}{2}\right), \quad \sin\theta_l = \frac{\gamma}{\sqrt{1+\gamma^2}}\sqrt{1-y-\frac{y^2 \gamma^2}{4}},   \\
\sin\theta_{l}^{\prime} &=&\frac{\sin\theta_{l}}{(1-y)}, \quad \cos\theta_{l}^{\prime} = \frac{\cos\theta_l + y\sqrt{1+\gamma^2}}{(1-y)}= - \displaystyle\frac{1-y - \displaystyle\frac{y \gamma^2}{2}}{(1-y)\sqrt{1+\gamma^2}}, 
\label{kkangles}
\end{eqnarray}
\end{subequations}
The beam energy, $k_o \equiv \mid \vec{k} \mid $, and final lepton energy, $k_o'\equiv \mid \vec{k}' \mid$ are, respectively,
\begin{equation}
\mid \vec{k} \mid  = \frac{Q}{\gamma y} \quad \quad \mid \vec{k}' \mid  = \mid \vec{k} \mid (1-y) = \frac{Q (1-y)}{\gamma y}.
\end{equation}
Finally, the angle between the two electrons, $\theta_{ee'}$, is defined by,
\begin{equation}
\cos \theta_{ee'} = \frac{1-y-\displaystyle\frac{y^2\gamma^2}{2}}{1-y}
\end{equation}

The outgoing photon angle, $\theta$, is obtained from the following equation that defines $t$, 
\begin{eqnarray}
t & = & (q-q^\prime)^2= -Q^2-2\nu \mid \vec{q}^\prime \mid (1+\sqrt{1+\gamma^2}\cos\theta), 
\label{cos} 
\end{eqnarray}
with,
\[\mid {q}^\prime \mid =   \nu + M - p_0^\prime  = \nu \left(1+\frac{t}{2M\nu}\right) = \frac{Q}{\gamma}\Big(1 + \frac{x_{Bj} t}{Q^2} \Big)
\label{q0} \]
so that,
\begin{eqnarray}
\cos\theta &=&  -\frac{1}{\sqrt{1+\gamma^2}} \left[ 1 + \frac{1}{2} \gamma^2 \frac{\left(1+\displaystyle\frac{t}{Q^2} \right)}{\left(1+\displaystyle\frac{x_{Bj} t}{Q^2}\right)}  \right].
\label{costheta} 
 \end{eqnarray}
Notice that the virtual photon is along the negative $z$ axis, therefore $q_3=-\mid \vec{q} \mid$.
Also that 
the specification of azimuthal angles does not change from the Laboratory to the Center of Mass (CoM) frame, since the ${{\gamma^*}}$ is in the same direction and the orientation of the planes is unchanged under the boost to the CoM. 

The allowed region of $t$ is given by varying $\cos\theta$ for each fixed $Q^2,\, \nu$, and $q_0^\prime$.  In the Laboratory frame this is equivalent to the elementary problem of finding either the minimum energy of the nucleon, $p_0^\prime$, or the maximum energy of the photon, $q_0^\prime$, that conserves the overall energy, $\nu+M$, and 3-momentum ${\vec q} = -\nu\sqrt{1+\gamma^2} {\hat z}$. 
Solving Eq.(\ref{cos}) for $q_0^\prime$ for $\cos\theta=1$, gives the minimum $t$, 
\begin{eqnarray}
t_0& = &-2M(\nu-q_{0\, max}^\prime)= \frac{Q^2\left(1-\sqrt{1+\gamma^2} + \frac{1}{2} \gamma^2\right)}{x_{Bj}\left( 1-\sqrt{1+\gamma^2} + \displaystyle\frac{1}{2 x_{Bj}} \gamma^2\right)} 
\approx - \frac{M^2x_B^2}{(1-x_B) } ] = - \frac{4 \xi^2 M^2}{1-\xi^2} 
\end{eqnarray}
so the minimum momentum transfer is equivalent to a target mass correction. 
The following relation holds in the given reference frame between $t$, $\xi$, $t_0$, and $\Delta_T$,
\footnote{In the Light Cone frame where we evaluate GPDs in what follows, 
the relation becomes, $t = t_0 - \frac{\Delta_T^2}{1-\xi^2}$. The relationship is gframe dependent because ${\bf \Delta}_T$ is not invariant under transverse boosts \cite{Diehl:2003ny}.}
\begin{equation}
t = t_0 - \Delta_T^2 \, \frac{1+\xi}{1-\xi} 
\end{equation}

An important variable that appears in all electroproduction processes is the ratio of longitudinal to transverse virtual photon flux,
\begin{equation}
\epsilon\equiv \frac{1-y-\frac{1}{4}y^2\gamma^2}{1-y+\frac{1}{2}y^2 +\frac{1}{4}y^2\gamma^2}
=\frac{\sum_h \mid A_h^0 \mid^2} 
{\sum_h \, \sum_{\Lambda_{{{\gamma^*}}} = \pm 1} \mid  A_h^{\Lambda_{{\gamma^*}}} \mid^2} 
\label{epsilon}
\end{equation}
where the functions $A_h^{\Lambda_\gamma}$ connect the lepton helicity to the virtual photon helicity. 

In summary, given the initial beam energy and momentum encoded in $k$, the initial proton energy and momentum, $p$, and $Q^2$, $x_{Bj}$, $t$, we can reconstruct all the components of the final particles four-vectors $k'$, $p'$, $q'$ as a function of the azymuthal angle $\phi$. 

\subsubsection{Phase dependence}
DVCS helicity amplitudes, Eq.(\ref{eq:DVCS1b}) are evaluated in the CoM frame of the final hadron system, which defines the hadron plane at an angle $\phi$ with respect to the lab lepton plane.
To evaluate the cross section one has to transform to the laboratory lepton frame by applying a rotation of $-\phi$ about the $z$ axis. Another way to express this is that the lepton produces a definite helicity virtual photon specified in the lepton $x-z$ plane, while the virtual photon's interaction with the hadrons occurs in the hadron plane which is rotated through an azimuthal angle $\phi$. Phases appear in the definition of the DVCS contribution to the cross section as a consequence of such a rotation about the axis where the virtual photon lies \cite{Dmitrasinovic:1989bf,Boffi:1993gs}. To implement this we first define the polarization vectors for the virtual photon of momentum $q$ along the {negative} $z$-axis in the laboratory frame,
\begin{eqnarray}
\label{eq:eps}
 \varepsilon^{\Lambda_{{\gamma^*}}=\pm 1}   & \equiv & \frac{1}{\sqrt{2}} (0; \mp 1, {i}, 0), \, \, \quad
\varepsilon^{\Lambda_{{\gamma^*}}=0} \equiv \frac{1}{Q} (\mid {\vec q} \mid; 0, 0 ,q_0)=\frac{1}{\gamma} (\sqrt{1+\gamma^2}; 0, 0, 1), 
\label{eps0} \nonumber \\
 \end{eqnarray}
The ejected (real) photon polarization vectors read,
\begin{equation}
\varepsilon^{\Lambda_\gamma^\prime = \pm 1}  \equiv \frac{1}{\sqrt{2}} ( 0; \mp \cos\theta \cos\phi +i \sin\phi,  \mp \cos\theta \sin\phi {+ i} \cos\phi , \pm \sin\theta), 
\label{eq:epsprime} 
\end{equation}
The outgoing photon polarization vectors obey the following completeness relation obtained summing over the physical (on-shell) states \cite{Gastmans},
\begin{eqnarray}
\label{eq:polLambda'}
\sum_{\Lambda_\gamma'} \left( \varepsilon^{\Lambda_\gamma^\prime}_\mu (q') \right)^* \varepsilon^{\Lambda_\gamma^\prime}_\nu(q') & = &  {-g_{\mu\nu}  }
\end{eqnarray}
%
One can see that
the $\phi$ rotation about the $z$-axis changes the phase of the transverse components, and leaves the longitudinal polarization vector unchanged. The transformed DVCS polarization vectors are,
\begin{eqnarray}
\label{eq:epsrot1}
\varepsilon^{\Lambda_{{\gamma^*}}   =  \pm 1} & \rightarrow & \frac{e^{-i\Lambda_{{\gamma^*}}\phi}}{\sqrt{2}} (0, \mp 1, {i}, 0) \\
\label{eps_hadron}
\varepsilon^{\Lambda_{\gamma'} =  \pm 1} 
& \rightarrow &
 \frac{e^{-i\Lambda_{\gamma}'\phi}}{\sqrt{2}} ( 0, \mp 1 , i,0)   + (0, 0_T, \pm \sin\theta)
\label{eq:epsprime1} 
\end{eqnarray} 
The dependence on the angle $\theta$ arises from the fact that the photon's momentum, $q'$, is produced at an angle with the $z$ axis. Eqs.(\ref{eq:epsprime},\ref{eq:epsprime1})  become the same as Eq.(\ref{eq:eps},\ref{eq:epsrot1}) in the forward ({\it i.e.} collinear with the virtual photon along the -$z$ direction) limit. 
From Eq.(\ref{costheta}) one can see that in the limit $t/Q^2 \rightarrow 0$, $\cos \theta \approx 1$ to order  ${\cal O}(1/Q^4)$ (given by $\gamma^4$), while $\sin \theta \approx 0$ to order ${\cal O}(1/Q^2)$ (or $\gamma^2$).

One can therefore display explicitly a phase term as shown in Eq.(\ref{eq:epsprime1}) in the $\theta \rightarrow 0$ limit. 

As we show in the following sections it is the incoming photon polarization vector, through Eq.(\ref{eq:epsrot1}), that characterizes the phase dependence of the DVCS contribution to the cross section. 

\subsection{Observables}
The helicity formalism allows us to identify polarization observables for the various beam and target configurations. 
%
The total number of twist two and twist three CFFs we wish to extract from the observables is 32=2$\times$16 (the factor 2 is from considering the $\Im {\rm m}$ and $\Re {\rm e}$ parts in each CFF), corresponding to 4 distinct GPDs in the quark twist-two sector, 8 twist-three quark GPDs (4 in the vector and axial vector sectors, respectively) and 4 transversity gluon GPDs (the explicit expressions for all of these quantities are given in Sec.\ref{sec:3}). The 4 twist-two and the 8 twist-three GPDs correspond to specific quark-proton polarization configurations listed in Table \ref{observ:tab} adopting the  symbolism where the first letter refers to the polarization of the quark, $P_q$, and the second to the polarization of the proton target, $P_p$. 

In the twist two sector, the following quark-proton polarizations contribute: $P_q P_p=UU, LL, UT, LT$. 
Along with the GPDs, we also list the Transverse Momentum Distrbutions (TMDs) corresponding to  the same spin configurations. Notice that the TMD $f_{1T}^\perp$ appears with an asterisk. This is to signify that while the $UT$ configuration is the same as for the GPD $E$, these two quantities have opposite behavior under $PT$ transformations, namely $E$ is naive T-even by definition, while $f_{1T}$ is naive T-odd: $E$ and $f_{1T}^\perp$ originate from the $\Re e$ and $\Im m$ parts of the same Generalized TMD (GTMD) \cite{Meissner:2009ww}.  
\begin{table}[htp]
\centering
\begin{tabular}{|c|c|c|c|c|c|}
\hline
GPD & Twist & $P_q P_p$ & TMD & $P_{Beam} P_{p}$ (DVCS) &  $P_{Beam}P_p$  ($\mathcal{I}$) \\
\hline 
${\bf H} + \displaystyle\frac{\xi^2}{1-\xi} E$ & 2 & $UU$ & $f_1$ & $UU$, $LL$, $UT^{\sin(\phi-\phi_s)}$, $LT^{\cos(\phi-\phi_s)}$ &    $UU^{\cos \phi}$,  $LU^{\sin \phi}$ \\
${\bf \widetilde{H}} + \displaystyle\frac{\xi^2}{1-\xi} \widetilde{E}$ & 2 & LL & $g_1$ &$UU$, $LL$, $UT^{\sin(\phi-\phi_s)}$, $LT^{\cos(\phi-\phi_s)}$ &  $UU^{\cos \phi}$, $UL^{\cos \phi}$ , $LU^{\sin \phi}$, $LL^{\sin \phi}$,  $UT^{\frac{\cos \phi}{\sin \phi}}$, $LT^{\cos \phi}$ \\
& & & & &\\
${\bf E}$ & 2 & UT & $f_{1T}^{\perp \, {\bf (*)}}$&  $UT^{\sin(\phi-\phi_s)}$, $LT^{\cos(\phi-\phi_s)}$& $UU^{\cos \phi}$, $LU^{\sin \phi}$, $UT$, $LT$, $UT^{\cos \phi}$, $UT^{\sin \phi}$ \\
& & & & &\\
${\bf \widetilde{E}}$ & 2 & LT & $g_{1T}$ &  $UT^{\sin(\phi-\phi_s)}$, $LT^{\cos(\phi-\phi_s)}$& $UL^{\sin \phi}$, $LL^{\cos \phi}$, $UT^{\cos \phi}$, $UT^{\sin \phi}$ \\
& & & & &\\
 {\bf H+E} & 2 & - & - & - & $UU^{\cos \phi}$,  $LU^{\sin \phi}$, $UL^{\sin \phi}$, $LL^{\cos \phi}$, $UT^{\cos \phi}$, $UT^{\sin \phi}$ \\
& & & & &\\
\hline 
& & & & &\\
${\bf 2\widetilde{H}_{2T} + E_{2T}} - \xi \widetilde{E}_{2T}$ & 3 & UU & $f^\perp$ & $UU^{\cos\phi}$, $LU^{\sin\phi}$ & $UU, LU$ \\
& & & & &\\
${\bf 2\widetilde{H}_{2T}' + E_{2T}'} - \xi \widetilde{E}_{2T}'$  & 3 & LL & $g_L^\perp$ & $UU^{\cos\phi}$, $LU^{\sin\phi}$ & $UU, LU$  \\
& & & & &\\
${\bf H_{2T} + \, \displaystyle\frac{t_o-t}{4M^2} \widetilde{H}_{2T}} $ & 3 & UT & $f_T^{\bf (*)}, f_{T}^{\perp \, {\bf (*)}}$ &$UU^{\cos\phi}$, $UL^{\cos\phi}$, $LU^{\sin\phi}$, $LL^{\cos\phi}$ &$UU, LU$  \\
& & & & &\\
${\bf H_{2T}' + \, \displaystyle\frac{t_o-t}{4 M^2} \widetilde{H}_{2T}'}$ & 3 & LT & $g_T', g_{T}^{\perp}$&
$UU^{\cos\phi}$,$UL^{\cos\phi}$, $LU^{\sin\phi}$, $LL^{\cos\phi}$ & $UU, LU$  \\
& & & & &\\
\hline
& & & & & \\
${\bf \widetilde{E}_{2T}} - \xi E_{2T}$ & 3 & UL & $f_L^{\perp \, {\bf (*)}}$ & $UU^{\cos\phi}$, $UL^{\cos\phi}$, $LU^{\sin\phi}$, $LL^{\cos\phi}$ & $UU, LU, UT$ \\
& & & && \\
${\bf \widetilde{E}_{2T}'} - \xi E_{2T}'$ & 3 & LU & $g^{\perp \, {\bf (*)}}$ &$UU^{\cos\phi}$, $UL^{\cos\phi}$,  $LU^{\sin\phi}$, $LL^{\cos\phi}$ & $UU, LU, UT$ \\
& & & & &\\
${\bf \widetilde{H}_{2T}}$ & 3 & UT$_x$ & $f_T^{\perp \, (*)}$ & $UU^{\cos\phi}$, $UL^{\cos\phi}$, $LU^{\sin\phi}$, $LL^{\cos\phi}$ & $UU, LU, UT$ \\
& & & & &\\
${\bf \widetilde{H}_{2T}'}$ & 3 & LT$_x$ & $g_T^\perp$ & $UU^{\cos\phi}$, $UL^{\cos\phi}$, $LU^{\sin\phi}$, $LL^{\cos\phi}$  &$UU, LU, UT$  \\
& & & & &\\
 \hline
\end{tabular}
\caption{Polarization observables for the DVCS and BH-DVCS intereference contributions to the $ep\rightarrow e'p'\gamma'$ cross section. The GPD content entering the complex CFFs for each polarization configuration is listed in the first column (we use boldface characters for the dominant terms); the corresponding twist is in the second column; the third column contains the polarization of the quark and proton; the fourth column indicates the TMDs with the same quark-proton polarization configuration; in the last two columns the observables' beam and target polarization configrations are displayed.}
\label{observ:tab}
\end{table}

In the twist three sector we find new relations: the combinations $2 \widetilde{H}_{2T} + E_{2T}$, and $2 \widetilde{H}_{2T}' + E_{2T}'$ encode quark-gluon-quark correlations that arise for an unpolarized quark in an unpolarized proton, and a longitudinally polarized quark proton configurations, respectively. These GPDs can be, therefore, described as twist-three correspondents of the GPDs $H$, and $\widetilde{H}$, respectively. The TMDs for the same configuration are also listed.
Similarly, the combination ${ H_{2T} + \, \displaystyle\frac{t_o-t}{4M^2} \widetilde{H}_{2T}} $ is the twist three correspondent of the GPD $E$.
Notice that the twist three TMDs with the same $UT$ polarization configuration are $f_T'$ and $f_T^\perp$, which are T-odd, similarly as for the twist two case of $f_{1T}^\perp$. The GPD
${H_{2T}' + \, \displaystyle\frac{t_o-t}{4M^2} \widetilde{H}_{2T}'}$ is the twist three correspondent to $\widetilde{E}$; it is an off-forward extension of $g_T'$ and $g_T^\perp$ ($g_T$, T-even) (see also \cite{Raja:2017xlo}).
The twist three distributions that cannot be associated with any of the twist two $P_q P_p$ polarization configurations are listed separately. These functions carry new physical information on the structure of the proton. Two new correlations involving longitudinal polarization correspond to the GPDs $\widetilde{E}_{2T}$, $\widetilde{E}_{2T}'$. These functions are particularly interesting because they single out the orbital component of angular momentum (OAM) \cite{Penttinen:2000dg,Hatta:2012cs,Courtoy:2013oaa,Rajan:2016tlg,Raja:2017xlo}. Notice that the careful analysis performed in this paper allows us to point out precisely which polarization configurations are sensitive to OAM and to, therefore, dispel the notion that OAM cannot be measured in DVCS.
Finally, the functions $\widetilde{H}_{2T}$, $\widetilde{H}_{2T}'$ involve ``in plane" transverse polarization. Their study will open up the way to understanding the contribution of transverse OAM. 

The various polarization configurations listed in the third column of Table \ref{observ:tab} enter the different beam polarizations, $P_{Beam}=U,L$, and target proton polarizations, $P_p=U,L,T$, in the DVCS and BH-DVCS interference (${\cal I}$) terms, which enter the cross section as listed below,
\begin{subequations}
\label{sig_pol}
\begin{align}
\sigma_{UU} &=  \sum_{h,\Lambda} \sigma_{h \Lambda} = \sigma_{UU}^{DVCS} + \sigma_{UU}^{BH} + \sigma_{UU}^{\cal I}\\
\sigma_{LU} &=  {\sum_\Lambda (\sigma_{+\Lambda} - \sigma_{- \Lambda})} =   \sigma_{LU}^{DVCS} +  \sigma_{LU}^{\cal I} 
\\
\sigma_{UL} &= {\sum_h (\sigma_{h+} - \sigma_{h-})}= \sigma_{UL}^{DVCS} +  \sigma_{UL}^{\cal I} \\
\sigma_{LL} &=  {(\sigma_{++} - \sigma_{+-}) - (\sigma_{-+} - \sigma_{--})} = \sigma_{LL}^{DVCS} + \sigma_{LL}^{BH} + \sigma_{LL}^{\cal I}  \\
\sigma_{UT} &= {\sum_h \left( \sigma^T_{h,+}-\sigma^T_{h,-} \right)}
 =  \sigma_{UT}^{DVCS} +  \sigma_{UT}^{\cal I}  \\
\sigma_{LT} &= {\sigma_{+\,+}^T - \sigma_{+-}^T -(\sigma_{-+}^T - \sigma_{-\,-}^T)} = \sigma_{LT}^{DVCS} + \sigma_{LT}^{BH} + \sigma_{LT}^{\cal I} ,
\end{align}
\end{subequations}
where $h$ and $\Lambda$ are the electron and target helicities, respectively. Eqs.(\ref{sig_pol}) can be used to navigate the last two columns in Table \ref{observ:tab}.
Notice that the various polarization observables in DVCS are interpreted differently than similar observables or helicity configurations in inclusive or semi-inclusive experiments. In DVCS the observables are bilinear forms that contain quadratic expressions of the CFFs (the interference term contains products of nucleon form factors and CFFs).
This makes it difficult to isolate specific GPDs within each observable, since summing terms with different polarizations does not produce cancellations like the ones appearing in inclusive scattering processes. For instance, unpolarized scattering  measures the PDF $f_1$ in inclusive scattering, while for DVCS the $UU$ term contains both the vector, $H$, and axial-vector, $\widetilde{H}$ GPDs. Equivalently, scattering of a longitudinally polarized electron from a longitudinally polarized target measures both vector and axial-vector GPDs in DVCS, while the axial vector component, $g_1$, can be singled out in the inclusive case. 
A clearer physics interpretation of the interference term is, however, attainable by formulating this contribution  according to the standard notation used for elastic electron-proton scattering in the one photon exchange approximation,  generalizing the Rosenbluth cross section \cite{Rosenbluth:1950yq} (Sec. \ref{sec:BHDVCS}). 

In conclusion, the physics picture summarized in Table \ref{observ:tab}, calls for a different approach than the standard analysis methods used so far to extract information from inclusive/semi-inclusive polarized scattering experiments. 
We, first of all, notice that by writing the cross section in a generalized Rosenbluth form, the GPD combination, $(H+E)$, appears naturally in the formalism as the coefficient of the magnetic form factor term. We, therefore, separately list this observable and the corresponding beam-target polarization configurations in Table \ref{observ:tab}. 
On more general grounds, in DVCS measurements one cannot rely on the dominance of any specific observable for any given polarization configuration, but multiple structure functions, translated into multiple CFFs, appear simultaneously in the cross section. 
%
Our easily readable formalism was constructed in such a way as to facilitate the analysis of these observables.  Numerical evaluations will be shown in a forthcoming publication.

\section{\label{sec:3}Deeply Virtual Compton Scattering Cross section}
In this Section we present the detailed structure of the cross section for  DVCS in terms of helicity amplitudes. Our formulation is consistent with the work in Refs.\cite{Arens:1996xw,Diehl:2005pc} where a general notation was introduced to describe the various beam and target polarization configurations for a wide variety of electron proton scattering processes from semi-inclusive deep inelastic scattering (SIDIS) to DVCS. While specifically for SIDIS a more detailed notation following Ref.\cite{Arens:1996xw,Diehl:2005pc} was developed subsequently in Ref.\cite{Bacchetta:2006tn}, an analogous complete description of the exclusive processes including DVCS, TCS and their respective background BH processes has been so far lacking.

The formalism presented here allows us to: 
\begin{itemize}
\item single out the various beam and target polarizations configurations contributing to the DVCS cross section with their specific dependence on the azimuthal angle $\phi$, between the lepton and hadron planes.  
\item describe observables including various beam and target asymmetries with GPDs up to twist three
\item in virtue of its covariant form, give a unified description that is readily usable for both fixed target and collider experimental setups.
\end{itemize}
%

%

\begin{figure}
\begin{center}
\includegraphics[width=12.cm]{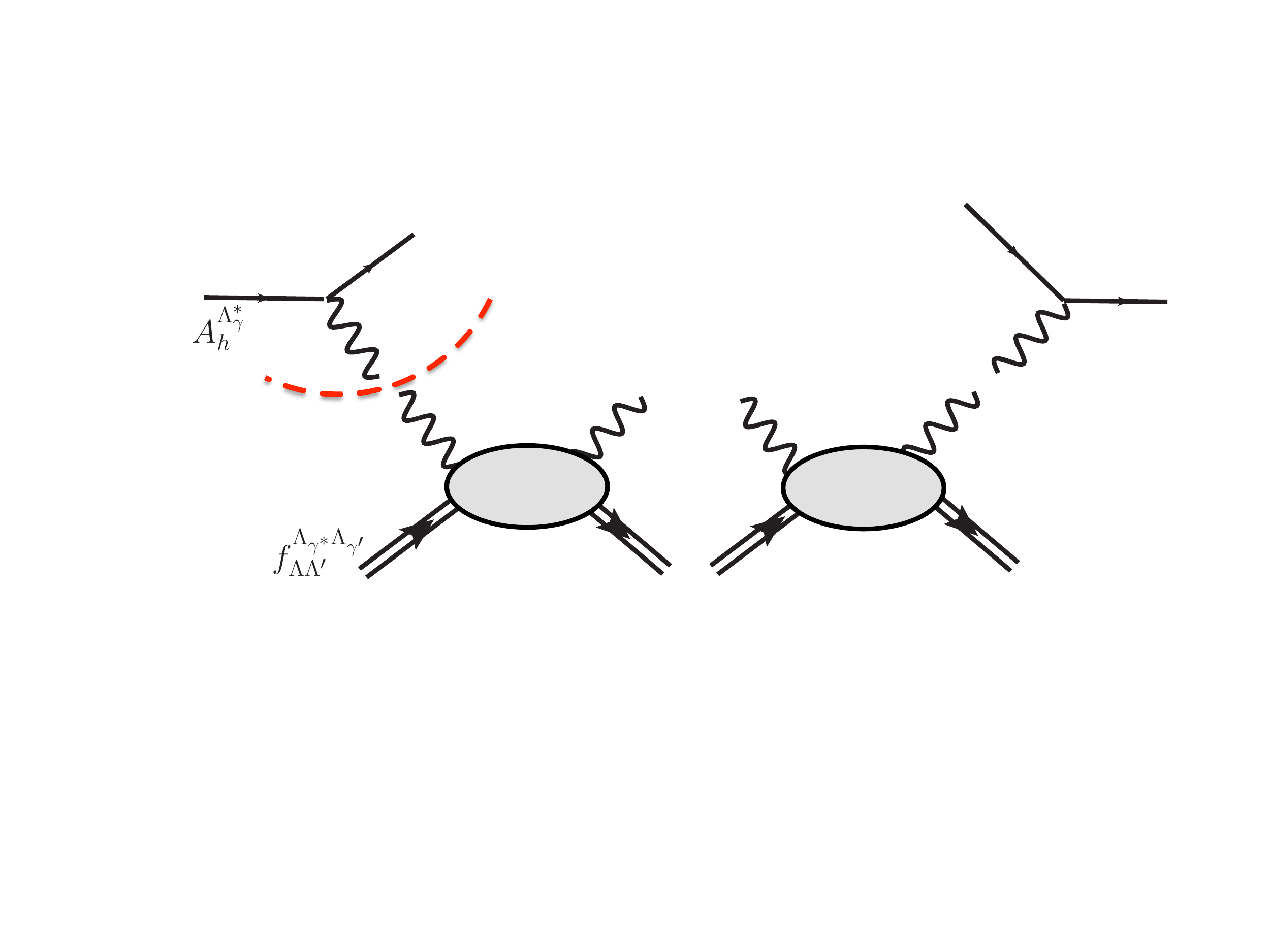}
\end{center}
\vspace{-3cm}
\caption{Factorization of the DVCS contribution to the cross section into leptonic and hadronic helicity amplitudes.}
\label{fig:DVCS}
\end{figure}
\subsection{General Formalism}
\label{sec:genform}
%
Following the formalism introduced in Refs.\cite{Arens:1996xw,Diehl:2005pc,Bacchetta:2006tn} one can derive the general expression describing all polarized and unpolarized contributions to the DVCS cross section, $\sigma \equiv d^5\sigma$ in Eq.(\ref{eq:coherent}), corresponding to the following configurations: unpolarized beam unpolarized target (UU), polarized beam unpolarized target (LU), unpolarized beam longitudinally polarized target (UL), polarized beam longitudinally polarized target (LL), unpolarized beam transversely polarized target (UT), polarized beam transversely polarized target (LT),
\footnote{Similarly to Ref.\cite{Bacchetta:2006tn}, while the first and second subscripts define the polarization of the beam and target, the third subscript in {\it e.g.} $F_{UU,T}$ specifies the polarization of the virtual photon. }
\begin{eqnarray}
\label{eq:xs5fold} 
\frac{d^5\sigma_{DVCS}}{d x_{Bj} d Q^2 d|t| d\phi d\phi_S } & = &
\Gamma
\big| T_{DVCS} \big|^2 \;
 \nonumber \\
& = & \frac{\Gamma}{Q^2 (1-\epsilon)}\left\{  F_{UU,T} + \epsilon F_{UU,L}+ \epsilon \cos 2\phi F_{UU}^{\cos 2 \phi} 
\right.  
+ \sqrt{\epsilon(\epsilon+1)} \cos \phi F_{UU}^{\cos \phi} \nonumber \\
& + & (2h)  \sqrt{2\epsilon(1-\epsilon)} \sin \phi F_{LU}^{\sin \phi}  \nonumber \\
& + & (2 \Lambda) \Big[ \sqrt{\epsilon(\epsilon+ 1)}\sin \phi  F_{UL}^{\sin \phi} +  \epsilon \sin 2 \phi F_{UL}^{\sin 2\phi} \nonumber \\
& +  & (2h)  \left( \sqrt{1-\epsilon^2} F_{LL}
+ 2  \sqrt{ \epsilon(1- \epsilon)} \cos \phi F_{LL}^{\cos \phi}  \right) \Big]\nonumber \\
& + & (2\Lambda_T) \left[\sin(\phi - \phi_S) \left( F_{UT,T}^{\sin(\phi - \phi_S) } +\epsilon F_{UT,L}^{\sin(\phi - \phi_S) }\right)\right.  \nonumber \\
&  & \left. + \quad   \epsilon \, \sin(\phi + \phi_S)  F_{UT}^{\sin(\phi + \phi_S) } +\epsilon \sin(3\phi-\phi_S)  F_{UT}^{\sin(3\phi - \phi_S) } \right. \nonumber \\
&& \left. + \sqrt{2\epsilon(1+\epsilon)}\left( \sin\phi_S F_{UT}^{\sin\phi_S} + \sin(2\phi-\phi_S) F_{UT}^{\sin(2\phi-\phi_S)} \right)\right] \nonumber \\
&  & + (2h) (2 \Lambda_T) \left[ \sqrt{1-\epsilon^2} \cos(\phi-\phi_S) F_{LT}^{\cos(\phi-\phi_S)} +\sqrt{2\epsilon (1-\epsilon)} \cos\phi_S F_{LT}^{\cos\phi_S}  \right. \nonumber \\
 & & \left. \left. \quad + \sqrt{2\epsilon(1-\epsilon)} \cos(2\phi-\phi_S) F_{LT}^{\cos(2\phi -\phi_S)} \right] \right\}
\end{eqnarray}
the longitudinal spin is $S_{||} \equiv \Lambda$,  while $\mid {\vec S}_T \mid \equiv \Lambda_T$, $S_T$ being the transverse proton spin at an angle $\phi_S$ with the lepton plane.
The dimensions of $T_{DVCS}$, Eq.(\ref{eq:TDVCScov}), are GeV$^{-1}$ (see Appendix \ref{appa}). 

In this respect, our formalism supersedes the work presented in Refs.\cite{Belitsky:1999sg,Belitsky:2001ns,Belitsky:2008bz,Belitsky:2010jw,Belitsky:2012ch} where the theoretical framework of DVCS on the proton was described by organizing the cross section in terms of ``angular harmonics" in the azymuthal angle, $\phi$. In 
our approach the various sources of both kinematical and dynamical dependence on the scale, $Q^2$ for the DVCS and BH processes can be readily singled out
and a more physical interpretation of the occurrence of twist two and twist three contributions appears from the combinations of the virtual photon polarizations, namely:
\vspace{0.5cm}

$\bullet$ Twist 2: $F_{UU,T}$, $F_{LL}$, $F_{UT,T}^{\sin (\phi-\phi_S)}$, $F_{LT}^{\cos (\phi-\phi_S)}$

$\bullet$ Twist 3: $F_{UU}^{\cos \phi}$, $F_{UL}^{\sin \phi}$, $F_{LU}^{\sin \phi}$, $F_{LL}^{\cos \phi}$, $F_{UT}^{\sin \phi_S}$, $F_{UT}^{\sin (2 \phi- \phi_S)}$, $F_{LT}^{\cos \phi_S}$, $F_{LT}^{\cos (2 \phi- \phi_S)}$

$\bullet$ Twist 4: $F_{UU,L}$, $F_{UT,L}^{\sin(\phi-\phi_s)}$

$\bullet$ Transverse gluons: $F_{UU}^{\cos 2\phi}$, $F_{UL}^{\sin 2\phi}$, $F_{UT}^{\sin (\phi+\phi_s)}$, $F_{UT}^{\sin (3\phi-\phi_s)}$.


\vspace{0.5cm}
Notice the striking similarity of the DVCS and SIDIS cross sections structure \cite{Bacchetta:2006tn}. We remark that despite the observables containining the same helicity structure, the helicity structure at the amplitude level is inherently different.
This is a consequence of the DVCS process being exclusive.

The amplitude, $T_{DVCS}$, was introduced in Section \ref{sec:2} factorized into its lepton and hadron contributions. 
We consider the structure of the cross section for the case in which the polarizations of the final photon, $\Lambda_\gamma'$, and nucleon, $\Lambda'$, are not detected while the initial nucleon and lepton have a definite longitudinal polarization, $h$ and $\Lambda$, respectively. 
For longitudinally polarized states we define,
\begin{eqnarray}
\sigma_{h \Lambda} &=& \Gamma \, \sum_{ \Lambda^\prime_\gamma, \, \Lambda^\prime}  \left( T^{h   \Lambda^\prime_\gamma}_{DVCS,  \Lambda \Lambda^\prime} \right)^* \, T^{h  \Lambda^\prime_\gamma}_{DVCS,\Lambda \Lambda^\prime}   =  
%
\Gamma
\sum_{\Lambda_{\gamma^*}^{(1)}, \Lambda_{\gamma^*}^{(2)}}  {\cal L}_h^{\Lambda_{\gamma^*}^{(1)}, \Lambda_{\gamma^*}^{(2)}} H^{\Lambda_{\gamma^*}^{(1)}, \Lambda_{\gamma^*}^{(2)}}_{\Lambda}
\label{eq:DVCS1b2}
\end{eqnarray}
where the lepton tensor contracted with the polarization vectors is defined as (Eq.\eqref{eq:A_intro}),
\begin{eqnarray}
\label{eq:lepton}
{\cal L}_h^{\Lambda_{\gamma^*}^{(1)} \Lambda_{\gamma^*}^{(2)}} =  \sum_{\Lambda_{\gamma^*}^{(1)}, \Lambda_{\gamma^*}^{(2)}} A_h^{\Lambda_{{\gamma^*}}^{(1)}} A_h^{\Lambda_{{\gamma^*}}^{(2)}} 
\end{eqnarray}
while the hadronic contribution written in terms of helicity amplitudes reads (Eq.\eqref{eq:famp}),
\begin{subequations}
\begin{eqnarray}
H^{\Lambda_{\gamma^*}^{(1)} \Lambda_{\gamma^*}^{(2)}}_{\Lambda} & = &  
\sum_{  \Lambda^\prime}  F^{\Lambda_{\gamma^*}^{(1)} \Lambda_{\gamma^*}^{(2)}}_{\Lambda \Lambda'}   \\
F^{\Lambda_{\gamma^*}^{(1)} \Lambda_{\gamma^*}^{(2)}}_{\Lambda \Lambda'} & = &   
\sum_{ \Lambda^\prime_\gamma}    \left[ f_{\Lambda, \Lambda'}^{\Lambda_{{\gamma^*}}^{(1)} ,\Lambda_\gamma^\prime}\right]^* f_{\Lambda, \Lambda'}^{\Lambda_{{\gamma^*}}^{(2)} ,\Lambda_\gamma^\prime} 
\end{eqnarray}
\end{subequations}
Separating out the leptonic and hadronic contributions we have, 
\begin{eqnarray}
&\sigma_{h \Lambda} = &  \sum_{ \Lambda^\prime_\gamma, \, \Lambda^\prime}   \sum_{\Lambda_{{\gamma^*}}^{(1)}} \left[A_h^{\Lambda_{{\gamma^*}}^{(1)}} f_{\Lambda, \Lambda'}^{\Lambda_{{\gamma^*}}^{(1)} ,\Lambda_\gamma^\prime}\right]^*\sum_{\Lambda_{{\gamma^*}}^{(2)}} A_h^{\Lambda_{{\gamma^*}}^{(2)}} f_{\Lambda, \Lambda'}^{\Lambda_{{\gamma^*}}^{(2)} ,\Lambda_\gamma^\prime}
 \nonumber \\  
& = &  \sum_{ \Lambda^\prime_\gamma, \, \Lambda^\prime}  \left(A_h^{1} f_{\Lambda, \Lambda^\prime}^{1,\Lambda_\gamma'} + A_h^{-1} f_{\Lambda, \Lambda^\prime}^{-1,\Lambda_\gamma'} + A_h^{0} f_{\Lambda, \Lambda^\prime}^{0,\Lambda_\gamma'} \right)^* \left(A_h^{1} f_{\Lambda, \Lambda^\prime}^{1,\Lambda_\gamma'} + A_h^{-1} f_{\Lambda, \Lambda^\prime}^{-1,\Lambda_\gamma'} + A_h^{0} f_{\Lambda, \Lambda^\prime}^{0,\Lambda_\gamma'} \right) \nonumber \\
& = & \sum_{ \Lambda^\prime_\gamma, \, \Lambda^\prime}%
(A_h^{1} )^2 \mid f_{\Lambda, \Lambda^\prime}^{1,\Lambda_\gamma'} \mid^2 + (A_h^{-1} )^2 \mid f_{\Lambda, \Lambda^\prime}^{-1,\Lambda_\gamma'} \mid^2 + ( A_h^{0} )^2 \mid f_{\Lambda, \Lambda^\prime}^{0,\Lambda_\gamma'} \mid^2  \nonumber \\
& + & A_h^{1} A_h^{0} \left[ \left(f_{\Lambda, \Lambda^\prime}^{1,\Lambda_\gamma'}  \right)^* f_{\Lambda, \Lambda^\prime}^{0,\Lambda_\gamma'} + \left(f_{\Lambda, \Lambda^\prime}^{0,\Lambda_\gamma'}  \right)^* f_{\Lambda, \Lambda^\prime}^{1,\Lambda_\gamma'} \right] +A_h^{-1}  A_h^{0} \left[  \left(  f_{\Lambda, \Lambda^\prime}^{-1,\Lambda_\gamma'} \right)^*f_{\Lambda, \Lambda^\prime}^{0,\Lambda_\gamma'} +   \left(f_{\Lambda, \Lambda^\prime}^{0,\Lambda_\gamma'}  \right)^* f_{\Lambda, \Lambda^\prime}^{-1,\Lambda_\gamma'}  \right]  \nonumber \\
&+ & A_h^{1} A_h^{-1} \left[ \left(f_{\Lambda, \Lambda^\prime}^{1,\Lambda_\gamma'} \right)^*f_{\Lambda, \Lambda^\prime}^{-1,\Lambda_\gamma'} + \left(f_{\Lambda, \Lambda^\prime}^{-1,\Lambda_\gamma'} \right)^*f_{\Lambda, \Lambda^\prime}^{1,\Lambda_\gamma'}  \right].
\label{eq:phase1}
\end{eqnarray}
Replacing the $A_h$ functions in the helicity structure module in Eq.(\ref{eq:phase1}) allows one to separate out the beam helicity, $h$, dependent terms as follows,
\begin{eqnarray}
&\sigma_{h \Lambda} = & 
\frac{1}{Q^2} \frac{1}{1-\epsilon} \Big\{2  \sum_{  \Lambda^\prime_\gamma, \, \Lambda^\prime}%
\left( \mid f_{\Lambda, \Lambda^\prime}^{1,\Lambda_\gamma'} \mid^2 +  \mid f_{\Lambda, \Lambda^\prime}^{-1,\Lambda_\gamma'} \mid^2 \right) + 2 \epsilon \sum_{  \Lambda^\prime_\gamma, \, \Lambda^\prime} \mid f_{\Lambda, \Lambda^\prime}^{0,\Lambda_\gamma'} \mid^2  \nonumber \\
& + & {\sqrt{2}} \sqrt{\epsilon(1+\epsilon)} \sum_{  \Lambda^\prime_\gamma, \, \Lambda^\prime}\left[ -\left(f_{\Lambda, \Lambda^\prime}^{1,\Lambda_\gamma'}  \right)^* f_{\Lambda, \Lambda^\prime}^{0,\Lambda_\gamma'} - \left(f_{\Lambda, \Lambda^\prime}^{0,\Lambda_\gamma'}  \right)^* f_{\Lambda, \Lambda^\prime}^{1,\Lambda_\gamma'}  +   \left(  f_{\Lambda, \Lambda^\prime}^{-1,\Lambda_\gamma'} \right)^*f_{\Lambda, \Lambda^\prime}^{0,\Lambda_\gamma'} +   \left(f_{\Lambda, \Lambda^\prime}^{0,\Lambda_\gamma'}  \right)^* f_{\Lambda, \Lambda^\prime}^{-1,\Lambda_\gamma'}  \right]  \nonumber \\
&- & 2 \epsilon \sum_{  \Lambda^\prime_\gamma, \, \Lambda^\prime}\left[ \left(f_{\Lambda, \Lambda^\prime}^{1,\Lambda_\gamma'} \right)^*f_{\Lambda, \Lambda^\prime}^{-1,\Lambda_\gamma'} + \left(f_{\Lambda, \Lambda^\prime}^{-1,\Lambda_\gamma'} \right)^*f_{\Lambda, \Lambda^\prime}^{1,\Lambda_\gamma'}  \right] \Big\}
\nonumber \\ 
& + & (2h) \Big\{ 2 \sqrt{1-\epsilon^2} \sum_{  \Lambda^\prime_\gamma, \, \Lambda^\prime}%
\left( \mid f_{\Lambda, \Lambda^\prime}^{1,\Lambda_\gamma'} \mid^2 -  \mid f_{\Lambda, \Lambda^\prime}^{-1,\Lambda_\gamma'} \mid^2 \right) \nonumber \\
& - & {\sqrt{2}} \sqrt{\epsilon(1-\epsilon)} \sum_{  \Lambda^\prime_\gamma, \, \Lambda^\prime}\left[ \left(f_{\Lambda, \Lambda^\prime}^{1,\Lambda_\gamma'}  \right)^* f_{\Lambda, \Lambda^\prime}^{0,\Lambda_\gamma'} + \left(f_{\Lambda, \Lambda^\prime}^{0,\Lambda_\gamma'}  \right)^* f_{\Lambda, \Lambda^\prime}^{1,\Lambda_\gamma'}  +   \left(  f_{\Lambda, \Lambda^\prime}^{-1,\Lambda_\gamma'} \right)^*f_{\Lambda, \Lambda^\prime}^{0,\Lambda_\gamma'} +   \left(f_{\Lambda, \Lambda^\prime}^{0,\Lambda_\gamma'}  \right)^* f_{\Lambda, \Lambda^\prime}^{-1,\Lambda_\gamma'}  \right] \Big\}, \nonumber \\
\label{eq:phase1p1} 
\end{eqnarray}
where the coefficients dependent on $\epsilon$, Eq.\eqref{epsilon}, are obtained from the various lepton tensor components in Section \ref{sec:leptensdvcs}. The transverse polarized case involves a different summation over the helicity states and is described separately in Section \ref{sec:DVCS_structfun}.

\subsubsection{Lepton Tensor}
\label{sec:leptensdvcs}
The bilinear terms defining the lepton tensor are given by, \begin{subequations}
\begin{eqnarray}
{\cal L}_h^{\pm 1 \pm 1}  =  A_h^{* \pm 1} A_h^{ \pm 1 } & = &  \frac{1}{Q^2} \frac{1}{1-\epsilon} \left[{2}  \pm 2(2h) \sqrt{(1+\epsilon)(1-\epsilon)} \right], \\    
{\cal L}_h^{0 0} =  A_h^{*0} A_h^{0} &  = &  \frac{1}{Q^2} \frac{2 \epsilon}{1-\epsilon}, \\ 
 {\cal L}_h^{\pm 1  \mp 1} =  A_h^{*1} A_h^{-1} &= & -   \frac{1}{Q^2} \frac{ 2 \epsilon}{1-\epsilon}, \\    
{\cal L}_h^{\pm 1  0} =   A_h^{*\pm1} A_h^{0} &  = &  \frac{1}{Q^2} \frac{{\sqrt{2}}}{1-\epsilon} \left[\mp  \sqrt{\epsilon(1+\epsilon)} - 2 h \sqrt{\epsilon(1-\epsilon)} \right]  .  
\end{eqnarray}
\end{subequations}
%

The helicity amplitudes, $A_h^{\Lambda_{\gamma^*}}$, for the transverse and longitudinal  virtual photon helicity components read, respectively, as,
\begin{subequations}
\label{A_h}
\begin{eqnarray}
 A_h^{\pm 1} & = & \mp \frac{1}{Q^2} \frac{1}{\sqrt{2}} \left[ \bar{u}(k^\prime,h)  \gamma_1 u(k,h)  \left(\varepsilon_1^{\pm 1} \right)^* \pm i  \bar{u}(k^\prime,h) \gamma_2 u(k,h)  \left(\varepsilon_2^{\pm 1} \right)^*  \right]  \nonumber \\ 
&  = &  
\frac{1}{\sqrt{Q^2}} \left( \mp \sqrt{\frac{1+\epsilon}{1-\epsilon}} - 2h \right) 
   \\
A_h^0  & = &  \bar{u}(k^\prime,h) \gamma_0 u(k,h) \varepsilon^0_0 - \bar{u}(k^\prime,h) \gamma_3 u(k,h) \varepsilon^0_3 = \frac{1}{\sqrt{Q^2}}\sqrt{\frac{2 \epsilon}{1-\epsilon}},  
\end{eqnarray}
\end{subequations}
Notice that the factor $1/Q^2$ in the first line of Eq.(\ref{A_h}) comes from the photon propagator. The rest of the matrix element depends on $\sqrt{Q^2}$ due to the lepton spinors normalization. As a result, the overall $Q^2$ dependence is $1/\sqrt{Q^2}$.
(the Dirac spinors and normalizations are shown in Appendix \ref{appa}). 
These terms satisfy the parity relations,
\begin{equation}
 A_h^{\Lambda_{{\gamma}}}=(-1)^{\Lambda_{\gamma}} \left[ A_{-h}^{-\Lambda_{\gamma}} \right]^*
 \end{equation}
No phase dependence appears here because the polarization vector is evaluated in the lepton plane. 

The $\epsilon$ dependent coefficients in Eq.(\ref{eq:xs5fold}) result from evaluating the lepton tensor, Eq.(\ref{eq:lepton}). 
For example, in the unpolarized case one has the following combinations,
\begin{subequations}
\begin{eqnarray}
F_{UU,T}   \rightarrow \sum_h {\cal L}_h^{\pm 1 \pm 1} & = &  \frac{1}{Q^2} \frac{1}{1-\epsilon} 4, \\    
F_{UU,L} \rightarrow \sum_h  {\cal L}_h^{0 0}  &  = &  \frac{1}{Q^2} \frac{4 \epsilon}{1-\epsilon}, 
\\ 
F_{UU}^{\cos 2 \phi}\rightarrow \sum_h  {\cal L}_h^{\pm 1  \mp 1} + {\cal L}_h^{\mp 1  \pm 1}  &= & -   \frac{1}{Q^2} \frac{ 4 \epsilon}{1-\epsilon}, \\    
F_{UU}^{\cos  \phi} \rightarrow \sum_h {\cal L}_h^{\pm 1  0} &  = & \mp  \frac{1}{Q^2} \frac{{\sqrt{2}}}{1-\epsilon}  \sqrt{\epsilon(1+\epsilon)}    .  
\end{eqnarray}
\end{subequations}
Notice that an overall factor of 2 was included in $\Gamma$ (Eq.\ref{eq:Gamma}) (see also Appendix \ref{appc}).

\subsubsection{Hadron Tensor}
The helicity amplitudes entering Eq.(\ref{eq:DVCS1b2}) are defined in terms of the hadron tensor as,
\begin{eqnarray}
f_{\Lambda \Lambda^\prime}^{{\Lambda_{\gamma^*} \Lambda_\gamma^\prime}}  & = & {\cal W}_{\mu\nu} \left[\varepsilon^{\Lambda_{\gamma^*}}(q)\right]^\mu [\varepsilon^{\Lambda_\gamma'}(q')]^{\nu *} 
\end{eqnarray}
where for $Q^2 >> M^2,(-t)$, $W_{\mu\nu}$ can be
written within the context of QCD factorization theorems as \cite{Collins:1998be}, 
\begin{eqnarray}
{\cal W}_{\mu \nu}&=&  \frac{1}{2} \left[-g_{T, \mu \nu} \, \int_{-1}^1 dx   \,  C^+(x,\xi) \, W^{[\gamma^+]}_{\Lambda \Lambda'}
 + i \epsilon_{T, \nu\mu}  \, \int_{-1}^1 dx  \, C^-(x,\xi) \, W^{[\gamma^+ \gamma_5]}_{\Lambda \Lambda'}  \right]
 \nonumber \\
& + & \frac{2Mx_{Bj}}{Q} (q+4 \xi P)_\mu \left[ g_{T, \nu i} \, \int_{-1}^1 dx  \,  C^+(x,\xi) W^{[\gamma^i]}_{\Lambda \Lambda'}
 + i \epsilon_{T, i \nu} \int_{-1}^1 dx  \, C^-(x,\xi) W^{[\gamma^i \gamma_5]}_{\Lambda \Lambda'}  
\right]
\label{tensor_trans}
\end{eqnarray}
with  \[\displaystyle  g^{\mu\nu} _T = g^{\mu\nu} - n_+^\mu n_-^\nu -  n_-^\nu n_+^\mu \equiv g^{ij}, \quad\quad \displaystyle  \epsilon^{\mu\nu} _T = \epsilon_{\alpha \beta \sigma \rho} g^{\alpha\mu} _T g^{\beta\nu} _T n_-^\rho 
n_+^\sigma \equiv \epsilon^{-+\mu\nu} \equiv \epsilon^{ij} \]  
$(i,j=1,2)$, $n_+$ and $n_-$ being unit light cone vectors. 
$W^{[\Gamma]}_{\Lambda \Lambda'}$ are the  quark-quark correlation functions,
\begin{eqnarray}
\label{eq:GPDcorr}
W_{\Lambda' \Lambda}^{[\Gamma]}(x,\xi,t) & = & \displaystyle\frac{1}{2}\int \frac{d z^-}{2 \pi} e^{ixP^+ z^-} \left. \langle p', \Lambda' \mid 
 \bar{\psi} \left(-\frac{z}{2}\right) \Gamma {\cal U} \psi\left(\frac{z}{2}\right)   \mid p, \Lambda \rangle \right|_{z^+=0,z_T=0} , 
\end{eqnarray} 
where ${\cal U}$ is the gauge link connection. For GPDs ${\cal U}$ is a straight link, implying that all GPDs are naive T-even.   
Writing out explicitly the proton helicity dependence, we have, for $\Gamma = \gamma^+, \gamma^+ \gamma^5$, the following vector and axial vector twist two correlation functions,
\begin{eqnarray}
\label{GPDvec}
W^{[\gamma^+]}_{\Lambda' \Lambda } &=&
\frac{1}{\sqrt{1-\xi^{2}}}\Bigg[H(1-\xi^{2})-\xi^{2}E \Bigg]\delta_{\Lambda,\Lambda'} +  \frac{\Lambda}{\sqrt{1-\xi^{2}}} \frac{\Delta^{1} + i \Lambda \Delta^{2}}{2M} E \, \delta_{\Lambda,-\Lambda'} \\
\label{GPDaxvec}
W^{[\gamma^+\gamma^5]}_{\Lambda' \Lambda } &=& 
\frac{1}{\sqrt{1-\xi^{2}}}\Bigg[\Lambda \widetilde{H}(1-\xi^{2}) - \Lambda \xi^{2}\widetilde{E} \Bigg]\delta_{\Lambda,\Lambda'} +
\frac{1}{\sqrt{1-\xi^{2}}} \frac{\Delta^{1}+i\Lambda \Delta^{2}}{2M} \xi \widetilde{E} \, \delta_{\Lambda,-\Lambda'},
\end{eqnarray}
whereas, for $\Gamma = \gamma^i, \gamma^i \gamma^5$ we adopt the following parameterization of the twist three correlation function \cite{Meissner:2009ww},
\begin{eqnarray}
\label{GPDvec3}
W^{\gamma^i}_{\Lambda \Lambda'} & = &
\frac{1}{\sqrt{1-\xi^2}} \Bigg[\frac{\Delta^{i}}{2P^+}\Big(2 \widetilde{H}_{2T}  + E_{2T} - \xi \widetilde{E}_{2T} \Big) + \frac{i\Lambda \epsilon^{ij}\Delta^{j}}{2P^+} \Big(\widetilde{E}_{2T} - \xi E_{2T} \Big)  
\Bigg] \delta_{\Lambda,\Lambda'} \nonumber \\ \\
& + & \sqrt{1-\xi^2} \Bigg[ -\frac{M}{P^+} ( \Lambda \delta_{i1} + i\delta_{i2} ) \Big( H_{2T} +  \frac{\xi}{1-\xi^2} \widetilde{E}_{2T} -\frac{\xi^2}{1-\xi^2} E_{2T} \Big) - \Lambda \frac{\Delta^{i}(\Delta^{1}+ i \Lambda\Delta^{2})}{2M P^+}\widetilde{H}_{2T} \Bigg] \delta_{\Lambda,-\Lambda'} \nonumber 
\\ 
\label{GPDaxvec3}
W^{\gamma^i\gamma^5}_{\Lambda \Lambda'} &=& {\frac{1}{\sqrt{1-\xi^{2}}}\Bigg[\frac{i\epsilon^{ij}\Delta^{j}}{2P^{+}}\Big( E_{2T}'- \xi \widetilde{E}_{2T}' \Big) 
+ \frac{i\epsilon^{ij}\Delta^{j}}{P^{+}}\widetilde{H}_{2T}' 
- \frac{\Lambda \Delta^{i}}{2P^{+}}\Big(\widetilde{E}_{2T}' - \xi E_{2T}' \Big) \Bigg]\delta_{\Lambda,\Lambda'} } \nonumber \\ 
&+&  {\frac{1}{\sqrt{1-\xi^{2}}}\Bigg[\frac{M(\delta_{i1} + i\Lambda \delta_{i2})}{P^{+}}\Big((1-\xi^{2})H_{2T}' + \xi \widetilde{E}_{2T}'- \xi^{2}E_{2T}'\Big) - \Lambda \frac{i\epsilon^{ij}\Delta^{j}(\Delta^{1}+ i\Lambda \Delta^{2})}{2MP^{+}}\widetilde{H}_{2T}'\Bigg]\delta_{\Lambda,-\Lambda'} } 
\end{eqnarray}
 \\
%
We adopt throughout the notation given in Ref.\cite{Meissner:2009ww} where for the chiral even twist two contributions,
as in the first parametrization introduced by Ji \cite{Ji:1996ek}, the letter $H$ signifies that in the forward limit these GPDs correspond to a PDF, while the ones denoted by $E$ are completely new functions. The tilde denotes the axial vector case \cite{Diehl:2001pm}.
Note that the matrix structures that enter the twist three vector $(\gamma^i)$ and axial vector $(\gamma^i\gamma^5)$ cases are identical to the ones occurring at the twist two level in the chiral-odd tensor sector. 
Hence, the GPDs are named using a similar notation: 
the corresponding twist three GPD occurring with the same matrix coefficient as the chiral-odd is named $F_{2T}$ in the vector case $\gamma^i$, and $F_{2T}'$ in the axial vector case $\gamma^i\gamma^5$ $F=H, E, \widetilde{H}, \widetilde{E})$.

The structure functions in Eq.(\ref{eq:xs5fold}) while reflecting the helicity structure of the various azymuthal angular modulations, are written in terms of complex valued Compton Form Factors (CFFs). CFFs are obtained as convolutions over the longitudinal momentum fraction, $x$, of the GPDs, $F(x,\xi,t)$, ($F=H, H, ...)$, and the Wilson coefficient functions, $C^\pm$,
\begin{eqnarray}
\mathcal{F}(\xi,t)  = {\cal C} \left( C^+ \, F\right) \equiv \int_{-1}^{1} dx   C^+(x,\xi) F(x,\xi,t), 
\quad\quad 
\mathcal{\widetilde{F}}(\xi,t) = {\cal C} \left( C^- \, \widetilde{F} \right) \equiv \int_{-1}^{1} dx   C^-(x,\xi) \widetilde{F}(x,\xi,t).
\label{CFF}
\end{eqnarray}
whereby, at leading order,
 \begin{equation}
 { C^\pm(x,\xi) = \frac{1}{x-\xi - i \epsilon} \mp \frac{1}{x+\xi - i \epsilon}, \quad \quad} 
 \end{equation}
and the real and imaginary parts of the CFFs are defined as,
\begin{subequations}
 \begin{eqnarray}
\mathcal{F} & = & \Re {\rm e} \mathcal{F} +i \Im {\rm m} \mathcal{F} = PV \int_{-1}^{1} dx  \left(\frac{1}{x-\xi}  {-} \frac{1}{x+\xi}\right) \, F(x,\xi,t) + i \pi \left[ F(\xi,\xi,t)  {-} F(-\xi,\xi,t) \right] , \\
  \widetilde{\mathcal{F}} & = & \Re {\rm e} \widetilde{\mathcal{F}} +i \Im {\rm m} \widetilde{\mathcal{F}} = PV \int_{-1}^{1} dx  \left(\frac{1}{x-\xi}  {+} \frac{1}{x+\xi}\right) \, \widetilde{F}(x,\xi,t) + i \pi \left[ \widetilde{F}(\xi,\xi,t)  {+} \widetilde{F}(-\xi,\xi,t) \right] .
 \end{eqnarray}
\end{subequations}

\subsection{Structure functions in terms of Compton Form Factors}
\label{sec:DVCS_structfun}
The structure functions appearing in Eq.(\ref{eq:xs5fold}) are given by quadratic terms in the CFFs multiplied by a kinematic factor, $w$, 
\[ w \, \Re{\rm e} \mathcal{F} \, \Re{\rm e}\mathcal{G}, \quad w \, \Im{\rm m} \mathcal{F} \, \Im{\rm m} \, \mathcal{G}, \quad w \, \Re{\rm e} \mathcal{F} \, \Im{\rm m} \, \mathcal{G}, \quad w \, \Im{\rm m} \mathcal{F} \, \Re{\rm e} \, \mathcal{G} ,\]
where we distinguish terms where both  $\mathcal{F}$ and ${\cal G}$ are of twist two, with $\mathcal{F,G}=\mathcal{H},\mathcal{E},\widetilde{\mathcal{H}},\widetilde{\mathcal{E}}$; terms where either $\mathcal{F(G)}$ is of twist two and ${\cal G(F)}$ is of twist three {\it i.e.}
$\mathcal{F,G}=\mathcal{H}_{2T},\mathcal{E}_{2T}$, $\widetilde{\mathcal{H}}_{2T},\widetilde{\mathcal{E}}_{2T},\mathcal{H}_{2T}',\mathcal{E}_{2T}'$ ,$ \widetilde{\mathcal{H}}_{2T}',\widetilde{\mathcal{E}}_{2T}'$; and finally, terms where both $\mathcal{F}$ and ${\cal G}$ are of twist three. Additional terms with transverse gluon polarization are also present. Their structure is described in Section \ref{sec:transvgluons}.
We have,

\vspace{0.5cm}
\noindent {\it Twist Two}
\begin{eqnarray}
\label{eq:FUUT0} 
F_{UU,T} & = & 4\Big[ (1-\xi^2)\Big[  (\Re e {\cal H})^2 +  (\Im m {\cal H})^2 + (\Re e \widetilde{\cal H})^2 + (\Im m \widetilde{\cal H})^2 \Big] + \displaystyle\frac{t_o-t}{2 M^2}  
\left[ (\Re e{\cal E})^2 + (\Im m{\cal E})^2  + 
 \, \xi^2 (\Re e\widetilde{\cal E})^2 +   \, \xi^2 (\Im m\widetilde{\cal E})^2 \right] \nonumber \\
&-&  \frac{2 \xi^2}{1-\xi^2} \,\left( \Re e {\cal H} \, \Re e {\cal E } + \Im m {\cal H} \Im m{\cal E } +  \Re e \widetilde{\cal H} \,  \Re e\widetilde{\cal E } + \Im m \widetilde{\cal H} \Im m \widetilde{\cal E }   \right) \Big] 
\\ \nonumber \\
F_{LL} & = & \displaystyle 4 \, \Big[ 2 (1-\xi^2) \, \Big( \Re e {\cal H} \,\Re e \widetilde{\cal H} + \Im m {\cal H} \,\Im m \widetilde{\cal H} \Big) + \, 2  \, \frac{t_o-t}{2 M^2} \,  \Big( \Re e{\cal E} \,(\xi \Re e\widetilde{\cal E}) 
+ \Im m{\cal E} \,(\xi \Im m \widetilde{\cal E}) \Big) \nonumber \\
&   + & \frac{2 \xi^2}{1-\xi^2} \left( \Re e {\cal H} \Re e\widetilde{\cal E } + \Im m {\cal H} \Im m \widetilde{\cal E} +  \Re e \widetilde{\cal H}  \Re e{\cal E } +  \Im m \widetilde{\cal H}  \Im m{\cal E }   \right)\Big] 
\label{eq:FUUL0} 
   \\ \nonumber \\
\label{eq:FUTT0} 
F_{UT,T}^{\sin(\phi - \phi_S) }&= &\frac{\sqrt{t_{0}-t}}{M}
\Big[ \Im  m {\cal H}  \Re  e {\cal E}  -  \Re e \mathcal{H}I\Im  m {\cal E} +    \Re e \widetilde{\cal H} \, (\xi \Im m \widetilde{\cal E}) - \Im  m \widetilde{\cal H}  \, (\xi \Re e \widetilde{\cal E}) \Big]
\\
\label{eq:FULT0} 
F_{LT}^{\cos(\phi-\phi_S)} & = & \frac{\sqrt{t_{0}-t}}{M}\Big[-\Re e \widetilde{\cal H} \Re  e {\cal E} -\Im  m \widetilde{\cal H}  \Im  m {\cal E} +   \Re e\mathcal{H} \,  (\xi \Re e \widetilde{\cal E}) + \Im  m {\cal H} \, (\xi \Im m \widetilde{\cal E})  + \frac{\xi^{2}}{1-\xi^{2}}\Big(  \Re e \widetilde{\mathcal{E}}  \Re  e {\cal E} + \Im m \widetilde{\cal E} \Im  m {\cal E} \Big)\Big]
\end{eqnarray}
At twist three, the structure functions listed below include the factor
\[ \frac{K}{\sqrt{Q^2}} =
\frac{\sqrt{t_0-t}}{\sqrt{Q^2}} x_{Bj} (1-\xi) ,\]
(see Eq.\eqref{eq:K} and Section \ref{sec:phasedep}),

\vspace{0.5cm}
\noindent {\it Twist three}
\begin{eqnarray}
F_{UU}^{\cos \phi} & = & - \frac{2 K}{\sqrt{Q^2}} \,  (1-\xi^2) \nonumber \\
& \times & \Re{\rm e} \Big\{  \left(2 \widetilde{\cal H}_{2T} + {\cal E}_{2T} + 2 \widetilde{\cal H}'_{2T} + {\cal E}'_{2T}\right)^{*}\Big( {\cal H}-\frac{\xi^2}{1-\xi^2} \mathcal{E} \Big)  \nonumber \\
&+& \Big(\mathcal{H}_{2T} + \frac{t_0-t}{4M^{2}} \widetilde{\mathcal{H}}_{2T} + \mathcal{H}_{2T}' +  \frac{t_0-t}{4M^{2}} \widetilde{\mathcal{H}}_{2T}'  \Big)^{*}\Big( \mathcal{E} - \xi \widetilde{\mathcal{E}} \Big)
\nonumber \\
&- & 2 \xi \,\left(\widetilde{\cal E}_{2T} + \widetilde{\cal E}'_{2T}\right)^{*} \Big(\widetilde{\cal H} - \frac{\xi^2}{1-\xi^2} \widetilde{\mathcal{E}}\Big) 
+ \frac{\xi}{1-\xi^2} \Big(\widetilde{\mathcal{E}}_{2T} - \xi \mathcal{E}_{2T} + \widetilde{\mathcal{E}}_{2T}' - \xi \mathcal{E}_{2T}' \Big)^{*} \Big( \mathcal{E} - \xi \widetilde{\mathcal{E}} \Big)
\nonumber \\
&+&  \frac{t_0-t}{16 M^2}   \Big(\mathcal{\widetilde{H}}_{2T} + \mathcal{\widetilde{H}}_{2T}'\Big)^{*} \Big( \mathcal{E} + \xi \widetilde{\mathcal{E}}\Big)   \Big\}
\\
F_{LU}^{\sin \phi} & = & - \frac{2 K}{\sqrt{Q^2}} \,  (1-\xi^2)\Im{\rm m} \Big\{  \left(2 \widetilde{\cal H}_{2T} + {\cal E}_{2T} + 2 \widetilde{\cal H}'_{2T} + {\cal E}'_{2T}\right)^{*}\Big( {\cal H}-\frac{\xi^2}{1-\xi^2} \mathcal{E} \Big)  \nonumber \\
&+& \Big(\mathcal{H}_{2T} + \frac{t_0-t}{4M^{2}} \widetilde{\mathcal{H}}_{2T} + \mathcal{H}_{2T}' +  \frac{t_0-t}{4M^{2}} \widetilde{\mathcal{H}}_{2T}'  \Big)^{*}\Big( \mathcal{E} - \xi \widetilde{\mathcal{E}} \Big)
\nonumber \\
&- & 2 \xi \,\left(\widetilde{\cal E}_{2T} + \widetilde{\cal E}'_{2T}\right)^{*} \Big(\widetilde{\cal H} - \frac{\xi^2}{1-\xi^2} \widetilde{\mathcal{E}}\Big) 
+ \frac{\xi}{1-\xi^2} \Big(\widetilde{\mathcal{E}}_{2T} - \xi \mathcal{E}_{2T} + \widetilde{\mathcal{E}}_{2T}' - \xi \mathcal{E}_{2T}' \Big)^{*} \Big( \mathcal{E} - \xi \widetilde{\mathcal{E}} \Big)
\nonumber \\
&+&  \frac{t_0-t}{16 M^2}   \Big(\mathcal{\widetilde{H}}_{2T} + \mathcal{\widetilde{H}}_{2T}'\Big)^{*} \Big( \mathcal{E} + \xi \widetilde{\mathcal{E}}\Big)   \Big\}
\\
F_{UL}^{\sin \phi} & = & - \frac{2 K}{\sqrt{Q^2}} \,  (1-\xi^2)\Im m \Big\{ \Big( {\cal E}_{2T} - \xi \widetilde{\cal E}_{2T} + {\cal E}'_{2T} - \xi \widetilde{\cal E}'_{2T} \Big)^{*}\Big( \widetilde{{\cal H}} -\frac{\xi^2}{1-\xi^2} \widetilde{\mathcal{E}} \Big) 
 \nonumber \\
&+ & \frac{t_0-t}{16 M^2}  \Big(\mathcal{\widetilde{H}}_{2T} + \mathcal{\widetilde{H}}_{2T}'\Big)^{*} \Big( \mathcal{E} + \xi \widetilde{\mathcal{E}}\Big)  
\nonumber \\
& +  & \Big(\mathcal{H}_{2T} + \frac{t_0-t}{4M^{2}} \widetilde{\mathcal{H}}_{2T} +  \mathcal{H}_{2T}' + \frac{t_0-t}{4M^{2}} \widetilde{\mathcal{H}}_{2T}' \Big)^{*}  \Big( \mathcal{E} - \xi \widetilde{\mathcal{E}} \Big) \nonumber \\
&+& \frac{\xi}{1-\xi^2} \Big(\widetilde{\mathcal{E}}_{2T} - \xi \mathcal{E}_{2T}+ \widetilde{\mathcal{E}}_{2T}' - \xi \mathcal{E}_{2T}'\Big)^{*}\Big( \mathcal{E} - \xi \widetilde{\mathcal{E}} \Big)
  \Big\} 
\\
F_{LL}^{\cos \phi} & =&  - \frac{2K}{\sqrt{Q^2}} \,  (1-\xi^2) \Re e \Big\{ \Big( {\cal E}_{2T} - \xi \widetilde{\cal E}_{2T} + {\cal E}'_{2T} - \xi \widetilde{\cal E}'_{2T} \Big)^{*}\Big( \widetilde{{\cal H}} -\frac{\xi^2}{1-\xi^2} \widetilde{\mathcal{E}} \Big) 
 \nonumber \\
&+ & \frac{t_0-t}{16 M^2}  \Big(\mathcal{\widetilde{H}}_{2T} + \mathcal{\widetilde{H}}_{2T}'\Big)^{*} \Big( \mathcal{E} + \xi \widetilde{\mathcal{E}}\Big)  
\nonumber \\
& +  & \Big(\mathcal{H}_{2T} + \frac{t_0-t}{4M^{2}} \widetilde{\mathcal{H}}_{2T} +  \mathcal{H}_{2T}' + \frac{t_0-t}{4M^{2}} \widetilde{\mathcal{H}}_{2T}' \Big)^{*}  \Big( \mathcal{E} - \xi \widetilde{\mathcal{E}} \Big) \nonumber \\
&+& \frac{\xi}{1-\xi^2} \Big(\widetilde{\mathcal{E}}_{2T} - \xi \mathcal{E}_{2T}+ \widetilde{\mathcal{E}}_{2T}' - \xi \mathcal{E}_{2T}'\Big)^{*}\Big( \mathcal{E} - \xi \widetilde{\mathcal{E}} \Big)
  \Big\} 
\\ 
F_{UT}^{\sin(\phi_S)} &= &\frac{K}{\sqrt{2 Q^2}}\sqrt{1-\xi^{2}} \Im m\Bigg[\Big(\mathcal{H}_{2T} + \mathcal{H}_{2T}' \Big) + \frac{t_{0}-t}{4M^{2}}\Big(\widetilde{\mathcal{H}}_{2T} + \widetilde{\mathcal{H}}_{2T}'  \Big) + \frac{\xi}{1-\xi^{2}}\Big(\widetilde{\mathcal{E}}_{2T} + \widetilde{\mathcal{E}}_{2T}' \Big) \nonumber \\&-& \frac{\xi^{2}}{1-\xi^{2}}\Big(\mathcal{E}_{2T} + \mathcal{E}_{2T}' \Big)  \Bigg] \Bigg(\mathcal{H} + \widetilde{\mathcal{H}} - \frac{\xi^{2}}{1-\xi^{2}}\Big(\mathcal{E} + \widetilde{\mathcal{E}} \Big) \Bigg)^{*} \nonumber \\ 
&+& \frac{K}{\sqrt{2}Q^{2}}  \frac{t_{0}-t}{4M^{2}}\Im m\Bigg[2\widetilde{\mathcal{H}}_{2T} + (1+\xi)\Big(\mathcal{E}_{2T} - \widetilde{\mathcal{E}}_{2T}\Big) + 2 \widetilde{\mathcal{H}}_{2T}' + (1+\xi)\Big(\mathcal{E}_{2T}' - \widetilde{\mathcal{E}}_{2T}'\Big) \Bigg] \Bigg(\mathcal{E}+\xi \widetilde{\mathcal{E}} \Bigg)^{*}
\\
F_{UT}^{\sin(2\phi - \phi_S) }& = &  -\frac{K}{\sqrt{2 Q^2}} \frac{t_{0}-t}{4M^{2}}
\Im m\Bigg[2\widetilde{\mathcal{H}}_{2T} + (1-\xi)\Big(\mathcal{E}_{2T} + \widetilde{\mathcal{E}}_{2T}\Big) + 2 \widetilde{\mathcal{H}}_{2T}' + (1-\xi)\Big(\mathcal{E}_{2T}' + \widetilde{\mathcal{E}}_{2T}'\Big) \Bigg]\nonumber  \\&\times&  \Bigg(\mathcal{E}-\xi \widetilde{\mathcal{E}} \Bigg)^{*} \nonumber \\
&-& \frac{K}{\sqrt{2}Q^{2}}\sqrt{1-\xi^{2}}\frac{t_{0}-t}{8M^{2}}
\Im m\Bigg[\widetilde{\mathcal{H}}_{2T} + \widetilde{\mathcal{H}}_{2T}' \Bigg]\Bigg(\mathcal{H} - \widetilde{\mathcal{H}} - \frac{\xi^{2}}{1-\xi^{2}}\Big(\mathcal{E} - \widetilde{\mathcal{E}} \Big) \Bigg)^{*}
\\
F_{LT}^{\cos(\phi_S)} & = & \frac{K}{\sqrt{2 Q^2}}\sqrt{1-\xi^{2}} \Re {\rm e} \Bigg[\Big(\mathcal{H}_{2T} + \mathcal{H}_{2T}' \Big) + \frac{t_{0}-t}{4M^{2}}\Big(\widetilde{\mathcal{H}}_{2T} + \widetilde{\mathcal{H}}_{2T}'  \Big) + \frac{\xi}{1-\xi^{2}}\Big(\widetilde{\mathcal{E}}_{2T} + \widetilde{\mathcal{E}}_{2T}' \Big) \nonumber \\&-& \frac{\xi^{2}}{1-\xi^{2}}\Big(\mathcal{E}_{2T} + \mathcal{E}_{2T}' \Big)  \Bigg] \Bigg(\mathcal{H} + \widetilde{\mathcal{H}} - \frac{\xi^{2}}{1-\xi^{2}}\Big(\mathcal{E} + \widetilde{\mathcal{E}} \Big) \Bigg)^{*} \nonumber \\ 
&+& \frac{K}{\sqrt{2}Q^{2}}  \frac{t_{0}-t}{4M^{2}}\Re {\rm e} \Bigg[2\widetilde{\mathcal{H}}_{2T} + (1+\xi)\Big(\mathcal{E}_{2T} - \widetilde{\mathcal{E}}_{2T}\Big) + 2 \widetilde{\mathcal{H}}_{2T}' + (1+\xi)\Big(\mathcal{E}_{2T}' - \widetilde{\mathcal{E}}_{2T}'\Big) \Bigg] \Bigg(\mathcal{E}+\xi \widetilde{\mathcal{E}} \Bigg)^{*}
\\ 
F_{LT}^{\cos(2\phi - \phi_S) } &= & \frac{K}{\sqrt{2 Q^2}} \frac{t_{0}-t}{4M^{2}}
\Re {\rm e} \Bigg[2\widetilde{\mathcal{H}}_{2T} + (1-\xi)\Big(\mathcal{E}_{2T} + \widetilde{\mathcal{E}}_{2T}\Big) + 2 \widetilde{\mathcal{H}}_{2T}' + (1-\xi)\Big(\mathcal{E}_{2T}' + \widetilde{\mathcal{E}}_{2T}'\Big) \Bigg]\nonumber  \\&\times&  \Bigg(\mathcal{E}-\xi \widetilde{\mathcal{E}} \Bigg)^{*} \nonumber \\
&+& \frac{K}{\sqrt{2}Q^{2}}\sqrt{1-\xi^{2}}\frac{t_{0}-t}{8M^{2}}
\Re {\rm e} \Bigg[\widetilde{\mathcal{H}}_{2T} + \widetilde{\mathcal{H}}_{2T}' \Bigg]\Bigg(\mathcal{H} - \widetilde{\mathcal{H}} - \frac{\xi^{2}}{1-\xi^{2}}\Big(\mathcal{E} - \widetilde{\mathcal{E}} \Big) \Bigg)^{*}
\end{eqnarray} 
The structure functions $F_{UU,L}$, $F_{UT,L}^{\sin(\phi - \phi_S) }$ are given by the product of two twist three CFFs. They enter the cross section suppressed by a $1/Q^2$ factor,
\begin{eqnarray}
F_{UU,L} &= & \frac{K^{2}}{Q^2}  \, \Bigg( \Big|2 \widetilde{\mathcal{H}}_{2T}  + (1-\xi)(\mathcal{E}_{2T} + \widetilde{\mathcal{E}}_{2T} ) + 2 \widetilde{\mathcal{H}}_{2T}' + (1-\xi)(\mathcal{E}_{2T}'   + \widetilde{\mathcal{E}}_{2T}') \Big|^{2}\nonumber \\
&+&  \Big|2 \widetilde{\mathcal{H}}_{2T}  + (1+\xi)(\mathcal{E}_{2T} - \widetilde{\mathcal{E}}_{2T} ) + 2 \widetilde{\mathcal{H}}_{2T}' + (1+\xi)(\mathcal{E}_{2T}'   - \widetilde{\mathcal{E}}_{2T}') \Big|^{2}\Bigg) \\
F_{UT,L} & = & \frac{K^{2}}{Q^2} \Bigg(\Big[2 \widetilde{\mathcal{H}}_{2T}  + (1-\xi)(\mathcal{E}_{2T} + \widetilde{\mathcal{E}}_{2T} ) + 2 \widetilde{\mathcal{H}}_{2T}' + (1-\xi)(\mathcal{E}_{2T}'   + \widetilde{\mathcal{E}}_{2T}') \Big]^{*}\Big[ (\mathcal{H}_{2T} + \mathcal{H}_{2T}') \nonumber \\&+& \frac{t_0-t}{4M^2}\Big(\mathcal{\widetilde{H}}_{2T} +  \mathcal{\widetilde{H}}_{2T}'\Big) + \frac{\xi}{1-\xi^{2}}\Big(\mathcal{\widetilde{E}}_{2T} + \mathcal{\widetilde{E}}_{2T}' \Big) 
- \frac{\xi^{2}}{1-\xi^{2}}\Big(\mathcal{E}_{2T} + \mathcal{E}_{2T}' \Big)  \Big] + \Big[ 2 \widetilde{ \mathcal{H}}_{2T} + (1+\xi)(\mathcal{E}_{2T} - \widetilde{\mathcal{E}}_{2T}) \nonumber \\ &+&  
2 \widetilde{\mathcal{H}}_{2T}' + (1+\xi)(\mathcal{E}_{2T}'  -  \widetilde{\mathcal{E}}_{2T}') \Big] \Big[\mathcal{ \widetilde{H}}_{2T} + \mathcal{\widetilde{H}}_{2T}'\Big]^{*}  \Bigg)
\end{eqnarray} 
The structure functions $F_{UU}^{\cos 2 \phi}$, $F_{UL}^{\sin 2 \phi}$ $F_{UT}^{\sin(\phi + \phi_S) },
F_{UT}^{\sin(3\phi - \phi_S) }$ involve two units of helicity flip. They are, therefore, described by transverse gluon GPDs (Section \ref{sec:transvgluons}).

The leading twist structure functions, Eqs.(\ref{eq:FUUT0}, \ref{eq:FUUL0}, \ref{eq:FUTT0}) display a similar content in terms of GPDs as in the expressions for the``Fourier coefficients"  in  Refs.\cite{Belitsky:2001ns,Belitsky:2010jw}. Notice, however, how, by expressing the kinematic coefficients in terms of the variable $\epsilon$ efficiently streamlines the formalism, avoiding any approximation. The twist three structure functions contain GPDs from the classification of  Ref.\cite{Meissner:2009ww}; they are entirely new, both in the GPD content, and in the kinematic coefficients.  

In order to pin down the twist two GPD structure of the nucleon, from this part of the cross section one would need 8 distinct measurements for the $\Re {\rm e}$ and $\Im {\rm m}$ parts of the 4 GPDs. $ep \rightarrow e' \gamma' p'$ from a polarized proton can provide only 4 of these measurements. Additional information can be obtained from related deeply virtual exclusive experiments. 



\subsection{Helicity Structure Functions}
The structure functions composition in terms of CFFs follows from the definition of the helicity amplitudes for the DVCS process given in Eq.(\ref{eq:famp}). In particular, we define the unpolarized components in Eq.(\ref{eq:xs5fold}) 
as,
\begin{subequations}
\label{FUU}
\begin{eqnarray}
\label{eq:FUUT}
F_{UU,T} &  = &  4 (\widetilde{F}_{+ +}^{1 1} + \widetilde{F}_{+ -}^{1 1} + \widetilde{F}_{- +}^{1 1} + \widetilde{F}_{- -}^{1 1}) ,\\ 
F_{UU,L} &  = & 2 \widetilde{F}^{00}_{++}  \\
 F_{UU}^{\cos   \phi}  &  = &  
- 2 \, \Re {\rm e} \left(\widetilde{F}_{++}^{01}+\widetilde{F}_{+-}^{01} +\widetilde{F}_{-+}^{01}+\widetilde{F}_{--}^{01} \right) \\
 F_{UU}^{\sin  \phi}  &  = &  2\Im m \Big( \widetilde{F}_{\Lambda+}^{01} + \widetilde{F}_{\Lambda -}^{01} + \widetilde{F}_{\Lambda +}^{0-1} +\widetilde{F}_{\Lambda -}^{0-1}\Big)=
 0 \\
 F_{UU}^{\cos  2 \phi}  &  = & 2 \,  \Re{\rm e}   \left(\widetilde{F}_{++}^{1-1}+\widetilde{F}_{+-}^{1-1}+\widetilde{F}_{-+}^{1-1}+\widetilde{F}_{--}^{1-1} \right) ,
\end{eqnarray}
\end{subequations}
while the structure functions involving longitudinal beam polarization are,
\begin{subequations}
\begin{eqnarray}
\label{FLU}
F_{LU} & = & 2(\widetilde{F}_{\Lambda +}^{11} + \widetilde{F}_{\Lambda -}^{11} - \widetilde{F}_{\Lambda +}^{-1-1} - \widetilde{F}_{\Lambda -}^{-1-1}) =
0, \\
 F_{LU}^{\sin   \phi}  &  = & -2 \, \Im {\rm m} \left(\widetilde{F}_{++}^{01}+\widetilde{F}_{+-}^{01} +\widetilde{F}_{-+}^{01}+\widetilde{F}_{--}^{01} \right) , \\
 F_{LU}^{\cos   \phi}  &  = & 2\Re e\Big( \widetilde{F}_{\Lambda+}^{01} + \widetilde{F}_{\Lambda -}^{01} + \widetilde{F}_{\Lambda +}^{0-1} +\widetilde{F}_{\Lambda -}^{0-1}\Big)=
 0 
 \end{eqnarray}
 \end{subequations}
the structure functions for longitudinal target polarization are,
\begin{subequations} 
\label{FUL}
\begin{eqnarray}
F_{UL}^{\sin  \phi} & = & 2\Im m\Big(\widetilde{F}_{++}^{01} + \widetilde{F}_{+-}^{01} - \widetilde{F}_{-+}^{01} - \widetilde{F}_{--}^{01} \Big)\\
F_{UL}^{\cos  \phi} & = & 2 \Re e \Big( \widetilde{F}_{\Lambda+}^{01} + \widetilde{F}_{\Lambda -}^{01} + \widetilde{F}_{\Lambda +}^{0-1} +\widetilde{F}_{\Lambda -}^{0-1}\Big)=
 0 
\end{eqnarray}
\end{subequations}
and finally the structure functions for both beam and target longitudinal polarization read, 
\begin{subequations}
\begin{eqnarray}
\label{eq:FLL}
F_{LL}  &  = &   2 (\widetilde{F}_{+ +}^{1 1} + \widetilde{F}_{+ -}^{1 1} - \widetilde{ F}_{- +}^{1 1} - \widetilde{F}_{- -}^{1 1} )   \\
 F_{LL}^{\sin \phi}  &  = & 2\Im m(\widetilde{F}_{\Lambda +}^{0 1} + \widetilde{F}_{\Lambda -}^{0 1} - \widetilde{F}_{\Lambda +}^{0 -1} - \widetilde{F}_{\Lambda -}^{0 -1}) =0 ,
  \\
 F_{LL}^{\cos \phi}  &  = & - 2 \, \Re {\rm e} \left(\widetilde{F}_{++}^{01}-\widetilde{F}_{+-}^{01} +\widetilde{F}_{-+}^{01}-\widetilde{F}_{--}^{01} \right)  
\end{eqnarray}
\end{subequations}
Note that  
$F_{LU}=F_{UL}=0$ from parity conservation (the properties of the structure functions under parity transformation are explained in Section \ref{sec:phasedep}).
Similarly, in the twist three case, $F_{UU}^{\sin \phi} = F_{LU}= F_{LU}^{\cos \phi} = 0$ as it would follow from parity conservation in the quark-proton scattering amplitude seen as a 2 body scattering process in the CoM. For instance, $F_{UU}^{\sin{\phi}}$ is given by the following combination of helicity amplitudes $ F_{++}^{01} + F_{+-}^{01} + F_{-+}^{01}+F_{--}^{01}+ F_{--}^{0-1} + F_{-+}^{0-1}+F_{+-}^{0-1} + F_{++}^{0-1}$ which is zero as if the 2-body scattering parity rules held. The scattering process for twist three objects cannot, however, be trivially reduced to a two body scattering process, thus implying that the gluon rescattering happens, in this case, in one plane.

For an initial target nucleon with definite transverse polarization, $\vec{S}_T$, with orientation
\begin{equation}
\vec{S}_T = S_T (\cos \phi_s, \sin\phi_s, 0)
\label{ST}
\end{equation}
in the target hadron rest frame with the target transverse spin relative to the lepton frame coordinates. 
%
The target polarization density matrix for longitudinal or transverse polarization, in the {helicity basis relative to the lepton frame coordinates}, is
\begin{equation}
\rho_{\Lambda'' \, \Lambda}^{\vec{S}_{T \, {\rm or} \, L}}=\frac{1}{2}\left(\begin{array}{cc}
1+S_L & S_T e^{+i\phi_S} \\
S_T e^{-i\phi_S} & 1-S_L
\end{array} \right)_{\Lambda''\, \Lambda},
\label{densitymatrix}
\end{equation}

The basis in which the transverse spin is diagonal is a transversity basis, $\Lambda_T$, wherein the state,
\begin{equation}
\mid \Lambda_T = \pm \frac{1}{2} \rangle = \frac{e^{-i \phi_s/2}}{\sqrt{2}} \left[ e^{+i\phi_s/2} \mid \Lambda
=+\frac{1}{2} \rangle  \pm e^{-i\phi_s/2} \mid \Lambda=-\frac{1}{2} \rangle \right]
\end{equation}
The amplitudes with definite transverse spin for the target will be given by linear combinations of helicity amplitudes,
\begin{equation}
{\widetilde T}^{h  \Lambda^\prime_\gamma}_{DVCS,  \Lambda_T \Lambda^\prime} =\frac{1}{\sqrt{2}} \left[ e^{+i\phi_s/2} \, T^{h  \Lambda^\prime_\gamma}_{DVCS,  \Lambda=+, \, \Lambda^\prime} + \Lambda_T e^{-i\phi_s/2} \, T^{h  \Lambda^\prime_\gamma}_{DVCS,  \Lambda=-, \, \Lambda^\prime} \right]
\end{equation}
where ${\widetilde T}$ is the amplitude with the target in the transversity basis.
One has for $S_T=\pm 1$ that the target is totally polarized in the $(\cos\phi_s, \sin \phi_s, 0)$ direction with transversity $\Lambda_T=\pm 1/2$. 

The $\Lambda$-dependence is defined through the spin four-vector,
\begin{eqnarray}
 S_{L \,\mu} &=& (2 \Lambda) \left(\frac{\mid{\vec p}\mid}{M}; {\bf 0}_T, -\frac{p^0}{M} \right)\,
 \rightarrow 
 - (2 \Lambda) \,  \hat{z}  \,\,\,\, {\rm rest \, frame} , 
 \\
S_{T \, \mu} &=& S_T\left(0; \cos \phi_s, \sin \phi_s,0\right) 
\label{SLcovariant}
\end{eqnarray}
where the minus sign in $\vec{S}_L$ follows the Trento convention \cite{Bacchetta:2004jz}. 


\vspace{0.3cm}
We distinguish the two cases of an unpolarized beam, $UT$,
\begin{subequations}
\label{eq:FUT}
\begin{eqnarray}
\label{eq:FUTT}
F_{UT,T}^{\sin(\phi - \phi_S) }&=& {\Im} m (\widetilde{F}_{T, + +}^{1  1} 
+ \widetilde{F}_{T, + -}^{1  1} )\\ F_{UT,L}^{\sin(\phi - \phi_S) }&=&2 {\Im} m (\widetilde{F}_{T, + +}^{0 \, 0} + \widetilde{F}_{T, +\, -}^{0 \, 0} )
\\ 
F_{UT}^{\sin(\phi + \phi_S) } &=& \frac{1}{2} {\Im}m(\widetilde{F}_{T, + +}^{1  -1} +\widetilde{F}_{T, +  -}^{1  -1}) 
\\
F_{UT}^{\sin(3\phi - \phi_S) }&=& \frac{1}{2}{\Im}m (\widetilde{F}_{T, - +}^{1  -1}+ \widetilde{F}_{T, -  -}^{1 -1})
\\
F_{UT}^{\sin(\phi_S)} &=& \frac{1}{\sqrt{2}} {\Im m} 
(\widetilde{F}_{T, + +}^{1 0}+ \widetilde{F}_{T, + -}^{10} ) 
\\ 
F_{UT}^{\sin(2\phi - \phi_S) }&=&\frac{1}{\sqrt{2}} {\Im} m (\widetilde{F}_{T, -  +}^{1 \, 0} + \widetilde{F}_{T, -  -}^{1  0})
\end{eqnarray}
\end{subequations}
and a polarized beam, $LT$,
\begin{subequations}
\label{eq:FLT}
\begin{eqnarray}
\label{eq:FLTc}
F_{LT}^{\cos(\phi-\phi_S)}&=&{\Re} e (\widetilde{F}^{1 \, 1}_{T, + \, +} + \widetilde{F}^{1 \, 1}_{T, + \, -})\\
F_{LT}^{\cos\phi_S}&=& -\frac{1}{\sqrt{2}} {\Re} e (\widetilde{F}_{T, + \, +}^{1 \, 0} + \widetilde{F}_{T, + \, -}^{1 \, 0} ) \\
F_{LT}^{\cos(2\phi-\phi_S)}&=&-\frac{1}{\sqrt{2}} {\Re} e (\widetilde{F}_{T,- \, +}^{1 \, 0} +\widetilde{F}_{T,- \, -}^{1 \, 0} )
\end{eqnarray}
\end{subequations}
Note that of the 6 imaginary parts, only 3 have corresponding real parts. That is a result of parity and Hermiticity of the transverse polarization dependent structure functions.

\subsubsection{$\gamma^* p \rightarrow \gamma' p'$ Helicity Amplitudes}
\label{sec:phasedep}
The structure functions appearing in Eq.(\ref{eq:xs5fold}) are constructed from  bilinear structures in the helicity basis (Eq.\ref{eq:famp}) for specific $\Lambda$ and summed over the final proton, $\Lambda'$ and photon, $\Lambda_{\gamma'}$, polarizations. We list below all the constructs appearing in the DVCS cross section. More details on  the cross section in terms of the helicity amplitudes are given in Appendix \ref{appc}.

The phase dependence of the various contributions to the cross section included in the bilinear forms listed above, derives from the helicity amplitudes property \cite{Jacob:1959at,Leader},
\begin{equation}
f_{\Lambda \Lambda^\prime}^{\Lambda_{\gamma^*} \Lambda_\gamma'} (\theta,\phi) = e^{-i(\Lambda_{\gamma^*} -\Lambda -\Lambda_\gamma'+ \Lambda')\phi} \,\widetilde{f}_{\Lambda \Lambda^\prime}^{\,\Lambda_{\gamma^*} \Lambda_\gamma'}(\theta),
\end{equation}
which follows from the definition of the rotated polarization vectors
in Eqs.(\ref{eq:epsrot1},\ref{eps_hadron}). The intermediate exchanged photon's phase not being measured generates phase ($\phi$) dependent configurations where $\Lambda_{\gamma^*}^{(1)} \neq \Lambda_{\gamma^*}^{(2)}$, at variance with the $\Lambda_\gamma'$, $\Lambda$, $\Lambda'$ terms where the $\phi$ dependence cancels out in the product of the $f$ functions times their conjugates. 

In a two-body scattering process the helicity amplitudes obey the following parity constraint,  
\begin{equation}
f^{-\Lambda_\gamma-\Lambda_\gamma^\prime}_{-\Lambda-\Lambda^\prime} = \eta (-1)^{\Lambda_{\gamma^*}-\Lambda-\Lambda_\gamma^\prime+\Lambda^\prime}
\left( f^{\Lambda_\gamma\Lambda_\gamma^\prime}_{\Lambda\Lambda^\prime}\right)^*
\label{eq:parity_f}
\end{equation}
with $\eta=+$ for photon or vector meson production and $\eta=-$ 
for pseudoscalar meson production. 

For longitudinal polarization the phase dependence of the structure functions is given by,
\begin{eqnarray}
F_{\Lambda\Lambda'}^{\Lambda_{\gamma^{*}}^{(1)} \Lambda_{\gamma^{*}}^{(2)}} = e^{i (\Lambda_{\gamma^*}^{(1)}-\Lambda_{\gamma^*}^{(2)}) \, \phi} \, \widetilde{F}_{\Lambda \Lambda'}^{\Lambda_{{\gamma^*}}^{(1)}\Lambda_{{\gamma^*}}^{(2)}} = 
e^{i (\Lambda_{\gamma^*}^{(1)}-\Lambda_{\gamma^*}^{(2)}) \, \phi} \, \sum_{\Lambda_{\gamma'}} \left( \widetilde{f}^{\Lambda_{{\gamma^*}}^{(1)} \Lambda_\gamma'}_{\Lambda \Lambda' } \right)^* \widetilde{f}^{\Lambda_{{\gamma^*}}^{(2)} \Lambda_\gamma^\prime}_{ \Lambda \Lambda'}  = 
\sum_{\Lambda_{\gamma'}} \left( f^{\Lambda_{{\gamma^*}}^{(1)} \Lambda_\gamma'}_{\Lambda \Lambda' } \right)^* f^{\Lambda_{{\gamma^*}}^{(2)} \Lambda_\gamma^\prime}_{ \Lambda \Lambda'}   
\label{eq:polarizedF}
\end{eqnarray}
whereas for the for the transverse case one has,


\begin{eqnarray}
&&F_{T, \Lambda\Lambda'}^{\Lambda_{\gamma^{*}}^{(1)} \Lambda_{\gamma^{*}}^{(2)}} = e^{i (\Lambda_{\gamma^*}^{(1)}-\Lambda_{\gamma^*}^{(2)}-2\Lambda) \, \phi} \, \widetilde{F}_{T,\Lambda \Lambda'}^{\Lambda_{{\gamma^*}}^{(1)}\Lambda_{{\gamma^*}}^{(2)}}  = \sum_{\Lambda_{\gamma'}} \left( f^{\Lambda_{{\gamma^*}}^{(1)} \Lambda_\gamma'}_{\Lambda \Lambda' } \right)^* f^{\Lambda_{{\gamma^*}}^{(2)} \Lambda_\gamma^\prime}_{ -\Lambda \Lambda'} = e^{i (\Lambda_{\gamma^*}^{(1)}-\Lambda_{\gamma^*}^{(2)}-2\Lambda) \, \phi} \, \sum_{\Lambda_{\gamma'}} \left( \widetilde{f}^{\Lambda_{{\gamma^*}}^{(1)} \Lambda_\gamma'}_{\Lambda \Lambda' } \right)^* \widetilde{f}^{\Lambda_{{\gamma^*}}^{(2)} \Lambda_\gamma^\prime}_{ -\Lambda \Lambda'} 
\label{eq:TpolarizedF}
\end{eqnarray}

The helicity structure functions which contain twist two quark GPDs are the ones with transverse $\gamma^*$.
\footnote{We disregard contributions of the type, $ \left( {f}^{1-1}_{\Lambda \Lambda'} \right)^* {f}^{1-1}_{ \Lambda \Lambda'}$ in these equations since they are suppressed at order $\alpha_S^2$.}
For an unpolarized or longitudinally polarized target one has,

\vspace{1cm}
\noindent {\it Twist Two: Unpolarized/Longitudinally Polarized}
\begin{subequations}
\begin{eqnarray}
F^{11}_{\Lambda \Lambda'} & = &    \left( {f}^{11}_{\Lambda \Lambda'} \right)^* {f}^{11}_{ \Lambda \Lambda'} = \mid {f}^{11}_{\Lambda \Lambda'} \mid^2 \\
F^{-1-1}_{\Lambda \Lambda'} & =  &   \left( {f}^{-1-1}_{\Lambda \Lambda'} \right)^* {f}^{-1-1}_{ \Lambda \Lambda'} = \mid {f}^{-1-1}_{\Lambda \Lambda'} \mid^2 .
\end{eqnarray}
\end{subequations}

\vspace{.1cm}
\noindent These helicity structure functions enter $F_{UU,T}$, Eq.(\ref{eq:FUUT}), and $F_{LL}$, Eq.(\ref{eq:FLL}). They obey the following parity relations,
\begin{eqnarray}
F^{\pm 1 \pm 1}_{\Lambda \Lambda'}  = F^{\mp 1 \mp 1}_{-\Lambda -\Lambda'}     
\end{eqnarray}

\noindent Because of these parity relations we have no leading twist $UL$ and $LU$ components in the DVCS contribution to the cross section. 
For transverse target spin the following amplitudes contribute to, $F_{UT,T}^{\sin (\phi-\phi_S)}$, Eq.(\ref{eq:FUTT}), and $F_{LT}^{\cos (\phi-\phi_S)}$, (\ref{eq:FLTc}), respectively.

\vspace{0.5cm}
\noindent {\it Twist Two: Transversely Polarized}
\begin{subequations}
\begin{eqnarray}
e^{2i\Lambda \phi} F^{11}_{T,\Lambda \Lambda'} & = &   \left( {\widetilde{f}}^{11}_{\Lambda \Lambda'} \right)^* {\widetilde{f}}^{11}_{ -\Lambda \Lambda'} 
\\
\label{trans_structa}
e^{2i\Lambda \phi}F^{-1-1}_{T,\Lambda \Lambda'} & =  &    \left( {\widetilde{f}}^{-1-1}_{\Lambda \Lambda'} \right)^* {\widetilde{f}}^{-1-1}_{ -\Lambda \Lambda'} 
\label{trans_structb}
\end{eqnarray}
\end{subequations}

\vspace{.5cm}
\noindent{\it Twist Two Transversity Gluons: Longitudinal Target}
\begin{subequations}
\begin{eqnarray}
e^{-i 2 \phi} \, F^{1-1}_{\Lambda \Lambda'} & =  & \left( {\widetilde{f}}^{11}_{\Lambda \Lambda'} \right)^* {\widetilde{f}}^{-11}_{ \Lambda \Lambda'} + \left( {\widetilde{f}}^{1-1}_{\Lambda \Lambda'} \right)^* {\widetilde{f}}^{-1-1}_{ \Lambda \Lambda'}
\label{transga}
\\
e^{i 2 \phi} \, F^{-11}_{\Lambda \Lambda'} &= & \left( {\widetilde{f}}^{-11}_{\Lambda \Lambda'} \right)^* {\widetilde{f}}^{11}_{ \Lambda \Lambda'} + \left( {\widetilde{f}}^{-1-1}_{\Lambda \Lambda'} \right)^* {\widetilde{f}}^{1-1}_{ \Lambda \Lambda'} .
\end{eqnarray}
\end{subequations}

\vspace{.5cm}
\noindent Notice that the double helicity flip terms  contribute at twist two: they involve transversity gluons in the term in ${f}^{-11}_{ \Lambda \Lambda'}$, which is suppressed by $\alpha_S$. 
We describe these contributions in Section \ref{sec:transvgluons}. 

\vspace{0.5cm}
\noindent{\it Twist Two Transversity Gluons: Transversely Polarized Target}
\begin{subequations}
\begin{eqnarray}
e^{-i (2 -2\Lambda )\phi} \, F^{1-1}_{T, \Lambda \Lambda'} & =  & \left( {\widetilde{f}}^{11}_{\Lambda \Lambda'} \right)^* {\widetilde{f}}^{-11}_{ -\Lambda \Lambda'} + \left( {\widetilde{f}}^{1-1}_{\Lambda \Lambda'} \right)^* {\widetilde{f}}^{-1-1}_{ -\Lambda \Lambda'}
\\
e^{i (2+2\Lambda )\phi} \, F^{-11}_{T, \Lambda \Lambda'} &= & \left( {\widetilde{f}}^{-11}_{\Lambda \Lambda'} \right)^* {\widetilde{f}}^{11}_{ -\Lambda \Lambda'} + \left( {\widetilde{f}}^{-1-1}_{\Lambda \Lambda'} \right)^* {\widetilde{f}}^{1-1}_{ -\Lambda \Lambda'} .
\label{transg2}
\end{eqnarray}
\end{subequations}

\vspace{.5cm} 
\noindent We list the  bilinear products of  helicity amplitudes involving the twist three GPDs: these contain a transversely polarized photon term, $\Lambda_{\gamma^*}=\pm 1$, multiplied by a longitudinally polarized photon term with $\Lambda_\gamma^*=0$. 

\vspace{0.5cm} 
\noindent{\it Twist Three: Longitudinally Polarized}
\begin{subequations}
\begin{eqnarray}
e^{i \phi} \, F^{01}_{\Lambda \Lambda'} & = &   \left( {\widetilde{f}}^{01}_{\Lambda \Lambda'} \right)^* {\widetilde{f}}^{11}_{ \Lambda \Lambda'}  \\
e^{-i \phi} \, F^{0-1}_{\Lambda \Lambda'} & =  &  \left( {\widetilde{f}}^{0-1}_{\Lambda \Lambda'} \right)^* {\widetilde{f}}^{-1-1}_{ \Lambda \Lambda'} 
\end{eqnarray}
\end{subequations}
with conjugates,
\begin{subequations}
\begin{eqnarray}
e^{-i \phi} \, \left(F^{01}_{\Lambda \Lambda'} \right)^*\equiv e^{-i \phi} \, F^{10}_{\Lambda \Lambda'} & = &   \left( {\widetilde{f}}^{11}_{\Lambda \Lambda'} \right)^* {\widetilde{f}}^{01}_{ \Lambda \Lambda'}  \\
e^{i \phi} \, \left(F^{0-1}_{\Lambda \Lambda'}\right)^* \equiv e^{i \phi} \, F^{-10}_{\Lambda \Lambda'}  & =  &    \left( {\widetilde{f}}^{-1-1}_{\Lambda \Lambda'} \right)^* {\widetilde{f}}^{0-1}_{ \Lambda \Lambda'} 
\end{eqnarray}
\end{subequations}
%

\vspace{0.5cm} 
\noindent{\it Twist Three: Transversely Polarized}
\begin{subequations}
\begin{eqnarray}
e^{i(1+2\Lambda)\phi} \, F^{01}_{T, \Lambda \Lambda'} & = &   \left( {\widetilde{f}}^{01}_{\Lambda \Lambda'} \right)^* {\widetilde{f}}^{11}_{ -\Lambda \Lambda'}  \\
e^{-i(1-2\Lambda) \phi} \, F^{0-1}_{T, \Lambda \Lambda'} & =  &  \left( {\widetilde{f}}^{0-1}_{\Lambda \Lambda'} \right)^* {\widetilde{f}}^{-1-1}_{ -\Lambda \Lambda'} 
\end{eqnarray}
\end{subequations}
with conjugates,
\begin{subequations}
\begin{eqnarray}
e^{-i(1+ 2 \Lambda) \phi} \, \left(F^{01}_{T, \Lambda \Lambda'} \right)^*\equiv e^{-i (1 + 2\Lambda )\phi} \, F^{10}_{T, -\Lambda \Lambda'} & = &   \left( {\widetilde{f}}^{11}_{-\Lambda \Lambda'} \right)^* {\widetilde{f}}^{01}_{ \Lambda \Lambda'}  \\
e^{i(1-2\Lambda) \phi} \, \left(F^{0-1}_{T, \Lambda \Lambda'}\right)^* \equiv e^{i(1 -2\Lambda ) \phi} \, F^{-10}_{T, -\Lambda \Lambda'}  & =  &    \left( {\widetilde{f}}^{-1-1}_{-\Lambda \Lambda'} \right)^* {\widetilde{f}}^{0-1}_{ \Lambda \Lambda'} 
\end{eqnarray}
\end{subequations}

\vspace{1cm}
\noindent The helicity amplitudes are written in terms of GPDs.
At twist two one has \cite{Diehl:2001pm}, 
\begin{subequations}
\label{helampftw2}
\begin{eqnarray}
f_{++}^{1 1}& = & \sqrt{1-\xi^2}\left(\mathcal{H} + \widetilde{\mathcal{H}} -\frac{\xi^2}{1-\xi^2} (\mathcal{E} + \widetilde{\mathcal{E}} ) \right)  \\
f_{--}^{1 1}&=&  \sqrt{1-\xi^2}\left(\mathcal{H} - \widetilde{\mathcal{H}} -\frac{\xi^2}{1-\xi^2} (\mathcal{E} - \widetilde{\mathcal{E}} ) \right)  \\
f_{+-} ^{1 1}&=& e^{i \phi} \frac{\sqrt{t_0 - t}}{2M} ( \mathcal{E} + \xi \widetilde{\mathcal{E}} ) \\
f_{- +} ^{1 1}&=&  - e^{-i \phi}  \frac{\sqrt{t_0 - t}}{2M} ( \mathcal{E} - \xi \widetilde{\mathcal{E}} ),
\end{eqnarray}
\end{subequations}
The GPD content of the helicity amplitudes in described in Section \ref{sec:parton}.
From these expressions one can see that by summing {\it e.g.} over the non-flip proton polarization (first two lines in Eqs.(\ref{helampftw2})), one would eliminate the axial vector CFF $\widetilde{\cal H}$ at the amplitude level. However, because the DVCS cross section involves the amplitudes modulus squared, there is no obvious simplification of results that allows us to interpret an observable (in this example the $UU$ one) in terms of specific GPDs: counter-intuitively the $UU$ observable contains both the vector and axial vector contributions.   

The twist three helicity amplitudes written in terms of twist three GPDs from the complete parametrization of the correlation function \cite{Meissner:2009ww} are presented for the first time here. They read,
\begin{subequations}
\label{eq:famps3}
\begin{eqnarray}
f_{++}^{01} &=& \! 
W^{\gamma^1+ i \gamma^2}_{++} + W^{(\gamma^1 + i \gamma^2)\gamma_5}_{++} =
\frac{K}{\sqrt{Q^2}} \, e^{i\phi} \Big[2 \widetilde{\mathcal{H}}_{2T}  + (1-\xi)(\mathcal{E}_{2T} + \widetilde{\mathcal{E}}_{2T} ) + 2 \widetilde{\mathcal{H}}_{2T}' + (1-\xi)(\mathcal{E}_{2T}'   + \widetilde{\mathcal{E}}_{2T}') \Big] \nonumber \\ \\
f_{--}^{01} &=&  \! W^{\gamma^1+ i \gamma^2}_{--} - W^{(\gamma^1 + i \gamma^2)\gamma_5}_{--}= \frac{K}{\sqrt{Q^2}}  \, e^{i\phi}  \Big[ 2 \widetilde{ \mathcal{H}}_{2T} + (1+\xi)(\mathcal{E}_{2T} - \widetilde{\mathcal{E}}_{2T}) +  
2 \widetilde{\mathcal{H}}_{2T}' + (1+\xi)(\mathcal{E}_{2T}'  -  \widetilde{\mathcal{E}}_{2T}') \Big] \nonumber \\ \\
f_{+-}^{01} &=& \! W^{\gamma^1+ i \gamma^2}_{+-} + W^{(\gamma^1 + i \gamma^2)\gamma_5}_{+-} =  - \frac{K}{\sqrt{Q^2}} \frac{\sqrt{t_0-t}}{2M }\,\, e^{2i\phi} \,  \Big[\mathcal{ \widetilde{H}}_{2T} + \mathcal{\widetilde{H}}_{2T}'\Big]\\
f_{-+}^{01} &=& \! W^{\gamma^1+ i \gamma^2}_{-+} - W^{(\gamma^1 + i \gamma^2)\gamma_5}_{-+} = \frac{K}{\sqrt{Q^2}}
\Big[ (\mathcal{H}_{2T} + \mathcal{H}_{2T}') + \frac{t_0-t}{4M^2}\Big(\mathcal{\widetilde{H}}_{2T} +  \mathcal{\widetilde{H}}_{2T}'\Big) + \frac{\xi}{1-\xi^{2}}\Big(\mathcal{\widetilde{E}}_{2T} + \mathcal{\widetilde{E}}_{2T}' \Big) \nonumber \\
&-& \frac{\xi^{2}}{1-\xi^{2}}\Big(\mathcal{E}_{2T} + \mathcal{E}_{2T}' \Big)  \Big]
\end{eqnarray}
\end{subequations}
where $K$ is obtained from the hard scattering amplitude (Sec. \ref{sec:parton}, Eqs.\eqref{eq:gtw3s},\eqref{eq:gtw3u}), using $\Delta_T^2= (t_0-t)(1-\xi^2)$,
\begin{equation}
\label{eq:K}
\frac{K}{\sqrt{Q^2}} = \frac{\sqrt{Q^{2}+\nu^{2}}-\nu}{\sqrt{Q^2}} \, \frac{\Delta_T}{2M} \frac{\sqrt{1-\xi^2}}{1+\xi} = \frac{\sqrt{1+\gamma^2}-1}{\gamma} \,    \frac{\sqrt{t_0-t}}{2M} (1-\xi) = 
\frac{1}{2}\frac{\sqrt{t_0-t}}{\sqrt{Q^2}} x_{Bj} (1-\xi) .
\end{equation}
%
The last expression in Eq.(\ref{eq:K}) was obtained for $\gamma^2 <<1$. 
Inserting the helicity amplitudes in the definitions for the bilinear structures we obtain for the twist two case,
\begin{subequations}
\begin{eqnarray}
F_{++}^{11} & = &  (1-\xi^2) \mid \mathcal{H} + \widetilde{\mathcal{H}}\mid^2 -  \frac{2 \xi^2}{1-\xi^2} \, \Re e \left[ (\mathcal{H} + \widetilde{\mathcal{H}})
 (\mathcal{E} + \widetilde{\mathcal{E}} ) \right]
   \\
F_{--}^{11} & = &   (1-\xi^2) \mid \mathcal{H} - \widetilde{\mathcal{H}}\mid^2 - \frac{2 \xi^2}{1-\xi^2} \, \Re e \left[ (\mathcal{H} - \widetilde{\mathcal{H}})  ( \mathcal{E} - \widetilde{\mathcal{E}} ) \right] \\
F_{+-}^{11} & = &  \frac{t_0 - t}{4M^2} \mid \mathcal{E} + \xi \widetilde{\mathcal{E}} \mid^2  \\
F_{-+}^{11} & = &  \frac{t_0 - t}{4M^2}  \mid \mathcal{E} - \xi \widetilde{\mathcal{E}} \mid^2   
\end{eqnarray}
\end{subequations}
where we have disregarded terms proportional to $\xi^4$. Notice that the phase dependence structure described in Section \ref{sec:phasedep} implies that the leading twist structure functions with longitudinal polarizations do not depend on $\phi$, despite a phase dependence appears in the helicity flip amplitudes in Eqs.(\ref{helampftw2}).

At twist three one has, 
\begin{subequations}
\begin{eqnarray}
F_{++}^{01} &=& \frac{K}{\sqrt{Q^2}}  \sqrt{1-\xi^2} \,  \Big( \Big(\mathcal{\widetilde{H}}_{2T}' +\frac{1-\xi}{2}\mathcal{E}_{2T}' \Big) + \Big(\mathcal{\widetilde{H}}_{2T}+\frac{1-\xi}{2}\mathcal{E}_{2T} \Big) + \frac{1-\xi}{2} \Big( \mathcal{\widetilde{E}}_{2T}+  \widetilde{\mathcal{E}}_{2T}' \Big ) \Big)^{*} \nonumber \\
& \times & \Big(\mathcal{H} + \widetilde{\mathcal{H}} -\frac{\xi^2}{1-\xi^2} (\mathcal{E} + \widetilde{\mathcal{E}} ) \Big)  
\\
F_{--}^{01} &=&  \frac{2K}{\sqrt{Q^2}} \, \sqrt{1-\xi^2}  \, \Big(\Big(\mathcal{\widetilde{H}}_{2T}' -\frac{1+\xi}{2}\mathcal{E}_{2T}' \Big) + \Big(\mathcal{\widetilde{H}}_{2T}-\frac{1+\xi}{2}\mathcal{E}_{2T} \Big)- \frac{1+\xi}{2} \Big( \mathcal{\widetilde{E}}_{2T}+  \widetilde{\mathcal{E}}_{2T}' \Big ) \Big)^* \nonumber \\
& \times & \Big(\mathcal{H} - \widetilde{\mathcal{H}} -\frac{\xi^2}{1-\xi^2} (\mathcal{E} - \widetilde{\mathcal{E}} ) \Big)\\
F_{+-}^{01} &=& - \frac{K}{\sqrt{Q^2}} \, \frac{t_0-t}{4 M^2}
\Big(\Big(\mathcal{\widetilde{H}}_{2T} + \mathcal{\widetilde{H}}_{2T}'\Big)^*\Big( \mathcal{E} + \xi \widetilde{\mathcal{E}}\Big)  \Big)\\
F_{-+}^{01} &=& - \frac{K}{\sqrt{Q^2}} \, 
\Big(\Big(\mathcal{H}_{2T} + \mathcal{H}_{2T}' + \frac{t_0-t}{4M^{2}}\Big( \widetilde{\mathcal{H}}_{2T} + \widetilde{\mathcal{H}}_{2T}' \Big) + \frac{\xi}{1-\xi^2} \Big(\widetilde{\mathcal{E}}_{2T} + \widetilde{\mathcal{E}}_{2T}' \Big)\nonumber \\ &-& \frac{\xi^2}{1-\xi^2} \Big(\mathcal{E}_{2T} + \mathcal{E}_{2T}' \Big) \Big)^* \Big( \mathcal{E} - \xi \widetilde{\mathcal{E}} \Big)  \Big)
\end{eqnarray}
\end{subequations}
The longitudinal structure function, $F_{UU,L}$ contains a ${\cal O}(1/Q^2)$ suppressed term which is bilinear in the twist three CFFs as it can be seen by examining the helicity structure,
\begin{subequations}
\begin{eqnarray}
F_{++}^{00} &=& \frac{K^{2}}{Q^{2}}  \, \Bigg( \Big|2 \widetilde{\mathcal{H}}_{2T}  + (1-\xi)(\mathcal{E}_{2T} + \widetilde{\mathcal{E}}_{2T} ) + 2 \widetilde{\mathcal{H}}_{2T}' + (1-\xi)(\mathcal{E}_{2T}'   + \widetilde{\mathcal{E}}_{2T}') \Big|^{2} \nonumber \\
&+&  \Big|2 \widetilde{\mathcal{H}}_{2T}  + (1+\xi)(\mathcal{E}_{2T} - \widetilde{\mathcal{E}}_{2T} ) + 2 \widetilde{\mathcal{H}}_{2T}' + (1+\xi)(\mathcal{E}_{2T}'   - \widetilde{\mathcal{E}}_{2T}') \Big|^{2}\Bigg)\\
F_{--}^{00} &=& \frac{K^{2}}{Q^{2}}  \, \Bigg( \Big|2 \widetilde{\mathcal{H}}_{2T}  + (1-\xi)(\mathcal{E}_{2T} + \widetilde{\mathcal{E}}_{2T} ) + 2 \widetilde{\mathcal{H}}_{2T}' + (1-\xi)(\mathcal{E}_{2T}'   + \widetilde{\mathcal{E}}_{2T}') \Big|^{2}\nonumber \\
&+&  \Big|2 \widetilde{\mathcal{H}}_{2T}  + (1+\xi)(\mathcal{E}_{2T} - \widetilde{\mathcal{E}}_{2T} ) + 2 \widetilde{\mathcal{H}}_{2T}' + (1+\xi)(\mathcal{E}_{2T}'   - \widetilde{\mathcal{E}}_{2T}') \Big|^{2}\Bigg) \\
F_{+-}^{00} &=& \frac{K^{2}}{Q^{2}} \Bigg(
\Big| (\mathcal{H}_{2T} + \mathcal{H}_{2T}') + \frac{t_0-t}{4M^2}\Big(\mathcal{\widetilde{H}}_{2T} +  \mathcal{\widetilde{H}}_{2T}'\Big) + \frac{\xi}{1-\xi^{2}}\Big(\mathcal{\widetilde{E}}_{2T} + \mathcal{\widetilde{E}}_{2T}' \Big) \nonumber \nonumber\\
&-& \frac{\xi^{2}}{1-\xi^{2}}\Big(\mathcal{E}_{2T} + \mathcal{E}_{2T}' \Big)  \Big|^{2} + \frac{(t_0-t)^{2}}{64M^4}\,\, \,  \Big|\mathcal{ \widetilde{H}}_{2T} + \mathcal{\widetilde{H}}_{2T}'\Big|^{2}\Bigg) \\
F_{-+}^{00} &=&  \frac{K^{2}}{Q^{2}}
\Bigg(\Big| (\mathcal{H}_{2T} + \mathcal{H}_{2T}') + \frac{t_0-t}{4M^2}\Big(\mathcal{\widetilde{H}}_{2T} +  \mathcal{\widetilde{H}}_{2T}'\Big) + \frac{\xi}{1-\xi^{2}}\Big(\mathcal{\widetilde{E}}_{2T} + \mathcal{\widetilde{E}}_{2T}' \Big) \nonumber \nonumber\\
&-& \frac{\xi^{2}}{1-\xi^{2}}\Big(\mathcal{E}_{2T} + \mathcal{E}_{2T}' \Big)  \Big|^{2} + \frac{(t_0-t)^{2}}{64M^4}\,\, \,  \Big|\mathcal{ \widetilde{H}}_{2T} + \mathcal{\widetilde{H}}_{2T}'\Big|^{2} \Bigg)
\end{eqnarray}
\end{subequations}

We write the structure functions for a transversely polarized target in an analogous way. They read,

\noindent {\em Twist Two}
\begin{subequations}
\begin{eqnarray}
F_{T,++}^{11} &=& -
\frac{\sqrt{t_{0}-t}}{2M}e^{-i\phi}
\Bigg[\mathcal{H} + \widetilde{\mathcal{H}} - \frac{\xi^{2}}{1-\xi^{2}}\Big(\mathcal{E} + \widetilde{\mathcal{E}}\Big) \Bigg]^{*}\Big(\mathcal{E} - \xi \widetilde{\mathcal{E}}\Big)\\
F_{T,--}^{11} &=&
\frac{\sqrt{t_{0}-t}}{2M}e^{i\phi}
\Bigg[\mathcal{H} - \widetilde{\mathcal{H}} - \frac{\xi^{2}}{1-\xi^{2}}\Big(\mathcal{E} - \widetilde{\mathcal{E}}\Big) \Bigg]^{*}\Big(\mathcal{E} + \xi \widetilde{\mathcal{E}}\Big) \\
F_{T,+-}^{11} &=&
\frac{\sqrt{t_{0}-t}}{2M}e^{-i\phi}
\Bigg[\mathcal{H} - \widetilde{\mathcal{H}} - \frac{\xi^{2}}{1-\xi^{2}}\Big(\mathcal{E} - \widetilde{\mathcal{E}}\Big) \Bigg]\Big(\mathcal{E} + \xi \widetilde{\mathcal{E}}\Big)^{*} \\
F_{T,-+}^{11} &=& -
\frac{\sqrt{t_{0}-t}}{2M}e^{i\phi}
\Bigg[\mathcal{H} + \widetilde{\mathcal{H}} - \frac{\xi^{2}}{1-\xi^{2}}\Big(\mathcal{E} + \widetilde{\mathcal{E}}\Big) \Bigg]\Big(\mathcal{E} - \xi \widetilde{\mathcal{E}}\Big)^{*}
\end{eqnarray}
\end{subequations}

\noindent{\em  Twist Three}
\begin{subequations}
\begin{eqnarray}
F_{T,++}^{01} &=& -
\frac{K}{\sqrt{Q^2}} \frac{\sqrt{t_{0}-t}}{2M}e^{-2i\phi}
\Bigg[2\widetilde{\mathcal{H}}_{2T} + (1-\xi)\Big(\mathcal{E}_{2T} + \widetilde{\mathcal{E}}_{2T}\Big) + 2 \widetilde{\mathcal{H}}_{2T}' + (1-\xi)\Big(\mathcal{E}_{2T}' + \widetilde{\mathcal{E}}_{2T}'\Big) \Bigg]^{*}  \Bigg(\mathcal{E}-\xi \widetilde{\mathcal{E}} \Bigg) \\
F_{T,--}^{01} &=& 
\frac{K}{\sqrt{Q^2}} \frac{\sqrt{t_{0}-t}}{2M}\Bigg[2\widetilde{\mathcal{H}}_{2T} + (1+\xi)\Big(\mathcal{E}_{2T} - \widetilde{\mathcal{E}}_{2T}\Big) + 2 \widetilde{\mathcal{H}}_{2T}' + (1+\xi)\Big(\mathcal{E}_{2T}' - \widetilde{\mathcal{E}}_{2T}'\Big) \Bigg]^{*}  \Bigg(\mathcal{E}+\xi \widetilde{\mathcal{E}} \Bigg) \\
F_{T,+-}^{01} &=& -
\sqrt{1-\xi^{2}} \frac{K}{\sqrt{Q^2}} \frac{\sqrt{t_{0}-t}}{2M}e^{-2i\phi}
\Bigg[\widetilde{\mathcal{H}}_{2T} + \widetilde{\mathcal{H}}_{2T}' \Bigg]^{*} \Bigg(\mathcal{H} - \widetilde{\mathcal{H}} - \frac{\xi^{2}}{1-\xi^{2}}\Big(\mathcal{E} - \widetilde{\mathcal{E}} \Big) \Bigg)\\
F_{T,-+}^{01} &=& 
\sqrt{1-\xi^{2}} \frac{2Mx_{Bj}}{\sqrt{Q^2}} \Bigg[\Big(\mathcal{H}_{2T} + \mathcal{H}_{2T}' \Big) + \frac{t_{0}-t}{4M^{2}}\Big(\widetilde{\mathcal{H}}_{2T} + \widetilde{\mathcal{H}}_{2T}'  \Big) + \frac{\xi}{1-\xi^{2}}\Big(\widetilde{\mathcal{E}}_{2T} + \widetilde{\mathcal{E}}_{2T}' \Big) \nonumber \\&-& \frac{\xi^{2}}{1-\xi^{2}}\Big(\mathcal{E}_{2T} + \mathcal{E}_{2T}' \Big)  \Bigg]^{*} \Bigg(\mathcal{H} + \widetilde{\mathcal{H}} - \frac{\xi^{2}}{1-\xi^{2}}\Big(\mathcal{E} + \widetilde{\mathcal{E}} \Big) \Bigg)
\end{eqnarray}
\end{subequations}

\subsubsection{Flavor composition}
In the previous Section we have omitted for simplicity an index, $q$, referring to the quark flavor of the various GPDs. The flavor composition of the proton GPDs, $F_p,F_{2T}^p,F_{2T}'^p$ is given by,
\begin{equation}
F_p =  \sum_q  e_q F_{q/p} = \sum_q  e_q F_q, 
\end{equation}
where $F_{q/p}\equiv F_q$ is the quark GPD in the proton, and $e_q$ is the quark charge. 
The neutron GPDs are obtained using isospin symmetry, namely $F_{u/n}=F_{d/p}$, $F_{d/n}=F_{u/p}$. 

%
%
\subsection{Partonic Structure of Polarized Structure Functions}
\label{sec:parton}
We now discuss the quark-gluon content of the various  configurations described by the helicity amplitudes $f_{\Lambda \Lambda'}^{\Lambda_\gamma \Lambda_\gamma^\prime}$ up to ${\cal O}(1/Q)$. Specific factorized formulations of the helicity amplitudes for deeply virtual exclusive processes were given in Ref.\cite{Collins:1998be} (see review in \cite{Diehl:2001pm})  at leading order. Twist three contributions were considered in Refs.\cite{Belitsky:2010jw,Belitsky:2012ch} for exclusive processes and in Ref.\cite{Bacchetta:2006tn} for SIDIS.
In this Section we show how these terms include GPDs from the general classification scheme of GPDs, GTMDs, and TMDs given in  Ref.\cite{Meissner:2009ww}.   

Considering a sufficiently large four-momentum transfer, $Q^2$, we can assume that perturbative QCD factorization works \cite{Radyushkin:2013nsa}. The twist two contribution, obtained taking the transverse polarization for both the incoming and outgoing photons, $\Lambda_{{\gamma^*}}=\Lambda_\gamma'= \pm 1$, reads ,
\footnote{Notice that the subscripts for the initial helicities and final helicities appear switched compared to the amplitudes, $f_{\Lambda\Lambda'}$. We stick to this notation in order to be consistent with the existing literature.}
\begin{eqnarray}
f_{\Lambda \Lambda^\prime}^{\pm 1 \pm 1} & =   & \sum_{\lambda,\lambda^\prime}  \, 
 g_{\lambda \lambda^\prime}^{\pm 1 \pm 1} (x,\xi,t; Q^2)  
 \otimes A_{\Lambda^\prime \lambda^\prime, \Lambda \lambda}(x,\xi,t), 
\label{facto}
\end{eqnarray}
where, the convolution integral is given by $\otimes \rightarrow \int_{-1}^1 d x$; $A_{\Lambda^\prime \lambda^\prime, \Lambda \lambda}$ \cite{Diehl:2003ny} is the quark-proton helicity amplitude describing the process, 
\[k(\lambda) + p (\Lambda)\rightarrow k'(\lambda') + p' (\Lambda') , \]
where $k (k')$ are the initial (final) quarks momenta, $p(p')$ are the initial (final) proton momenta, namely
\begin{eqnarray}
A_{\Lambda' \lambda', \Lambda \lambda} = \int \frac{d z^- \, d^2{\bf z}_T}{(2 \pi)^3} e^{ixP^+ z^- - i \bar{\bf k}_T\cdot {\bf z}_T} \left. \langle p', \Lambda' \mid {\cal O}_{\lambda' \lambda}(z) \mid p, \Lambda \rangle \right|_{z^+=0} .
\end{eqnarray} 
At twist two, the bilocal quark field operators are written as,
\begin{eqnarray}
\label{tw2_operator}
{\cal O}_{\pm \pm}(z) & = & \bar{\psi}\left(-\frac{z}{2}\right) \gamma^+(1 \pm \gamma_5)  \psi\left(\frac{z}{2}\right) \equiv \phi_\pm^\dagger \phi_\pm ,
\end{eqnarray}
defining the chiral even transitions between quark $\pm, \pm$ helicity states; in the equation we have also defined $\phi_\pm= \gamma^+\gamma^- \psi$, which correspond to the ``good" components of $\psi$, {\it i.e.} its independent degrees of freedom obtained from the QCD equations of motion  \cite{Jaffe:1997yz}.
We connect specifically with the correlation function parametrization displayed in Eqs.(\ref{GPDvec},\ref{GPDaxvec}) by noticing that,
\begin{eqnarray}
W^{[\gamma^+]}_{\Lambda\Lambda'} = \frac{1}{2} \left[A_{\Lambda' +, \Lambda +} +  A_{\Lambda' -, \Lambda -} \right], \quad\quad W^{[\gamma^+\gamma_5]}_{\Lambda\Lambda'} = \frac{1}{2} \left[A_{\Lambda' +, \Lambda +} -  A_{\Lambda' -, \Lambda -} \right]
\end{eqnarray}
The expressions of the quark-proton helicity amplitudes in terms of GPDs read \cite{Diehl:2003ny},
\begin{subequations}
\label{GPDeven}
\begin{eqnarray}
A_{++,++}  & = & \sqrt{1-\xi^2} \left( \frac{H + \widetilde{H}}{2} - \frac{\xi^2}{1-\xi} \frac{E + \widetilde{E}}{2} \right)  \\
A_{+-,+-}  & = & \sqrt{1-\xi^2} \left( \frac{H - \widetilde{H}}{2} - \frac{\xi^2}{1-\xi} \frac{E - \widetilde{E}}{2} \right)  \\
A_{++,-+} & = & - \frac{\Delta_1 -  i \Delta_2}{2M \sqrt{1-\xi^2}} \,  \frac{E - \xi \widetilde{E}}{2} = - e^{-i\phi} \, \frac{\sqrt{t_0 - t}}{2M} \frac{E - \xi \widetilde{E}}{2} \\
A_{-+,++} & = &  \frac{\Delta_1 +  i \Delta_2}{2M \sqrt{1-\xi^2}} \,  \frac{E + \xi \widetilde{E}}{2} = e^{i\phi} \, \frac{\sqrt{t_0 - t}}{2M} \frac{E + \xi \widetilde{E}}{2}
\end{eqnarray}
\end{subequations}
These amplitudes are calculated in the CoM frame of the outgoing photon and proton, with $\vec{q}$ in the negative z-direction. We write the relevant four vectors components as,
\footnote{Here we use the notation, $v\equiv(v^+,v^-,{\bf v}_T)$.}
\begin{subequations}
\begin{eqnarray}
k & \equiv & \left((x+\xi)P^+,k^-,{\bf k}_T + \frac{{\bf \Delta}_{T}}{2}\right) \\
k' & \equiv & \left((x-\xi)P^+,k'^-,{\bf k}_T- \frac{{\bf \Delta}_{T}}{2} \right) \\
p & \equiv & \left((1+\xi)P^+,\frac{M^2+ \Delta_T^2/4}{2(1+\xi)P^+},\frac{{\bf \Delta}_T}{2}\right) \\
p' & \equiv & \left((1-\xi)P^+,\frac{M^2+ \Delta_T^2/4}{2(1-\xi)P^+},-\frac{{\bf \Delta}_T}{2}\right) \\
\Delta & \equiv & \left(- 2 \xi P^+,\frac{\xi\left(M^2+ \Delta_T^2/4 \right)}{P^+(1-\xi^2)},-{\bf \Delta}_T \right)
\end{eqnarray}
\end{subequations}
where the momentum fraction, $x$, in the symmetric frame defined by the reference vector, $P$, \cite{Belitsky:2005qn}, 
\[x= \frac{k^++k'^+}{P^+} .\]

%
The $\gamma^* q \rightarrow \gamma' q$ scattering amplitude $g_{\lambda\lambda^\prime}^{\Lambda_{\gamma^*} \Lambda_\gamma^\prime}$ reads,
\begin{eqnarray} 
g_{\lambda \lambda^\prime}^{\Lambda_{\gamma^*}  \Lambda_\gamma^\prime}&  =  &  \bar{u}(k^\prime,\lambda^\prime) \gamma^\mu \gamma^+ \gamma^\nu u(k,\lambda)    \left[\frac{ (\epsilon_\mu^{\Lambda_\gamma^\prime})^* \epsilon_\nu^{\Lambda_{\gamma^*}} }{\hat{s} - i \epsilon } + \frac{ (\epsilon_\nu^{\Lambda_\gamma^\prime})^*  \epsilon_\mu^{ \Lambda_{\gamma^*}} }{\hat{u} -i \epsilon} \right] \, q^- .
\label{ampg}
\end{eqnarray}
From parity conservation one obtains the following relations \cite{Diehl:2003ny},
\begin{equation}
A_{-\Lambda' -\lambda', -\Lambda -\lambda} = (-1)^{\Lambda' -\lambda'- \Lambda +\lambda} \left(A_{\Lambda' \lambda', \Lambda \lambda}\right)^* \quad\quad\quad g_{\lambda\lambda^\prime}^{\Lambda_{\gamma^*} \Lambda_\gamma^\prime} = (-1)^{\lambda -\lambda^\prime + \Lambda_{\gamma^*} -\Lambda_\gamma^\prime} g_{-\lambda-\lambda^\prime}^{-\Lambda_{\gamma^*} -\Lambda_\gamma^\prime}.
\end{equation}
The allowed helicity combinations are dictated by parity conservation, namely, the only allowed, independent amplitudes are: $g^{11}_{++}$ in the $s$ channel, and $g^{11}_{--}$ in the $u$ channel. The allowed helicity and longitudinal spin configurations are summarized in Table \ref{helicity:tab}.
By using the following relations for the invariants,
\begin{eqnarray}
\label{eq:s_hat}
\hat{s} & = & (k+q)^2 \approx \frac{Q^2}{2\xi}(x-\xi), \quad\quad
\hat{u}  =  (k^\prime -q)^2 \approx \frac{Q^2}{2 \xi} (x+\xi), \quad\quad
q^-  \approx  (Pq)/P^+  =  \frac{Q^2}{2(2 \xi) P^+},
\end{eqnarray}
we obtain the following expressions,
\begin{subequations}
\label{eq:g11}
\begin{eqnarray}
g_{++}^{11} & = & \sqrt{x^2-\xi^2} \, \frac{1}{x-\xi + i\epsilon}
 \, \quad\quad {\rm s-channel} \\
g_{--}^{11} & = &  \sqrt{x^2-\xi^2} \, \frac{1}{x+\xi + i\epsilon} \, \quad\quad {\rm u-channel}  ,  
\end{eqnarray}
\end{subequations}
Notice that the parity transformation is carried out independently from the complex denominators in Eq.(\ref{eq:g11}) where the $i \epsilon$ term follows from the analytic properties of the amplitudes and is, therefore, not a feature of the helicity configurations.  Moreover, the  $\approx$ sign in Eq.(\ref{eq:s_hat}) signifies that target mass corrections of the same type as the ones appearing in deep inelastic scattering were disregarded (see {\it e.g.} Ref.\cite{Schienbein:2007gr} for a review).
The convolution in Eq.(\ref{facto}) for the leading order helicity amplitudes can then be written as,
\begin{eqnarray}
\label{flambda}
f_{\Lambda \Lambda'}^{1 1}& = &  g_{++}^{1  1}  
 \otimes A_{\Lambda^\prime +, \Lambda +} + g_{--}^{ 1 1}  
 \otimes A_{\Lambda^\prime -, \Lambda -} 
  =  \left[ \frac{1}{x- \xi + i \epsilon } 
 \otimes A_{\Lambda^\prime +, \Lambda +} +   \frac{1}{x + \xi + i \epsilon}  
 \otimes A_{\Lambda^\prime -, \Lambda -}  \right] 
\nonumber \\
\end{eqnarray}
where the normalization factor, $\sqrt{x^2 - \xi^2}$ is inserted in the quark-parton structures $A_{\Lambda^\prime \lambda', \Lambda \lambda}$. f$_{\Lambda \Lambda'}^{-1 -1}$ is obtained through the parity transformation in Eq.(\ref{eq:parity_f}).
By inserting the expressions for $A_{\Lambda^\prime +, \Lambda +}$ and $A_{\Lambda^\prime -, \Lambda -}$ from Eqs.(\ref{GPDeven}) 
 and using the parity relations for the $f$ amplitudes, 
one recovers the expressions for the photon-proton helicity amplitudes given in Eqs.(\ref{helampftw2}) in terms of Compton Form Factors.

\begin{table}[htp]
\centering
\begin{tabular}{|c|ccccccc|}
\hline 
 & $\;\; \Lambda_\gamma$ & $\lambda$ & $\;\; \Lambda_\gamma'$ & $\lambda'$ &  $\;\; S_z^{\gamma+q}$ & $\epsilon_i^{+ *} \epsilon_i^+$ & $\epsilon_i^{- *} \epsilon_i^{-}$ \\
 \hline 
s-channel & 1 & $\frac{1}{2}$ & 1 & $\frac{1}{2}$ & $-\frac{1}{2}$ & 1 & 0 \\
u-channel & -1 & $\frac{1}{2}$ & -1 & $\frac{1}{2}$ & $-\frac{1}{2}$ & 0 & 1\\
\hline
\end{tabular}
\caption{Helicity and longitudinal spin configurations at twist two. The polarization vectors indices are $i=1,2$.}
\label{helicity:tab}
\end{table}

At twist three a similar factorized form for $f_{\Lambda \Lambda'}^{\Lambda_\gamma \Lambda_\gamma^\prime}$ holds \cite{Anikin:2000em}.
It is important to specify precisely our definition of ``twist" which is given here by the order in $1/{P^+}$ at which the matrix elements corresponding to the field operators ${\cal O}_{\lambda' \lambda}(z)$ in the correlation function (see {\it e.g.} Eq.(\ref{tw2_operator})), contribute to the amplitude. The ${\cal O}(1/{P^+})$, twist three matrix elements for the chiral even operators are defined by the following operators, 
\begin{eqnarray}
\label{tw3_operator}
{\cal O}_{\pm \pm}(z) & = & \bar{\psi}\left(-\frac{z}{2}\right) (\gamma^1 \pm i \gamma^2)(1 \pm \gamma_5)  \psi\left(\frac{z}{2}\right). 
\end{eqnarray}
We adopt the same notation as in Ref.\cite{Jaffe:1997yz} to identify the composite quark-gluon fields in the helicity amplitudes we write an asterisk on  
the helicity label for the quark within the bad component; for instance, in $g^{01}_{-+^*}$ in the s-channel, the initial quark has helicity $\lambda=-1/2$, the final quark has also helicity $\lambda'=-1/2$, and the final gluon has helicity $\lambda_g=+1$, so that the total longitudinal spin is conserved counting the gluon as part of the final state.  

As a consequence, at twist three we obtain twice as many expressions for the quark-proton helicity amplitudes in terms of GPDs. 
For the proton helicity conserving terms one has,  
\begin{subequations}
\label{eq:GPDnonfliptw3}
\begin{eqnarray}
A_{++,+-^*}  &=& \sqrt{1-\xi^2} \frac{\Delta_{T}}{4 M}e^{i\phi}\left[\widetilde{H}_{2T} + (1-\xi) \frac{E_{2T} + \widetilde{E}_{2T}}{2} + \widetilde{H}_{2T}' + (1-\xi)\frac{E_{2T}'+  \widetilde{E}_{2T}'}{2} \right]\\
A_{++^*,+-}  & = & -\sqrt{1-\xi^2} \frac{\Delta_{T}}{4 M}e^{i\phi}\Big[\widetilde{H}_{2T}+ (1-\xi) \frac{E_{2T} + \widetilde{E}_{2T}}{2}- \widetilde{H}_{2T}' -  (1-\xi)\frac{E_{2T}'+  \widetilde{E}_{2T}'}{2}  \Big] \\
A_{-+,--^*} & = & \sqrt{1-\xi^2} \frac{\Delta_{T}}{4 M}e^{i\phi}\Big[\widetilde{H}_{2T}+ (1+\xi) \frac{E_{2T} - \widetilde{E}_{2T}}{2} +  \widetilde{H}_{2T}' +  (1+\xi)\frac{E_{2T}'-  \widetilde{E}_{2T}'}{2}\Big]\\
A_{-+^*,--} & = & -\sqrt{1-\xi^2} \frac{\Delta_T}{4 M} e^{i\phi}\Big[\widetilde{H}_{2T}+ (1+\xi) \frac{E_{2T} - \widetilde{E}_{2T}}{2} -  \widetilde{H}_{2T}' -  (1+\xi)\frac{E_{2T}'-  \widetilde{E}_{2T}'}{2} \Big]
\end{eqnarray}
\end{subequations}
while for the proton helicity flip terms the amplitudes are, 
\begin{subequations}
\label{eq:GPDfliptw3}
\begin{eqnarray}
A_{++,--^*}  & = & \frac{\sqrt{1-\xi^{2}}}{2} \Bigg[H_{2T} + \frac{t_0-t}{4M^2}\widetilde{H}_{2T} - \frac{\xi^{2}}{1-\xi^{2}}E_{2T}  +  \frac{\xi}{1-\xi^{2}}\widetilde{E}_{2T} + H_{2T}' + \frac{t_0-t}{4M^2} \widetilde{H}_{2T}' -  \frac{\xi^{2}}{1-\xi^{2}}E_{2T}' +  \frac{\xi}{1-\xi^{2}} \widetilde{E}_{2T}' 
\Bigg] \nonumber \\ \\
A_{++^*,--}  & = & -\frac{\sqrt{1-\xi^{2}}}{2} \Bigg[H_{2T} + \frac{t_0-t}{4M^2}\widetilde{H}_{2T} - \frac{\xi^{2}}{1-\xi^{2}}E_{2T}  +  \frac{\xi}{1-\xi^{2}}\widetilde{E}_{2T} - H_{2T}' - \frac{t_0-t}{4M^2} \widetilde{H}_{2T}' +  \frac{\xi^{2}}{1-\xi^{2}}E_{2T}' -  \frac{\xi}{1-\xi^{2}} \widetilde{E}_{2T}'
\Bigg] \nonumber \\ \\
A_{-+,+-^*} & = &  -\frac{\sqrt{1-\xi^{2}}}{2} \, \frac{t_0-t}{4M^{2}} \, e^{2i\phi}\Big(\widetilde{H}_{2T} + \widetilde{H}_{2T}' \Big)\\
A_{-+^*,+-} & = &\frac{\sqrt{1-\xi^{2}}}{2} \, \frac{t_0-t}{4M^{2}} \, e^{2i\phi}\Big(\widetilde{H}_{2T} - \widetilde{H}_{2T}' \Big)
\end{eqnarray}
\end{subequations}
The helicity amplitudes read 
\begin{subequations}
\begin{eqnarray}
f_{++}^{01}& = & g_{-^{*},+}^{0,+1}\otimes A_{++,+-^{*}} + g_{-,+^{*}}^{0,+1} \otimes A_{++^{*},+-}
  \\
f_{--}^{01}&=& g_{-^{*},+}^{0,+1}\otimes A_{-+,--^{*}} + g_{-,+^{*}}^{0,+1} \otimes A_{-+^{*},--}
\\
f_{+-} ^{0 1}&=& g_{-^{*},+}^{0,+1}\otimes A_{-+,+-^{*}} + g_{-,+^{*}}^{0,+1} \otimes A_{-+^{*},+-} 
  \\
f_{- +} ^{0 1}&=& g_{-^{*},+}^{0,+1}\otimes A_{++,--^{*}} +g_{-,+^{*}}^{0,+1} \otimes A_{++^{*},--}, 
\end{eqnarray}
\end{subequations}
where we used the allowed by parity conservation helicity values which in the $s$ channel are given  by,
\begin{equation}
\label{eq:gtw3s}
g_{\lambda,\lambda'}^{0,\Lambda_{\gamma}'}= -\frac{e^{2}}{\sqrt{2}}\frac{1}{x-\xi+i\epsilon}\Bigg(\frac{\sqrt{Q^{2}+\nu^{2}}-\nu}{2P^{+}Q} \Bigg)
\overline{u}(k+\frac{\Delta}{2},\lambda')\Bigg[\Lambda_{\gamma}'\gamma^{1} +i\Lambda_{\gamma}'\gamma^{2}\gamma_{5}+ i\gamma^{2}+\gamma^{1}\gamma_{5}  \Bigg]u(k-\frac{\Delta}{2},\lambda)
\end{equation}
and in the $u$ channel by, 
\begin{equation}
\label{eq:gtw3u}
g_{\lambda,\lambda'}^{0,\Lambda_{\gamma}'} = e^{2}\frac{1}{x+\xi+i\epsilon}\Bigg(\frac{\sqrt{Q^{2}+\nu^{2}}-\nu}{2P^{+}Q} \Bigg)\overline{u}(k+\frac{\Delta}{2},\lambda')\Bigg[-\Lambda_{\gamma}'\gamma^{1}+i\Lambda_{\gamma}'\gamma^{2}\gamma_{5}-i\gamma^{2}+\gamma^{1}\gamma^{5} \Bigg]u(k-\frac{\Delta}{2},\lambda)
\end{equation}

Notice that the parity relations for the twist three amplitudes, $f_{\Lambda,\Lambda'}^{\Lambda_{\gamma^{*}},\Lambda_{\gamma}'}$, are the same as for the two body scattering processes. 
%
The DVCS cross section does not allow us to directly disentangle specific GPDs by appropriately choosing the beam and target polarizations because these appear in the cross section embedded in bilinear expressions of the CFFs. 
Notwithstanding, as explained throughout this paper, each GPD/CFF or GPD linear combination can be identified with specific polarization observables. 

\subsection{Transverse Gluon Amplitudes}
\label{sec:transvgluons} 
Up to this point we have ignored the contributions to the DVCS cross sections from gluons, noting that the virtual and real photons do not interact directly with the gluon content of the nucleons. That interaction occurs through quark loops, that suppresses the amplitudes by order $\alpha_{EM}$ while still contributing at leading twist. However, for double photon helicity flip, the leading contribution to the DVCS cross sections are from gluon double helicity flip or gluon transversity. 
The analog of the CFFs that connect the gluon GPDs to the double helicity flip $\gamma^* \, + N \rightarrow \gamma \, + \, N$ are obtained from the gluon helicity double flip amplitudes convoluted with the sum over quark loops.

The double helicity flip structure functions, $F_{UU}^{\cos 2 \phi}$, $F_{UL}^{\sin 2 \phi}$, $F_{UT}^{\sin(\phi + \phi_S) },
F_{UT}^{\sin(3\phi - \phi_S) }$, involve gluon transversity GPDs. They are given as:

\begin{eqnarray}
F_{UU}^{\cos 2 \phi} & = &  -2\frac{\alpha_{S}}{2\pi}\sqrt{1-\xi^{2}}\frac{t_{0}-t}{4M^{2}} \,  \Re{\rm e} \Bigg[\sqrt{1-\xi^{2}}\Big(\widetilde{\mathcal{H}}_{T}^{g} + (1-\xi)\frac{\mathcal{E}_{T}^{g} + \widetilde{\mathcal{E}}_{T}^{g}}{2} \Big) \Big(\mathcal{H} + \widetilde{\mathcal{H}} - \frac{\xi^{2}}{1-\xi^{2}}(\mathcal{E} + \widetilde{\mathcal{E}} \Big)^{*} \nonumber \\&+& \sqrt{1-\xi^{2}}\Big(\widetilde{\mathcal{H}}_{T}^{g} + (1+\xi)\frac{\mathcal{E}_{T}^{g} - \widetilde{\mathcal{E}}_{T}^{g}}{2}  \Big)\Big(\mathcal{H} - \widetilde{\mathcal{H}} - \frac{\xi^{2}}{1-\xi^{2}}(\mathcal{E} + \widetilde{\mathcal{E}} \Big)^{*} \nonumber \\&+& 
\frac{\sqrt{t_{0}-t}}{2M}\Big(\widetilde{\mathcal{H}}_{T}^{g} + (1+\xi)\frac{\mathcal{E}_{T}^{g} - \widetilde{\mathcal{E}}_{T}^{g}}{2} \Big)\Big(\mathcal{E} + \xi \widetilde{\mathcal{E}}\Big)^{*} \nonumber \\&-& \sqrt{1-\xi^{2}}\Big(\mathcal{H}_{T}^{g} + \frac{t_{0}-t}{M^{2}}\widetilde{\mathcal{H}}_{T}^{g} - \frac{\xi^{2}}{1-\xi^{2}}\mathcal{E}_{T}^{g} + \frac{\xi}{1-\xi^{2}}\widetilde{\mathcal{E}}_{T}^{g} \Big)\Big(\mathcal{E} - \xi \widetilde{\mathcal{E}} \Big)^{*}\Bigg] \\
F_{UL}^{\sin 2 \phi} & =& -2\frac{\alpha_{S}}{2\pi}\sqrt{1-\xi^{2}}\frac{t_{0}-t}{4M^{2}} \,  \Im{\rm m} \Bigg[\sqrt{1-\xi^{2}}\Big(\widetilde{\mathcal{H}}_{T}^{g} + (1-\xi)\frac{\mathcal{E}_{T}^{g} + \widetilde{\mathcal{E}}_{T}^{g}}{2} \Big) \Big(\mathcal{H} + \widetilde{\mathcal{H}} - \frac{\xi^{2}}{1-\xi^{2}}(\mathcal{E} + \widetilde{\mathcal{E}} \Big)^{*} \nonumber \\&+& \sqrt{1-\xi^{2}}\Big(\widetilde{\mathcal{H}}_{T}^{g} + (1+\xi)\frac{\mathcal{E}_{T}^{g} - \widetilde{\mathcal{E}}_{T}^{g}}{2}  \Big)\Big(\mathcal{H} - \widetilde{\mathcal{H}} - \frac{\xi^{2}}{1-\xi^{2}}(\mathcal{E} + \widetilde{\mathcal{E}} \Big)^{*} \nonumber \\&+& 
\frac{\sqrt{t_{0}-t}}{2M}\Big(\widetilde{\mathcal{H}}_{T}^{g} + (1+\xi)\frac{\mathcal{E}_{T}^{g} - \widetilde{\mathcal{E}}_{T}^{g}}{2} \Big)\Big(\mathcal{E} + \xi \widetilde{\mathcal{E}}\Big)^{*} \nonumber \\&-& \sqrt{1-\xi^{2}}\Big(\mathcal{H}_{T}^{g} + \frac{t_{0}-t}{M^{2}}\widetilde{\mathcal{H}}_{T}^{g} - \frac{\xi^{2}}{1-\xi^{2}}\mathcal{E}_{T}^{g} + \frac{\xi}{1-\xi^{2}}\widetilde{\mathcal{E}}_{T}^{g} \Big)\Big(\mathcal{E} - \xi \widetilde{\mathcal{E}} \Big)^{*}\Bigg]\\
F_{UT}^{\sin(3\phi - \phi_S) } & = &  -4\frac{\alpha_{S}}{2\pi}\sqrt{1-\xi^{2}}\frac{\sqrt{t_{0}-t}^{3}}{8M^{3}}\Im m\Bigg[(1-\xi^{2})\Big(\widetilde{\mathcal{H}}_{T}^{g} \Big)\Big(\mathcal{H} - \widetilde{\mathcal{H}} - \frac{\xi^{2}}{1-\xi^{2}}(\mathcal{E} - \widetilde{\mathcal{E}})\Big)^{*} \nonumber \\ &+& \Big(\widetilde{\mathcal{H}}_{T}^{g} + (1-\xi)\frac{\mathcal{E}_{T}^{g} + \widetilde{\mathcal{E}}_{T}^{g}}{2} \Big) \Big(\mathcal{E} - \xi \widetilde{\mathcal{E}} \Big)^{*} \Bigg]
\end{eqnarray}

They correspond to the following helicity structures,
\begin{subequations}
\begin{eqnarray}
F_{UU}^{\cos 2 \phi} & = &  2 \,  \Re{\rm e}   \left(\widetilde{F}_{++}^{1-1}+\widetilde{F}_{+-}^{1-1} \right)\\
F_{UL}^{\sin 2 \phi} & = &2 \,  \Im{\rm m}   \left(\widetilde{F}_{++}^{1-1}+\widetilde{F}_{+-}^{1-1} \right) \\
F_{UT}^{\sin(\phi + \phi_S) } & = & 2 \Im {\rm m} \{\widetilde{F}^{+1 -1}_{T, + +} +\widetilde{F}^{+1 -1}_{T, + -} \} =0\\
F_{UT}^{\sin(3\phi - \phi_S) } & = &  2 \Im {\rm m} \{\widetilde{F}^{-1 +1}_{T, + +} +\widetilde{F}^{-1 +1}_{T, + -} \} = 2\Im {\rm m} \{\widetilde{F}_{T,--}^{1-1} - \widetilde{F}_{T,-+}^{1-1} \}
\end{eqnarray}
\end{subequations}

We write the double flip helicity structure functions in terms of the gluon transversity GPDs:

\begin{subequations}
\begin{eqnarray}
F_{++}^{1-1} &=& -\frac{\alpha_{s}}{2\pi}(1-\xi^{2})\frac{t_{0}-t}{4M^{2}}e^{2i\phi}\Bigg[\Big(\widetilde{\mathcal{H}}_{T}^{g} + (1-\xi)\frac{\mathcal{E}_{T}^{g} + \widetilde{\mathcal{E}}_{T}^{g}}{2} \Big) \Big(\mathcal{H} + \widetilde{\mathcal{H}} - \frac{\xi^{2}}{1-\xi^{2}}(\mathcal{E} + \widetilde{\mathcal{E}} \Big)^{*} \nonumber \\&+& \Big(\widetilde{\mathcal{H}}_{T}^{g} + (1+\xi)\frac{\mathcal{E}_{T}^{g} - \widetilde{\mathcal{E}}_{T}^{g}}{2}  \Big)\Big(\mathcal{H} - \widetilde{\mathcal{H}} - \frac{\xi^{2}}{1-\xi^{2}}(\mathcal{E} + \widetilde{\mathcal{E}} \Big)^{*} \Bigg]\\
F_{+-}^{1-1} &=& -\frac{\alpha_{s}}{2\pi}\sqrt{1-\xi^{2}}\frac{t_{0}-t}{4M^{2}}e^{2i\phi}\Bigg[\frac{\sqrt{t_{0}-t}}{2M}\Big(\widetilde{\mathcal{H}}_{T}^{g} + (1+\xi)\frac{\mathcal{E}_{T}^{g} - \widetilde{\mathcal{E}}_{T}^{g}}{2} \Big)\Big(\mathcal{E} + \xi \widetilde{\mathcal{E}}\Big)^{*} \nonumber \\&-& \sqrt{1-\xi^{2}}\Big(\mathcal{H}_{T}^{g} + \frac{t_{0}-t}{M^{2}}\widetilde{\mathcal{H}}_{T}^{g} - \frac{\xi^{2}}{1-\xi^{2}}\mathcal{E}_{T}^{g} + \frac{\xi}{1-\xi^{2}}\widetilde{\mathcal{E}}_{T}^{g} \Big)\Big(\mathcal{E} - \xi \widetilde{\mathcal{E}} \Big)^{*} \Bigg]
\end{eqnarray}
\end{subequations}
for the longitudinal/unpolarized target and,
\begin{subequations}
\begin{eqnarray}
F_{T,++}^{1-1} &=& -\frac{\alpha_{s}}{2\pi}\sqrt{1-\xi^{2}}\frac{\sqrt{t_{0}-t}}{2M}e^{i\phi}\Bigg[(1-\xi^{2})\Big(\mathcal{H}_{T}^{g} + \frac{t_{0}-t}{M^{2}}\widetilde{\mathcal{H}}_{T}^{g} - \frac{\xi^{2}}{1-\xi^{2}}\mathcal{E}_{T}^{g} + \frac{\xi}{1-\xi^{2}}\widetilde{\mathcal{E}}_{T}^{g} \Big) \Big(\mathcal{H} + \widetilde{\mathcal{H}} - \frac{\xi^{2}}{1-\xi^{2}}(\mathcal{E} + \widetilde{\mathcal{E}}) \Big)^{*} \nonumber\\&-& \frac{t_{0}-t}{4M^{2}}\Big(\widetilde{\mathcal{H}}_{T}^{g} + (1+\xi)\frac{\mathcal{E}_{T}^{g} - \widetilde{\mathcal{E}}_{T}^{g}}{2} \Big) \Big(\mathcal{E} + \xi \widetilde{\mathcal{E}} \Big)^{*}  \Bigg] \\
F_{T, +-}^{1-1} &=& \frac{\alpha_{s}}{2\pi}\sqrt{1-\xi^{2}}\frac{\sqrt{t_{0}-t}}{2M}e^{i\phi}\Bigg[(1-\xi^{2})\Big(\mathcal{H}_{T}^{g} + \frac{t_{0}-t}{M^{2}}\widetilde{\mathcal{H}}_{T}^{g} - \frac{\xi^{2}}{1-\xi^{2}}\mathcal{E}_{T}^{g} + \frac{\xi}{1-\xi^{2}}\widetilde{\mathcal{E}}_{T}^{g} \Big) \Big(\mathcal{H} + \widetilde{\mathcal{H}} - \frac{\xi^{2}}{1-\xi^{2}}(\mathcal{E} + \widetilde{\mathcal{E}}) \Big)^{*} \nonumber\\&-& \frac{t_{0}-t}{4M^{2}}\Big(\widetilde{\mathcal{H}}_{T}^{g} + (1+\xi)\frac{\mathcal{E}_{T}^{g} - \widetilde{\mathcal{E}}_{T}^{g}}{2} \Big) \Big(\mathcal{E} + \xi \widetilde{\mathcal{E}} \Big)^{*}  \Bigg] \\
F_{T,--}^{1-1} &=& -\frac{\alpha_{s}}{2\pi}\sqrt{1-\xi^{2}}\frac{\sqrt{t_{0}-t}^{3}}{8M^{3}}e^{3i\phi}\Bigg[(1-\xi^{2})\Big(\widetilde{\mathcal{H}}_{T}^{g} \Big)\Big(\mathcal{H} - \widetilde{\mathcal{H}} - \frac{\xi^{2}}{1-\xi^{2}}(\mathcal{E} - \widetilde{\mathcal{E}})\Big)^{*} \nonumber \\ &+& \Big(\widetilde{\mathcal{H}}_{T}^{g} + (1-\xi)\frac{\mathcal{E}_{T}^{g} + \widetilde{\mathcal{E}}_{T}^{g}}{2} \Big) \Big(\mathcal{E} - \xi \widetilde{\mathcal{E}} \Big)^{*} \Bigg]\\
F_{T,-+}^{1-1} &=&\frac{\alpha_{s}}{2\pi}\sqrt{1-\xi^{2}}\frac{\sqrt{t_{0}-t}^{3}}{8M^{3}}e^{3i\phi}\Bigg[(1-\xi^{2})\Big(\widetilde{\mathcal{H}}_{T}^{g} \Big)\Big(\mathcal{H} - \widetilde{\mathcal{H}} - \frac{\xi^{2}}{1-\xi^{2}}(\mathcal{E} - \widetilde{\mathcal{E}})\Big)^{*} \nonumber \\ &+& \Big(\widetilde{\mathcal{H}}_{T}^{g} + (1-\xi)\frac{\mathcal{E}_{T}^{g} + \widetilde{\mathcal{E}}_{T}^{g}}{2} \Big) \Big(\mathcal{E} - \xi \widetilde{\mathcal{E}} \Big)^{*} \Bigg]
\end{eqnarray}
\end{subequations}
for the transverse target polarization.

We use equations~\eqref{transga} to ~\eqref{transg2} to define the double flip helicity structure functions through their subsequent proton/photon amplitudes,
\begin{subequations}
\begin{eqnarray}
F_{++}^{1-1} & = & e^{i 2\phi}\left[\left( {\widetilde{f}}^{11}_{+ +} \right)^* {\widetilde{f}}^{-11}_{ + +} + \left( {\widetilde{f}}^{1-1}_{+ +} \right)^* {\widetilde{f}}^{-1-1}_{ + +}\right]\\
F_{+-}^{1-1} & = & e^{i 2\phi}\left[\left( {\widetilde{f}}^{11}_{+ -} \right)^* {\widetilde{f}}^{-11}_{ + -} + \left( {\widetilde{f}}^{1-1}_{+ -} \right)^* {\widetilde{f}}^{-1-1}_{ + -}\right]
\end{eqnarray}
\end{subequations}
for longitudinal target polarization and by,
\begin{subequations}
\begin{eqnarray}
F_{T,++}^{1-1} & = &  e^{i \phi}\left[\left( {\widetilde{f}}^{11}_{+ +} \right)^* {\widetilde{f}}^{-11}_{ - +} + \left( {\widetilde{f}}^{1-1}_{+ +} \right)^* {\widetilde{f}}^{-1-1}_{ - +}\right]\\
F_{T,+-}^{1-1} & = &  e^{i \phi}\left[\left( {\widetilde{f}}^{11}_{+ -} \right)^* {\widetilde{f}}^{-11}_{ - -} + \left( {\widetilde{f}}^{1-1}_{+ -} \right)^* {\widetilde{f}}^{-1-1}_{ - -}\right]   \\
F_{T,--}^{1-1} & = &  e^{i 3\phi}\left[\left( {\widetilde{f}}^{11}_{--} \right)^* {\widetilde{f}}^{-11}_{ +-} + \left( {\widetilde{f}}^{1-1}_{--} \right)^* {\widetilde{f}}^{-1-1}_{ +-}\right]\\
F_{T,-+}^{1-1} & = &  e^{i 3\phi}\left[\left( {\widetilde{f}}^{11}_{-+} \right)^* {\widetilde{f}}^{-11}_{ ++} + \left( {\widetilde{f}}^{1-1}_{-+} \right)^* {\widetilde{f}}^{-1-1}_{ ++}\right]  
\end{eqnarray}
\end{subequations}
for transverse target polarization.
The leading contributions in $\alpha_s$ have the simple form for the helicity amplitudes, in which the $\gamma$ helicities match the gluon helicities\cite{Diehl:2001pm}. The helicity amplitudes read,
\begin{equation}
f_{\Lambda \Lambda^\prime}^{g\, \Lambda_\gamma  -\Lambda_\gamma^\prime}(\xi,t) =-\frac{\alpha_s}{2\pi}  \int_{-1}^{+1} \frac{dx}{x} \left(\frac{1}{\xi -x -i\epsilon} - \frac{1}{\xi +x -i\epsilon} \right) \, A^g_{\Lambda^\prime \Lambda_g'=\Lambda_\gamma', \Lambda \Lambda_g = \Lambda_\gamma^*}(x,\xi,t) 
\label{CFFgluon}
\end{equation}
Writing explicitly the helicty values, one has,
\begin{subequations}
\begin{eqnarray}
f_{++}^{g\,-+} &=& -e^{2i\phi}\sqrt{1-\xi^2}\, \frac{t_0 -t}{4M^2} \frac{\alpha_s}{2\pi}  \Big(\widetilde{\mathcal{H}}_T^g+(1-\xi)\frac{\mathcal{E}_T^g+\widetilde{\mathcal{E}}_T^g}{2}\Big) \\
f_{--}^{g\, -+} &=& -e^{2i\phi}\sqrt{1-\xi^2}\, \frac{t_0 -t}{4M^2}\frac{\alpha_s}{2\pi}   \Big(\widetilde{\mathcal{H}}_T^g+(1+\xi)\frac{\mathcal{E}_T^g-\widetilde{\mathcal{E}}_T^g}{2}\Big) \\
f_{-+}^{g\,-+} &=& -e^{i\phi} (1-\xi^2) \frac{\sqrt{t_0-t}}{2M} \frac{\alpha_s}{2\pi} \Big( \mathcal{H}_T^g +\frac{t_0-t}{M^2} \widetilde{\mathcal{H}}_T^g - \frac{\xi^2}{1-\xi^2} \mathcal{E}_T^g +  \frac{\xi}{1-\xi^2} \,\widetilde{\mathcal{E}}_T^g\Big) \\
f_{+-}^{g\,-+} &=& - e^{3i\phi} (1-\xi^2) \frac{\sqrt{t_0-t}^3}{8M^3}\frac{\alpha_s}{2\pi}  \widetilde{\mathcal{H}}_T^g,
\label{gluonflip2}
\end{eqnarray}
\end{subequations}

The GPD content of the amplitudes is obtained through the following gluon-proton helicity amplitudes 
\begin{subequations}
\begin{eqnarray}
A_{++,+-} &=& e^{2i\phi}\sqrt{1-\xi^2}\, \frac{t_0 -t}{4M^2} \Big({\tilde H}_T^g+(1-\xi)\frac{E_T^g+\tilde{E}_T^g}{2}\Big) \\
A_{-+,--} &=& e^{2i\phi}\sqrt{1-\xi^2}\, \frac{t_0 -t}{4M^2} \Big({\tilde H}_T^g+(1+\xi)\frac{E_T^g-\tilde{E}_T^g}{2}\Big) \\
A_{++,--} &=& e^{i\phi} (1-\xi^2) \frac{\sqrt{t_0-t}}{2M} \Big( H_T^g +\frac{t_0-t}{M^2} \tilde{H}_T^g - \frac{\xi^2}{1-\xi^2} E_T^g +  \frac{\xi}{1-\xi^2} \,\tilde{E}_T^g\Big) \\
A_{-+,+-} &=& - e^{3i\phi} (1-\xi^2) \frac{\sqrt{t_0-t}^3}{8M^3}\tilde{H}_T^g,
\label{gluonflip}
\end{eqnarray}
\end{subequations}
Notice that a similar structure appears for the quark helicity flip amplitudes \cite{Diehl:2003ny}.
\footnote{The phases and signs are in agreement with Ref.~\cite{Diehl:2003ny}, wherein each helicity flip amplitude for quarks is multiplied here by the complex conjugate of that reference's overall factor $e^{+i\phi} \sqrt{1-\xi^2}\sqrt{t_0-t}/2M$.}


\section{Bethe Heitler Cross Section}
\label{sec:BH} 
Similarly to DVCS, the BH contribution to the cross section defined in Section \ref{sec:2} in terms of helicity amplitudes can also be cast in a form emphasizing the various beam and target polarization configurations.  In what follows we present a covariant form of the cross section. 

\subsection{General Structure}
The cross section reads,
\begin{eqnarray}
\label{eq:xs5foldbh}
\frac{d^5\sigma_{BH}}{d x_{Bj} d Q^2 d|t| d\phi d\phi_S }  = 
\Gamma \, \big| T_{BH} \big|^2 
 =\frac{\Gamma}{t} \Big\{F_{UU}^{BH} +  (2 \Lambda)(2h) F_{LL}^{BH} + (2 \Lambda_T) (2h) F_{LT}^{BH}  \Big\}.
\end{eqnarray}
where $\Gamma$ was defined in Eq.\eqref{eq:xs5foldgeneral}. Notice that we do not consider in the cross section the terms $F_{LU}^{BH}$ and $F_{UL}^{BH}$, $F_{UT}^{BH}$, where either the target or beam are polarized, since they involve a $Z^o$ exchange and they are therefore suppressed.

The helicity structure of the amplitude $T_{BH}$ defined in  Eq.(\ref{eq:BH1b}), is given by,
\begin{eqnarray}
T_{BH, \Lambda\Lambda'}^{h \Lambda_\gamma^\prime} & = &  \left[ B_{h \Lambda_\gamma^\prime}(k,  k^\prime, q^\prime) \right]^\nu \left[J_{\Lambda \Lambda^\prime}(p, p^\prime) \right]_\nu ,
\label{eq:BH0}
\end{eqnarray}
where the matrix element of the lepton part, $B_{h \,\Lambda_\gamma^\prime }^\nu$, is,
\begin{eqnarray}
B_{h \,\Lambda_\gamma^\prime }^\nu & = &  \frac{1}{\Delta^2} L_h^{\mu\,\nu} \epsilon_\mu^{*\Lambda_\gamma^\prime}(q^\prime)  \nonumber \\
& = & \frac{1}{\Delta^2}  {\bar u}(k^\prime, h) \left[ \gamma^\mu ({\slas k}^\prime + {\slas q} ^\prime) \gamma^\nu \frac{1}{(k^\prime + q^\prime)^2} +\gamma^\nu ({\slas k} - {\slas q} ^\prime) \gamma^\mu \frac{1}{(k- q^\prime)^2} \right] u(k, h) 
\epsilon_\mu^{*\Lambda_\gamma^\prime}(q^\prime) .
\label{BHamp}
\end{eqnarray}
The (massless) leptons conserve helicity 
\footnote{We use $\gamma^5 \, u(k,h) = 2\,h\, u(k,h)$, where $h$ = lepton helicity = $\pm \frac{1}{2}$, throughout.}
in the EM process and satisfy the Dirac equation (${\slas k} u(k) = 0$), with normalization $\bar{u}u=2m$. $L_h^{\mu\,\nu}$ satisfies the gauge conditions, 
\begin{eqnarray}
\label{lep_gaugeinv1}
L_h^{\mu\nu} q^\prime_\mu =  L_h^{\mu \nu} \Delta_\nu =0.
\end{eqnarray}
The nucleon matrix elements of the EM current operator is defined (using the Gordon Identity) as,
\begin{eqnarray}
\label{eq:elastic}
\left[J_{\Lambda \Lambda^\prime}\right]_\nu &=& 
 \overline{U}(p',\Lambda')\left[ \left(F_1+ F_2\right)\gamma_\nu 
  -\frac{(p+p')_\nu}{2M} F_2 \right]U(p,\Lambda),
\end{eqnarray}
$F_1$  and $F_2$ being the Dirac and Pauli form factors, ($F_1+F_2 = G_M$). Note that in Eq.\eqref{eq:elastic} we explicitly consider hadronic helicity states with $\Lambda=\pm1/2$.
%

\begin{figure}
\begin{center}
\includegraphics[width=12.cm]{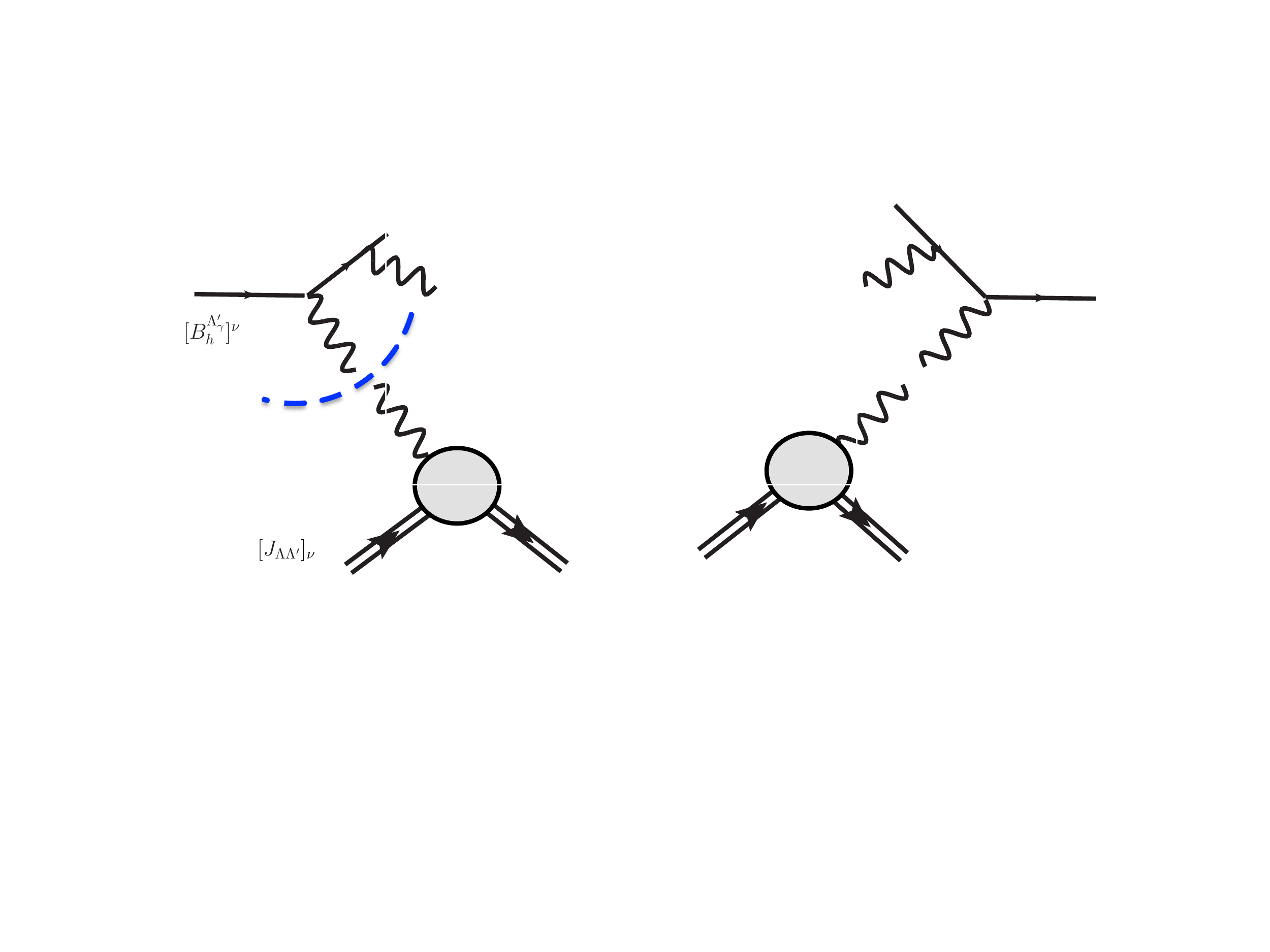}
\end{center}
\vspace{-3cm}
\caption{Factorization of the BH contribution to the cross section into its leptonic and hadronic components.}
\label{fig:BH}
\end{figure}

The BH matrix element modulus squared entering Eq.(\ref{eq:xs5foldbh}) can be written as the product of a leptonic tensor, 
\begin{equation}
\label{lepBH}
\mathcal{L}_{h \Lambda_{\gamma'}}
^{\nu \rho}= B_{h \,\Lambda_\gamma^\prime}^{\nu} B_{h \,\Lambda_\gamma^\prime}^{\rho  \, *},  
\end{equation}
and a hadronic tensor which reads,
\begin{equation}
\left[W^{BH}_{\Lambda \Lambda'}\right]_{\nu \rho} = \left[ J_{\Lambda\Lambda^\prime}\right]_\nu \, \left[ J_{\Lambda \Lambda^\prime}\right]_{\rho}^*  ,
\label{WBH} 
\end{equation}
when the target and recoil nucleon are either longitudinally polarized or unpolarized (when helicities are summed over), and  
\begin{eqnarray}
\left[W^{BH}_{\Lambda_T}\right]_{\nu \rho} &=& \sum_{\Lambda,\Lambda^\prime, \Lambda^{\prime \prime}}\rho_{\Lambda'' \, \Lambda}^{\vec{S}_\perp}\left[ J_{\Lambda\Lambda^\prime}\right]_\nu \, \left[ J_{\Lambda^{\prime \prime} \Lambda^\prime}\right]_{\rho}^* =S_T \sum_{\Lambda^\prime} \left[ J_{\Lambda_T \Lambda^\prime}\right]_\nu \, \left[ J_{\Lambda_T \Lambda^\prime}\right]_{\rho}^*, \\
&=& \Lambda_T
\left[e^{+i(\phi_s)}  \sum_{\Lambda^\prime} \left[ J_{- \Lambda^\prime}\right]_\nu \, \left[ J_{+ \Lambda^\prime}\right]_{\rho}^*  + e^{-i(\phi_s)}  \sum_{\Lambda^\prime } \left[ J_{+ \Lambda^\prime}\right]_\nu \, \left[ J_{- \Lambda^\prime}\right]_{\rho}^*, \right]
\label{WBHT}
\end{eqnarray}
for a transverse polarized target, written in terms of
the target spin density matrix, Eq.(\ref{densitymatrix}). 
The factorization of the BH cross section into its lepton and hadron/nucleon parts is sketched in Fig.\ref{fig:BH}.
Only two types of (parity conserving) contributions describe the BH process, which involve either an unpolarized or a polarized (longitudinally or transversely) target. As we show in what follows,  the unpolarized term is generated by multiplying the symmetric part of the lepton tensor with respect to the Lorentz indices, with the symmetric part of the proton current, whereas the polarized terms are given by the product of the respective antisymmetric components of the lepton and hadron tensors. The product of the antisymmetric contribution to the lepton/hadron tensors with the symmetric/antisymmetric tensor would yield the $UL, UT$ and $LU$ beam-target spin correlations. The latter correspond to parity violating terms and we do not consider them in this paper. For all the allowed polarization configurations the cross section will depend on only two distinct structure functions which are quadratic in the form factors, and which get multiplied by different kinematic coefficients. This form of the cross section is defined as ``Rosenbluth-type". Further details on the contraction of the lepton and hadron tensors are given below, in Sections \ref{sec:lepBH} and \ref{sec:hadBH}, respectively.

\subsubsection{Unpolarized Target: $F_{UU}^{BH}$}
The unpolarized amplitude squared can be computed evaluating the following contraction of the lepton and hadron tensors,
\begin{equation}
\mid T^{BH}_{unpol}  \mid^2 \,
= \, 
\frac{1}{2\,t^2} \left[W^{BH}_{unpol}\right]_{\nu \rho} \,  \sum_{h \Lambda_\gamma'} 
\mathcal{L}^{\nu \rho }_{h \Lambda_\gamma'}
\equiv \, \frac{1}{t} F_{UU}^{BH}
\label{TBH2_unpol}
\end{equation}
After performing the sum over the physical photon polarizations, and summing all contributions,
we obtain for the unpolarized structure function:
\begin{align}
F^{BH}_{UU}  \,= &\,
\frac{1}{t}\,\frac{ {8}\,M^2}{\myprod{k}{q'} \myprod{k'}{q'}}
\Bigg[
 2 \tau G_M^2
 \Big(
 \myprod{k}{\Delta}^2
 +
 \myprod{k'}{\Delta}^2
\Big)
 \nonumber\\& \hspace{50pt}
+ (F_1^2 + \tau F_2^2)
\Big( 
4 \tau 
 \big(
 \myprod{k}{P}^2
 +
 \myprod{k'}{P}^2
 \big)
 -(\tau +1)
\big(
 \myprod{k}{\Delta}^2
 +
 \myprod{k'}{\Delta}^2
\big)
\Big)
\Bigg]\, ,
\label{eq:BHunpolF}
\end{align}
where  $\tau=-t/(4M^2)$.
Notice the dimensions of the differential cross section is (nb/GeV$^4$) $\rightarrow$ GeV$^{-6}$ 
so that $F^{BH}_{UU}$ is dimensionless,
as required by the formulation in Eq.(\ref{eq:BH0}).

One can write the contribution to the cross section in Eq.(\ref{eq:xs5foldbh}) in the Rosenbluth type form,
\begin{equation}
\label{eq:BHunpol}
\frac{d^5\sigma^{BH}_{unpol}}{d x_{Bj} d Q^2 d|t| d\phi d\phi_S } \equiv \frac{\Gamma}{t} F_{UU}^{BH} 
=
\frac{\Gamma}{t} \left[ A(y,x_{Bj},t,Q^2,\phi) \left(F_1^2 + \tau F_2^2 \right)+ B(y,x_{Bj},t,Q^2,\phi) \tau G_M^2(t)  \right]
\end{equation}
with,
\begin{align}
A   = &\frac{{8}\,M^2}{t \myprod{k}{q'} \myprod{k'}{q'}}
\Bigg[  
4 \tau 
 \Big(
 \myprod{k}{P}^2
 +
 \myprod{k'}{P}^2
 \Big)
 -(\tau +1)
\Big(
 \myprod{k}{\Delta}^2
 +
 \myprod{k'}{\Delta}^2
\Big)
\Bigg] 
\label{eq:Aunpol}\\
B = & 
\frac{{16} \,M^2}{t \myprod{k}{q'} \myprod{k'}{q'}}
\Big[
 \myprod{k}{\Delta}^2
 +
 \myprod{k'}{\Delta}^2
\Big]  \,,
\label{eq:Bunpol}
\end{align}
where $\Gamma$ is the factor multiplying the amplitude squared in Eq.\eqref{eq:xs5foldgeneral}.
Eqs.\eqref{eq:BHunpol}-\eqref{eq:Bunpol} agree with the expression for the unpolarized BH cross section given in Ref.\cite{Kroll:1995pv}.

By writing all the covariant products appearing in the expressions in terms of $t$, $Q^2$, $x_{Bj}$, $y$, $(k\,\Delta)$, one has,
\begin{align}
\label{eq:inv1}
(k\,q^\prime)
= &
-\frac{Q^2}{2}- (k\,\Delta\,) = -\frac{Q^2}{2}\left(1 + \frac{2(k\,\Delta\,)}{Q^2} \right)
 \\
\label{eq:inv2}
(k^\prime\,q^\prime)
= &
\, \frac{t}{2}- (k\,\Delta) = \frac{Q^2}{2}\left(\frac{t}{Q^2} - \frac{2(k\,\Delta\,)}{Q^2} \right)
\\
\label{eq:inv3}
 \myprod{k}{\Delta}^2
 +
 \myprod{k'}{\Delta}^2
= &
\, \, Q^4\left[
\frac{1}{4}\left(1-\frac{t}{Q^2}\right)^2
+
\left(1-\frac{t}{Q^2}
+
\frac{2\,(k\,\Delta)}{Q^2}\right)\,
\frac{(k\,\Delta)}{Q^2}
\right]
\hspace{110pt}
\\
\label{eq:inv4}
\myprod{k}{P}^2
 +
 \myprod{k'}{P}^2
 =& \frac{Q^4}{4} \left\{\frac{1}{4}
 \Bigg[
 \frac{2-x_{Bj}}{x_{Bj}}
 +
\frac{t}{Q^2}
 \Bigg]\right.^2
 \nonumber \\
-
&
\left.
\Bigg[
\frac{2-x_{Bj}}{x_{Bj}}
+
\frac{t}{Q^2}
-
2\left(
\frac{1}{x_{Bj}\,y}
+
\frac{(k\,\Delta)}{Q^2}
\right)
\Bigg]
\left(
\frac{1}{x_{Bj}\,y}
+
\frac{(k\,\Delta)}{Q^2}
\right)
\right\}
\,.
\end{align}
The formulation above is useful for numerical studies since it enables us to disentangle the $Q^2$ dominant terms from the subdominant ones ({\it e.g.} $t/Q^2$), and the $Q^2$ independent ones ({\it e.g.}$(2-x_{Bj})/x_{Bj}$). 
The kinematic $\phi$ dependence in the target rest frame is entirely described by $(k \Delta)$. Notice that this 
invariant is of order $Q^2$, so that the term $(k\Delta)/Q^2$ is only mildly $Q^2$ dependent.

\subsubsection{Polarized Target: $F_{LL}^{BH}$ and $F_{LT}^{BH}$\label{subsec:BHpol}}
When the target is polarized either longitudinally or transversely,
one may write
\begin{equation}
\mid T^{BH}_{pol}  \mid^2 \,
=\, \frac{1}{t}  \sum_{\Lambda_\gamma'}\,
\left[W^{BH}_{pol}\right]_{\nu \rho} \, \mathcal{L}^{\nu \rho }_{h \Lambda_\gamma'}\nonumber \\
\equiv \, \frac{1}{t}  F_{LL}^{BH} \; {\rm or} \; \frac{1}{t}  F_{LT}^{BH}
\label{TBH2_pol}\,,
\end{equation}
After summing over polarizations of the final photon, and after several simplifications, the polarized BH structure functions read,
\begin{align}
{
F^{BH}_{LL(LT)} \,
}
=& \frac{2 h}{t^2} \, \frac{8\,M^2 }{ \myprod{k}{q'} \myprod{k'}{q'}}\nonumber\\
\times&\Bigg\{
\hspace{10pt}G_M^2
\Big[
\frac{\myprod{{p'}}{  S_{{L(T)}}  }}{M}
\Big(
\myprod{k'}{\Delta}^2
-\myprod{k}{\Delta}^2
\Big)
+
\frac{\myprod{k}{S_{{L(T)}}}}{M}
\Big(
\Delta^2\myprod{k}{\Delta}
\Big)
-
\frac{\myprod{k'}{S_{{L(T)}}}}{M}
\Big(
\Delta^2\myprod{k'}{\Delta}
\Big)
\Big] 
\nonumber \\ &\hspace{5pt}
-
F_2 \,G_M
\Big[
\frac{\myprod{{p'}}{S_{{L(T)}}}}{M}
\Big(
\big(
\myprod{k'}{\Delta}^2
-\myprod{k}{\Delta}^2
\big)
-
2\tau
\big(
\myprod{k'}{\Delta}\myprod{k'}{{p}}
-\myprod{k}{\Delta}\myprod{k}{{p}}
\big)
\Big)
\nonumber \\ &\hspace{70pt}
+
\frac{\myprod{k}{S_{{L(T)}}}}{M}
\Big(
 (1+\tau) \Delta^2\myprod{k}{\Delta}
\Big)
-
\frac{\myprod{k'}{S_{{L(T)}}}}{M}
\Big(
  (1+\tau ) \Delta^2\myprod{k'}{\Delta}
\Big)
\Big]
\hspace{10pt}
\Bigg\}\,,
\label{eq:BHpolF}
\end{align}
The $\Lambda${($\Lambda_T$)}-dependence is given through the spin four-vector{s $S_{L(T)}$}, defined in Section \ref{sec:3}.

The contribution to the cross section, Eq.(\ref{eq:xs5foldbh}) is,
\begin{equation}
\label{eq:BHpol}
\frac{d^5\sigma^{BH}_{pol}}{d x_{Bj} d Q^2 d|t| d\phi d\phi_S } \equiv \frac{\Gamma}{{t}} F_{{LL(LT)}}^{BH} 
=
\frac{\Gamma}{{t}} \left[\frac{1}{{t}} \widetilde{A}_{{L(T)}}(y,x_{Bj},t,Q^2,\phi) F_2 \, G_M + \frac{1}{{t}} \widetilde{B}_{{L(T)}}(y,x_{Bj},t,Q^2,\phi)  \, G_M^2  \right],
\end{equation}
where we defined,
\begin{eqnarray}
\widetilde{A}_{{L(T)}} &=& {-} \frac{(2 h) \, 8\,M^2 }{ \myprod{k}{q'} \myprod{k'}{q'}}\Big[
\frac{\myprod{{p'}}{S_L}}{M}
\Big(
\big(
\myprod{k'}{\Delta}^2
-\myprod{k}{\Delta}^2
\big)
-
2\tau
\big(
\myprod{k'}{\Delta}\myprod{k'}{{p}}
-\myprod{k}{\Delta}\myprod{k}{{p}}
\big)
\Big)
\nonumber \\ 
&& \hspace{20pt} + \, 
t \, \frac{\myprod{k}{S_L}}{M}
 (1+\tau) \myprod{k}{\Delta}
- t \, 
\frac{\myprod{k'}{S_L}}{M}
(1+\tau)  \myprod{k'}{\Delta} \Big]
\label{eq:Apol}
\\ 
\widetilde{B}_{{L(T)}} &=& \frac{(2 h) \, 8\,M^2 }{ \myprod{k}{q'} \myprod{k'}{q'}}\Big[
\frac{\myprod{{p'}}{S_L}}{M}
\Big(
\myprod{k'}{\Delta}^2
-\myprod{k}{\Delta}^2
\Big)
+ \, t \, 
\frac{\myprod{k}{S_L}}{M}
\myprod{k}{\Delta}
- \, t \, 
\frac{\myprod{k'}{S_L}}{M}
\myprod{k'}{\Delta}
\Big]
\label{eq:Bpol}
\end{eqnarray}
In addition to the  invariants in Eqs.~\eqref{eq:inv1} to \eqref{eq:inv4}, the following spin-independent 
four-vector products are needed to compute the polarized Bethe-Heitler cross section, 
\begin{align}
(k'\,\Delta)
=&
Q^2\left[
\frac{1}{2}\left(1-\frac{t}{Q^2}\right)+\frac{(k\,\Delta)}{Q^2}
\right]
\label{eq:inv5}
\\
\myprod{k'}{\Delta}^2
-\myprod{k}{\Delta}^2
=&
Q^4\left[
\frac{1}{4}\left(1-\frac{t}{Q^2}\right)
+
\frac{\,k\,\Delta}{Q^2}
\right]
\left(1-\frac{t}{Q^2}\right)
\label{eq:inv6}
\\
\myprod{k'}{\Delta}\myprod{k'}{p}
-\myprod{k}{\Delta}\myprod{k}{p}
=
&
\frac{Q^4}{2 x_{Bj}} 
\left(
\frac{1-y}{2\,y}
\left( 1-\frac{t}{Q^2}\right)
-\frac{\myprod{k}{\Delta}}{Q^2}
\right)
\label{eq:inv7}
\,.
\end{align}
Note that the formulation for $F^{BH}_{UU}$ and $F^{BH}_{LL(LT)}$ presented so far is given in terms of Lorentz-invariant quantities. In order
to perform explicit calculations, the invariants $(k \Delta)$, $(p'S)$, $(kS)$ and $(k'S)$ must be computed in a given reference frame.

{

To evaluate the Bethe Heitler polarized cross section, 
all that is needed is the invariants involving the spin fourvector $S_{L(T)}$.
Recalling the conventions in the target rest frame of Fig.~\ref{fig:kinematics}
\begin{align}
S^\alpha_L&=2\Lambda\,\,\,(0,0,0,1)\nonumber\\
S^\alpha_T&=2\Lambda_T(0,\cos\phi_S,\sin\phi_S,0)\,,\nonumber
\end{align}
}
one gets
\begin{align}
(p^\prime\,S_L) \,
=&
\frac{ (2\,\Lambda) \,M}{\sqrt{1+\gamma^2}}
\left[
x_{Bj}\left(1-\frac{t}{Q^2}\right)
-
\frac{t}{2\,M^2}
\right]
\\
(k\,S_L) \,
=&
\frac{(2\,\Lambda) \,Q^2}{\sqrt{1+\gamma^2}}
\left(1 + \frac{y \gamma^2}{2}\right)
\frac{1}{2 M x_{Bj} y}
\\
(k^\prime\,S_L) \,
=&
\frac{(2\,\Lambda) \,Q^2}{\sqrt{1+\gamma^2}}
\left(
1-y-\displaystyle\frac{y \gamma^2}{2}
\right)
\frac{1}{2 M x_{Bj} y}\,,
\end{align}
for the longitudinal polarization case, and
\begin{align}
({p'}\,S_T) \,
=&\,
2\,\Lambda_T
\frac{\left[-t (1-x_{Bj})\left(1+x_{Bj}\displaystyle\frac{t}{Q^2}\right) -M^2 x_{Bj}^2-\displaystyle\frac{M^2 x_{Bj}^2 t}{Q^2}\left(2+\displaystyle\frac{t}{Q^2}\right)\right]^{1/2}}{\sqrt{1+\gamma^2}}
{\cos(\phi-\phi_s)}
\\
(k\,S_T) \,
=&\,
(k'\,S_T) \,
=\,
{-}
\frac{2\,\Lambda_T\,Q}{\sqrt{1+\gamma^2}}
\,
\frac{1}{y}
\,
\left(1-y-\frac{y^2 \gamma^2}{4}\right)^{1/2}
{\cos(\phi_s)}\,,
\end{align}
for the transverse polarization case.
We remark that in our treatment $\Lambda(\Lambda_T )=\pm1/2$. 
In the following sections we describe in detail the structure of the lepton (Sec.\ref{sec:lepBH}) and hadron, (Sec.\ref{sec:hadBH}) tensors. 

\subsection{Structure of Lepton Tensor}
\label{sec:lepBH}
The structure of the lepton tensor is obtained by first writing the BH amplitude in Eq.(\ref{BHamp}) using the Dirac equation and the Lorentz condition, $q^\prime \cdot \epsilon(q^\prime) = 0$, along with the reduction of products of three Dirac matrices into single Dirac matrices and kinematic scalars, 
\begin{eqnarray}
B_{h,\,\Lambda_\gamma^\prime}^\nu&=& \frac{1}{\Delta^2}{\bar u}(k^\prime, h) \left[  (\epsilon_{\Lambda_\gamma^\prime}^{\, \, *} \cdot (k^\prime+k) ) \, \gamma^\nu \, D_+  +   (\epsilon_{\Lambda_\gamma^\prime}^* \cdot (k^\prime-k) ) \, \gamma^\nu \, D_-  \right.  \nonumber \\
& & \left. \quad \quad \quad \quad + (q^{\prime \, \nu} {\slas \epsilon}^{\, \prime \, *}  - \epsilon^{\,\prime \,* \,\nu}{\slas q}^{\, \prime } ) D_- +\,i\, 2 h\, \epsilon^{\alpha \beta \nu \rho} \epsilon^{\Lambda_\gamma' \, *}_\alpha \, q^\prime_\beta  \, \gamma_\rho \, D_+ \right] u(k,h)\,,
\label{BH3}
\end{eqnarray}
where, disregarding the electron mass, we have defined,
\begin{subequations}
\label{eq:Dplus}
\begin{eqnarray}
D_+ &=& \frac{1}{(k^\prime + q^\prime)^2} + \frac{1}{(k - q^\prime)^2} = \frac{1}{2(k'q')} - \frac{1}{2(kq')} = - \frac{1}{4(kq') (k'q')} 
\left(Q^2 +t  \right) \\
D_- &=& -\frac{1}{(k^\prime + q^\prime)^2} - \frac{1}{(k - q^\prime)^2} =- \frac{1}{2(k'q')} - \frac{1}{2(kq')} = \frac{1}{4(kq') (k'q')} 
\left(Q^2 - t + 4 (k \Delta) \right).
\end{eqnarray}
\end{subequations}
 We write the lepton amplitude in terms of three structures,
 \begin{equation}
B_{h \,\Lambda_\gamma^\prime}^{\nu} = B_{h \,\Lambda_\gamma^\prime}^{\nu \, (1)}  + B_{h \,\Lambda_\gamma^\prime}^{\nu \, (2)}  + B_{h \,\Lambda_\gamma^\prime}^{\nu \, (3)} ,
 \end{equation}
where
\begin{subequations}
\label{eq:Bdecomp}
\begin{align}
B_{h \,\Lambda_\gamma^\prime}^{\nu \, (1)}
=&  \, {\bar u}(k^\prime, h)  \gamma^\nu  u(k,h) \left\{ \left[  (\epsilon_{\Lambda_\gamma^\prime}^{\, \, *}  k^\prime) + (\epsilon_{\Lambda_\gamma^\prime}^{\, \, *} k) \right] D_+  + \left[ (\epsilon_{\Lambda_\gamma^\prime}^{\, \, *} k^\prime) - (\epsilon_{\Lambda_\gamma^\prime}^{\, \, *} k) \right]  D_-   \right\} 
\label{Bnew1} \\
B_{h \,\Lambda_\gamma^\prime}^{\nu (2)}
=&  {\bar u}(k^\prime, h)  \gamma^\mu u(k,h)  \left[ q^{\prime \, \nu}   \epsilon_{\Lambda_\gamma^\prime \,  \mu}^{\, \, *}  - \epsilon_{\Lambda_\gamma^\prime}^{\, \, * \nu} \,  q'_\mu  \right]  \,  D_-  
\label{Bnew2}\\
B_{h,\,\Lambda_\gamma^\prime}^{\nu (3)}
=&  \,i\, 2 h \, \, {\bar u}(k^\prime, h) \gamma_\sigma  u(k,h)  \,\,  \epsilon^{\alpha \beta \, \nu \, \sigma}  \epsilon^* _{\Lambda_\gamma^\prime, \alpha} \, q^\prime_\beta \, D_+ 
\label{Bnew3}
\end{align}
\end{subequations}
This decomposition is purely practical, as the number of independent helicity amplitudes for the BH process is 4 (once summed over $\Lambda_{\gamma'}$, see Ref.\cite{Gastmans}) while for our calculation we grouped the first two into one single term.
Following Eq.\eqref{eq:Bdecomp}, the BH lepton tensor, $\mathcal{L}_{h \Lambda_{\gamma'}}$, Eq.(\ref{lepBH}), is obtained as the sum of six terms displayed below,
\begin{equation}
\mathcal{L}_{h \Lambda_{\gamma'}}
^{\nu \rho}= B_{h \,\Lambda_\gamma^\prime}^{\nu} B_{h \,\Lambda_\gamma^\prime}^{\rho  \, *} = \sum_{a,b} \mathcal{L}_{h \Lambda_{\gamma'}}
^{\nu \rho (a,b)},  
\end{equation}
where $\mathcal{L}_{h \Lambda_\gamma'}^{\nu \rho (a,b)}$, $(a,b =1,2,3)$ read,
 \begin{align}
 \label{Lcal11}
t^2 \Lcal{1}{1}=&2 C C^*\Big[-{g}^{\nu \rho } \myprod{k}{k'}-2 i h \epsilon ^{{k}k'\nu \rho }+{k}^{\rho } k'{}^{\nu }+{k}^{\nu } k'{}^{\rho }\Big]  \\ 
 \label{Lcal12}
t^2 \Lcal{1}{2}=&2 C D_-\Big[{q'}^{\rho } k'{}^{\nu } \myprod{k}{\realeps{}{}}-{q'}^{\rho } {\realeps{}{}}^{\nu } \myprod{k}{k'}+{k}^{\nu } {q'}^{\rho } \myprod{k'}{\realeps{}{}}-{\realeps{}{}}^{\rho }
 k'{}^{\nu } \myprod{k}{q'}
 \nonumber \\ & \hspace{30pt} 
+{q'}^{\nu } {\realeps{}{}}^{\rho } \myprod{k}{k'}
-{k}^{\nu } {\realeps{}{}}^{\rho } \myprod{k'}{q'}-2 i h {q'}^{\rho } \epsilon ^{{k}k'\nu {\realeps{}{}}}+2 i h {\realeps{}
{}}^{\rho } \epsilon ^{ {k} k'\nu  {q'}}\Big]  \\ 
 \label{Lcal13}
t^2 \Lcal{1}{3}=&2 C D_+\Big[\myprod{k}{\realeps{}{}} \left({g}^{\nu \rho } \myprod{k'}{q'}-{q'}^{\nu } k'{}^{\rho }\right)+\myprod{k'}{\realeps{}{}} \left({k}^{\rho } {q'}^{\nu }-{g}^{\nu \rho } \myprod{k}
{q'}\right)
 \nonumber \\ & \hspace{30pt} 
+2 i h \left(\myprod{k}{k'} \epsilon ^{{q'}\nu \rho {\realeps{}{}}}+{k}^{\nu } \epsilon ^{{q'}\rho k'{\realeps{}{}}}+k'{}^{\nu } \epsilon ^{{q'}\rho {k}{\realeps{}{}}}\right)+{\realeps{}
{}}^{\nu } \left(k'{}^{\rho } \myprod{k}{q'}-{k}^{\rho } \myprod{k'}{q'}\right)
\Big]  \\ 
\label{Lcal22}
t^2 \Lcal{2}{2}=&2 D_-^2\Big[-{q'}^{\rho } {\realeps{*}{}}^{\nu } \myprod{k}{\realeps{}{}} \myprod{k'}{q'}+{q'}^{\rho } \myprod{k'}{\realeps{}{}} \left({q'}^{\nu } \myprod{k}{\realeps{*}{}}-{\realeps{*}{}
}^{\nu } \myprod{k}{q'}\right)
 \nonumber \\ & \hspace{30pt} 
+{q'}^{\nu } {q'}^{\rho } \myprod{k}{\realeps{}{}} \myprod{k'}{\realeps{*}{}}+2 {\realeps{}{}}^{\rho } {\realeps{*}{}}^{\nu } \myprod{k}{q'} \myprod{k'}{q'}-{q'}^{\nu } {\realeps{}
{}}^{\rho } \myprod{k}{\realeps{*}{}} \myprod{k'}{q'}
 \nonumber \\ & \hspace{30pt} 
-{q'}^{\nu } {\realeps{}{}}^{\rho } \myprod{k}{q'} \myprod{k'}{\realeps{*}{}}+{q'}^{\nu } {q'}^{\rho } \myprod{k}{k'}
 \nonumber \\ & \hspace{30pt} 
-2 i h {q'}^{\nu } {q'}
^{\rho } \epsilon ^{{k}k'{\realeps{*}{}}{\realeps{}{}}}+2 i h {q'}^{\rho } {\realeps{*}{}}^{\nu } \epsilon ^{{k}k'{q'}{\realeps{}{}}}-2 i h {q'}^{\nu } {\realeps{}{}}^{\rho } \epsilon ^{ {k}
 k' {q'} {\realeps{*}{}}}
 \Big]  \\ 
\label{Lcal23}
t^2 \Lcal{2}{3}=&2 D_- D_+\Big[2 i h \left({q'}^{\nu } \myprod{k}{\realeps{*}{}}-{\realeps{*}{}}^{\nu } \myprod{k}{q'}\right) \epsilon ^{\rho k'{q'}{\realeps{}{}}}+2 i h {q'}^{\nu } \myprod{k}{k'} \epsilon ^{\rho {q'}
{\realeps{*}{}}{\realeps{}{}}}-2 i h {\realeps{*}{}}^{\nu } \myprod{k'}{q'} \epsilon ^{\rho {k}{q'}{\realeps{}{}}}
 \nonumber \\ & \hspace{30pt} 
+2 i h {q'}^{\nu } \myprod{k'}{\realeps{*}{}} \epsilon ^{\rho {k}{q'}
{\realeps{}{}}}
+\myprod{k}{q'} \left({q'}^{\rho } {\realeps{*}{}}^{\nu }-{q'}^{\nu } {\realeps{*}{}}^{\rho }\right) \myprod{k'}{\realeps{}{}}-{q'}^{\rho } {\realeps{*}{}}^{\nu } \myprod{k}{\realeps{}{}
} \myprod{k'}{q'}
 \nonumber \\ & \hspace{30pt} 
+{q'}^{\nu } {\realeps{*}{}}^{\rho } \myprod{k}{\realeps{}{}} \myprod{k'}{q'}-{q'}^{\nu } k'{}^{\rho } \myprod{k}{q'}+{k}^{\rho } {q'}^{\nu } \myprod{k'}{q'}
\Big]  \\ 
\label{Lcal33}
t^2 \Lcal{3}{3}=&2 D_+^2\Big[\myprod{k'}{q'} \left(-2 \myprod{k}{q'} \left({g}^{\nu \rho }+{\realeps{}{}}^{\nu } {\realeps{*}{}}^{\rho }\right)+\myprod{k}{\realeps{*}{}} \left({q'}^{\rho } {\realeps{}{}}
^{\nu }-2 i h \epsilon ^{\nu \rho {q'}{\realeps{}{}}}\right)+{k}^{\nu } {q'}^{\rho }\right)
 \nonumber \\ & \hspace{30pt} 
+\myprod{k}{q'} \myprod{k'}{\realeps{*}{}} \left({q'}^{\rho } {\realeps{}{}}^{\nu }+2 i h \epsilon ^{\nu \rho {q'}
{\realeps{}{}}}\right)+2 i h \epsilon ^{\rho {q'}{\realeps{*}{}}{\realeps{}{}}} \left( {k}^{\nu } \myprod{k'}{q'}-k'{}^{\nu } \myprod{k}{q'}\right)
 \nonumber \\ & \hspace{30pt} 
+{q'}^{\nu } \Bigg(\myprod{k'}{q'} \left({\realeps{*}
{}}^{\rho } \myprod{k}{\realeps{}{}}+{k}^{\rho }\right)+\myprod{k}{q'} \left({\realeps{*}{}}^{\rho } \myprod{k'}{\realeps{}{}}+k'{}^{\rho }\right)
 \nonumber \\ & \hspace{50pt} 
-{q'}^{\rho } \left(\myprod{k}{\realeps{*}{}
} \myprod{k'}{\realeps{}{}}+\myprod{k}{\realeps{}{}} \myprod{k'}{\realeps{*}{}}+\myprod{k}{k'}\right)\Bigg)
 \nonumber \\ & \hspace{30pt} 
+{q'}^{\rho } k'{}^{\nu } \myprod{k}{q'}\Big]  
\end{align}
To obtain the expressions above we used
$u(k,h) \bar{u}(k,h)=\not\!{k} (1+ h\gamma_5)/2\,,
\,\,\bar{u} u = 2m \rightarrow 0\,
$
and defined
\begin{eqnarray}
\label{eq:C}
C & = &  \left[  (\epsilon_{\Lambda_\gamma^\prime}^{\, \, *}  k^\prime) + (\epsilon_{\Lambda_\gamma^\prime}^{\, \, *} k) \right] D_+  + \left[ (\epsilon_{\Lambda_\gamma^\prime}^{\, \, *} k^\prime) - (\epsilon_{\Lambda_\gamma^\prime}^{\, \, *} k) \right]  D_-  \approx  \frac{(\epsilon_{\Lambda_\gamma^\prime}^{\, \, *}  k^\prime)}{(k'q')} -  \frac{(\epsilon_{\Lambda_\gamma^\prime}^{\, \, *} k)}{(kq')} .
\end{eqnarray}
Summing over the photon polarization we find,
\begin{align}
 {\cal L}^{\nu \rho (1,1)}_h=&2 \Big[\frac{k'^2}{\myprod{k'}{q'}^2}+\frac{k^2}{\myprod{k}{q'}^2}-\frac{2\myprod{k}{k'}}{\myprod{k}{q'}\myprod{k'}{q'}}
 \Big]\Big[-{g}^{\nu \rho } \myprod{k}{k'}-2 i h \epsilon ^{{k}k'\nu \rho }+{k}^{\rho } k'{}^{\nu }+{k}^{\nu } k'{}^{\rho }\Big]  \\ 
{\cal L}_h^{\nu \rho (1,2)}= &2 D_-\Big[\frac{{q'}^{\rho } k{}^{\nu } k'^2-{k'}^{\rho } {k'}^{\nu } \myprod{k}{q'}-{k}^{\nu } {k'}^{\rho } \myprod{k'}{q'}+2ih k'^{\rho} \epsilon^{kk'\nu q'}}{\myprod{k'}{q'}}
 \nonumber \\ & \hspace{30pt} 
+\frac{k'^\nu k^\rho \myprod{k}{q'}-q'^\rho k'^\nu k^2-q'^\nu k^\rho \myprod{k}{k'}-2ih{}k^\rho \epsilon^{kk'\nu q'} }{\myprod{k}{q'}}\Big]  \\ 
{\cal  L}_h^{\nu \rho (1,3)}=&2D_+\Big[\frac{1}{\myprod{k'}{q'}}\Big( \myprod{k}{k'} \left({g}^{\nu \rho } \myprod{k'}{q'}-{q'}^{\nu } k'{}^{\rho }\right)+k'^2 \left({k}^{\rho } {q'}^{\nu }-{g}^{\nu \rho } \myprod{k}{q'}\right)+k'^\nu k'^\rho \myprod{q'}{k}
 \nonumber \\ & \hspace{30pt} 
-k'^\nu k^\rho \myprod{k'}{q'}+2ih{}(\epsilon^{q'\nu \rho k'}+k'^\nu \epsilon ^{q' \rho kk'})\Big)+\frac{1}{\myprod{k}{q'}}\Big( \myprod{k}{k'} \left({g}^{\nu \rho } \myprod{k}{q'}-{q'}^{\nu } k{}^{\rho }\right)
 \nonumber \\ & \hspace{30pt} 
+k^2 \left({k'}^{\rho } {q'}^{\nu }-{g}^{\nu \rho } \myprod{k}{q'}\right)+k^\nu k'^\rho \myprod{q'}{k}-k^\nu k^\rho \myprod{k'}{q'}+2ih{}(\epsilon^{q'\nu \rho k}+k^\nu \epsilon ^{q' \rho k'k}) \Big)\Big] \\
{\cal L}_h^{\nu\rho(2,3)}=&2 D_- D_+\Big[\myprod{k'}{q'}\Big(2{q'}^{\nu } {k}^{\rho }-{q'}^{\rho } {k}^{\nu } \Big)+\myprod{k}{q'}\Big({q'}^{\rho }{k'}^{\nu}-2{q'}^{\nu }{k'}^{\rho}\Big)
 \nonumber \\ & \hspace{30pt} 
+2 i h \myprod{k}{q'} \epsilon ^{{\nu}{\rho}{k'}{q'}}+2 i h \myprod{k'}{q'}  \epsilon ^{ {\nu}
 {\rho} {k}{q'}} 
\Big]  \\
{\cal L}_h^{\nu\rho(2,2)}= &2D_-^2 \Big[-\myprod{k'}{q'}\Big({q'}^{\rho } {k}^{\nu }+{q'}^{\nu } {k}^{\rho } \Big)-\myprod{k}{q'}\Big({q'}^{\rho }{k'}^{\nu}+{q'}^{\nu }{k'}^{\rho}\Big) 
 \nonumber \\ & \hspace{30pt} 
+2{ }{g}^{\rho \nu }\myprod{k}{q'} \myprod{k'}{q'}+3{} {q'}^{\nu } {q'}^{\rho } \myprod{k}{k'}
 \nonumber \\ & \hspace{30pt} 
+2 i h {q'}^{\rho } \epsilon ^{{k}k'{q'}{\nu}}-2 i h {q'}^{\nu }  \epsilon ^{ {k}
 {k'} {q'}\rho} \Big] \\
{\cal  L}_h^{\nu\rho(3,3)}=&2 D_+^2\Big[2\myprod{k'}{q'} \Big({k}^{\nu}{q'}^{\rho}+{k}^{\rho}{q'}^{\nu}\Big)+2\myprod{k}{q'} \Big({k}^{\nu}{q'}^{\rho}+{k}^{\rho}{q'}^{\nu}\Big)-4{g}^{\nu \rho}\myprod{k'}{q'}\myprod{k}{q'}-3{} {q'}^{\nu } {q'}^{\rho } \myprod{k}{k'}
 \nonumber \\ & \hspace{30pt} 
+2 i h  \epsilon ^{{\nu}{\rho}{q'}{k}}-2 i h \epsilon ^{ {\nu}
 {\rho} {q'}{k'}} \Big]
 \end{align}
We notice that in order to derive the final expression for the cross section, one needs the following relations among the antisymmetric symbols, 
\begin{eqnarray}
&&\epsilon^{\alpha \beta \gamma \delta} \epsilon_{\alpha \beta \gamma \delta} = -4!,  \;\;\; 
\epsilon^{\mu \beta \gamma \delta} \epsilon_{\alpha \beta \gamma \delta} = -3! \, \delta_\alpha^\mu,  \;\;\;
\epsilon^{\mu \nu \gamma \delta} \epsilon_{\alpha \beta \gamma \delta} = -2! \delta^{\mu \nu}_{\alpha \beta} 
= -2 (\delta_\alpha^\mu \delta_\beta^\nu - \delta_\alpha^\nu \delta_\beta^\mu), \;\;\; \nonumber \\
&& \epsilon^{\mu \nu \zeta \delta} \epsilon_{\alpha \beta \gamma \delta} = - \delta^{\mu \nu \zeta}_{\alpha \beta \gamma} 
= - \delta^\mu_\alpha \delta^{\nu \zeta}_{\beta \gamma} - \delta^\nu_\alpha \delta^{\zeta \mu}_{\beta \gamma} - \delta^\zeta_\alpha \delta^{\mu \nu}_{\beta \gamma},
\label{epsilons}
\end{eqnarray}

\subsection{Structure of Hadron Tensor}
\label{sec:hadBH}
The unpolarized and polarized hadronic tensors for BH are obtained from the following helicity combinations of Eq.(\ref{WBH}),
\begin{align}
\left[W^{BH}_{unpol}\right]^{\nu \rho}  
=&
\,\frac{1}{2}\sum_{\Lambda \Lambda\prime}\,\,\,\,\big[W^{BH}_{\Lambda \Lambda'}\big]^{\nu \rho} 
%
\\
\left[W^{BH}_{pol}\right]^{\nu \rho}  
=&
\,\frac{1}{2}\sum_{\Lambda\prime} \big[W^{BH}_{\Lambda \Lambda'}\big]^{\nu \rho}
-
\big[W^{BH}_{-\Lambda \Lambda'}\big]^{\nu \rho}
\label{BHWtensors}
\end{align}
Working in helicity basis we write the polarization combinations for the product of currents in Eq.(\ref{WBH}), $\left[ J_{\Lambda''\Lambda^\prime}\right]_\nu \, \left[ J_{\Lambda \Lambda^\prime}\right]_{\rho}^* $ as,
\begin{align}
\left[W^{BH}_{unpol}\right]^{\nu \rho} 
+&
\left[W^{BH}_{pol}\right]^{\nu \rho} 
=
\nonumber \\
\frac{1}{2}
&{\text Tr}\left\{\Big({\slas p'} +M\Big)\left(G_M\gamma^\nu-F_2\frac{{P}^\nu}{M}\right)\Big(1+\gamma_5 {\slas S}_L\Big)\Big({\slas p}+M\Big)\left(G_M\gamma^\rho-F_2\frac{{P}^\rho}{M}\right)\right\}\,,
\label{WBHcovariant}
\end{align}
Notice that the longitudinal polarized proton case is obtained by inserting $(\gamma^5 \slas{S})_{\Lambda \Lambda''}$.

The unpolarized and polarized hadronic tensors correspond respectivelly to the symmetric and antisymmetric parts of the RHS of Eq.\eqref{WBHcovariant}. 
From Eq.\eqref{WBHcovariant} the symmetric and antisymmetric parts can be readily extracted \cite{Sofiatti:2011yi},
\begin{align}
\left[W^{BH}_{unpol}\right]^{\nu \rho}  =&
\,\,\,4 M^2 \left[ - \tau G_M^{\,2} \left(g^{\nu \rho} - \frac{\Delta^\nu \Delta^\rho}{\Delta^2}\right) +
\frac{(F_1^2+ \tau F_2^2)}{M^2} {P}^\nu {P}^\rho
\right],
\label{WBH_unpol} \\
\left[W^{BH}_{pol}\right]^{\nu \rho} =& -i\, 2 M
\Bigg[  G_M^{\,2}\, \epsilon^{\nu\rho\alpha'\beta'} S_L{\,}_{\alpha'} \Delta_{\beta'}   
  +   
\frac{F_2\, G_M}{\,M^2} \left( \epsilon^{\nu\alpha'\beta'\gamma'} {P}^\rho - {P}^\nu \epsilon^{\rho\alpha'\beta'\gamma'}  \right) S_L{\,}_{\alpha'} {P}_{\beta'} \Delta_{\gamma'} 
\Bigg],
\label{WBH_pol} 
\end{align}
with $\tau = -t/4M^2$. We remark that our choice for the proton spinor normalizations, $\bar{U}U=2M$, brings extra factors of the proton mass in Eqs. \eqref{WBH_unpol} and \eqref{WBH_pol}, with respect to some of the expressions given in the literature ({\it e.g.} \cite{Sofiatti:2011yi}). 

For transverse target polarization, as above, one may simply replace $S_{L \,\alpha}\to S_{T \,\alpha}$. 
We remark that in our treatment $\Lambda(\Lambda_T )=\pm1/2$.
\footnote{The substitution of $S_L$ is equivalent to keeping helicity labels. 
 It can be seen from the density matrix Eq.(\ref{densitymatrix}) that for $S_{L \, (or \, T)} = \pm 1$, the target is totally polarized in the longitudinal (or transverse) direction with helicity (or transversity) $\Lambda=\pm 1/2$ (or $\Lambda_T =\pm 1/2$). }
The cross section contributions, Eqs.(\ref{eq:BHunpol},\ref{eq:BHpol}),  are obtained by contracting the lepton structures ${\cal L}^{\nu \rho (a,b)}_h$, $(a,b=1,2,3)$ with the corresponding hadronic tensor components,  $[W^{BH}_{unpol(pol)}]_{\nu \rho}$, and summing over the 6 different structures obtained for $(a,b)=11,12,13,22,23,33$,using the symmetry between $a$ and $b$. Detailed formulae for the intermediate calculation are given in Appendix \ref{app:BH}.

\section{BH-DVCS Interference}
\label{sec:BHDVCS} 
In this Section we present the interference term between the BH and DVCS helicity amplitudes appearing in Eq.(\ref{eq:xs5foldgeneral}). Its general structure reads,
\begin{equation}
\sum_{\Lambda_{\gamma^*}} \mathcal{I}_{\Lambda \Lambda'}^{h, \Lambda_{\gamma^*} \Lambda_{\gamma'}}  =  \left( T_{BH, \, \Lambda \Lambda'} ^{h \Lambda_{\gamma'} \, }\right)^* T_{DVCS, \, \Lambda \Lambda'}^{h \Lambda_{\gamma'}}
+ \left( T_{DVCS, \, \Lambda \Lambda'}^{h \Lambda_{\gamma'}} \right)^* T_{BH, \, \Lambda \Lambda'} ^{h \Lambda_{\gamma'} \, }
\label{eq:Isec5}
\end{equation}
where
$T^{h \Lambda_{\gamma'}}_{DVCS,\Lambda \Lambda'}$ and 
$T^{h \Lambda_{\gamma'}}_{BH,\Lambda \Lambda'}$ are defined, respectively, in Sections \ref{sec:3} and \ref{sec:BH}. 
We present a formulation of this term that allows us to follow as closely as possible the helicity formalism displayed in Section \ref{sec:3} and in  Refs.\cite{Bacchetta:2006tn,Sofiatti:2011yi}. The phase structure of the cross section is however, more elaborate, as we explain in detail below, since we are dealing with two different virtual photons, one with momentum $q$ for DVCS, and one with momentum $\Delta$, for  BH. Because the latter is tilted relatively to the $z$-axis, the kinematical coefficients for the interference contributions to the cross section are given by more complex expressions. 

\subsection{General Formalism}
\label{sec:Int_A}
In what follows,similarly to the pure DVCS, Eq.\eqref{eq:xs5fold}, and BH, Eq.(\ref{eq:BH0}) contributions, we write a master formula organized according to the beam and target polarization configurations, 
{
\begin{eqnarray}
\label{eq:int5fold} 
&& \frac{d^5\sigma_{{\cal I}}}{d x_{Bj} d Q^2 d|t| d\phi d\phi_S }  = e_l
\Gamma
\left( T_{BH}^* T_{DVCS}
+ T_{DVCS}^* T_{BH} \right) \;
 \nonumber \\
& = & e_l \, \frac{\Gamma}{ Q^2 \mid t \mid} \Big\{F_{UU}^{\cal I}  + (2h) F_{LU}^{\cal I} 
+ (2 \Lambda) F_{UL}^{\cal I}   + (2h)(2 \Lambda)  F_{LL}^{\cal I}  +  (2 \Lambda_T) F_{UT}^{\cal I} + (2h)(2 \Lambda_T) F_{LT}^{\cal I}  \Big\}, 
\end{eqnarray}
where the electron/positron beam charge is $e_l=\pm 1$.
Notice that differently from the BH cross section, the terms $F_{UL}^{\cal I} $, $F_{LU}^{\cal I} $ and $F_{UT}^{\cal I} $ are present since in this case they are not  parity violating. }

\vspace{0.5cm}
Eq.(\ref{eq:int5fold}) can be factorized into its lepton, $L$, and hadron, $H$, components, 
\begin{equation}
\label{interf2}
{\cal I}_{\Lambda \Lambda'}^{h, \Lambda_{\gamma^*}} =  \sum_{\Lambda_\gamma'} (L_{{\cal I}, \, h}^{ \Lambda_{\gamma^*} \Lambda_\gamma'})^{* \, \rho} (H_{{\cal I}, \, \Lambda\Lambda'}^{\Lambda_{\gamma^*}\Lambda_\gamma'})_\rho  + (L_{{\cal I}, \, h}^{ \Lambda_{\gamma^*} \Lambda_\gamma'})^\rho (H_{{\cal I}, \, \Lambda, \Lambda'}^{\Lambda_{\gamma^*} \Lambda_\gamma'})^*_\rho.
\end{equation}
The phase structure of the BH-DVCS interference term arises similarly to the DVCS contribution (Sec.\ref{sec:3}) where 
the $\phi$ dependence is determined by the exchanged photon polarization vectors (Eqs.(\ref{eq:phase1})). 
{
However, differently from the pure DVCS term where the azimuthal angular dependence resides entirely in the phase factors (Eq.\eqref{eq:DVCS1b}), the BH-DVCS contribution contains an additional $\phi$ dependence of kinematical origin. The kinematical $\phi$ dependence arises in the same way as on the BH cross section.  In previous literature by introducing  an expansion in Fourier harmonics, the distinction between $\phi$ dependence from the ``phase" and $\phi$ dependence from the ``kinematics" has not been made clear. Here we give an exact treatment indicating the two different sources if azimuthal angular dependence.  }

The phase dependence for the lepton and hadron amplitudes can be summarized schematically as follows,
\begin{eqnarray}
\left(L_{{\cal I}, \, h}^{\Lambda_{\gamma^*} \Lambda_\gamma'} \right)^\rho & = &  B^{\rho}_{h,\Lambda_\gamma'} \left(A_h^{\Lambda_\gamma^*}\right)^* 
\rightarrow e^{-i \Lambda_\gamma'\phi} \\
\left(H_{{\cal I} \, \Lambda \Lambda'}^{\Lambda_{\gamma^*},\Lambda_\gamma'} \right)_\rho & = &   
 \left(J_{\Lambda \Lambda'}\right)_\rho \, f_{\Lambda \Lambda'}^{\Lambda_{\gamma^*} \Lambda_\gamma'} \rightarrow e^{-i( \Lambda - \Lambda')\phi}  \, e^{-i( \Lambda_{\gamma^*} - \Lambda - \Lambda_\gamma' +\Lambda')\phi}  \equiv e^{-i( \Lambda_{\gamma^*} - \Lambda_\gamma')\phi}
 \\
 && \left(L_{{\cal I}, \, h}^{\Lambda_{\gamma^*} \Lambda_\gamma'} \right)^\rho \left(H_{{\cal I} \, \Lambda \Lambda'}^{\Lambda_{\gamma^*},\Lambda_\gamma'} \right)^*_\rho \rightarrow  e^{i \Lambda_{\gamma^*} \phi}
\label{eq:phasedep}
\end{eqnarray}
Notice that the DVCS amplitude for the lepton part,  $A_h^{\Lambda_{\gamma^*}}$, does not carry any angular dependence since it is defined entirely in the lepton plane.

The form of the lepton and hadron tensors are given in what follows,  respectively,   in Sections \ref{sec:DVCSBH_Lepton}  and \ref{sec:DVCSBH_Hadron}.  

\subsubsection{Lepton Tensor}
\label{sec:DVCSBH_Lepton}
\begin{figure}
\label{BH-DVCSlephad}
\begin{center}
\includegraphics[width=12.cm]{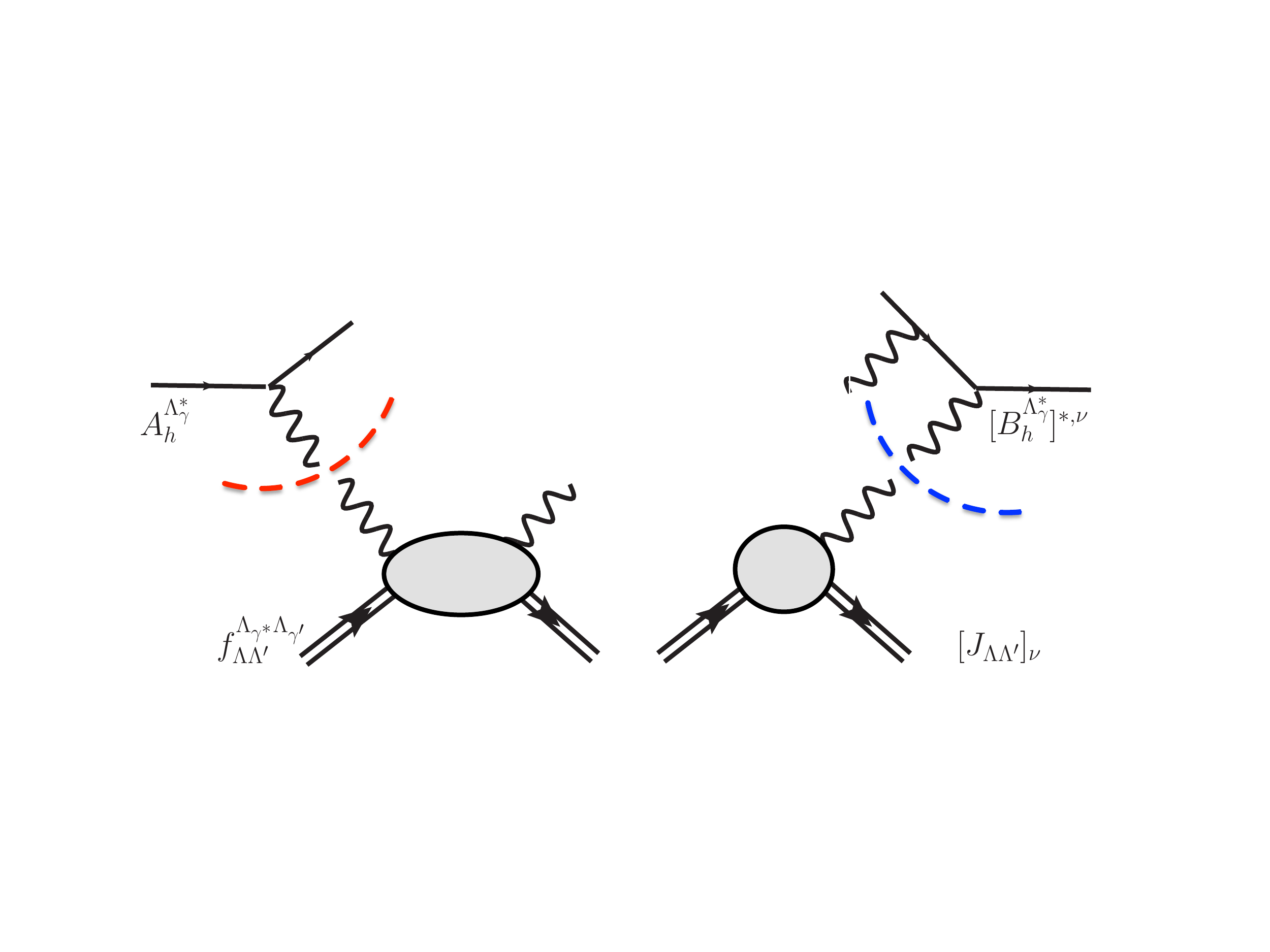}
\end{center}
\vspace{-3cm}
\caption{Factorization of the BH-DVCS interference contributions to the cross section into their respective leptonic and hadronic helicity amplitudes.}
\end{figure}

The BH-DVCS interference lepton tensor is defined as,
\begin{eqnarray}
\label{eq:int_lep2}
\left(L_{{\cal I}, \, h}^{\Lambda_{\gamma^*} \Lambda_\gamma'} \right)^\rho & = &  B^{\rho}_{h,\Lambda_\gamma'} \left(A_h^{\Lambda_\gamma^*}\right)^*
\nonumber \\ 
&=& \left(\sum_{a=1,3}B^{\rho\,(a)}_{h,\Lambda_\gamma'} \right) \Bigg( \frac{1}{Q^2} \bar{u}(k',h) \gamma^\nu u(k,h) \left(\epsilon_\nu^{\Lambda_{\gamma^*}}\right)^*\Bigg)^*  
=
\frac{1}{t \, Q^2} \frac{\epsilon^{\Lambda_{\gamma^*}}_\nu}{C^*}
 \left({\cal L}_{h \Lambda_\gamma'}^{\rho \nu \, (1,1) \, *} + {\cal L}_{h \Lambda_\gamma'}^{\rho \nu \, (1,2)\, *} + {\cal L}_{h \Lambda_\gamma'}^{ \rho \nu \, (1,3) \, *}  \right), \nonumber \\
\label{eq:Int_lep_amp}
\end{eqnarray}
The expressions for the DVCS lepton amplitude, $A_h^{\Lambda_{\gamma^*}}$, and the BH lepton amplitude, $B_{h,\Lambda_\gamma'}^\rho$ are given in Eq.(\ref{eq:A_intro}), and Eq.(\ref{eq:B_intro}), respectively; the terms ${\cal L}_{h \Lambda_\gamma'}^{\nu \rho \, (1,a)}, a=1,2,3$ are defined in Eqs.(\ref{Lcal11},\ref{Lcal12},\ref{Lcal13}), whereas the coefficient and $C$ is given in  Eq.(\ref{eq:C}). Notice that the factor $t$ arises from writing the amplitude $A_h^{\Lambda_\gamma^*}$ in terms of $B_{h , \Lambda_\gamma'}^{(1)}$. The dimensions of the lepton tensor, $L_{\cal I}^\rho$, are GeV$^{-3}$ (see also Appendix \ref{appa}). 
Writing out explicitly the dependence on the polarization vectors one has,
\begin{eqnarray}
\label{eq:Int_lep_amp2}
 \left(L_{{\cal I}, \, h}^{\Lambda_{\gamma^*} \Lambda_\gamma'}\right)^{* \, \rho}
 & = &  \frac{2}{t \, Q^2} \Bigg\{ \Big[ \left( k\epsilon^{\Lambda_{\gamma^*}}\right)^* k'^\rho + \left(k'\epsilon^{\Lambda_{\gamma^*}}\right)^* k^\rho - (\epsilon^{\Lambda_{\gamma^*}\rho})^* (kk')  
 - {2}i h k_\alpha^\prime k_\beta \left(\epsilon_\nu^{\Lambda_{\gamma^*}} \right)^* \epsilon^{\alpha \nu \beta \rho } \Big]  \Bigg(\frac{(\epsilon^{\Lambda_\gamma'} k')}{(k'q')} - \frac{(\epsilon^{\Lambda_\gamma'} k)}{(kq')} \Bigg)  \nonumber \\
%
 & + &  \Big[   \left( k\epsilon^{\Lambda_{\gamma^*}}\right)^* q^{\prime \, \rho} (\epsilon^{\Lambda_\gamma'} k^\prime) + \left(k'\epsilon^{\Lambda_{\gamma^*}}\right)^* q^{\prime \, \rho} ( \epsilon^{\Lambda_\gamma^\prime} k) -[q^{\prime \rho} (\epsilon^{\Lambda_\gamma^\prime} \epsilon^{\Lambda_{\gamma^*}}) - \left(q'\epsilon^{\Lambda_{\gamma^*}}\right)^* \epsilon^{\Lambda_\gamma' \, \rho}](k k^\prime)   
\nonumber \\
& &   {-\left(k\epsilon^{\Lambda_{\gamma^*}} \right)^* \epsilon^{\Lambda_\gamma' \, \rho} (k' q') -\left(k'\epsilon^{\Lambda_{\gamma^*}}\right)^* \epsilon^{\Lambda_\gamma' \, \rho} (k q')} - {2} ihk^\prime_\alpha k_\beta \epsilon_\nu^{\Lambda_{\gamma^*}} \epsilon^{\alpha \nu \beta \delta} 
\left(q^{\prime \rho} \epsilon^{\Lambda_\gamma'}_\delta - q^{\prime}_\delta \epsilon^{{\Lambda_\gamma'} \rho} \right) \Big] 
\nonumber \\
& \times & \left(\frac{1}{(k'q')} + \frac{1}{(kq')} \right) \, \nonumber \\
%
& + & 
 \Big[  \left(q'\epsilon^{\Lambda_{\gamma^*}}\right)^*  k^\rho (\epsilon^{\Lambda_\gamma^\prime}k')- \left(q'\epsilon^{\Lambda_{\gamma^*}}\right)^*  k'^\rho (\epsilon^{\Lambda_\gamma^\prime}k)  - (\epsilon^{\Lambda_\gamma^\prime} \epsilon^{\Lambda_{\gamma^*} \, *}) k^\rho (k' q') 
\nonumber \\ 
& + & (\epsilon^{\Lambda_\gamma'} \epsilon^{\Lambda_{\gamma^*} \, *}) k'^\rho (k q')  
 +  \left( \epsilon^{\Lambda_{\gamma^*} \, \rho} \right)^* ((k'q')(\epsilon^{\Lambda_\gamma^\prime} k) - (k q')(\epsilon^{\Lambda_\gamma'} k') )   
\nonumber \\
& + & 
{2} ih  \left( \left(k'\epsilon^{\Lambda_{\gamma^*}}\right)^* k_\delta + \left(k\epsilon^{\Lambda_{\gamma^*}}\right)^* k'_\delta - \epsilon^{\Lambda_{\gamma^*} \, *}_\delta (kk') \right)    \epsilon^{\alpha \beta \, \rho\, \delta}  \epsilon^{\Lambda_\gamma'}_\alpha \, q^\prime_\beta 
\Big]  \left(\frac{1}{(k'q')} - \frac{1}{(kq')} \right)\Bigg\}
\end{eqnarray}
Reorganizing the terms by grouping them under the same four-vector index $\rho$, and separating the unpolarized lepton from the polarized lepton one has,
\vspace{0.5cm}

\begin{eqnarray}
\label{Interf_lep}
\left(L_{{\cal I}, \, h}^{\Lambda_{\gamma^*} \Lambda_\gamma'}\right)^{* \, \rho}
 & = &   \left(L_{{\cal I}, \, S}^{\Lambda_{\gamma^*} \Lambda_\gamma'}\right)^{* \, \rho} 
 +  2 i h \, \left(L_{{\cal I}, \, A}^{\Lambda_{\gamma^*} \Lambda_\gamma'}\right)^{* \, \rho}
\end{eqnarray}
where,
\begin{subequations}
\begin{eqnarray}
\label{Interf_lep2}
\left(L_{{\cal I}, \, S}^{\Lambda_{\gamma^*} \Lambda_\gamma'}\right)^{* \, \rho} & = & \frac{1}{t \, Q^2} \Big(k'^\rho A_{k'}^{\Lambda_{\gamma^*} \Lambda_\gamma'} + k^\rho A_k^{\Lambda_{\gamma^*} \Lambda_\gamma'} + q'^\rho A_{q'}^{\Lambda_{\gamma^*} \Lambda_\gamma'} + \epsilon_{\Lambda_{\gamma^*}}^\rho A_{\epsilon^*}^{\Lambda_{\gamma^*} \Lambda_\gamma'} + \epsilon_{\Lambda_\gamma'}^\rho A_{\epsilon'}^{\Lambda_{\gamma^*} \Lambda_\gamma'} \Big) \\
\label{eq:L_int_anti}
\left(L_{{\cal I}, \, A}^{\Lambda_{\gamma^*} \Lambda_\gamma'}\right)^{* \, \rho} &= & \frac{1}{t \, Q^2} \Big[ k_\alpha' k_\beta \left( \epsilon_\nu^{\Lambda_{\gamma^*}} \right)^*  
  \left(\frac{(\epsilon_{\Lambda_\gamma'} k')}{(k'q')} - \frac{(\epsilon_{\Lambda_\gamma'} k)}{(kq')} \right)
\epsilon^{\alpha \nu \beta \rho} \nonumber \\
&- & k_\alpha' k_\beta \left( \epsilon_\nu^{\Lambda_{\gamma^*}} \right)^*   \left( q'^\rho \epsilon^{\Lambda_\gamma'}_\delta - q'_\delta \epsilon^{\Lambda_\gamma' \rho} \right) 2D^-  \epsilon^{\alpha \nu \beta \delta} \nonumber \\
&+ &  q^\prime_\beta \left( \left(k'\epsilon^{\Lambda_{\gamma^*}}\right)^* k_\delta + \left(k\epsilon^{\Lambda_{\gamma^*}}\right)^* k'_\delta - \epsilon^{\Lambda_{\gamma^*} \, *}_\delta (kk') \right) \epsilon^{\Lambda_\gamma'}_\alpha \,   2D^+ \epsilon^{\alpha \beta \, \rho\, \delta}   \Big]
\end{eqnarray}
\end{subequations}
with coefficients, $A_{k'}^{\Lambda_{\gamma^*} \Lambda_\gamma'}, A_{k}^{\Lambda_{\gamma^*} \Lambda_\gamma'}, A_{q'}^{\Lambda_{\gamma^*} \Lambda_\gamma'}, A_{\epsilon^*}^{\Lambda_{\gamma^*} \Lambda_\gamma'}, A_{\epsilon'}^{\Lambda_{\gamma^*} \Lambda_\gamma'}$ defined by:
\begin{subequations}
\label{lep_Acoeff}
\begin{eqnarray}
A_{k'}^{\Lambda_{\gamma^*} \Lambda_\gamma'} & = & 2 \frac{(k \epsilon^*)(k' \epsilon')}{(k'q')} - 2 \frac{(k \epsilon^*)(k \epsilon')}{(kq')} - \frac{(q' \epsilon^*)(k \epsilon')}{(k'q')} + \frac{(q' \epsilon^*)(k \epsilon')}{(kq')} + \frac{(\epsilon^* \epsilon')(kq')}{(k'q')} - (\epsilon^* \epsilon') \nonumber \\
& = & {4 D_+  (k \epsilon^*)(k \epsilon')  - 2 D_+ {(q' \epsilon^*)(k \epsilon')}  + 2 D_+ {(\epsilon^* \epsilon')(kq')} }\\ \nonumber  \\ 
A_k^{\Lambda_{\gamma^*} \Lambda_\gamma'} & = & 2 \frac{(k' \epsilon^*)(k' \epsilon')}{(k'q')} - 2 \frac{(k' \epsilon^*)(k \epsilon')}{(kq')} + \frac{(q' \epsilon^*)(k' \epsilon')}{(k'q')} - \frac{(q' \epsilon^*)(k' \epsilon')}{(kq')} + \frac{(\epsilon^* \epsilon')(k'q')}{(kq')} - (\epsilon^* \epsilon') \nonumber \\
& = & {4 D_+ {(k' \epsilon^*)(k \epsilon')} + 2 D_+ {(q' \epsilon^*)(k' \epsilon')}  - 2 D_+ {(\epsilon^* \epsilon')(k'q')} }\\ \nonumber  \\
A_{q'}^{\Lambda_{\gamma^*} \Lambda_\gamma'}& = & \frac{(k \epsilon^*)(k' \epsilon')}{(k'q')} + \frac{(k \epsilon^*)(k' \epsilon')}{(kq')} + \frac{(q' \epsilon^*)(k' \epsilon')}{(k'q')} + \frac{(q' \epsilon^*)(k' \epsilon')}{(kq')}- \frac{(\epsilon^* \epsilon')(kk')}{(k'q')} - \frac{(\epsilon^* \epsilon')(kk')}{(kq')} \nonumber \\
& = & 2 D_- {(k \epsilon^*)(k' \epsilon')} + 2 D_- {(q' \epsilon^*)(k' \epsilon')}  - 2 D_- {(\epsilon^* \epsilon')(kk')} \\ \nonumber \\
A_{\epsilon^*}^{\Lambda_{\gamma^*} \Lambda_\gamma'}& = & - 2 \frac{(k k')(k' \epsilon')}{(k'q')} + 2 \frac{(k k')(k \epsilon')}{(kq')} + (k \epsilon') - \frac{(k'q')(k \epsilon')}{(kq')}- \frac{(kq')(k'\epsilon')}{(k'q')} + (k'\epsilon') \nonumber \\
& = & {- 4 D_+ {(k k')(k \epsilon')} + 2 D_+ {(k'q')(k \epsilon')}  - 2 D_+ {(k'\epsilon')(kq')}} \\ \nonumber \\
A_{\epsilon'}^{\Lambda_{\gamma^*} \Lambda_\gamma'}& = & \frac{(k k')(q' \epsilon^*)}{(k'q')} + \frac{(k k')(q' \epsilon^*)}{(kq')} - (k \epsilon^*) - \frac{(k'q')(k \epsilon^*)}{(kq')}- \frac{(kq')(k'\epsilon^*)}{(k'q')} - (k'\epsilon^*) \nonumber \\
& = & 2 D_- {(k k')(q' \epsilon^*)} - 2 D_- {(k'q')(k \epsilon^*)}  - 2 D_- {(k'\epsilon^*)(kq')} \\ \nonumber
\end{eqnarray}
\end{subequations}
$D^+$ and $D^-$ where defined for the BH cross section in  {Eq.(\ref{eq:Dplus}) as,
\[ D^+ = \frac{1}{2(k'q')} - \frac{1}{2(kq')} \;\;\;\; \quad D^- = -\frac{1}{2(k'q')} - \frac{1}{2(kq')} \] . }
Notice the structure of the lepton tensor: it depends on both the outgoing photon polarization vector $\epsilon^{\Lambda_\gamma'}$ and the exchanged DVCS virtual photon polarization vector $\epsilon^{\Lambda_{\gamma^*}}$. These two polarization vectors will be combined 
with the polarization vectors for the twist two and twist three contributions to the hadronic tensor in Sections \ref{sec:DVCSBH_tw2} and \label{sec:DVCSBH_tw3}, respectively. 
As we explain below, this will determine the phase dependence of the DVCS-BH interference term.  

\vspace{0.5cm}
\subsubsection{Hadron Tensor}
\label{sec:DVCSBH_Hadron}
The hadron contribution to BH-DVCS is defined as the product of the proton current, Eq.(\ref{eq:elastic}),
\begin{equation}
 \left[J_{\Lambda \Lambda^\prime}\right]_\rho = 
 \overline{U}(p',\Lambda')\left[ G_M \gamma_\rho 
  -\frac{(p+p')_\rho}{2M} F_2 \right]U(p,\Lambda)
\end{equation}
with $G_M=F_1+ F_2$, and the hadron tensor contracted with the photon polarization vectors,  Eq.(\ref{tensor_trans}),
\begin{eqnarray}
\label{eq:Interf_had}
\left(H_{{\cal I} \, \Lambda \Lambda'}^{\Lambda_{\gamma^*},\Lambda_\gamma'} \right)_\rho & = &   \left( J_{\Lambda \Lambda'} \right)_\rho \left[ \epsilon^{\Lambda_{{\gamma^*}}}_{\beta}(q) W^{\beta \alpha}_{\Lambda \Lambda'} \left(\epsilon^{\Lambda_\gamma'}_{\alpha}(q')\right)^*  \right]   
\end{eqnarray}
We now separate the structure of $W_{\Lambda\Lambda'}$ into its twist two and twist three components. For twist two, using the definitions in Eqs.(\ref{GPDvec},\ref{GPDaxvec}), one has,
\begin{eqnarray}
 \left(H_{{\cal I} \, \Lambda \Lambda'}^{\Lambda_{\gamma^*},\Lambda_\gamma'} \right)_\rho  =  \left( J_{\Lambda \Lambda'} \right)_\rho  \left(-g_T^{\alpha \beta} \mathcal{F}^S_{\Lambda \Lambda'} + i \epsilon_T^{\beta\alpha} \mathcal{F}^A_{\Lambda \Lambda'} \right) \left(\epsilon^{\Lambda_\gamma' }_{\alpha}(q')\right)^* \epsilon^{\Lambda_{{\gamma^*}}}_{\beta}(q) ,
\end{eqnarray} 
\noindent and correspondingly: $\Lambda_{\gamma^*}=\pm 1$, $\Lambda_{\gamma'}=\pm 1$.
We introduced the notation, ${\cal F}_S$ and ${\cal F}_A$ for the symmetric (S), and antisymmetric (A) components of the hadron tensor, 
\begin{subequations}
\begin{eqnarray}
\label{FS}
\mathcal{F}^S_{\Lambda\Lambda'} & = & \frac{1}{2} \, \int_{-1}^1 dx  \, C^+(x,\xi) \, W^{[\gamma^+]}_{\Lambda \Lambda'}  \\
\label{FA}
\mathcal{F}^A_{\Lambda\Lambda'} & = & \frac{1}{2} \, \int_{-1}^1 dx  \,
 C^-(x,\xi) \, W^{[\gamma^+ \gamma_5]}_{\Lambda \Lambda'} 
 \end{eqnarray}
\end{subequations}
with,
 \begin{equation*}
 C^\pm(x,\xi) = \frac{1}{x-\xi - i \epsilon} \pm \frac{1}{x+\xi - i \epsilon}.
 \end{equation*}

To understand the polarization configurations of the products of the currents we need to understand the projection of the spin. 
In our notation we define our helicity spinors by separating out the momentum from the helicity dependent part.
\begin{equation}
U(p,\Lambda) = \frac{\slashed{p} + M}{\sqrt{p^{0} + M}}U(0,\Lambda) \qquad \overline{U}(p',\Lambda') = \overline{U}(0,\Lambda')\frac{\slashed{p} + M}{\sqrt{p^{0} + M}}
\end{equation}
where the helicity dependent pieces obey our general rule for the overlap of two different states.
\begin{equation}
U(0,\Lambda)\overline{U}(0,\Lambda') = \frac{1}{4}(1+\gamma^{0})(1+\Lambda \gamma^{3}\gamma_{5})\delta_{\Lambda,\Lambda'} + \frac{1}{4}(1+\gamma^{0})(\gamma^{1}+ i\Lambda \gamma^{2})\gamma_{5}\delta_{\Lambda,-\Lambda'}
\end{equation}
so that the general expression for the overlap of two different states with different helicity and momentum is given by
\begin{equation}
\overline{U}(p',\Lambda') U(p,\Lambda) = \frac{1}{4\sqrt{(p^{0}+M)(p^{\prime 0}+M}} {\rm Tr} \Big\{ (1+\gamma^{0})(1+\Lambda \gamma^{3}\gamma_{5})(\slashed{p'}+M)(\slashed{p}+M) \Big\} \delta_{\Lambda,\Lambda'}
\end{equation}
This general result is useful for the calculation of amplitudes of the overlap of two momentum states; however, in the calculation of the interference term we are calculating the overlap of two currents with the same momentum. Thus simplifications can be made.

The interference term can be outlined as follows for states of the same helicity (no transverse spin polarization states):
\begin{eqnarray}
J_{\Lambda,\Lambda' \, \rho}\mathcal{F}_{\Lambda,\Lambda'}^{S,A} &=& \overline{U}(p',\Lambda')\Big[\text{BH} \Big]_{\rho}U(p,\Lambda)\overline{U}(p,\Lambda)\Big[\text{DVCS} \Big]U(p',\Lambda') \nonumber \\
&=& {\rm Tr} \Big\{ U(p',\Lambda')\overline{U}(p',\Lambda')\Big[\text{BH} \Big]_{\rho}U(p,\Lambda)\overline{U}(p,\Lambda)\Big[\text{DVCS} \Big]^{S,A} \Big\},
\end{eqnarray}
{where 
\begin{subequations}
\begin{eqnarray}
\Big[\text{BH} \Big]_{\rho} & = & \left[ G_M \gamma_\rho 
  -\frac{(p+p')_\rho}{2M} F_2 \right] \\
\Big[\text{DVCS} \Big]^S & = &  H \gamma^+ + E \frac{\sigma^{+i} \Delta_i}{2M}, \quad\quad
\Big[\text{DVCS} \Big]^A  =   \tilde{H} \gamma^+\gamma_5 + \tilde{E} \frac{\Delta^+}{M}\gamma_5.
\end{eqnarray}
\end{subequations}
}
Since we are summing over the polarization of the final state proton we can use the relation $\sum_{\Lambda'}U(p',\Lambda')\overline{U}(p',\Lambda') = \slashed{p}'+M$, to obtain,
\begin{eqnarray}
&=& {\rm Tr} \Big\{(\slashed{p}'+M)\Big[\text{BH} \Big]_{\rho}U(p,\Lambda)\overline{U}(p,\Lambda)\Big[\text{DVCS} \Big] \Big\} \nonumber \\
&=& {\rm Tr}\Big\{(\slashed{p}'+M)\Big[\text{BH} \Big]_{\rho}(1+\gamma^{0})(1+\Lambda \gamma^{3}\gamma_{5})(\slashed{p}+M)(\slashed{p}+M)\Big[\text{DVCS} \Big] \Big\}
\end{eqnarray}
{This expression can be simplified using the covariant form of the spin vector.
We define the spin vector in its usual form with $S^{2} = -1$ and $(pS)=0$, $\{p,S\}=0$.}
We can use the spin vector to make our gamma matrix structure above covariant using the rest frame in which $S^{0} = 0$.
\begin{eqnarray}
1+ \gamma^{0} + \Lambda \gamma^{3}\gamma_{5} + \Lambda \gamma^{0}\gamma^{3}\gamma_{5} &=& 1 + S^{0}\gamma^{0} + \Lambda S^{3}\gamma^{3} \gamma_{5} + \Lambda S^{0}\gamma^{0}\gamma^{3}\gamma^{5} \nonumber \\
&=& 1 + \Lambda \gamma_{5}\nonumber \\
&=& 1 + \slashed{S}\gamma_{5}
\end{eqnarray}
and since $S,p$ anti commute like $\gamma_{5}$ the two of them together commute, therefore our final result for the outline is as follows
\begin{eqnarray}
{\rm Tr} \Big\{(\slashed{p}' +M)\Big[\text{BH} \Big]_{\rho}(\slashed{p}+M)(1+\slashed{S}\gamma_{5})\Big[\text{DVCS} \Big] \Big\}
\end{eqnarray}
where $\slashed{S} = \Lambda$.

Similarly for a transversely polarized target we must involve the case where the DVCS process no longer conserves helicity. Therefore we can use the operator $(\gamma^{1} + i \Lambda \gamma^{2})\gamma_{5}$ in which we see that when we make the expression covariant using the spin vector we obtain,
\begin{equation}
{\rm Tr} \Big\{(\slashed{p}' + M)\Big[ \text{BH}\Big]_{\rho}(\slashed{p} + M)(1+\slashed{S}_{T}\gamma_{5})\Big[\text{DVCS} \Big] \Big\}
\end{equation}
where $\slashed{S}_{T} = \Lambda_{T}\gamma^{i}$. Notice the similarity between the interference expressions cast with our formalism and Eq.\eqref{WBHcovariant}. 

\vspace{0.5cm}
\noindent The covariant expressions for the products $J^\rho_{\Lambda \Lambda'} {\cal F}^{S,A}_{\Lambda \Lambda'}$ are given by polarization configuration below.

\vspace{1cm}\noindent
{\em Unpolarized Target}
 \begin{eqnarray}
\label{FSunpol}
\sum_{\Lambda,\Lambda'} (J_{\Lambda \Lambda'})_\rho \, \mathcal{F}^S_{\Lambda \Lambda'} &= &2 {P}_{\rho}\left(F_{1} {\cal H} + \tau F_{2}{\cal E} \right)  
+  \xi \Delta_\rho  G_M \left({\cal H}+ {\cal E} \right) \\
\label{FAunpol}
\sum_{\Lambda,\Lambda'}(J_{\Lambda \Lambda'})_\rho \, \mathcal{F}^A_{\Lambda \Lambda'} &= &  \frac{1}{(2 {P}^+)} i \epsilon^{\mu\sigma\nu +}g_{\sigma \rho} \tilde{t}_{\mu\nu} G_M \widetilde{\cal H} 
\end{eqnarray}

\vspace{1cm}
\noindent {\em Longitudinally Polarized Target}
\begin{eqnarray}
\label{FSpol}
(J_{++})_\rho \, \mathcal{F}^S_{++} - (J_{--})_\rho \, \mathcal{F}^S_{--}  &=&- \frac{i}{P^{+}}\epsilon^{\mu \sigma \nu +}g_{\sigma \rho} \tilde{t}_{\mu\nu}  G_M \left({\cal H}+ {\cal E} \right) 
\\
\label{FApol}
(J_{++})_\rho \, \mathcal{F}^A_{++} - (J_{--})_\rho \, \mathcal{F}^A_{--} & = & -2P_{\rho}\left( F_{1} \widetilde{\cal H} + \tau F_{2} \widetilde{\cal E} -  F_{1} \xi \widetilde{\cal E} \right) 
- \xi \Delta_{\rho} G_M \widetilde{\cal H}, 
\end{eqnarray}

\vspace{1cm}
\noindent 
{\em In Plane Transversely Polarized Target}
\begin{eqnarray}
\Big[(J_{\Lambda \Lambda'})_\rho \, \mathcal{F}_{S, \, \Lambda \Lambda'} \Big]^1 & = &\frac{i}{M}\epsilon^{\mu \sigma \nu 1}g_{\sigma \rho}\Big[P_{\mu}P_{\nu} - \frac{1}{2}\widetilde{t}_{\mu \nu} \Big] G_M \mathcal{E} + \frac{i}{2MP^{+}}\epsilon^{\mu \nu 1 +}P_{\rho}\widetilde{t}_{\mu\nu}F_{2} \Big(\mathcal{H}+\mathcal{E} \Big) +\frac{iM}{P^{+}}\epsilon^{ \mu \sigma 1 +}g_{\sigma \rho}\Delta_{\mu} G_M \Big( \mathcal{H} + \mathcal{E} \Big) \nonumber \\ \\
\Big[ (J_{\Lambda \Lambda'})_\rho \, \mathcal{F}_{A,\Lambda \Lambda'} \Big]^1
 &=& -\frac{M}{P^{+}}\Big[g^{ 1}_{\rho}\Delta^{+} - \Delta^{1}g^{ +}_{\rho} \Big] G_M \widetilde{\cal H} - \frac{1}{M}\Delta^{1}P_{\rho}F_{2}\widetilde{\cal H} 
 - \frac{\xi}{2M} \Delta^{1} \Delta_{\rho} G_M \widetilde{\cal E} 
\end{eqnarray}

\vspace{.5 cm}
\noindent
{\em Out of Plane Transversely Polarized Target}
\begin{eqnarray}
\Big[(J_{\Lambda \Lambda'})_\rho \, \mathcal{F}_{S, \, \Lambda \Lambda'} \Big]^2 & = &\frac{i}{M}\epsilon^{\mu \sigma \nu 2}g_{\sigma \rho}\Big[P_{\mu}P_{\nu} - \frac{1}{2}\widetilde{t}_{\mu \nu} \Big] G_M \mathcal{E} + \frac{i}{2MP^{+}}\epsilon^{\mu \nu 2 +}P_{\rho}\widetilde{t}_{\mu\nu}F_{2} \Big(\mathcal{H}+\mathcal{E} \Big) +\frac{iM}{P^{+}}\epsilon^{ \mu \sigma 2 +}g_{\sigma \rho}\Delta_{\mu} G_M \Big( \mathcal{H} + \mathcal{E} \Big) \nonumber \\ \\
\Big[ (J_{\Lambda \Lambda'})_\rho \, \mathcal{F}_{A,\Lambda \Lambda'} \Big]^2
 &=& -\frac{M}{P^{+}}\Big[g^{ 2}_{\rho}\Delta^{+} - \Delta^{2}g^{ +}_{\rho} \Big] G_M \widetilde{\cal H} - \frac{}{M}\Delta^{2}P_{\rho}F_{2}\widetilde{\cal H} 
 - \frac{\xi}{2M} \Delta^{2} \Delta_{\rho} G_M \widetilde{\cal E} 
\end{eqnarray}
where,
\begin{equation}
\label{tmunu}
\tau=-\Delta^2/4M^2 ; \quad \quad \quad 
\tilde{t}_{\mu\nu} = P_{\mu}\Delta_{\nu} - P_{\nu} \Delta_{\mu} + \frac{1}{2}\Delta_{\mu}\Delta_{\nu} .    
\end{equation} 
Notice that, differently from the BH and DVCS contributions, the unpolarized hadronic tensor has now an antisymmetric part that appears at twist two from the product of the proton current and the axial vector component from the GPD correlation function. Although this term is generated in an analogous way as in the parity violating contributions to elastic scattering \cite{Sofiatti:2011yi}, it is parity conserving in exclusive photon production. 

To evaluate the twist three contribution we multiply the proton current, Eqs.(\ref{eq:elastic}), with the twist three components of the hadronic tensor, Eqs.(\ref{tensor_trans},,\ref{GPDvec3},\ref{GPDaxvec3}), 
\begin{eqnarray}
\label{eq:Interf_had_three}
 \left(H_{{\cal I} \, \Lambda \Lambda'}^{\Lambda_{\gamma^*},\Lambda_\gamma'} \right)_\rho 
& = & \left( J_{\Lambda \Lambda'} \right)_\rho 
(q+4\xi P)^{\beta} \Big( g^{\alpha j} \mathcal{F}_{\Lambda\Lambda', j}^{S} +i \epsilon^{j \alpha} \mathcal{F}_{\Lambda\Lambda', j}^{A}\Big)\left(\epsilon^{\Lambda_\gamma' }_{\alpha}(q')\right)^* \epsilon^{\Lambda_{{\gamma^*}}}_{\beta}(q) 
\end{eqnarray} 
\noindent where now: $\Lambda_{\gamma^*}=0$, $\Lambda_{\gamma'}=\pm 1$.

${\cal F}_S^{j}$ and ${\cal F}_A^{j}$ are the symmetric and antisymmetric components of tensor, respectively  defined in terms of CFFs in Section \ref{sec:3} as, 
\begin{eqnarray}
\label{FS_2}
\mathcal{F}^{S,j}_{\Lambda\Lambda'} & = & \frac{2Mx_{Bj}}{Q} \, \int_{-1}^1 dx  \, C^+(x,\xi) \, W^{[\gamma^j]}_{\Lambda \Lambda'}  
\\
\label{FA_2}
\mathcal{F}^{A,j}_{\Lambda\Lambda'} & = & \frac{2Mx_{Bj}}{Q}\, \int_{-1}^1 dx  \,
 C^-(x,\xi) \, W^{[\gamma^j  \gamma_5]}_{\Lambda \Lambda'}
 \end{eqnarray}
By inserting the expressions of the correlation functions in terms of CFFs, we can evaluate Eqs.(\ref{eq:Interf_had},\ref{eq:Interf_had_three}) for specific target polarizations. For the product  $J^\rho_{\Lambda \Lambda'} {\cal F}^{S,A}_{\Lambda \Lambda', j}$ we find,

\vspace{0.5cm}
\noindent
{\em Unpolarized Target}
\begin{eqnarray}
\sum_{\Lambda,\Lambda'} (J_{\Lambda \Lambda'})_\rho \, \mathcal{F}^{S}_{ \Lambda \Lambda',j} & = & 
4 {P}_{\rho} \frac{\Delta_{j}}{{P}^{+}}\Big[F_{1} (2\widetilde{\cal H}_{2T} + \mathcal{E}_{2T}) - F_{2} (\mathcal{H}_{2T}+ \tau \widetilde{\mathcal{H}}_{2T})  \Big] \nonumber \\
&+& \nonumber g_{\rho j} \frac{2M^2}{P^+} G_M \Big[ \tau (\xi \mathcal{E}_{2T} - \widetilde{\cal E}_{2T}) + 4 \xi  \mathcal{H}_{2T} \Big] \nonumber
\\
&-& \Delta_\rho \frac{\Delta_{j}}{P^+} G_M  \widetilde{\cal E}_{2T} 
\\
\sum_{\Lambda,\Lambda'} (J_{\Lambda \Lambda'})_\rho \, \mathcal{F}^{A}_{ \Lambda \Lambda', j} & = & 
4 {P}_{\rho} \frac{-i\epsilon_{ij}\Delta^{i}}{{P}^{+}}\Big[F_{1} (2\widetilde{\cal H}_{2T}' + \mathcal{E}_{2T}') - F_{2} (\mathcal{H}_{2T}' + \tau \widetilde{\cal H}_{2T}')  \Big] \nonumber \\
&-& \nonumber i\epsilon_{ij} g^{ i}_{\rho} \frac{2M^2}{P^+} G_M \Big[ \tau (\xi \mathcal{E}_{2T}' - \widetilde{\cal E}_{2T}') + 4 \xi \mathcal{H}_{2T}' \Big] \nonumber
\\
&+& \Delta_\rho \frac{i\epsilon_{ij}\Delta^i}{P^+} G_M \widetilde{\cal E}_{2T}' 
\end{eqnarray}

\noindent 
{\em Longitudinally Polarized Target}
\begin{eqnarray}
(J_{\Lambda \Lambda})_\rho \, \mathcal{F}^{S}_{ \Lambda \Lambda, j}  &=& \frac{4M^{2}}{(P^{+})^{2}}i\epsilon^{\mu \sigma + k}g_{\sigma \rho}g_{jk}\Delta_{\mu} G_M  \mathcal{H}_{2T} 
- \frac{2}{(P^{+})^{2}}P_{\rho}i\epsilon^{\mu \nu + k} g_{jk}\tilde{t}_{\mu\nu} F_{2} \mathcal{H}_{2T}  \nonumber \\
&+& \frac{i\Delta_{j}}{(P^{+})^{2}}\epsilon^{\mu \sigma \nu +}g_{\sigma \rho}\tilde{t}_{\mu\nu} G_M  \mathcal{E}_{2T}
+ \frac{4i}{P^{+}}\epsilon^{\mu \sigma \nu k}g_{jk}g_{\sigma \rho}\Big(P_{\mu}P_{\nu} + \frac{1}{2}\tilde{t}_{\mu\nu} \Big) G_M (\xi \mathcal{E}_{2T} - \widetilde{\cal E}_{2T}) \\
(J_{\Lambda \Lambda})_\rho \, \mathcal{F}^{A}_{ \Lambda \Lambda,j}  &=&  \frac{4M^{2}}{(P^{+})^{2}}\epsilon_{ij}\epsilon^{\mu \sigma + i}g_{\sigma \rho}\Delta_{\mu} G_M  \mathcal{H}_{2T} 
+ \frac{2}{(P^{+})^{2}}P_{\rho}\epsilon_{ij}\epsilon^{\mu \nu + i} \tilde{t}_{\mu\nu} F_{2} \mathcal{H}_{2T}  \nonumber \\
&+& \frac{\epsilon_{ij}\Delta^{i}}{(P^{+})^{2}}\epsilon^{\mu \sigma \nu +}g_{\sigma \rho}\tilde{t}_{\mu\nu} G_M  \mathcal{E}_{2T}
+ \frac{4}{P^{+}}\epsilon_{ij}\epsilon^{\mu \sigma \nu i}g_{\sigma \rho}\Big(P_{\mu}P_{\nu} + \frac{1}{2}\tilde{t}_{\mu\nu} \Big) G_M (\xi \mathcal{E}_{2T} - \widetilde{\cal E}_{2T})
\end{eqnarray}

\noindent 
{\em In Plane Transversely Polarized Target}
\begin{eqnarray}
(J_{\Lambda \Lambda'})_\rho \, \mathcal{F}^S_{\Lambda \Lambda', j}
 &=& -\frac{8M}{P^{+}}\epsilon^{- 1 + j}P_{\rho}F_{2}H_{2T} \nonumber \\
 &-&\frac{M}{(P^{+})^{2}}\Big(4\epsilon^{\mu \sigma \nu +}g_{\sigma \rho}g^{1j}(P_{\mu}P_{\nu} - \frac{1}{2}P_{\mu}\Delta_{\nu} + \frac{1}{2}\Delta_{\mu}P_{\nu} - \frac{1}{4}\Delta_{\mu}\Delta_{\nu}) \nonumber\\
 &+& \epsilon^{1 + j \sigma}(2P_{\rho}P_{\sigma}-\frac{1}{2}\Delta_{\rho}\Delta_{\sigma}) \Big)(F_{1}+F_{2})H_{2T}
 \nonumber\\ &+& \frac{2iM}{(P^{+})^{2}}\epsilon^{\mu \sigma 1 +}g_{\sigma \rho}\Delta_{\mu}\Delta^{j}(F_{1}+F_{2})E_{2T} + \frac{4i\xi M}{P^{+}}\epsilon^{\mu \sigma 1 j}g_{\sigma \rho}\Delta_{\mu}(F_{1}+F_{2})E_{2T}\nonumber \\ &-& \frac{i}{M(P^{+})^{2}}\epsilon^{\mu \nu 1 +}\Delta^{j}( \Delta_{\mu}P_{\nu}- P_{\mu}\Delta_{\nu}  - \frac{1}{2}\Delta_{\mu}\Delta_{\nu})P_{\rho}F_{2}E_{2T}\nonumber\\&-& 
 \xi\frac{4i}{MP^{+}}\epsilon^{\mu \nu 1 j}(P_{\mu}P_{\nu} - \frac{1}{2}P_{\mu}\Delta_{\nu} + \frac{1}{2}\Delta_{\mu}P_{\nu} - \frac{1}{4}\Delta_{\mu}\Delta_{\nu})P_{\rho}F_{2}E_{2T}\nonumber\\
 &-& \frac{4i}{MP^{+}}\epsilon^{\mu \sigma \nu 1 }g_{\sigma \rho}\Delta^{j}(P_{\mu}P_{\nu} - \frac{1}{2}P_{\mu}\Delta_{\nu} + \frac{1}{2}\Delta_{\mu}P_{\nu} - \frac{1}{4}\Delta_{\mu}\Delta_{\nu})(F_{1}+F_{2})\widetilde{H}_{2T} \nonumber\\
 &+& \frac{4i}{MP^{+}}\epsilon^{\mu \nu 1 j}(P_{\mu}P_{\nu} - \frac{1}{2}P_{\mu}\Delta_{\nu} + \frac{1}{2}\Delta_{\mu}P_{\nu} - \frac{1}{4}\Delta_{\mu}\Delta_{\nu})P_{\rho}F_{2}\widetilde{E}_{2T}
 \nonumber\\
 &-& \frac{4iM}{P^{+}}\epsilon^{\mu \sigma 1 j}g_{\sigma \rho}\Delta_{\mu}(F_{1}+F_{2})\widetilde{E}_{2T}
\end{eqnarray}

\noindent 
{\em Out of Plane Transversely Polarized Target}
\begin{eqnarray}
(J_{\Lambda \Lambda'})_\rho \, \mathcal{F}^S_{\Lambda \Lambda', j}
 &=& -\frac{8M}{P^{+}}\epsilon^{- 2 + j}P_{\rho}F_{2}H_{2T} \nonumber \\
 &-&\frac{M}{(P^{+})^{2}}\Big(4\epsilon^{\mu \sigma \nu +}g_{\sigma \rho}g^{2j}(P_{\mu}P_{\nu} - \frac{1}{2}P_{\mu}\Delta_{\nu} + \frac{1}{2}\Delta_{\mu}P_{\nu} - \frac{1}{4}\Delta_{\mu}\Delta_{\nu})\nonumber \\
 &+& \epsilon^{2 + j \sigma}(2P_{\rho}P_{\sigma}-\frac{1}{2}\Delta_{\rho}\Delta_{\sigma}) \Big)(F_{1}+F_{2})H_{2T}
 \nonumber\\ &+& \frac{2iM}{(P^{+})^{2}}\epsilon^{\mu \sigma 2 +}g_{\sigma \rho}\Delta_{\mu}\Delta^{j}(F_{1}+F_{2})E_{2T} + \frac{4i\xi M}{P^{+}}\epsilon^{\mu \sigma 2 j}g_{\sigma \rho}\Delta_{\mu}(F_{1}+F_{2})E_{2T}\nonumber \\ &-& \frac{i}{M(P^{+})^{2}}\epsilon^{\mu \nu 2 +}\Delta^{j}( \Delta_{\mu}P_{\nu}- P_{\mu}\Delta_{\nu}  - \frac{1}{2}\Delta_{\mu}\Delta_{\nu})P_{\rho}F_{2}E_{2T}\nonumber\\&-& 
 \xi\frac{4i}{MP^{+}}\epsilon^{\mu \nu 2 j}(P_{\mu}P_{\nu} - \frac{1}{2}P_{\mu}\Delta_{\nu} + \frac{1}{2}\Delta_{\mu}P_{\nu} - \frac{1}{4}\Delta_{\mu}\Delta_{\nu})P_{\rho}F_{2}E_{2T}\nonumber\\
 &-& \frac{4i}{MP^{+}}\epsilon^{\mu \sigma \nu 2 }g_{\sigma \rho}\Delta^{j}(P_{\mu}P_{\nu} - \frac{1}{2}P_{\mu}\Delta_{\nu} + \frac{1}{2}\Delta_{\mu}P_{\nu} - \frac{1}{4}\Delta_{\mu}\Delta_{\nu})(F_{1}+F_{2})\widetilde{H}_{2T} \nonumber\\
 &+& \frac{4i}{MP^{+}}\epsilon^{\mu \nu 2 j}(P_{\mu}P_{\nu} - \frac{1}{2}P_{\mu}\Delta_{\nu} + \frac{1}{2}\Delta_{\mu}P_{\nu} - \frac{1}{4}\Delta_{\mu}\Delta_{\nu})P_{\rho}F_{2}\widetilde{E}_{2T}
 \nonumber\\
 &-& \frac{4iM}{P^{+}}\epsilon^{\mu \sigma 2 j}g_{\sigma \rho}\Delta_{\mu}(F_{1}+F_{2})\widetilde{E}_{2T}
\end{eqnarray}



Notice that by specifying the twist two or twist three structures, we make a choice in the polarization vectors components that in this case, at variance with the DVCS squared contribution, is frame specific. 

We contract the lepton and hadron tensors while preserving the helicity and phase structure,
keeping in mind that the hadronic plane is rotated compared to the leptonic plane, or that the polarization vector $\epsilon_{\mu}^{\Lambda_{\gamma}^{*}}$ comes with a phase of $e^{-i\Lambda_{\gamma}^{*}\phi}$ (see Section \ref{sec:2}).  One has the following two configurations, for a transversely and longitudinally polarized virtual photon, respectively, defining the twist two and twist three structures,
\begin{eqnarray}
twist \quad 2 \rightarrow \sum_{\Lambda_{\gamma}^{*}  =   \pm 1} \Big(\epsilon_{\mu}^{\Lambda_{\gamma}^{*}}\Big)^{*}\epsilon_{\nu}^{\Lambda_{\gamma}^{*}} & = & \cos{\phi} \, g^{T}_{\mu \nu} - \sin{\phi} \, \epsilon^{T}_{\mu \nu}
\\
twist \quad 3 \rightarrow  \Big(\epsilon_{\mu}^{\Lambda_{\gamma}^{*}=0}\Big)^{*} \epsilon_{\nu}^{\Lambda_{\gamma}^{*}=0} & = & g_{\mu \nu}^L  
\end{eqnarray}
where the non zero components for longitudinal photon polarization are: $g_{\mu \nu}^L$ are $g_{0 0}^L  = 1+ \nu^2/Q^2$, $g_{0 3}^L = g_{3 0}^L = \sqrt{\nu^2 + Q^2}\nu/Q^2$, and  $g_{3 3}^L = \nu^2/Q^2$.

We can also perform a similar procedure for the transverse polarization of the target. Using our covariant expression previously derived for the interference term after summation over final hadron state polarization. We find two configurations $(\gamma^{1}+i\gamma^{2})\gamma_{5}$ corresponding to a $\Lambda = +1$ and $(\gamma^{1}-i\gamma^{2})\gamma_{5}$ corresponding to a $\Lambda = -1$. Therefore we can use the spin density matrix to see what phase of the spin vector these polarization states correspond with while simultaneously summing over $\Lambda_{\gamma}^{*}$ as we did in our previous longitudinal case.

\begin{eqnarray}
\gamma^{1}\gamma^{5} - i \gamma^{2}\gamma_{5} &\rightarrow& \sum_{\Lambda_{\gamma}^{*} = \pm 1} \Big(\epsilon_{\mu}^{\Lambda_{\gamma}^{*}} \Big)^{*}\epsilon_{\nu}^{\Lambda_{\gamma}^{*}}e^{i\phi_{S}}\\
\gamma^{1}\gamma^{5} + i \gamma^{2}\gamma_{5} &\rightarrow& \sum_{\Lambda_{\gamma}^{*} = \pm 1} \Big(\epsilon_{\mu}^{\Lambda_{\gamma}^{*}} \Big)^{*}\epsilon_{\nu}^{\Lambda_{\gamma}^{*}}e^{-i\phi_{S}}
\end{eqnarray}

Solving this system of equations for ``in-plane" polarization or polarization along the $1$-direction, and ``out of plane" polarization or polarization along the $2$-direction gives us

\begin{eqnarray}
\gamma^{1}\gamma_{5} &\rightarrow& \Big[\cos{(\phi + \phi_{S})} + \cos{(\phi - \phi_{S})} \Big]g_{\mu \nu}^{T} - \Big[\sin{(\phi + \phi_{S})} + \sin{(\phi - \phi_{S})} \Big]\epsilon_{\mu \nu}^{T} \\
\gamma^{2}\gamma_{5} &\rightarrow& \Big[-\sin{(\phi + \phi_{S})} + \sin{(\phi - \phi_{S})} \Big]g_{\mu \nu}^{T} - \Big[\cos{(\phi + \phi_{S})} - \cos{(\phi - \phi_{S})} \Big]\epsilon_{\mu \nu}^{T}
\end{eqnarray}

since we are using a specific orientation for $\phi_{S}$ we can then utilize this to simplify our expressions.

\begin{eqnarray}
\phi_{S} \to 0 \implies \gamma^{1}\gamma_{5} &\rightarrow& \cos{(\phi)}g_{\mu \nu}^{T} - \sin{(\phi)}\epsilon_{\mu \nu }^{T} \\
\phi_{S} \to \frac{\pi}{2} \implies \gamma^{2}\gamma_{5} &\rightarrow& -\cos{(\phi)}g_{\mu \nu}^{T} + \sin{(\phi)}\epsilon_{\mu \nu }^{T}
\end{eqnarray}

\subsection{Polarization Configurations}
\label{sec:DVCSBH_tw2}
In what follows we organize the cross section into its twist two and twist three contributions. 

\subsubsection{$F_{UU}^{\cal I}$: Unpolarized Beam, Unpolarized Target}
For the unpolarized beam, unpolarized target contribution to the cross section we have,
\begin{eqnarray}
F_{UU}^{\cal I}  &= &  F^{{\cal I}, tw 2}_{UU}  + \frac{K}{\sqrt{Q^2}} F_{UU}^{{\cal I}, tw 3} 
\end{eqnarray}
with, 

\vspace{0.5cm}
\noindent {\em Twist  two}
\begin{subequations}
\begin{eqnarray}
\label{eq:Int_FUU}
F_{UU}^{{\cal I}, tw 2} &=&   A_{UU}^{\cal I}  \Re e \left(F_1 \mathcal{H} + \tau F_2  \mathcal{E} \right)   + B_{UU}^{\cal I}    G_M \Re e \left( \mathcal{H}+ \mathcal{E} \right)
 + C_{UU}^{\cal I}   
G_M \Re e\mathcal{ \widetilde{H}}    
\end{eqnarray}

\vspace{0.5cm}
\noindent {\em Twist  three}
\begin{eqnarray}
\label{eq:Int_FUU3}
F_{UU}^{{\cal I}, tw 3} &= &  \Re e \Bigg\{ A^{ (3) \cal I}_{UU}  \Big[ F_{1}(2\mathcal{\widetilde{H}}_{2T} + \mathcal{E}_{2T}) + F_{2}( \mathcal{H}_{2T} + \tau \mathcal{\widetilde{H}}_{2T}) \Big]
 \nonumber \\
 && + B^{(3) \cal I}_{UU} G_M \, \widetilde{E}_{2T}   + C^{(3) \cal I}_{UU}  G_M \, \Big[2\xi  H_{2T} - \tau( \widetilde{E}_{2T} -\xi E_{2T} )  \Big]
\Bigg\} \nonumber \\
&+&  \Re e \Bigg\{ \widetilde{A}^{(3) \cal I}_{UU}  \Big[F_{1}(2\widetilde{H}'_{2T} + E_{2T}') + F_{2}(H_{2T}' + \tau\widetilde{H}_{2T}' )\Big] \nonumber \\  && + \widetilde{B}^{(3) \cal I}_{UU}  G_M \widetilde{E}_{2T}' +
 \widetilde{C}^{(3) \cal I}_{UU}  G_M \Big[2\xi  H_{2T}' - \tau( \widetilde{E}_{2T}' -\xi E_{2T}' )  \Big]\Bigg\}
\end{eqnarray}
\end{subequations}
Both the twist-two and twist-three contributions are organized as a sum of terms each one including:
\begin{itemize}
    \item kinematic coefficients: $A_{UU}^{\cal I}(y,x_{Bj},t,Q^2,\phi), B_{UU}^{\cal I}(y,x_{Bj},t,Q^2,\phi), C_{UU}^{\cal I}(y,x_{Bj},t,Q^2,\phi)$. The latter are obtained by contracting the lepton tensor, Eq.\eqref{Interf_lep}, with the four-vectors, $P_\rho, \Delta_\rho$, from the symmetric component of the hadronic tensor, ${\cal F}_S$, Eq.\eqref{FSunpol}, and  $i \epsilon_{\mu\rho\nu +} \tilde{t}^{\mu\nu}$ from the anti-symmetric component ${\cal F}_A$, Eq.\eqref{FAunpol}.  
    \item products of the nucleon electromagnetic form factors and the Compton form factors. 
\end{itemize}

As explained in Section \ref{sec:Int_A}, the phase dependence is determined by the DVCS virtual photon polarization vector: the twist two term is associated with transverse virtual photon polarization, generating the $\cos \phi$ term in  $F_{UU}^{{\cal I}, tw 2}$ (Eq.\eqref{eq:Int_FUU}), whereas the twist three term is associated with longitudinal virtual photon polarization which carries no $\phi$ dependence (Eq.\eqref{eq:Int_FUU3}). 

The expressions for $A_{UU}^{\cal I}$, $B_{UU}^{\cal I}$, $C_{UU}^{\cal I}$ are given by

\begin{subequations}
\label{unpol_coeff}
\begin{eqnarray}
A^{\cal I}_{UU} &=&\frac{1}{(kq')(k'q')}\Bigg\{ (Q^{2}+t)\bigg[\big((kq')_T - 2(kk)_T - 2(kq')\big)(Pk') + \big(2(k'q') - 2(k'k)_T - (k'q')_T\big)(Pk) + (k'q')(kP)_T \nonumber \\
&+& (kq')(k'P)_T - 2(kk')(kP)_T\bigg] - (Q^{2}-t+4(k\Delta))\bigg[\big(2(kk') - (k'q')_T - (kk')_T\big)(Pq') + 2(kk')(Pq')_T \nonumber \\
&-& (k'q')(kP)_T - (kq')(k'P)_T\bigg]\Bigg\} \cos{\phi}\\
B^{\cal I}_{UU} &=& \frac{\xi}{2(kq')(k'q')} \Bigg\{ (Q^{2}+t)\bigg[\big((kq')_T - 2(kk)_T - 2(kq')\big)(\Delta k') + \big(2(k'q') - 2(k'k)_T - (k'q')_T\big)(\Delta k) + (k'q')(k\Delta)_T \nonumber \\
&+& (kq')(k'\Delta)_T - 2(kk')(k\Delta)_T\bigg] - (Q^{2}-t +4(k\Delta))\bigg[\big(2(kk') - (k'q')_T - (kk')_T\big)(\Delta q') + 2(kk')(\Delta q')_T \nonumber \\
&-& (k'q')(k\Delta)_T - (kq')(k'\Delta)_T\bigg] \Bigg \} \cos{\phi}\\
C^{\cal I}_{UU}&=& \frac{1}{2(kq')(k'q')}\Bigg\{ (Q^{2}+t)\bigg[ 2(kk')(k\Delta)_{T} - (k'q')(k\Delta)_{T} - (kq')(k'\Delta)_{T}
+ 4\xi (kk')(kP)_{T} - 2 \xi (k'q')(kP)_{T} \nonumber \\ &-& 2 \xi (kq')(k'P)_{T}\bigg]
- (Q^{2}-t+4(k\Delta))\bigg[(kk')(\Delta q')_{T} - (k'q')(\Delta k)_{T} - (kq')(\Delta k')_{T} + 2 \xi (kk')(P q')_{T} - 2\xi (k'q')(P k)_{T} \nonumber \\
&-& 2\xi (kq')(P k')_{T}\bigg] \Bigg \} \cos{\phi}
\end{eqnarray}
\end{subequations}

with the transverse components defined by invariant quantities in the laboratory frame as,
\begin{eqnarray*}
(kk)_{T} &=& (kk')_{T} = \frac{1}{2}\frac{\epsilon}{1-\epsilon}Q^{2}  \quad\quad (q' \Delta)_T = -\Delta_T^2 = -(1-\xi^2)(t_0-t)    \\
(kq')_{T} &=&  (k'q')_{T} = -(k\Delta)_T = - (k' \Delta)_T = \frac{Q^{2}}{\gamma \sqrt{1+\gamma^{2}}}\sqrt{\frac{\epsilon}{2(1-\epsilon)}} \left(1+ \frac{x t}{Q^2} \right)\sin \theta \cos{\phi}
\end{eqnarray*}
For comparison we re-write below the unpolarized BH coefficients, $A_{UU}^{BH}$, $B_{UU}^{BH}$, (Eq.\eqref{eq:BHunpol}), to underline both the similarities and differences in their structure, 
\begin{eqnarray*}
A^{BH}_{UU}  & = & \frac{8\,M^2}{\myprod{k}{q'} \myprod{k'}{q'} }
\Bigg[  4 \tau  \Big( \myprod{k}{P}^2 +\myprod{k}{P}^2 \Big) -(\tau +1)
\Big( \myprod{k}{\Delta}^2 + \myprod{k'}{\Delta}^2 \Big) \Bigg]  \\
B^{BH}_{UU} & = & 
\frac{16 \,M^2}{\myprod{k}{q'} \myprod{k'}{q'}}
\Big[ \myprod{k}{\Delta}^2+ \myprod{k'}{\Delta}^2\Big]  \,
\end{eqnarray*}
Eqs.\eqref{unpol_coeff} are obtained evaluating the following four-vector products involving the momenta, $P_\rho$, $\Delta_\rho$, and the tensor structure $\tilde{t}_{\mu\nu}$, Eq.\eqref{tmunu}, 
\begin{eqnarray}
 A_{UU}^{\cal I}  & = & 2 (P \Sigma_{S})  \qquad
 B_{UU}^{\cal I} =   \xi (\Delta \Sigma_{S}) \qquad
 C_{UU}^{\cal I}  =  \frac{1}{2P^{+}}\epsilon^{\mu \sigma \nu +}g_{\sigma \rho}\widetilde{t}_{\mu \nu}\Sigma_{A}^{ \rho},
\end{eqnarray}
where, 
\begin{eqnarray}
\label{SigmaS}
\Sigma^{\rho}_{S} &=& k^{\prime \rho}\widetilde{A}_{k'}^{S} + k^{\rho}\widetilde{A}_{k}^{S} + q^{\prime \rho}\widetilde{A}_{q'}^{S} + 
\widetilde{A}_{\epsilon^{*}}^{S \rho} + 
\widetilde{A}_{\epsilon'}^{S \rho} \\
\label{SigmaA}
\Sigma^{\rho}_{A} &=& k^{\prime \rho}\widetilde{A}_{k'}^{A} + k^{\rho}\widetilde{A}_{k}^{A} + q^{\prime \rho}\widetilde{A}_{q'}^{A} + 
\widetilde{A}_{\epsilon^{*}}^{A \rho} + 
\widetilde{A}_{\epsilon'}^{A \rho}.
\end{eqnarray}
The $\widetilde{A}$ coefficients were evaluated by taking the coefficients labeled $A$ in the lepton tensor, Eqs.\eqref{lep_Acoeff}, which depend on the kinematic variables $k,k',q'$ 
and multiplying them by the photon polarization vectors, $\epsilon_\alpha^{\Lambda_{\gamma'}}$ and $\epsilon_\beta^{\Lambda_{\gamma^*}}$ from the hadronic tensor, Eq.\eqref{eq:Interf_had}, and by either the symmetric ($S$) or the anti-symmetric ($A$) tensors ($g_{T}^{\alpha \beta}$ and $ \epsilon_{T}^{\alpha \beta}$) to give us the $S$, and $A$ part of the $\widetilde{A}$. 
The twist-two contribution is finally obtained by summing over the transverse polarizations of the photons ($\Lambda_{\gamma'} =\Lambda_{\gamma^*}= \pm 1$), using the summations rules listed. The first terms in Eq.\eqref{SigmaS} and Eq.\eqref{SigmaA}, are respectively given by,
\begin{eqnarray}
\label{Atilde_S}
k^{\rho}\widetilde{A}_{k}^{S} &=& -g_{T}^{\alpha \beta}\sum_{\Lambda_{\gamma}',\Lambda_{\gamma}^{*}} A_{k}^{\Lambda_{\gamma}^{*},\Lambda_{\gamma}'}k^{\rho}\Big(\epsilon_{\alpha}^{\Lambda_{\gamma}'}\Big)^{*} \epsilon_{\beta}^{\Lambda_{\gamma}^{*}}\\
\label{Atilde_A}
k^{\rho}\widetilde{A}_{k}^{A} &=& -\epsilon_{T}^{\alpha \beta}\sum_{\Lambda_{\gamma}',\Lambda_{\gamma}^{*}} A_{k}^{\Lambda_{\gamma}^{*},\Lambda_{\gamma}'}k^{\rho}\Big(\epsilon_{\alpha}^{\Lambda_{\gamma}'}\Big)^{*} \epsilon_{\beta}^{\Lambda_{\gamma}^{*}} .
\end{eqnarray}
The other terms in Eqs.(\ref{SigmaS},\ref{SigmaA}) are obtained similarly.
Calculating explicitly Eqs.(\ref{Atilde_S},\ref{Atilde_A}) we obtain,
\begin{subequations}
\begin{eqnarray}
\widetilde{A}_{k'}^{S} &=& -2D_{+}\Bigg[2(kk)_{T} -(kq')_{T} + 2(kq')\Bigg]\cos{\phi}  + 2D_{+}\Bigg[(k \times q^{'})_{T}\Bigg]\sin{\phi}\\
\widetilde{A}_{k}^{S}  &=& -2D_{+}\Bigg[2(k'k)_{T} +(k'q')_{T} - 2(k'q')\Bigg]\cos{\phi}  - 2D_{+}\Bigg[ (k'\times q')_{T}  \Bigg]\sin{\phi}\\
\widetilde{A}_{q'}^{S}  &=& -2D_{-}\Bigg[(kk')_{T} + (k'q')_{T} - 2(kk')\Bigg]\cos{\phi} - 2D_{-}\Bigg[(k' \times q')_{T} \Bigg]\sin{\phi}\\
\widetilde{A}_{\epsilon^{*}}^{S \rho}  &=& 2D_{+}\Bigg[2(kk')k^{\rho}_{T} - (k'q')k^{\rho}_{T} - (kq')k^{\prime \rho}_{T} \Bigg]\cos{\phi} 
- 2D_{+} \Bigg[2(kk')k_{T}^{\beta}\epsilon^{\rho}_{T \beta} - (k'q')k_{T}^{\beta}\epsilon^{\rho}_{T\beta} - (kq')k^{\prime \beta}\epsilon^{\rho}_{T \beta} \Bigg]\sin{\phi} \nonumber \\ \\
\widetilde{A}_{\epsilon'}^{S \rho} &=& -2D_{-}\Bigg[(kk')q^{\prime \rho}_{T} - (k'q')k^{\rho}_{T}-(kq')k^{\prime \rho}_{T} \Bigg]\cos{\phi}  + 2D_{-}\Bigg[ (kk')q^{\prime \mu}\epsilon_{\mu \rho}^{T}- (k'q')k^{\mu}\epsilon_{\mu \rho}^{T}- (kq')k^{\prime \mu}\epsilon_{\mu \rho}^{T} \Bigg]\sin{\phi}.
\end{eqnarray}
\end{subequations}
Note that this derivation is for the term: $T_{BH}^*T_{DVCS}$ in Eq.\eqref{eq:int5fold}: taking the complex conjugate and summing, we obtain specific cancellations of the $\phi$ dependence based on the polarization, {\it i.e.} the $UU$ polarization listed above has a cancellation of the $\sin{\phi}$ component of the coefficient. Similarly, the $LL$ components contain only the $\cos \phi $ terms. The $UL$ and $LU$ polarizations will, on the contrary, have a cancellation of the $\cos{\phi}$ dependent piece.

To project out the twist-three components we start from expresssions analogous to Eqs.(\ref{SigmaS},\ref{SigmaA}), namely
\begin{eqnarray}
\Sigma^{\rho j}_{S} &=& k^{\prime \rho}\widetilde{A}_{k',S}^{j}+  k^{ \rho}\widetilde{A}_{k,S}^{j} +  q^{\prime \rho}\widetilde{A}_{q',S}^{j} +  \widetilde{A}_{\epsilon^{*},S}^{ \rho j}+  \widetilde{A}_{\epsilon',S}^{ \rho j} \\ 
\Sigma^{\rho j}_{A} &=& k^{\prime \rho}\widetilde{A}_{k',A}^{j}+  k^{ \rho}\widetilde{A}_{k,A}^{j} +  q^{\prime \rho}\widetilde{A}_{q',A}^{j} +  \widetilde{A}_{\epsilon^{*},A}^{ \rho j}+  \widetilde{A}_{\epsilon',A}^{ \rho j} ,
\end{eqnarray}
where the $\tilde{A}$ coefficients are obtained by summing over the photon polarizations with $\Lambda_{\gamma}^{*} = 0$. The contraction with the symmetric and anti-symmetric components involves, in this case, a transverse index that is eventually contracted with the transverse index from the twist-three correlation function (see Section \ref{sec:DVCSBH_tw3}),
\begin{eqnarray}
k^{\rho}\widetilde{A}_{k,S}^{j} &=& (q+4\xi P)^{\beta}g_{T}^{\alpha j}\sum_{\Lambda_{\gamma}'} A_{k}^{\Lambda_{\gamma}^{*}=0,\Lambda_{\gamma}'}k^{\rho}\Big(\epsilon_{\alpha}^{\Lambda_{\gamma}'}\Big)^{*} \epsilon_{\beta}^{\Lambda_{\gamma}^{*}=0}\\
k^{\rho}\widetilde{A}_{k,A}^{j} &=&(q+4\xi P)^{\beta}\epsilon_{T}^{\alpha j}\sum_{\Lambda_{\gamma}'} A_{k}^{\Lambda_{\gamma}^{*}=0,\Lambda_{\gamma}'}k^{\rho} \Big(\epsilon_{\alpha}^{\Lambda_{\gamma}'}\Big)^{*} \epsilon_{\beta}^{\Lambda_{\gamma}^{*}=0}
\end{eqnarray}
Note that the longitudinally polarized photon which governs the twist three structure involves only the $0,3$ components of the momenta, $k, k', q'$. The twist three coefficients are, therefore, given by,
\begin{subequations}
\begin{eqnarray}
A_{UU}^{(3) \mathcal{I}} &=& -\frac{\xi }{2(kq')(k'q')}\frac{1}{P^{+}}\Bigg \{ (Q^{2}+t)\Big(2(k\Delta)_{T}(kP)_{L}(k'P) + (k'\Delta)_{T}(q'P)_{L}(k'P) + 2(k\Delta)_{T}(k'P)_{L}(kP) \nonumber \\&\qquad&+ (k'\Delta)_{T}(q'P)_{L}(kP)  - 2(kk')(k\Delta)_{T}(PP)_{L} + (k'q')(k\Delta)_{T}(PP)_{L} - (kq')(k'\Delta)_{T}(PP)_{L} \Big) \nonumber\\ &-& (Q^{2}-t+4(k\Delta))\Big((k'\Delta)_{T}(kP)_{L}(q'P) + (k'\Delta)_{T}(q'P)_{L}(q'P) + (kk')(q'P)_{L}(P\Delta)_{T} - (k'q')(kP)_{L}(P\Delta)_{T} \nonumber\\ &\qquad& - (kq')(k'P)_{L}(P\Delta)_{T} \Big) \Bigg\}\\
B_{UU}^{(3) \mathcal{I}} &=& -\frac{ \xi M^{2}}{(kq')(k'q')}\frac{1}{P^{+}} \Bigg\{ (Q^{2}+t) \Big(2(kP)_{L}(kk')_{T} + (Pq')_{L}(k'k')_{T} + 2(k'P)_{L}(kk)_{T} + (Pq')_{L}(kk')_{T} \Big) \nonumber\\
&-& (Q^{2}-t+4(k\Delta))\Big((kP)_{L}(k'q')_{T} + (Pq')_{L} (q'k')_{T} + 2((kk')(Pq')_L - (k'q')(Pk)_L - (kq')(Pk')_L)\Big)\Bigg\} \\
C_{UU}^{(3) \mathcal{I}} &=& \frac{\xi }{2(kq')(k'q')}\frac{1}{P^{+}}\Bigg\{(Q^{2}+t) \Big(2(k\Delta)_{T}(kP)_{L}(k'\Delta) + (k'\Delta)_{T}(q'P)_{L}(k'\Delta) + 2(k\Delta)_{T}(k'P)_{L}(k\Delta) \nonumber \\ &\qquad& + (k'\Delta)_{T}(q'P)_{L}(k\Delta) -  2(kk')(k\Delta)_{T}(P\Delta)_{L} + (k'q')(k\Delta)_{T}(P\Delta)_{L} - (kq')(k'\Delta)_{T}(P\Delta)_{L} \Big)\nonumber \\
&-&  (Q^{2}-t+4(k\Delta))\Big((k'\Delta)_{T}(kP)_{L}(q'\Delta) + (k'\Delta)_{T}(q'P)_{L}(q'\Delta) + (kk')(Pq')_{L}(\Delta\Delta)_{T} - (k'q')(kP)_{L}(\Delta\Delta)_{T}\nonumber  \\&\qquad& - (kq')(k'P)_{L}(\Delta\Delta)_{T}\Big)\Bigg\}
\end{eqnarray}
\end{subequations}
\begin{eqnarray}
\widetilde{A}_{UU}^{(3) \mathcal{I}} &=& - A_{UU}^{(3)\mathcal{I}}, \qquad
\widetilde{B}_{UU}^{(3) \mathcal{I}} = - B_{UU}^{(3)\mathcal{I}}, \qquad
\widetilde{C}_{UU}^{(3) \mathcal{I}}- C_{UU}^{(3)\mathcal{I}} 
\end{eqnarray}


\begin{eqnarray}
(AB)_{L} = A^{\mu}B^{\nu}g_{\mu \nu}^{L} = A^{0}B^{0}(1 + \frac{\nu^{2}}{Q^{2}}) + (A^{0}B^{3}+ A^{3}B^{0})\sqrt{\nu^2 + Q^2}\nu/Q^2 + A^{3}B^{3}\nu^2/Q^2
\end{eqnarray}
where $g_{\mu \nu}^{L}$ is defined following Eq. (229).

\subsubsection{$F_{LU}^{\cal I}$: Longitudinally Polarized Beam, Polarized Target}

\noindent For a longitudinally polarized beam and unpolarized target,
\begin{eqnarray}
F_{LU}^{\cal I} &= & F^{{\cal I}, tw 2}_{LU}  +  \frac{K}{\sqrt{Q^2}} F^{{\cal I}, tw 3}_{LU} ,
\end{eqnarray}
we obtain a structure analogous to the unpolarized case, where  the $\Re e$ parts of the CFFs are replaced with with the $\Im m$ parts,
namely,

\vspace{0.5cm}
\noindent{\em Twist Two}
\begin{subequations}
\begin{eqnarray}
F^{{\cal I} , tw2 }_{LU} 
 & = & A_{LU}^{\cal I} \Im m  \left(F_1 \mathcal{H} + \tau F_2  \mathcal{E} \right)  + B_{LU}^{\cal I}  G_M \, \Im m \left( \mathcal{H}+ \mathcal{E} \right)
 + C_{LU}^{\cal I} G_M \Im m \mathcal{ \widetilde{H}}
\end{eqnarray}

\vspace{0.5cm}
\noindent{\em Twist Three}
\begin{eqnarray}
F^{{\cal I} , tw3 }_{LU} & = & \Im m \Bigg\{ A^{ (3) \cal I}_{LU}  \Big[ F_{1}(2\mathcal{\widetilde{H}}_{2T} + \mathcal{E}_{2T}) + F_{2}( \mathcal{H}_{2T} + \tau \mathcal{\widetilde{H}}_{2T}) \Big]
 \nonumber \\
 && + B^{(3) \cal I}_{LU} G_M \, \widetilde{E}_{2T}   
+ C^{(3) \cal I}_{LU}  G_M \, \Big[2\xi  H_{2T} - \tau( \widetilde{E}_{2T} -\xi E_{2T} )  \Big]
\Bigg\} \nonumber \\
&+& \Im m \Bigg\{ \widetilde{A}^{(3) \cal I}_{LU}  \Big[F_{1}(2\widetilde{H}'_{2T} + E_{2T}') + F_{2}(H_{2T}' + \tau\widetilde{H}_{2T}' )\Big] \nonumber \\  && + \widetilde{B}^{(3) \cal I}_{LU}  G_M \widetilde{E}_{2T}'  +
 \widetilde{C}^{(3) \cal I}_{LU}  G_M \Big[2\xi  H_{2T}' - \tau( \widetilde{E}_{2T}' -\xi E_{2T}' )  \Big]\Bigg\}
\end{eqnarray}
\end{subequations}
The coefficients are obtained by contracting the antisymmetric part of the lepton tensor, $L_{{\cal I}, \, A}^{\Lambda_{\gamma^*} \Lambda_\gamma'}$ (Eq.\eqref{eq:L_int_anti}) 

\begin{equation}
A_{LU}^{\cal I}  = 2 (P D_{S})  \qquad
B_{LU}^{\cal I}   = \xi (\Delta D_{S} )\qquad
C_{LU}^{\cal I}  = \frac{1}{2P^{+}}\epsilon^{\mu \sigma \nu +}g_{\sigma \rho}\widetilde{t}_{\mu \nu}D_{A}^{ \rho}
\end{equation}

\begin{eqnarray}
A_{LU}^{(3)\cal I} &=& 4P_{\rho}\frac{\Delta_{j}}{P^{+}}\mathcal{D}_{S}^{\rho j} \qquad B_{LU}^{(3)\cal I} = g_{\rho j}\frac{2M^{2}}{P^{+}}\mathcal{D}_{S}^{\rho j}\qquad C_{LU}^{(3)\cal I} = - \Delta_{\rho}\frac{\Delta_{j}}{P^{+}}\mathcal{D}_{S}^{\rho j} \nonumber \\
\widetilde{A}_{LU}^{(3)\cal I} &=& -4P_{\rho}\frac{\epsilon_{ij}\Delta^{i}}{P^{+}}\mathcal{D}_{A}^{\rho j} \qquad \widetilde{B}_{LU}^{(3)\cal I} = -\epsilon_{ij}g_{\rho}^{i}\frac{2M^{2}}{P^{+}}\mathcal{D}_{A}^{\rho j} \qquad \widetilde{C}_{LU}^{(3)\cal I} =  \Delta_{\rho}\frac{\epsilon_{ij}\Delta^{i}}{P^{+}}\mathcal{D}_{A}^{\rho j}
\end{eqnarray}
Similarly with the anti-symmetric lepton tensor we get:

\begin{eqnarray}
\mathcal{D}_{S}^{ \rho j}&=& (q+4\xi P)^{\beta}g_{T}^{\alpha j} \sum_{\Lambda_{\gamma}'}L_{{\cal I}, \, A}^{\Lambda_{\gamma^*}=0 \, \Lambda_\gamma' \, \rho}\Big(\epsilon_{\alpha}^{\Lambda_{\gamma}'}\Big)^{*} \epsilon_{\beta}^{\Lambda_{\gamma}^{*}=0} \\
\mathcal{D}_{A}^{\rho j}&=&(q+4\xi P)^{\beta}\epsilon_{T}^{\alpha j} \sum_{\Lambda_{\gamma}'}L_{{\cal I}, \, A}^{\Lambda_{\gamma^*}=0 \, \Lambda_\gamma' \, \rho}\Big(\epsilon_{\alpha}^{\Lambda_{\gamma}'}\Big)^{*} \epsilon_{\beta}^{\Lambda_{\gamma}^{*}=0}
\end{eqnarray}

and for the antisymmetric terms, respectively,
\begin{subequations}
\begin{eqnarray}
\widetilde{A}_{k'}^{A} &=& 2D_{+}\Bigg[ (k \times q')_{T} \Bigg]\cos{\phi} + 2D_{+}\Bigg[2(kk)_{T} -(kq')_{T} + 2(kq') \Bigg]\sin{\phi}\\
\widetilde{A}_{k}^{A} &=& -2D_{+}\Bigg[ (k'\times q')_{T} \Bigg]\cos{\phi}  + 2D_{+}\Bigg[2(kk')_{T} + (k'q')_{T} - 2(k'q') \Bigg]\sin{\phi}\\
\widetilde{A}_{q'}^{A} &=& -2D_{-}\Bigg[ (k' \times q')_{T} \Bigg]\cos{\phi} + 2D_{-}\Bigg[(kk')_{T}+(k'q')_{T} - 2(kk') \Bigg]\sin{\phi}\\
\widetilde{A}_{\epsilon^{*}}^{A \rho} &=&   2D_{+}\Bigg[2(kk')k_{\alpha}\epsilon_{T}^{\alpha \rho} - (k'q')k_{\alpha}^{T}\epsilon_{T}^{\alpha \rho} - (kq')k^{\prime T}_{\alpha}\epsilon_{T}^{\alpha \rho} \Bigg]\cos{\phi}\nonumber \\
&-&2D_{+}\Bigg[2(kk')k^{\rho}_{T} - (k'q')k^{\rho}_{T} - (kq')k^{\prime \rho}_{T} \Bigg]\sin{\phi}\\
\widetilde{A}_{\epsilon '}^{A \rho}&=& -2D_{-}\Bigg[ (kk')q^{\prime T}_{\beta}\epsilon_{T}^{\rho \beta} - (k'q')k^{T}_{\beta}\epsilon_{T}^{\rho \beta} - (kq')k^{\prime T}_{\beta}\epsilon_{T}^{\rho \beta}\Bigg]\cos{\phi}\nonumber  \\ &+& 2D_{-}\Bigg[(kk')q^{\prime \rho}_{T} - (k'q')k^{\rho}_{T} - (kq')k^{\prime \rho}_{T} \Bigg]\sin{\phi} \, .
\end{eqnarray}
\end{subequations}

A similar derivation for the anti-symmetric part of the lepton tensor 
(denoted by the lepton helicity $h$ in the lower index of the $\widetilde{A}$), 
yields,
\begin{eqnarray}
D_{S}^{\rho} &=& -g_{T}^{\alpha \beta} \sum_{\Lambda_{\gamma}', \Lambda_{\gamma}^{*}}L_{{\cal I}, \, A}^{\Lambda_{\gamma^*} \Lambda_\gamma' \rho}\Big(\epsilon_{\alpha}^{\Lambda_{\gamma}'}\Big)^{*} \epsilon_{\beta}^{\Lambda_{\gamma}^{*}} \\
D_{A}^{\rho} &=&-\epsilon_{T}^{\alpha \beta} \sum_{\Lambda_{\gamma}', \Lambda_{\gamma}^{*}}L_{{\cal I}, \, A}^{\Lambda_{\gamma^*} \Lambda_\gamma' \rho }\Big(\epsilon_{\alpha}^{\Lambda_{\gamma}'}\Big)^{*} \epsilon_{\beta}^{\Lambda_{\gamma}^{*}} 
\end{eqnarray}

\subsubsection{$F_{UL}^{\cal I}$: Unpolarized Beam, Longitudinally Polarized Target}

\noindent Contracting the lepton tensor with the  symmetric, Eq.(\ref{FSpol}), and anti-symmetric, Eq.(\ref{FApol})), hadronic tensor components  for a longitudinally polarized proton yields,
\begin{eqnarray}
\label{eq:IntLU}
F_{UL}^{\cal I } &= &  F^{{\cal I }, tw 2}_{UL}    + \frac{K}{\sqrt{Q^2}} F^{{\cal I}, tw 3}_{UL} 
\end{eqnarray}
with,

\noindent{\em Twist Two}
\begin{subequations}
\begin{eqnarray}
 F_{UL}^{{\cal I}, tw 2} &=&  A_{UL}^{\cal I}  \Im {\rm m} \Big( F_{1} (\mathcal{\widetilde{H}} - \xi  \mathcal{\widetilde{E}} ) + \tau F_{2} \mathcal{\widetilde{E}} \Big) 
  + B_{UL}^{\cal I}   G_M \Im m \mathcal{\widetilde{H}}
 + C_{UL}^{\cal I}  
G_M \Im {\rm m} \left(\mathcal{H}+ \mathcal{E}\right)
\end{eqnarray}

\vspace{0.5cm}
\noindent{\em Twist Three}
\begin{eqnarray}
F_{UL}^{{\cal I}, tw 3} &=& \Im m \Big\{A_{UL}^{(3)\cal I}G_{M}\mathcal{H}_{2T} + B_{UL}^{(3)\cal I}F_{2}\mathcal{H}_{2T} \nonumber \\&& + C_{UL}^{(3) \cal I}G_{M}\mathcal{E}_{2T} + D_{UL}^{(3)\cal I}G_{M}(\xi \mathcal{E}_{2T} - \widetilde{\mathcal{E}}_{2T}) \Big \}\nonumber \\
&+&  \Im m \Big\{ \widetilde{A}_{UL}^{(3)\cal I}G_{M}\mathcal{H}_{2T}' + \widetilde{B}_{UL}^{(3)\cal I}F_{2}\mathcal{H}_{2T}' \nonumber  \\&& + \widetilde{C}_{UL}^{(3) \cal I}G_{M}\mathcal{E}_{2T}' + \widetilde{D}_{UL}^{(3)\cal I}G_{M}(\xi \mathcal{E}_{2T}' - \widetilde{\mathcal{E}}_{2T}') \Big \}
\end{eqnarray}
\end{subequations}

\begin{equation*}
A_{UL}^{\cal I}  = -2(P \Sigma_{A})  \qquad
B_{UL}^{\cal I}   = -\xi (\Delta \Sigma_{A}) \qquad
C_{UL}^{\cal I}  = -\frac{1}{2P^{+}}\epsilon^{\mu \sigma \nu +}g_{\sigma \rho}\widetilde{t}_{\mu \nu}\Sigma_{S}^{ \rho}
\end{equation*}

\begin{eqnarray*}
A_{UL}^{(3) \cal I} = \frac{4M^{2}}{(P^{+})^{2}}\epsilon^{\mu \sigma + k}g_{\sigma \rho}g_{jk}\Delta_{\mu}\Sigma^{ \rho j}_{S}\qquad
B_{UL}^{(3) \cal I} &=& - \frac{2}{(P^{+})^{2}}P_{\rho}\epsilon^{\mu \nu + k}g_{jk}\widetilde{t}_{\mu \nu}\Sigma^{\rho j}_{S} \qquad 
C_{UL}^{(3) \cal I} = \frac{\Delta_{j}}{(P^{+})^{2}}\epsilon^{\mu \sigma \nu +}g_{\sigma \rho}\widetilde{t}_{\mu \nu}\Sigma^{\rho j}_{S} \\
D_{UL}^{(3) \cal I} &=& \frac{4}{P^{+}}\epsilon^{\mu \sigma \nu k}g_{jk}g_{\sigma \rho}(P_{\mu}P_{\nu} + \frac{1}{2}\widetilde{t}_{\mu \nu})\Sigma^{\rho j}_{S} \\
\widetilde{A}_{UL}^{(3) \cal I} = \frac{4M^{2}}{(P^{+})^{2}}\epsilon_{ij}\epsilon^{\mu \sigma + i}g_{\sigma \rho}\Delta_{\mu} \Sigma^{ \rho j}_{A}\qquad
 \widetilde{B}_{UL}^{(3) \cal I} &=& - \frac{2}{(P^{+})^{2}}P_{\rho}\epsilon_{ij}\epsilon^{\mu \nu + i}\widetilde{t}_{\mu \nu} \Sigma^{ \rho j}_{A}
\qquad
\widetilde{C}_{UL}^{(3) \cal I} = \frac{\epsilon_{ij}\Delta^{i}}{(P^{+})^{2}}\epsilon^{\mu \sigma \nu +}g_{\sigma \rho}\widetilde{t}_{\mu \nu}\Sigma^{ \rho j}_{A}\\
\widetilde{D}_{UL}^{(3) \cal I} &=& \frac{4}{P^{+}}\epsilon_{ij}\epsilon^{\mu \sigma \nu i}g_{\sigma \rho}(P_{\mu}P_{\nu} + \frac{1}{2}\widetilde{t}_{\mu \nu})\Sigma^{\rho j}_{A}
\end{eqnarray*}

\subsubsection{$F_{LL}^{\cal I}$ Longitudinally Polarized Beam, Longitudinally Polarized Target}

For both longitudinally polarized beam and target we have a similar structure to Eq.(\ref{eq:IntLU}), where now the $\Im m$ part of the CFFs is replaced by their $\Re e$ part,
\begin{eqnarray}
F_{LL} &= & F^{{\cal I} \, tw 2}_{LL} + \frac{K}{\sqrt{Q^2}} F^{{\cal I}, tw 3}_{LL} 
\end{eqnarray}
with,

\vspace{0.5cm}
\noindent{\em Twist Two}
\begin{subequations}
\begin{eqnarray}
F_{LL}^{{\cal I}, tw2} & = & 
A_{LL}^{\cal I}  \Re {\rm e} \Big( F_{1} (\mathcal{\widetilde{H}} - \xi  \mathcal{\widetilde{E}} ) + \tau F_{2} \mathcal{\widetilde{E}} \Big) 
  + B_{LL}^{\cal I}   G_M \Re e \mathcal{\widetilde{H}}
 + C_{LL}^{\cal I}   
G_M \Re {\rm e} \left(\mathcal{H}+ \mathcal{E}\right)
 \end{eqnarray}
 
 \vspace{0.5cm}
\noindent{\em Twist Three}
\begin{eqnarray}
F_{LL}^{{\cal I},tw3} & = &  \Re e \Big\{A_{LL}^{(3)\cal I}G_{M}\mathcal{H}_{2T} + B_{LL}^{(3)\cal I}F_{2}\mathcal{H}_{2T} \nonumber\\&& + C_{LL}^{(3) \cal I}G_{M}\mathcal{E}_{2T} + D_{LL}^{(3)\cal I}G_{M}(\xi \mathcal{E}_{2T} - \widetilde{\mathcal{E}}_{2T}) \Big \}\nonumber\\
&+& \Re e \Big\{ \widetilde{A}_{LL}^{(3)\cal I}G_{M}\mathcal{H}_{2T}' + \widetilde{B}_{LL}^{(3)\cal I}F_{2}\mathcal{H}_{2T}' \nonumber\\&& + \widetilde{C}_{LL}^{(3) \cal I}G_{M}\mathcal{E}_{2T}' + \widetilde{D}_{LL}^{(3)\cal I}G_{M}(\xi \mathcal{E}_{2T}' - \widetilde{\mathcal{E}}_{2T}') \Big \}
\end{eqnarray}
\end{subequations}

\begin{equation}
A_{LL}^{\cal I} = -2(PD_{A}) \qquad
B_{LL}^{\cal I}=  - \xi( \Delta D_{A})\qquad
C_{LL}^{\cal I} = -\frac{1}{P^{+}}\epsilon^{\mu \sigma \nu +}g_{\sigma \rho}\widetilde{t}_{\mu \nu} D_{S}^{\rho}
\end{equation}

\begin{eqnarray}
A_{LL}^{(3) \cal I} = \frac{4M^{2}}{(P^{+})^{2}}\epsilon^{\mu \sigma + k}g_{\sigma \rho}g_{jk}\Delta_{\mu}\mathcal{D}_{S}^{\rho j}\qquad
B_{LL}^{(3) \cal I} &=& - \frac{2}{(P^{+})^{2}}P_{\rho}\epsilon^{\mu \nu + k}g_{jk}\widetilde{t}_{\mu \nu}\mathcal{D}_{S}^{\rho j} \qquad
C_{LL}^{(3) \cal I} = \frac{\Delta_{j}}{(P^{+})^{2}}\epsilon^{\mu \sigma \nu +}g_{\sigma \rho}\widetilde{t}_{\mu \nu} \mathcal{D}_{S}^{\rho j} \nonumber \\
D_{LL}^{(3) \cal I} &=& \frac{4}{P^{+}}\epsilon^{\mu \sigma \nu k}g_{jk}g_{\sigma \rho}(P_{\mu}P_{\nu} + \frac{1}{2}\widetilde{t}_{\mu \nu})\mathcal{D}_{S}^{\rho j} \nonumber \\
\widetilde{A}_{LL}^{(3) \cal I} = \frac{4M^{2}}{(P^{+})^{2}}\epsilon_{ij}\epsilon^{\mu \sigma + i}g_{\sigma \rho}\Delta_{\mu} \mathcal{D}_{A}^{\rho j} \qquad
 \widetilde{B}_{LL}^{(3) \cal I} &=& - \frac{2}{(P^{+})^{2}}P_{\rho}\epsilon_{ij}\epsilon^{\mu \nu + i}\widetilde{t}_{\mu \nu} \mathcal{D}_{A}^{\rho j}
\qquad
\widetilde{C}_{LL}^{(3) \cal I} = \frac{\epsilon_{ij}\Delta^{i}}{(P^{+})^{2}}\epsilon^{\mu \sigma \nu +}g_{\sigma \rho}\widetilde{t}_{\mu \nu}\mathcal{D}_{A}^{\rho j}\nonumber\\
\widetilde{D}_{LL}^{(3) \cal I} &=& \frac{4}{P^{+}}\epsilon_{ij}\epsilon^{\mu \sigma \nu i}g_{\sigma \rho}(P_{\mu}P_{\nu} + \frac{1}{2}\widetilde{t}_{\mu \nu})\mathcal{D}_{A}^{\rho j}
\end{eqnarray}

\subsubsection{$F_{UT}^{\cal I}$: Unpolarized Beam, Transversely Polarized Target, UT}

For an in plane transverse target polarization we have, for an unpolarized beam,

 \begin{eqnarray}
F_{UT_{x}}^{\cal I} &=&  \Big(\Sigma^{\rho}_{S} \cos \phi  + \Sigma^{\rho}_{A}  \sin \phi \Big) \Big\{  A_{T_{x}, \rho}^{\mathcal{I}} G_M \Im m \mathcal{E}  
+ B_{T_{x} \, \rho}^{\mathcal{I}} F_{2}\Im m( \mathcal{H} +  \mathcal{E}) 
 + C_{T_{x} \, \rho}^{\mathcal{I}} G_M \Im m ( \mathcal{H}+ \mathcal{E}) \nonumber \\&&  
 + \widetilde{A}_{T_{x}}^{\cal I}  G_M \Im m \mathcal{\widetilde{H}}  + \widetilde{B}_{T_{x}}^{\cal I} F_{2}\Im m \mathcal{\widetilde{H}}   + \widetilde{C}_{T_{x}}^{\cal I}  G_M \Im m \mathcal{\widetilde{E}}\Big \}
 \end{eqnarray}
 
 \begin{equation}
 \begin{split}
 A_{T_{x} \, \rho}^{\mathcal{I}}&= \frac{4}{M}\epsilon^{\mu \sigma \nu 1}g_{\sigma \rho}\Big[P_{\mu}P_{\nu}-\frac{1}{2}\widetilde{t}_{\mu \nu} \Big]\\
 B_{T_{x} \, \rho}^{\mathcal{I}} &= \frac{2}{MP^{+}}\epsilon^{\mu \nu 1 +}P_{\rho}\widetilde{t}_{\mu \nu}\\
 C_{T_{x} \, \rho}^{\mathcal{I}} &= \frac{4M}{P^{+}}\epsilon^{\mu \sigma 1 +}g_{\sigma \rho}\Delta_{\mu}
 \end{split}
 \qquad \qquad
 \begin{split}
 \widetilde{A}_{T_{x}}^{\mathcal{I}}&= -\frac{4M}{P^{+}}\Big[g^{1}_{\rho}\Delta^{+} - \Delta^{1}g_{\rho}^{+} \Big]\\
 \widetilde{B}_{T_{x}}^{\mathcal{I}} &= -\frac{4}{M}\Delta^{1}P_{\rho}\\
 \widetilde{C}_{T_{x}}^{\mathcal{I}} &= -\frac{2\xi}{M}\Delta^{1}\Delta_{\rho}
 \end{split}
 \end{equation}
 
 For an out of plane transverse target polarization we have, for an unpolarized beam,
 
 \begin{eqnarray}
F_{UT_{y}}^{\cal I} &=&  -\Big(\Sigma^{\rho}_{S} \cos \phi  + \Sigma^{\rho}_{A}  \sin \phi \Big) \Big\{  A_{T_{y}, \rho}^{\mathcal{I}} G_M \Im m \mathcal{E}  
+ B_{T_{y} \, \rho}^{\mathcal{I}} F_{2}\Im m( \mathcal{H} +  \mathcal{E}) 
 + C_{T_{y} \, \rho}^{\mathcal{I}} G_M \Im m ( \mathcal{H}+ \mathcal{E}) \nonumber \\&&  
 + \widetilde{A}_{T_{y}}^{\cal I}  G_M \Im m \mathcal{\widetilde{H}}  + \widetilde{B}_{T_{y}}^{\cal I} F_{2}\Im m \mathcal{\widetilde{H}}   + \widetilde{C}_{T_{y}}^{\cal I}  G_M \Im m \mathcal{\widetilde{E}}\Big \}
 \end{eqnarray}

 \begin{equation}
 \begin{split}
 A_{T_{y}}^{\mathcal{I}}&= \frac{4i}{M}\epsilon^{\mu \sigma \nu 2}g_{\sigma \rho}\Big[P_{\mu}P_{\nu}-\frac{1}{2}\widetilde{t}_{\mu \nu} \Big]\\
 B_{T_{y}}^{\mathcal{I}}&= \frac{2i}{MP^{+}}\epsilon^{\mu \nu 2 +}P_{\rho}\widetilde{t}_{\mu \nu}\\
 C_{T_{y}}^{\mathcal{I}}&= \frac{4iM}{P^{+}}\epsilon^{\mu \sigma 2 +}g_{\sigma \rho}\Delta_{\mu}
 \end{split}
 \qquad \qquad
 \begin{split}
 \widetilde{A}_{T_{y}}^{\mathcal{I}}&= -\frac{4M}{P^{+}}\Big[g^{2}_{\rho}\Delta^{+} - \Delta^{2}g_{\rho}^{+} \Big]\\
 \widetilde{B}_{T_{y}}^{\mathcal{I}} &= -\frac{4}{M}\Delta^{2}P_{\rho}\\
 \widetilde{C}_{T_{y}}^{\mathcal{I}} &= -\frac{2\xi}{M}\Delta^{2}\Delta_{\rho}
 \end{split}
 \end{equation}

\subsubsection{$F_{LT}^{\cal I}$: Longitudinally Polarized Beam, Transversely Polarized Target, LT}
For an in plane transverse target polarization we have, for a longitudinally polarized beam,

\begin{eqnarray}
F_{LT_{x}}^{\cal I} &=&  \Big(D^{\rho}_{S} \cos \phi  + D^{\rho}_{A}  \sin \phi \Big) \Big\{  A_{T_{x}, \rho}^{\mathcal{I}} G_M \Im m \mathcal{E}  
+ B_{T_{x} \, \rho}^{\mathcal{I}} F_{2}\Im m( \mathcal{H} +  \mathcal{E}) 
 + C_{T_{x} \, \rho}^{\mathcal{I}} G_M \Im m ( \mathcal{H}+ \mathcal{E}) \nonumber \\&&  
 + \widetilde{A}_{T_{x}}^{\cal I}  G_M \Im m \mathcal{\widetilde{H}}  + \widetilde{B}_{T_{x}}^{\cal I} F_{2}\Im m \mathcal{\widetilde{H}}   + \widetilde{C}_{T_{x}}^{\cal I}  G_M \Im m \mathcal{\widetilde{E}}\Big \}
\end{eqnarray}

For an out of plane transverse target polarization we have, for a longitudinally polarized beam,

\begin{eqnarray}
F_{LT_{y}}^{\cal I} &=&  -\Big(D^{\rho}_{S} \cos \phi  + D^{\rho}_{A}  \sin \phi \Big) \Big\{  A_{T_{y}, \rho}^{\mathcal{I}} G_M \Im m \mathcal{E}  
+ B_{T_{y} \, \rho}^{\mathcal{I}} F_{2}\Im m( \mathcal{H} +  \mathcal{E}) 
 + C_{T_{y} \, \rho}^{\mathcal{I}} G_M \Im m ( \mathcal{H}+ \mathcal{E}) \nonumber \\&&  
 + \widetilde{A}_{T_{y}}^{\cal I}  G_M \Im m \mathcal{\widetilde{H}}  + \widetilde{B}_{T_{y}}^{\cal I} F_{2}\Im m \mathcal{\widetilde{H}}   + \widetilde{C}_{T_{y}}^{\cal I}  G_M \Im m \mathcal{\widetilde{E}}\Big \}
\end{eqnarray}


\subsection{Discussion of Results}
A few remarks are in order:
\begin{enumerate}
    \item 
 For each polarization configuration, the structure of the cross section is clearly separated into its twist two and twist three contributions. The distinction between the two types of contributions is defined both by their different phase structure and by the $1/\sqrt{Q^2}$ suppression of the twist three term. 
The phase structure is such that all twist two terms appear with either a $\cos \phi$ or a $\sin \phi$ factor, while the twist three appears with no phase. In addition to the phase dependence on $\phi$, all coefficients depend on $\phi$ as dictated by the kinematics. This is at variance with the pure DVCS contributions (Sec.\ref{sec:3}).

\item The unpolarized target contribution to the cross section displays a generalized Rosenbluth form
 similar to the one describing the BH term.
Considering, for instance,  Eqs.\eqref{eq:Int_FUU},\eqref{eq:Int_FUU3} 
one can single out the electric and magnetic type contributions,
\[ (F_1 \, {\cal H} + \tau F_2 {\cal E } ) \leftrightarrow  (F_1^2 + \tau  F_2^2 ) \equiv G_E^2 \quad  \quad \quad (F_1 + F_2)({\cal H} + {\cal E}) \leftrightarrow  (F_1+F_2)^2 \equiv G_M^2. \]
However, differently from the elastic  and  BH scattering processes, a term containing the axial vector GPD, 
\[ \propto (F_1 + F_2) \, \widetilde{\cal H}\] 
is also present, as it is allowed by the DVCS helicity structure. It is interesting to notice that a contribution with the same structure ($G_M G_A$) appears both in elastic and BH scattering as a parity violating term. The GPD $\widetilde{E}$ decouples from the unpolarized measurement since it involves longitudinally polarized quarks in a transversely polarized target (Table \ref{observ:tab}). 

\item The unpolarized twist three structure constructed according to the scheme displayed in Table \ref{observ:tab}, contains the following terms:
\[ (2 \widetilde{H}_{2T} + E_{2T} ), \quad ({H}_{2T} + \tau \widetilde{H}_{2T} )  \quad \quad {\rm and} \quad\quad \widetilde{E}_{2T}' , \quad  ( {H}_{2T}' + \tau \widetilde{H}_{2T}' ). \]
$(2 \widetilde{H}_{2T} + E_{2T} )$ is the unpolarized twist three GPD and it  can be seen as the generalization of the Cahn TMD, $f_\perp$; analogously, $({H}_{2T} + \tau \widetilde{H}_{2T} )$ is the twist three contribution corresponding to the same polarization configuration of the GPD $E$. The latter is key in understanding final state interactions. In the axial vector sector, $\widetilde{E}_{2T}'$ corresponds to the twist two GPD $\widetilde{H}$, and ${H}_{2T}' + \tau \widetilde{H}_{2T}'$ to the GPD  $\widetilde{E}$.

\item Of particular interest are the GPDs 

\[ \widetilde{E}_{2T}, \quad\quad(2 \widetilde{H}_{2T}' + E_{2T}' ) \]

which are direct measurements of the quark contribution to the proton Orbital Angular Momentum, $L_z$, and Spin-Orbit interaction term,  $(L \cdot S)$, respectively \cite{Courtoy:2013oaa,Raja:2017xlo,Rajan:2016tlg}.

\end{enumerate}
Based on our formalism, other polarization configurations can be analyzed using a similar scheme as the one described in detail for the unpolarized case.

\section{Conclusions and Outlook}
\label{sec:conc}
DVCS gives unique access to the three-dimensional picture of the quarks and gluons lending great potential to unveil much more about the spin structure of the nucleon.  But of course this is only true if there is a more-less universal approach to the physical interpretation to leading power accuracy, and that the strategies for phenomenological analysis are practical with well defined uncertainties.   Here we have taken the fist steps on a long path to improve the prospects of information extraction and optimized data acquisition.

We have presented a generalized and comprehensive description of the cross section for DVCS scattering in terms of the helicity amplitude structures up to twist three accuracy in a covariant form.  Our transpicuous layout intends 
to help both devising theoretical constraints and implementing them in the analysis framework towards the goal 
of more accurately and completely extracting information at the amplitude level from both fixed target and collider data. 
We believe our presentation of the various beam target polarization contributions beyond leading twist with full azimuthal angular dependence allows for a more concise and intelligible representation in comparison to the ensconced Fourier harmonics.
The most important and direct consequence of our approach is that it allows us to organize both the BH and DVCS contributions to the cross section according to a generalized Rosenbluth formulation. The corresponding experimental extraction technique, the Rosenbluth separation method, has been used as a standard procedure for the extraction of the proton’s electromagnetic form factors since the inception, providing the highest precision determinations in Ref.\cite{Qattan:2004ht,Yurov}. Our approach opens up the possibility of analyzing DVCS data with a similar technique, unveiling terms that were previously disregarded, and that, as we explained herewith, measure directly the angular momentum contribution to the DVCS cross section.  {In this respect, notice that radiative corrections, namely two photon exchanges which have been advocated to play a role in elastic $ep$ scattering processes \cite{Afanasev:2017gsk}, might contribute and are a subject for future studies}. 

The step in thoroughness in the description of the DVCS observables is, therefore, critical for moving forward. There exists a distinct necessity for not only higher twist contributions to fully describe the experimental data, but also a transparent depiction of the phase dependence and interference terms at higher twist to obtain a more complete set of constraints.  We expect that the additional constraints combined with modern computational tools will provide the essential architecture to more accurately extract the Compton form factors.

The observables evaluated are presented in an exact treatment in all contributions apart from the operator product expansion in the hadronic tensor where dynamic twist three GPDs are considered, including kinematic power corrections.  Much attention was also given to the dependence on the azimuthal angle, $\phi$, of the different contributions to the cross section, disentangling the phase dependence resulting from the DVCS virtual photon polarization vectors from the pure kinematic $\phi$ dependence.  In this way the formalism presented in this paper offers the advantage of giving a transparent representation of both the phase structure and the $Q^2$ dependence of the cross section. In particular, the order in inverse powers of $Q$, corresponds to GPDs of different twist.

We intend for our formalism to be useful in both large and small-$x$ physics, enabling a more detailed study of the differences between forward and non-forward distributions over the full range of distance scales.  With consideration of kinematic suppression and observable interdependence as well as the relationship processes we expect the quality of nucleon imaging to radically improve in years to come.  To this extent the experimental and phenomenological approach must evolve together and take full advantage of machine intelligence.  Additional investigations in future publications including numerical analyses, and an extension of our approach to consider both recoil polarization in deeply virtual exclusive electron scattering, as well as timelike Compton scattering, will help to reveal this potential in more detail. 

\acknowledgements
This work was funded by DOE grants DOE grant DE-SC0016286 and in part by the DOE Topical Collaboration on TMDs (B.K. and S.L.), DOE grant DE-FG02-96ER40950 (L.C., D.K. and A.M.), and the SURA Center for Nuclear Femtohgraphy (CNF) grant. We thank Mauro Anselmino, Stepan Stepanian, Raphael Dupr\'e, Whitney Armstrong, Charles Hyde for comments and stimulating discussions and to Abha Rajan for participating in the early stages of this work.
\appendix
\section{Details of general definitions}
\label{appa}
\subsection{Dimensions}
The cross sections dimensions are in nb/GeV$^4$. The matrix elements modulus squared, $\mid T\mid^2$, carries dimensions of GeV$^{-2}$, or
$(\hbar c)^2 \times 10$ nb. 

A summary of the dimensions of the various contributions is presented in Table \ref{table:dim}.
In particular, the dimensions of $T_{DVCS}$ in GeV$^{-1}$ 
are carried by the factor $1/Q$ in  Eq.(\ref{eq:DVCS1b}). The lepton amplitude, $A_h^{\Lambda_{\gamma^*}}$ has dimension GeV (linear in $Q$), while the hadron amplitudes are dimensionless; the photon propagator carries dimensions GeV$^{-2}$, through the $1/Q^2$ factor, yielding GeV$^{-1}$ for the product. 
%

\begin{table}
\begin{tabular}{|c|c|c|c|c|}
\hline 
Process/Type & Lepton & Hadron & $\gamma$ propagator & Total \\
\hline 
$\mid BH \mid^2$ & 0 & $M^2$ &  $M^{-4}$ & $M^{-2}$ \\
\hline  
$\mid DVCS \mid^2$ & $M^2$ & 0 & $M^{-4}$ & $M^{-2}$ \\
\hline 
BH-DVCS & M & M & $M^{-4}$ & $M^{-2}$ \\
\hline
\end{tabular}
\caption{Dimensions of the lepton, hadron and photon propagator contributions to the cross section. The spinors are normalized to $2M$.}
\label{table:dim}
\end{table}


\section{GPDs in terms of helicity dependent correlation functions}
\label{appb}
By inverting Eqs.(\ref{helampftw2},\ref{eq:famps3}) we can identify the combination of proton helicities that describes each GPD.
\vspace{.5cm}

Twist-2 GPDs
\begin{eqnarray*}
\frac{1}{\sqrt{1-\xi^{2}}}\frac{\Delta_{T}}{M}E &=& e^{-i\phi}W_{+-}^{\gamma^{+}} - e^{i\phi}W_{-+}^{\gamma^{+}}\\
\frac{1}{\sqrt{1-\xi^{2}}}2(1-\xi^{2})H &=& W_{++}^{\gamma^{+}} + W_{--}^{\gamma^{+}}+ 2\xi^{2}\frac{M}{\Delta_{T}}\Big(e^{-i\phi}W_{+-}^{\gamma^{+}} - e^{i\phi}W_{-+}^{\gamma^{+}} \Big)\\
\frac{\xi}{\sqrt{1-\xi^{2}}}\frac{\Delta_{T}}{M}\widetilde{E} &=& e^{-i\phi}W_{+-}^{\gamma^{+}\gamma_{5}}+e^{i\phi}W_{-+}^{\gamma^{+}\gamma_{5}}\\
\frac{1}{\sqrt{1-\xi^{2}}}2(1-\xi^{2})\widetilde{H} &=& W_{++}^{\gamma^{+}\gamma_{5}} - W_{--}^{\gamma^{+}\gamma_{5}} + 2\xi \frac{M}{\Delta_{T}}\Big(e^{-i\phi}W_{+-}^{\gamma^{+}\gamma_{5}} + e^{i\phi}W_{-+}^{\gamma^{+}\gamma_{5}} \Big)
\end{eqnarray*}
\vspace{.5cm}
Twist-3 GPDs
\begin{eqnarray*}
\frac{1}{\sqrt{1-\xi^{2}}}\frac{\Delta_{T}}{P^{+}}(\widetilde{E}_{2T} - \xi E_{2T}) &=& \Big(W_{++}^{\gamma^{1}}+iW_{++}^{\gamma^{2}} \Big)e^{-i\phi} - \Big( W_{--}^{\gamma^{1}}+iW_{--}^{\gamma^{2}}\Big)e^{-i\phi}\\
\frac{1}{\sqrt{1-\xi^{2}}}\frac{\Delta_{T}}{P^{+}}(E_{2T} - \xi \widetilde{E}_{2T}) &=& \Big(W_{++}^{\gamma^{1}}+iW_{++}^{\gamma^{2}} \Big)e^{-i\phi} + \Big( W_{--}^{\gamma^{1}}+iW_{--}^{\gamma^{2}}\Big)e^{-i\phi}\\
&-&\frac{M}{\Delta_{T}} \Bigg(\Big(W_{-+}^{\gamma^{1}} - iW_{-+}^{\gamma^{2}} \Big)e^{2i\phi} - \Big(W_{+-}^{\gamma^{1}}+iW_{+-}^{\gamma^{2}} \Big)e^{-2i\phi}\Bigg) \\
\frac{1}{\sqrt{1-\xi^{2}}}\frac{\Delta_{T}^{2}}{MP^{+}}2\widetilde{H}_{2T} &=& \Big(W_{-+}^{\gamma^{1}} - iW_{-+}^{\gamma^{2}} \Big)e^{2i\phi} - \Big(W_{+-}^{\gamma^{1}}+iW_{+-}^{\gamma^{2}} \Big)e^{-2i\phi} \\
-\frac{1}{\sqrt{1-\xi^{2}}}\frac{2M}{P^{+}}(1-\xi^{2})2H_{2T} &=& \Big(W_{+-}^{\gamma^{1}} - i W_{+-}^{\gamma^{2}} \Big) - \Big(W_{-+}^{\gamma^{1}}+iW_{-+}^{\gamma^{2}} \Big) \\&-& 4\frac{M}{\Delta_{T}}e^{-i\phi}\Bigg(\Big(W_{++}^{\gamma^{1}}+iW_{++}^{\gamma^{2}}\Big) - \Big(W_{--}^{\gamma^{1}}+iW_{--}^{\gamma^{2}}  \Big)\Bigg) \\&+& \frac{1}{2}\Bigg( \Big(W_{-+}^{\gamma^{1}} - i W_{-+}^{\gamma^{2}} \Big)e^{2i\phi} - \Big(W_{+-}^{\gamma^{1}} + i W_{+-}^{\gamma^{2}} \Big)e^{-2i\phi} \Bigg)\\
\frac{1}{\sqrt{1-\xi^{2}}}\frac{\Delta_{T}}{P^{+}}(E_{2T}' - \xi \widetilde{E}_{2T}') &=& e^{-i\phi}\Big(W_{++}^{\gamma^{1}\gamma_{5}}+iW_{++}^{\gamma^{2}\gamma_{5}}\Big) +e^{-i\phi}\Big(W_{--}^{\gamma^{1}\gamma_{5}} + i W_{--}^{\gamma^{2}\gamma_{5}} \Big)\\
&+& \frac{M}{\Delta_{T}}\Bigg(\Big(W_{+-}^{\gamma^{1}\gamma_{5}}+ i W_{+-}^{\gamma^{2}\gamma_{5}} \Big)e^{-2i\phi} + \Big(W_{-+}^{\gamma^{1}\gamma_{5}} - i W_{-+}^{\gamma^{2}\gamma_{5}} \Big)e^{2i\phi} \Bigg)\\
\frac{1}{\sqrt{1-\xi^{2}}}\frac{\Delta_{T}}{P^{+}}(\widetilde{E}_{2T}' - \xi E_{2T}') &=& e^{-i\phi}\Big( W_{++}^{\gamma^{1}\gamma_{5}}+iW_{++}^{\gamma^{2}\gamma_{5}}\Big) - e^{-i\phi}\Big(W_{--}^{\gamma^{1}\gamma_{5}} + i W_{--}^{\gamma^{2}\gamma_{5}} \Big)\\
-\frac{1}{\sqrt{1-\xi^{2}}}\frac{\Delta_{T}^{2}}{MP^{+}}2\widetilde{H}_{2T}' &=& \Big(W_{+-}^{\gamma^{1}\gamma_{5}}+ i W_{+-}^{\gamma^{2}\gamma_{5}} \Big)e^{-2i\phi} + \Big(W_{-+}^{\gamma^{1}\gamma_{5}} - i W_{-+}^{\gamma^{2}\gamma_{5}} \Big)e^{2i\phi}
\end{eqnarray*}

\section{Detailed Lepton and Hadron Helicity Structure of the DVCS Cross Section}
\label{appc}
We write the cross section in Eq.\eqref{eq:phase1p1}
in terms of the helicity dependent structure functions,
\begin{eqnarray}
\label{eq:phase3}
\sigma_{h\Lambda}
& = & \frac{1}{Q^2} \frac{1}{1-\epsilon} \Big\{4 (F_{\Lambda +}^{1 1} + F_{\Lambda -}^{1 1} +  F_{\Lambda +}^{-1 -1} + F_{\Lambda -}^{-1 -1} )+ 
2 \epsilon  (F_{\Lambda +}^{0 0} + F_{\Lambda -}^{0 0} )
\nonumber \\
&+&   \sqrt{2} \sqrt{\epsilon(1+\epsilon)}  \,  2 \Big[ \cos \phi \, \Re {\rm e}   (- F_{\Lambda +}^{0 1} - F_{\Lambda -}^{0 1} +  F_{\Lambda +}^{0 -1} + F_{\Lambda -}^{0 -1}  ) \nonumber \\
&+& \sin \phi \, \Im {\rm m}   (-F_{\Lambda +}^{0 1} - F_{\Lambda -}^{0 1} -  F_{\Lambda +}^{0 -1} - F_{\Lambda -}^{0 -1}  ) \Big]
\nonumber \\
 &- & 2 \epsilon\,\left[ \cos 2 \phi  \, \Re {\rm e}    (F_{\Lambda +}^{1 -1} + F_{\Lambda -}^{1 -1}) - \sin 2 \phi  \, \Im {\rm m}   (F_{\Lambda +}^{1 -1} + F_{\Lambda -}^{1 -1}) \right] \Big\}
\nonumber \\ 
&+& 2 (2h) \Big\{ \sqrt{1-\epsilon^2} \, (F_{\Lambda +}^{1 1} + F_{\Lambda -}^{1 1} -  F_{\Lambda +}^{-1 -1} - F_{\Lambda -}^{-1 -1} ) 
\nonumber \\ 
& - &   \sqrt{\epsilon (1-\epsilon)} \,  \Big[ \cos \phi \, \Re {\rm e}   (F_{\Lambda +}^{0 1} + F_{\Lambda -}^{0 1} + F_{\Lambda +}^{0 -1} + F_{\Lambda -}^{0 -1}  )     
\nonumber 
\\
& + &  \sin \phi \, \Im {\rm m}   (F_{\Lambda +}^{0 1} + F_{\Lambda -}^{0 1} - F_{\Lambda +}^{0 -1} - F_{\Lambda -}^{0 -1}
   \,\Big] \Big\} .
\end{eqnarray}

\noindent 
{\it Transverse polarization}

\noindent To obtain a similar structure for the transversely polarized target in the $\phi_S$ direction 
using the density matrix in Eq.\eqref{densitymatrix}, we define the polarized cross section as, 
\begin{eqnarray}
\frac{1}{2}(\sigma_{h,\Lambda_T=+}^T -\sigma_{h,\Lambda_T=-}^T)&=& 
e^{+i(\phi_s)}  \sum_{\Lambda^\prime,  \Lambda^\prime_\gamma} \left( T^{h  \Lambda^\prime_\gamma}_{DVCS, + \Lambda^\prime} \right)^* \, T^{h  \Lambda^\prime_\gamma}_{DVCS,- \Lambda^\prime} + e^{-i(\phi_s)}  \sum_{\Lambda^\prime,  \Lambda^\prime_\gamma} \left( T^{h  \Lambda^\prime_\gamma}_{DVCS,  - \Lambda^\prime} \right)^* \, T^{h  \Lambda^\prime_\gamma}_{DVCS, +\Lambda^\prime}  \nonumber \\
&=&
2\Re {\rm e} {\Big\{e^{+i(\phi_s)}  \sum_{\Lambda^\prime,  \Lambda^\prime_\gamma} \left( T^{h  \Lambda^\prime_\gamma}_{DVCS, + \Lambda^\prime} \right)^* \, T^{h  \Lambda^\prime_\gamma}_{DVCS,- \Lambda^\prime} }\Big\} \equiv \sigma^T_{h,\Delta\Lambda_T}
\end{eqnarray}
Working out the terms, one has,
\begin{eqnarray}
&\sigma_{h \Lambda_T=+}^{(1)} = & \sum_{\Lambda^\prime,  \Lambda^\prime_\gamma} \left( T^{h  \Lambda^\prime_\gamma}_{DVCS, + \Lambda^\prime} \right)^* \, T^{h  \Lambda^\prime_\gamma}_{DVCS,- \Lambda^\prime} =   \sum_{ \Lambda^\prime_\gamma, \, \Lambda^\prime}  \sum_{\Lambda_{{\gamma^*}}^{(1)}} \left[ A_h^{\Lambda_{{\gamma^*}}^{(1)}} f_{+, \Lambda'}^{\Lambda_{{\gamma^*}}^{(1)} ,\Lambda_\gamma^\prime}\right]^*\sum_{\Lambda_{{\gamma^*}}^{(2)}} A_h^{\Lambda_{{\gamma^*}}^{(2)}} f_{-, \Lambda'}^{\Lambda_{{\gamma^*}}^{(2)} ,\Lambda_\gamma^\prime}
 \nonumber \\  
& = &  \sum_{ \Lambda^\prime_\gamma, \, \Lambda^\prime}  \left(A_h^{1} f_{+, \Lambda^\prime}^{1,\Lambda_\gamma'} + A_h^{-1} f_{+, \Lambda^\prime}^{-1,\Lambda_\gamma'} + A_h^{0} f_{+, \Lambda^\prime}^{0,\Lambda_\gamma'} \right)^* \left(A_h^{1} f_{-, \Lambda^\prime}^{1,\Lambda_\gamma'} + A_h^{-1} f_{-, \Lambda^\prime}^{-1,\Lambda_\gamma'} + A_h^{0} f_{-, \Lambda^\prime}^{0,\Lambda_\gamma'} \right) \nonumber \\
& = & \sum_{ \Lambda^\prime_\gamma, \, \Lambda^\prime}%
(A_h^{1} )^2  \left(f_{+, \Lambda^\prime}^{1,\Lambda_\gamma'}  \right)^* f_{-, \Lambda^\prime}^{1,\Lambda_\gamma'}  + (A_h^{-1} )^2 \left(f_{+, \Lambda^\prime}^{-1,\Lambda_\gamma'}  \right)^* f_{-, \Lambda^\prime}^{-1,\Lambda_\gamma'} + ( A_h^{0} )^2 \left(f_{+, \Lambda^\prime}^{0,\Lambda_\gamma'}  \right)^* f_{-, \Lambda^\prime}^{0,\Lambda_\gamma'}  \nonumber \\
& + & A_h^{1} A_h^{0} \left[ \left(f_{+, \Lambda^\prime}^{1,\Lambda_\gamma'}  \right)^* f_{-, \Lambda^\prime}^{0,\Lambda_\gamma'} + \left(f_{+, \Lambda^\prime}^{0,\Lambda_\gamma'}  \right)^* f_{-, \Lambda^\prime}^{1,\Lambda_\gamma'} \right] +A_h^{-1}  A_h^{0} \left[  \left(  f_{+, \Lambda^\prime}^{-1,\Lambda_\gamma'} \right)^*f_{-, \Lambda^\prime}^{0,\Lambda_\gamma'} +   \left(f_{+, \Lambda^\prime}^{0,\Lambda_\gamma'}  \right)^* f_{-, \Lambda^\prime}^{-1,\Lambda_\gamma'}  \right]  \nonumber \\
&+ & A_h^{1} A_h^{-1} \left[ \left(f_{+, \Lambda^\prime}^{1,\Lambda_\gamma'} \right)^*f_{-, \Lambda^\prime}^{-1,\Lambda_\gamma'} + \left(f_{+, \Lambda^\prime}^{-1,\Lambda_\gamma'} \right)^*f_{-, \Lambda^\prime}^{1,\Lambda_\gamma'}  \right],
\label{eq:phase2}
\end{eqnarray}
Evaluating the lepton process amplitudes, $A_h^{\Lambda_{\gamma^*}}$ we have,
\begin{eqnarray}
&\sigma_{h \Lambda_T=+}^{(1)} = & 
\frac{1}{Q^2} \frac{1}{1-\epsilon} \Big\{2  \sum_{  \Lambda^\prime_\gamma, \, \Lambda^\prime}%
\left[ (f_{+ \Lambda^\prime}^{1 \Lambda_\gamma'})^* f_{- \Lambda^\prime}^{1 \Lambda_\gamma'} +  (f_{+ \Lambda^\prime}^{-1 \Lambda_\gamma'})^* f_{- \Lambda^\prime}^{-1 \Lambda_\gamma'} \right] 
+ 2 \epsilon \sum_{  \Lambda^\prime_\gamma, \, \Lambda^\prime} (f_{+ \Lambda^\prime}^{0\Lambda_\gamma'})^*f_{- \Lambda^\prime}^{0\Lambda_\gamma'}  \nonumber \\
& + & \sqrt{2}\sqrt{\epsilon(1+\epsilon)} \sum_{  \Lambda^\prime_\gamma, \, \Lambda^\prime}\left[ -\left(f_{+ \Lambda^\prime}^{1,\Lambda_\gamma'}  \right)^* f_{- \Lambda^\prime}^{0,\Lambda_\gamma'} - \left(f_{+ \Lambda^\prime}^{0,\Lambda_\gamma'}  \right)^* f_{- \Lambda^\prime}^{1,\Lambda_\gamma'}  +   \left(  f_{+ \Lambda^\prime}^{-1,\Lambda_\gamma'} \right)^*f_{- \Lambda^\prime}^{0,\Lambda_\gamma'} +   \left(f_{+ \Lambda^\prime}^{0,\Lambda_\gamma'}  \right)^* f_{-\Lambda^\prime}^{-1,\Lambda_\gamma'}  \right]  \nonumber \\
&- & 2 \epsilon \sum_{  \Lambda^\prime_\gamma, \, \Lambda^\prime}\left[ \left(f_{+ \Lambda^\prime}^{1,\Lambda_\gamma'} \right)^*f_{- \Lambda^\prime}^{-1,\Lambda_\gamma'} + \left(f_{+ \Lambda^\prime}^{-1,\Lambda_\gamma'} \right)^*f_{-\Lambda^\prime}^{1,\Lambda_\gamma'}  \right] \Big\}
\nonumber \\ 
& + & (2h) \Big\{ 2 \sqrt{1-\epsilon^2} \sum_{  \Lambda^\prime_\gamma, \, \Lambda^\prime}%
\left( (f_{+ \Lambda^\prime}^{1 \Lambda_\gamma'})^* f_{- \Lambda^\prime}^{1 \Lambda_\gamma'} -  (f_{+ \Lambda^\prime}^{-1 \Lambda_\gamma'})^* f_{- \Lambda^\prime}^{-1 \Lambda_\gamma'} \right) \nonumber \\
& - & \sqrt{2} \sqrt{\epsilon(1-\epsilon)} \sum_{  \Lambda^\prime_\gamma, \, \Lambda^\prime}\left[ \left(f_{+ \Lambda^\prime}^{1,\Lambda_\gamma'}  \right)^* f_{- \Lambda^\prime}^{0,\Lambda_\gamma'} + \left(f_{+ \Lambda^\prime}^{0,\Lambda_\gamma'}  \right)^* f_{- \Lambda^\prime}^{1,\Lambda_\gamma'}  +   \left(  f_{+ \Lambda^\prime}^{-1,\Lambda_\gamma'} \right)^*f_{- \Lambda^\prime}^{0,\Lambda_\gamma'} +   \left(f_{+ \Lambda^\prime}^{0,\Lambda_\gamma'}  \right)^* f_{-\Lambda^\prime}^{-1,\Lambda_\gamma'}  \right] \Big\} \nonumber \\
\end{eqnarray}
Similarly to the longitudinally polarized case we now write the cross section in terms of structure functions. We have, 

\begin{eqnarray}
\sigma^{(1)}_{h \Lambda_{T}=+} &=& \frac{1}{Q^{2}}\frac{1}{1-\epsilon} e^{-i\phi}\Big \{ 2 \Big(\widetilde{F}_{++,T}^{11}  + \widetilde{F}_{+-,T}^{11} +\widetilde{F}_{++,T}^{-1-1}  + \widetilde{F}_{+-,T}^{-1-1} \Big) + 2\epsilon \Big(\widetilde{F}_{++,T}^{00}  + \widetilde{F}_{+-,T}^{00}\Big)  \nonumber \\ 
 &+& \sqrt{2}\sqrt{\epsilon(1+\epsilon)}\Big(-e^{i\phi}\widetilde{F}_{++,T}^{10} -e^{i\phi}\widetilde{F}_{+-,T}^{10} + e^{i\phi}\widetilde{F}_{++,T}^{0 -1} + e^{i\phi}\widetilde{F}_{+-,T}^{0 -1} \nonumber \\&-& e^{-i\phi}\widetilde{F}_{++,T}^{01} - e^{-i\phi}\widetilde{F}_{+-,T}^{01} + e^{-i\phi}\widetilde{F}_{++,T}^{-1 0} + e^{-i\phi}\widetilde{F}_{+-,T}^{-1 0} \Big)
\nonumber \\
 &-& 2\epsilon \Big(e^{2i\phi}\widetilde{F}_{++,T}^{1-1 } +e^{2i\phi}\widetilde{F}_{+-,T}^{1-1} + e^{-2i\phi}\widetilde{F}_{++,T}^{-11}+e^{-2i\phi}\widetilde{F}_{+-,T}^{-11} \Big)
 \nonumber  \\&+& (2h)
 \Big [ 2\sqrt{1-\epsilon^{2}}\Big(\widetilde{F}_{++,T}^{11} +\widetilde{F}_{+-,T}^{11}  - \widetilde{F}_{++,T}^{-1-1} -\widetilde{F}_{+-,T}^{-1-1} \Big) \nonumber \\
 &-& \sqrt{2}\sqrt{\epsilon(1-\epsilon)}\Big(e^{i\phi}\widetilde{F}_{++,T}^{10} + e^{i\phi}\widetilde{F}_{+-,T}^{10}+ e^{i\phi}\widetilde{F}_{++,T}^{0-1}+e^{i\phi}\widetilde{F}_{+-,T}^{0-1} \nonumber \\&+& e^{-i\phi}\widetilde{F}_{++,T}^{-10}+e^{-i\phi}\widetilde{F}_{+-,T}^{-10} + e^{-i\phi}\widetilde{F}_{++,T}^{01} +e^{-i\phi}\widetilde{F}_{+-,T}^{01} \Big) \Big ] \Big \}
\end{eqnarray}
Now with the terms organized with common phase factors,using parity and Hermitian conjugation, all the transverse target structure functions can be written in terms of the virtual photon helicities with $\Lambda_\gamma*^{(1)} = +1$ or 0 and $\Lambda_\gamma*^{(2)} =+1, 0 , -1$.   Then the single $S_T$ terms become $2i \mathcal{I}$m($\widetilde{F}_{\Lambda, \Lambda^\prime,T}^{\Lambda_\gamma*^{(1)}, \Lambda_\gamma*^{(2)}})$, while the double polarization terms reduce to $2 \mathcal{R}$e($\widetilde{F}_{\Lambda, \Lambda^\prime,T}^{\Lambda_\gamma*^{(1)}, \Lambda_\gamma*^{(2)})}$, 
\begin{eqnarray}
\sigma^T_{h,\Delta\Lambda_T}&=& 
2\Re {\rm e} {\Big\{e^{+i(\phi_s)}  \sum_{\Lambda^\prime,  \Lambda^\prime_\gamma} \left( T^{h  \Lambda^\prime_\gamma}_{DVCS, + \Lambda^\prime} \right)^* \, T^{h  \Lambda^\prime_\gamma}_{DVCS,- \Lambda^\prime} }\Big\} \nonumber \\
&=&2\Re {\rm e} {\Big\{e^{+i(\phi_s)} \sigma_{h,\Lambda_T=+}^{(1)}} \Big \}  \nonumber \\
&=& \frac{1}{Q^{2}}\frac{2}{1-\epsilon}\Big\{2\epsilon\cos{(\phi_{S}-\phi)}\Big(\Re e \widetilde{F}_{++,T}^{00} +\Re e \widetilde{F}_{+-,T}^{00}  \Big) \nonumber\\&-& 2\epsilon \sin{(\phi_{S}-\phi)}\Big( \Im m\widetilde{F}_{++,T}^{00} +\Im m\widetilde{F}_{+-,T}^{00} \Big) \nonumber \\
&+& 2 \cos{(\phi_{S}-\phi)}\Big( \Re e\widetilde{F}_{++,T}^{11} +\Re e\widetilde{F}_{+-,T}^{11} \Big) \nonumber
\\
&-& 2 \sin{(\phi_{S}-\phi)}\Big( \Im m\widetilde{F}_{++,T}^{11} +\Im m\widetilde{F}_{+-,T}^{11} \Big) \nonumber
\\
&+& 2 \cos{(\phi_{S}-\phi)}\Big( \Re e\widetilde{F}_{++,T}^{-1-1} +\Re e\widetilde{F}_{+-,T}^{-1-1} \Big) \nonumber
\\
&-& 2 \sin{(\phi_{S}-\phi)}\Big( \Im m\widetilde{F}_{++,T}^{-1-1} +\Im m\widetilde{F}_{+-,T}^{-1-1} \Big) \nonumber
\\
&-&2\epsilon \cos{(\phi + \phi_{S})}\Big(\Re e \widetilde{F}_{++,T}^{1-1} + \Re e \widetilde{F}_{+-,T}^{1-1}  \Big) \nonumber
\\&+&2\epsilon \sin{(\phi + \phi_{S})}\Big( \Im m \widetilde{F}_{++,T}^{1-1} + \Im m \widetilde{F}_{+-,T}^{1-1}  \Big) \nonumber
\\&-&2\epsilon \cos{(-3\phi + \phi_{S})}\Big( \Re e \widetilde{F}_{++,T}^{-11} + \Re e \widetilde{F}_{+-,T}^{-11} \Big) \nonumber
\\&+&2\epsilon \sin{(-3\phi + \phi_{S})}\Big(-\Im m \widetilde{F}_{++,T}^{-11} - \Im m \widetilde{F}_{+-,T}^{-11} \Big) \Big\} \nonumber
\\&+& (2h) \Big \{ 
2\sqrt{1-\epsilon^{2}}\cos{(\phi_{S}-\phi)}\Big(2\Re e \widetilde{F}_{++,T}^{11} + 2\Re e \widetilde{F}_{+-,T}^{11}  \Big)  \nonumber
\\&+& 2\sqrt{1-\epsilon^{2}}\sin{(\phi_{S}-\phi)}\Big(-2\Im m \widetilde{F}_{++,T}^{11} -2\Im m \widetilde{F}_{+-,T}^{11} \Big) \nonumber
\\&-& \sqrt{2}\sqrt{\epsilon(1-\epsilon)} \cos{(\phi_S -2 \phi)} \Big(2\Re e \widetilde{F}_{++,T}^{01} + 2\Re e \widetilde{F}_{+-,T}^{01}  \Big) \nonumber
\\&-& \sqrt{2}\sqrt{\epsilon(1-\epsilon)} \sin{(\phi_S -2 \phi)} \Big( - 2\Im m \widetilde{F}_{++,T}^{01} - 2\Im m \widetilde{F}_{+-,T}^{01}   \Big)\nonumber
\\&-& \sqrt{2}\sqrt{\epsilon(1-\epsilon)} \cos{(\phi_S)} \Big( 2\Re e \widetilde{F}_{++,T}^{10} + 2\Re e \widetilde{F}_{+-,T}^{10}  \Big) \nonumber
\\&-& \sqrt{2}\sqrt{\epsilon(1-\epsilon)} \sin{( \phi_{S})} \Big(2\Im m  \widetilde{F}_{++,T}^{10} + 2\Im m \widetilde{F}_{+-,T}^{10} \Big) 
\Big\}
\end{eqnarray}
Imposing as parity conservation and Hermiticity 
one finds that the single transverse polarization involves the imaginary parts of the bilinear product of helicity amplitudes while the beam and target polarizations involve the real parts. The final cross section expression reads,
\begin{eqnarray}
\sigma_{h,\Lambda_T=+}^T 
&=& \frac{1}{Q^{2}}\frac{2}{1-\epsilon}\Big\{-2\epsilon \sin{(\phi_{S}-\phi)}\Big( \Im m\widetilde{F}_{++,T}^{00} +\Im m\widetilde{F}_{+-,T}^{00} \Big) \nonumber \\
&-& 2 \sin{(\phi_{S}-\phi)}\Big( \Im m\widetilde{F}_{++,T}^{11} +\Im m\widetilde{F}_{+-,T}^{11} \Big) \nonumber \\
&+&2\epsilon \sin{(\phi + \phi_{S})}\Big( \Im m \widetilde{F}_{++,T}^{1-1} + \Im m \widetilde{F}_{+-,T}^{1-1}  \Big) \nonumber
\\&+&2\epsilon \sin{(-3\phi + \phi_{S})}\Big(-\Im m \widetilde{F}_{++,T}^{-11} - \Im m \widetilde{F}_{+-,T}^{-11} \Big)  \nonumber \\
&-& \sqrt{2}\sqrt{\epsilon(1+\epsilon)} \sin{(\phi_S -2 \phi)} \Big(2\Im m \widetilde{F}_{++,T}^{01} + 2\Im m \widetilde{F}_{+-,T}^{01}  \Big) \nonumber
\\&-& \sqrt{2}\sqrt{\epsilon(1+\epsilon)} \sin{(\phi_S)} \Big( 2\Im m \widetilde{F}_{++,T}^{10} + 2\Im m \widetilde{F}_{+-,T}^{10}  \Big) \Big\} \nonumber
\\&+& (2h) \Big \{ 
2\sqrt{1-\epsilon^{2}}\cos{(\phi_{S}-\phi)}\Big(2\Re e \widetilde{F}_{++,T}^{11} + 2\Re e \widetilde{F}_{+-,T}^{11}  \Big)  \nonumber
\\&-& \sqrt{2}\sqrt{\epsilon(1-\epsilon)} \cos{(\phi_S -2 \phi)} \Big(2\Re e \widetilde{F}_{++,T}^{01} + 2\Re e \widetilde{F}_{+-,T}^{01}  \Big) \nonumber
\\&-& \sqrt{2}\sqrt{\epsilon(1-\epsilon)} \cos{(\phi_S)} \Big( 2\Re e \widetilde{F}_{++,T}^{10} + 2\Re e \widetilde{F}_{+-,T}^{10}  \Big) \Big \}
\end{eqnarray}
the terms in this expression 
can be read directly to provide the transverse structure functions in section III, Eq.\eqref{eq:xs5fold}. 

\section{Details of the BH cross section calculation for specific $\Lambda_\gamma'$ polarization}
\label{app:BH}
For future applications, in this section we provide the contributions to the BH cross section for an unpolarized outgoing photon. 
\subsection{Unpolarized Target}
The contraction in Eq.(\ref{TBH2_unpol}), results in the following terms,
\begin{align}
\sigmaunpol{1}{1}=&
\hspace{12pt}
16\,M^2 |C|^2 
\hspace{10pt}
\Big\{W_1 \Big[ 2 \myprod{k}{k'}\Big] +  \frac{W_2}{M^2} \Big[2 \myprod{k}{\bar{P}} \myprod{k'}{\bar{P}}-M^2 (\tau +1) \myprod{k}{k'}\Big] \Big\}\\ 
\sigmaunpol{1}{2}=&
\hspace{12pt}
16\,M^2 C D_- 
\hspace{6pt}
\Big\{\frac{W_2}{M^2} \Big[\myprod{\bar{P}}
{q'} \left(\myprod{k}{\bar{P}} \myprod{k'}{\realeps{}{}}+\myprod{k}{\realeps{}{}} \myprod{k'}{\bar{P}}\right)
\nonumber \\ & \hspace{72pt} 
-\myprod{\bar{P}}{\realeps{}{}} \left(\myprod{k}{\bar{P}} \myprod{k'}{q'}+\myprod{k}{q'} \myprod{k'}{\bar{P}}\right)\Big]
\Big\}\\ 
\sigmaunpol{1}{3}=&
\hspace{12pt}
16\,M^2 C D_+ 
\hspace{6pt}
\Big\{
 W_1 \Big[ 2 \myprod{k}{q'} \myprod{k'}{\realeps{}{}}- 2 \myprod{k}{\realeps{}{}} \myprod{k'}{q'}\Big]
\nonumber \\ & \hspace{58pt} 
-\frac{W_2}{M^2}\Big[\myprod{\bar{P}}{q'} \left(\myprod{k}
{\realeps{}{}} \myprod{k'}{\bar{P}}-\myprod{k}{\bar{P}} \myprod{k'}{\realeps{}{}}\right)
\nonumber \\ & \hspace{80pt} 
+\myprod{\bar{P}}{\realeps{}{}} \left(\myprod{k}{\bar{P}} \myprod{k'}{q'}-\myprod{k}{q'} \myprod{k'}{\bar{P}
}\right)
\nonumber \\ & \hspace{80pt} 
-M^2 (1+\tau) \left(\myprod{k}{\realeps{}{}} \myprod{k'}{q'}-\myprod{k}{q'} \myprod{k'}{\realeps{}{}}\right)\Big]
\Big\}\\ 
\sigmaunpol{2}{2}=&
\hspace{12pt}
16\,M^2  D_-^2 
\hspace{14pt}
\Big\{
W_1 \Big[2 \myprod{k}{q'} \myprod{k'}{q'}\Big] 
\nonumber \\ & \hspace{58pt} 
- \frac{W_2}{M^2}\Big[
\myprod{k}{\realeps{*}{}} \myprod{\bar{P}}{q'} \myprod{\bar{P}}{\realeps{}{}} \myprod{k'}{q'}+\myprod{k}{q'} \myprod{\bar{P}}
{q'} \myprod{\bar{P}}{\realeps{}{}} \myprod{k'}{\realeps{*}{}}
\nonumber \\ & \hspace{80pt} 
+\myprod{k}{\realeps{}{}} \myprod{\bar{P}}{q'} \myprod{\bar{P}}{\realeps{*}{}} \myprod{k'}{q'}
+\myprod{k}{q'} \myprod{\bar{P}}{q'} \myprod{\bar{P}
}{\realeps{*}{}} \myprod{k'}{\realeps{}{}}
\nonumber \\ & \hspace{80pt} 
-\myprod{k}{\realeps{*}{}} \myprod{\bar{P}}{q'}{}^2 \myprod{k'}{\realeps{}{}}-\myprod{k}
{\realeps{}{}} \myprod{\bar{P}}{q'}{}^2 \myprod{k'}{\realeps{*}{}}
\nonumber \\ & \hspace{80pt} 
-2 \myprod{k}{q'} \myprod{\bar{P}}{\realeps{}{}} \myprod{\bar{P}}{\realeps{*}{}} \myprod{k'}{q'}-\myprod{k}{k'} \myprod{\bar{P}}{q'}{}^2\Big]
\Big\}\\ 
\sigmaunpol{2}{3}=
&- 16\,M^2 D_+ D_-   \Big\{ \frac{W_2}{M^2}\Big[\myprod{\bar{P}}{q'} \left(\myprod{k}{q'} \myprod{k'}{\bar{P}}-\myprod{k}{\bar{P}} \myprod{k'}{q'}\right)\Big]\Big\}\\ 
\sigmaunpol{3}{3}=&\hspace{12pt}
16\,M^2 D_+^2 \hspace{14pt}
\Big\{ 
W_1 \Big[2\myprod{k}{q'} \myprod{k'}{q'}\Big]
\nonumber \\ & \hspace{58pt} 
+ \frac{W_2}{M^2}
\Big[
\myprod{k}{\realeps{*}{}} \myprod{\bar{P}}{q'} \myprod{\bar{P}}{\realeps{}{}} \myprod{k'}{q'}+\myprod{k}{q'} \myprod{\bar{P}}{q'}
 \myprod{\bar{P}}{\realeps{}{}} \myprod{k'}{\realeps{*}{}}
\nonumber \\ & \hspace{80pt} 
+\myprod{k}{\realeps{}{}} \myprod{\bar{P}}{q'} \myprod{\bar{P}}{\realeps{*}{}} \myprod{k'}{q'}
+\myprod{k}{q'} \myprod{\bar{P}}{q'} \myprod{\bar{P}
}{\realeps{*}{}} \myprod{k'}{\realeps{}{}}
\nonumber \\ & \hspace{80pt} 
-\myprod{k}{\realeps{*}{}} \myprod{\bar{P}}{q'}{}^2 \myprod{k'}{\realeps{}{}}-\myprod{k}
{\realeps{}{}} \myprod{\bar{P}}{q'}{}^2 \myprod{k'}{\realeps{*}{}}
\nonumber \\ & \hspace{80pt} 
-2 \myprod{k}{q'} \myprod{\bar{P}}{\realeps{}{}} \myprod{\bar{P}}{\realeps{*}{}} \myprod{k'}{q'}-\myprod{k}{k'} \myprod{\bar{P}}{q'}{}^2
\nonumber \\ & \hspace{80pt} 
+2 \myprod{k}{\bar{P}
} \myprod{\bar{P}}{q'} \myprod{k'}{q'}+2 \myprod{k}{q'} \myprod{\bar{P}}{q'} \myprod{k'}{\bar{P}}
-2M^2 \left( \tau  +1\right) \myprod{k}{q'} \myprod{k'}{q'}\Big]
\Big\}
\end{align}
where: $W_1\equiv\tau G_M^2$, $W_2\equiv (F_1^2 + \tau F_2^2)$ and ${P}\equiv(p+p')/2$.
The expressions above involve the polarization vector for the outgoing photon, $\epsilon_{\Lambda_\gamma'}$. 
\subsection{Target proton polarization}
\begin{align}
\sigmapol{1}{1}=&(2\,h)\,8 M |C|^2 \nonumber \\
\times&\Big\{
G_M^2\,
\Big[ \myprod{k}{S} \myprod{k'}{\Delta }-\myprod{k}{\Delta } \myprod{k'}{S}\Big] \nonumber \\ 
&
-(1+\tau)G_M\,F_2
\Big[\myprod{k}{S} \myprod{k'}{\Delta } 
-\myprod{k}{\Delta } \myprod{k'}{S} \nonumber \\ 
&\hspace{30pt}-
\frac{\myprod{P'}{S}}{2\,M^2(1+\tau)} 
\left(\myprod{k}{\bar{P}} \myprod{k'}{\Delta }-\myprod{k}{\Delta } \myprod{k'}{\bar{P}}\right)
\Big]
\Big\}
\label{sigmapol11}
\end{align}
\begin{align}
\sigmapol{1}{2}=&(2\,h)\,4 M C D_-  \nonumber \\
\times&
\Big\{
G_M^2
\Big[
\myprod{S}{\realeps{}{}} \left(\myprod{k}{\Delta } \myprod{k'}{q'}-\myprod{k}{q'} \myprod{k'}{\Delta }\right) \nonumber \\ 
&+\myprod{k}{S} \left(-\myprod{\Delta }{\realeps{}{}} \myprod{k'}{q'}+\myprod{q'}{\Delta } \myprod{k'}{\realeps{}{}}-\myprod{k}{\Delta } \myprod{k'}{\realeps{}{}}+\myprod{k}{\realeps{}{}} \myprod{k'}{\Delta }\right) \nonumber \\ 
&+\myprod{k'}{S} \left(-\myprod{k}{\realeps{}{}} \left(\myprod{k'}{\Delta }+\myprod{q'}{\Delta }\right)+\myprod{k}{\Delta } \myprod{k'}{\realeps{}{}}+\myprod{k}{q'} \myprod{\Delta }{\realeps{}{}}\right) \nonumber \\ 
&+\myprod{P'}{S} \left(\myprod{k}{\Delta } \myprod{k'}{\realeps{}{}}-\myprod{k}{\realeps{}{}} \myprod{k'}{\Delta }\right)
\Big] \nonumber \\ 
&- (1+\tau)F_2 G_M
\Big[
\myprod{S}{\realeps{}{}} \left(\myprod{k}{\Delta } \myprod{k'}{q'}-\myprod{k}{q'} \myprod{k'}{\Delta }\right) \nonumber \\ 
&+  \myprod{k}{S} \left(-\myprod{\Delta }{\realeps{}{}} \myprod{k'}{q'}+\myprod{q'}{\Delta } \myprod{k'}{\realeps{}{}}-\myprod{k}{\Delta } \myprod{k'}{\realeps{}{}}+\myprod{k}{\realeps{}{}} \myprod{k'}{\Delta }\right) \nonumber \\ 
&-  \myprod{k'}{S} \left(\myprod{k}{\realeps{}{}} \left(\myprod{k'}{\Delta }+\myprod{q'}{\Delta }\right)-\myprod{k}{\Delta } \myprod{k'}{\realeps{}{}}-\myprod{k}{q'} \myprod{\Delta }{\realeps{}{}}\right) \nonumber \\ 
&+\frac{\myprod{P'}{S}}{2\,M^2(1+\tau)} 
\Big(
-\myprod{k}{\realeps{}{}} 
\left(\myprod{k'}{\Delta } \left(\myprod{\bar{P}}{q'}+2 M^2 (\tau +1)\right)-\myprod{q'}{\Delta } \myprod{k'}{\bar{P}}\right)\nonumber\\
&+\myprod{k}{\Delta } 
\Big(\myprod{k'}{\realeps{}{}} \left(\myprod{\bar{P}}{q'}+2 M^2 (\tau +1)\right)-\myprod{\bar{P}}{\realeps{}{}} \myprod{k'}{q'}\Big)\nonumber\\
&+\myprod{\Delta }{\realeps{}{}} \myprod{k}{\bar{P}} \myprod{k'}{q'}
-\myprod{q'}{\Delta } \myprod{k}{\bar{P}} \myprod{k'}{\realeps{}{}}\nonumber \\
&-\myprod{k}{q'} \myprod{\Delta }{\realeps{}{}} \myprod{k'}{\bar{P}}
+\myprod{k}{q'} \myprod{\bar{P}}{\realeps{}{}} \myprod{k'}{\Delta }
\Big)
\Big]
\Big\}
\label{sigmapol12}
\end{align}
\begin{align}
\sigmapol{1}{3}=&(2\,h) 8 M C D_+\nonumber\\
\times&
\Big\{
G_M^2\Big[ 
\myprod{S}{\realeps{}{}} \left(\myprod{k}{\Delta } \myprod{k'}{q'}+\myprod{k}{q'} \myprod{k'}{\Delta }\right) \nonumber \\ 
&-\myprod{k}{S} \left(\myprod{\Delta }{\realeps{}{}} \myprod{k'}{q'}-\myprod{q'}{\Delta } \myprod{k'}{\realeps{}{}}+\myprod{k}{\Delta } \myprod{k'}{\realeps{}{}}+\myprod{k}{\realeps{}{}} \myprod{k'}{\Delta }\right) \nonumber \\ 
&+\myprod{k'}{S} \left(\myprod{k}{\realeps{}{}} \left(\myprod{k'}{\Delta }+\myprod{q'}{\Delta }\right)+\myprod{k}{\Delta } \myprod{k'}{\realeps{}{}}-\myprod{k}{q'} \myprod{\Delta }{\realeps{}{}}\right) \nonumber \\ 
&+\myprod{P'}{S} \left(\myprod{k}{\Delta } \myprod{k'}{\realeps{}{}}+\myprod{k}{\realeps{}{}} \myprod{k'}{\Delta }\right)
\Big] \nonumber \\ 
&-F_2 G_M (1+\tau)
\Big[
\myprod{S}{\realeps{}{}} \left(\myprod{k}{\Delta } \myprod{k'}{q'}+\myprod{k}{q'} \myprod{k'}{\Delta }\right) \nonumber \\ 
&-\myprod{k}{S} \left(\myprod{\Delta }{\realeps{}{}} \myprod{k'}{q'}-\myprod{q'}{\Delta } \myprod{k'}{\realeps{}{}}+\myprod{k}{\Delta } \myprod{k'}{\realeps{}{}}+\myprod{k}{\realeps{}{}} \myprod{k'}{\Delta }\right) \nonumber \\ 
&+\myprod{k'}{S} \left(\myprod{k}{\realeps{}{}} \left(\myprod{k'}{\Delta }+\myprod{q'}{\Delta }\right)+\myprod{k}{\Delta } \myprod{k'}{\realeps{}{}}-\myprod{k}{q'} \myprod{\Delta }{\realeps{}{}}\right) \nonumber \\ 
&+\frac{\myprod{P'}{S}}{2 M^2(1+\tau)}\Big(\myprod{k}{\realeps{}{}} \left(\myprod{k'}{\Delta } \left(\myprod{\bar{P}}{q'}+2 M^2 (\tau +1)\right)-\myprod{q'}{\Delta } \myprod{k'}{\bar{P}}\right)\nonumber \\
&\hspace{20pt}+\myprod{k}{\Delta } \left(\myprod{k'}{\realeps{}{}} \left(\myprod{\bar{P}}{q'}+2 M^2 (\tau +1)\right)-\myprod{\bar{P}}{\realeps{}{}} \myprod{k'}{q'}\right)\nonumber \\
&\hspace{20pt}+\myprod{\Delta }{\realeps{}{}} \myprod{k}{\bar{P}} \myprod{k'}{q'}-\myprod{q'}{\Delta } \myprod{k}{\bar{P}} \myprod{k'}{\realeps{}{}}\nonumber \\
&\hspace{20pt}+\myprod{k}{q'} \myprod{\Delta }{\realeps{}{}} \myprod{k'}{\bar{P}}-\myprod{k}{q'} \myprod{\bar{P}}{\realeps{}{}} \myprod{k'}{\Delta }
\Big)
\Big] 
\Big\}
\label{sigmapol13}
\end{align}
\begin{align}
\sigmapol{2}{2}=&(2\,h) 8 M D_-^2 \nonumber\\
\times\,{\cal R}e&\Big\{
-G_M^2 \Big[
\hspace{20pt}\myprod{q'}{\Delta } \myprod{S}{\realeps{*}{}} \left(\myprod{k}{q'} \myprod{k'}{\realeps{}{}}-\myprod{k}{\realeps{}{}} \myprod{k'}{q'}\right) \nonumber \\ 
&\hspace{50pt}+\myprod{k}{S} \Big(
\myprod{k'}{q'} \left(
\myprod{k}{\realeps{*}{}} \myprod{\Delta }{\realeps{}{}}
- \myprod{q'}{\Delta }
+ \myprod{k}{\Delta }\right)\nonumber\\
&\hspace{80pt}-\myprod{k}{q'} \left(
\myprod{\Delta }{\realeps{*}{}} \myprod{k'}{\realeps{}{}}
+ \myprod{k'}{\Delta }\right)\Big)\nonumber\\
&\hspace{50pt}+\myprod{k'}{S}
\Big(-\myprod{k'}{q'} 
\left(
\myprod{k}{\realeps{*}{}} \myprod{\Delta }{\realeps{}{}}
+\myprod{k}{\Delta }
\right)\nonumber \\
&\hspace{80pt}+\myprod{k}{q'} 
\left(
\myprod{\Delta }{\realeps{*}{}} \myprod{k'}{\realeps{}{}}
+ \myprod{k'}{\Delta }+ \myprod{q'}{\Delta }\right)\Big) \nonumber \\ 
&\hspace{50pt}+\myprod{P'}{S} \Big(
-\myprod{k'}{q'} 
\left(
\myprod{k}{\realeps{*}{}} \myprod{\Delta }{\realeps{}{}}
+\myprod{k}{\Delta }\right)\nonumber\\
&\hspace{80pt}+\myprod{k}{q'}
\left(
\myprod{\Delta }{\realeps{*}{}} \myprod{k'}{\realeps{}{}}
+ \myprod{k'}{\Delta }
\right)
\Big)
\Big] \nonumber \\ 
&+\frac{G_M F_2 }{M^2}
\Big[
\hspace{15pt}\myprod{S}{\realeps{*}{}} \Big(\myprod{k}{\realeps{}{}} \myprod{\bar{P}}{q'} \left(\myprod{\bar{P}}{q'} \myprod{k'}{\Delta }-\myprod{q'}{\Delta } \myprod{k'}{\bar{P}}\right)\nonumber\\
&\hspace{80pt}+\myprod{q'}{\Delta } \myprod{k}{\bar{P}} \myprod{\bar{P}}{q'} \myprod{k'}{\realeps{}{}}-\myprod{k}{q'} \myprod{\bar{P}}{q'} \myprod{\bar{P}}{\realeps{}{}} \myprod{k'}{\Delta }\nonumber\\
&\hspace{80pt}+\myprod{k}{\Delta } \myprod{\bar{P}}{q'} \left(\myprod{\bar{P}}{\realeps{}{}} \myprod{k'}{q'}-\myprod{\bar{P}}{q'} \myprod{k'}{\realeps{}{}}\right)\nonumber\\
&\hspace{80pt}-\myprod{q'}{\Delta } \myprod{k}{\bar{P}} \myprod{\bar{P}}{\realeps{}{}} \myprod{k'}{q'}+\myprod{k}{q'} \myprod{q'}{\Delta } \myprod{\bar{P}}{\realeps{}{}} \myprod{k'}{\bar{P}}\Big) \nonumber \\ 
&\hspace{50pt}+\myprod{k}{S} 
\Big(
\myprod{\Delta }{\realeps{*}{}} \myprod{\bar{P}}{q'}{}^2 \myprod{k'}{\realeps{}{}}
+\myprod{k}{\realeps{*}{}} \myprod{\Delta }{\realeps{}{}} \myprod{\bar{P}}{q'} \myprod{k'}{\bar{P}}\nonumber \\
&\hspace{80pt}-\myprod{\Delta }{\realeps{*}{}} \myprod{\bar{P}}{q'} \myprod{\bar{P}}{\realeps{}{}} \myprod{k'}{q'}
-\myprod{k}{\realeps{*}{}} \myprod{\bar{P}}{q'} \myprod{\bar{P}}{\realeps{}{}} \myprod{k'}{\Delta }\nonumber \\
&\hspace{80pt}-\myprod{q'}{\Delta } \myprod{\bar{P}}{q'} \myprod{\bar{P}}{\realeps{*}{}} \myprod{k'}{\realeps{}{}}
-\myprod{k}{q'} \myprod{\Delta }{\realeps{*}{}} \myprod{\bar{P}}{\realeps{}{}} \myprod{k'}{\bar{P}}\nonumber \\
&\hspace{80pt}+ \myprod{q'}{\Delta } \myprod{\bar{P}}{\realeps{}{}} \myprod{\bar{P}}{\realeps{*}{}} \myprod{k'}{q'}
+ \myprod{k}{q'} \myprod{\bar{P}}{\realeps{}{}} \myprod{\bar{P}}{\realeps{*}{}} \myprod{k'}{\Delta }\nonumber\\
&\hspace{80pt}+\myprod{k}{\bar{P}} 
\left(
-\myprod{\Delta }{\realeps{*}{}} \myprod{\bar{P}}{q'} \myprod{k'}{\realeps{}{}}
+\myprod{\Delta }{\realeps{*}{}} \myprod{\bar{P}}{\realeps{}{}} \myprod{k'}{q'}
- \myprod{\bar{P}}{q'} \myprod{k'}{\Delta }
\right)\nonumber\\
&\hspace{80pt}+\myprod{k}{\Delta } 
\left(
\myprod{\bar{P}}{q'}\myprod{\bar{P}}{\realeps{*}{}} \myprod{k'}{\realeps{}{}}
- \myprod{\bar{P}}{\realeps{}{}}\myprod{\bar{P}}{\realeps{*}{}} \myprod{k'}{q'}
+ \myprod{\bar{P}}{q'} \myprod{k'}{\bar{P}}
\right)\nonumber\\
&\hspace{80pt}+ \myprod{\bar{P}}{q'}{}^2 \myprod{k'}{\Delta }
- \myprod{q'}{\Delta } \myprod{\bar{P}}{q'} \myprod{k'}{\bar{P}}
\Big) \nonumber \\ 
&\hspace{50pt}+\myprod{k'}{S} 
\Big(
-\myprod{k}{\realeps{*}{}} \myprod{\Delta }{\realeps{}{}} \myprod{\bar{P}}{q'} \myprod{k'}{\bar{P}}
+\myprod{k}{\realeps{*}{}} \myprod{\bar{P}}{q'} \myprod{\bar{P}}{\realeps{}{}} \myprod{k'}{\Delta }\nonumber \\
&\hspace{80pt}+\myprod{k}{q'} \myprod{\Delta }{\realeps{*}{}} \myprod{\bar{P}}{\realeps{}{}} \myprod{k'}{\bar{P}}
- \myprod{k}{q'} \myprod{\bar{P}}{\realeps{}{}} \myprod{\bar{P}}{\realeps{*}{}} \myprod{k'}{\Delta}\nonumber \\
&\hspace{80pt}+\myprod{k}{\bar{P}} 
\Big(
\myprod{\bar{P}}{q'} 
\left(
\myprod{\Delta }{\realeps{*}{}} \myprod{k'}{\realeps{}{}}
+ \myprod{q'}{\Delta }
+ \myprod{k'}{\Delta }
\right)-\myprod{k'}{q'}
\myprod{\Delta }{\realeps{*}{}} \myprod{\bar{P}}{\realeps{}{}}
\Big)\nonumber\\
&\hspace{80pt}-\myprod{k}{\Delta }
\Big(
\myprod{\bar{P}}{q'} 
\myprod{\bar{P}}{\realeps{}{}} \myprod{k'}{\realeps{*}{}}
- \myprod{\bar{P}}{\realeps{}{}} \myprod{\bar{P}}{\realeps{*}{}} \myprod{k'}{q'}
+ \myprod{\bar{P}}{q'} \myprod{k'}{\bar{P}}
+ \myprod{\bar{P}}{q'}{}^2
\Big)\nonumber \\
&\hspace{80pt}+\myprod{k}{q'} \myprod{\Delta }{\realeps{*}{}} \myprod{\bar{P}}{q'} \myprod{\bar{P}}{\realeps{}{}}
+\myprod{k}{\realeps{*}{}} \myprod{q'}{\Delta } \myprod{\bar{P}}{q'} \myprod{\bar{P}}{\realeps{}{}}\nonumber\\
&\hspace{80pt}-\myprod{k}{\realeps{*}{}} \myprod{\Delta }{\realeps{}{}} \myprod{\bar{P}}{q'}{}^2
- \myprod{k}{q'} \myprod{q'}{\Delta } \myprod{\bar{P}}{\realeps{}{}} \myprod{\bar{P}}{\realeps{*}{}}
\Big) \nonumber \\ 
&\hspace{50pt}+\myprod{P'}{S}
\Big(
-\myprod{k}{\realeps{*}{}} \myprod{\Delta }{\realeps{}{}} \myprod{\bar{P}}{q'} \myprod{k'}{\bar{P}}
+\myprod{k}{\realeps{*}{}} \myprod{\bar{P}}{q'} \myprod{\bar{P}}{\realeps{}{}} \myprod{k'}{\Delta }\nonumber\\
&\hspace{80pt}+\myprod{k}{q'} \myprod{\Delta }{\realeps{*}{}} \myprod{\bar{P}}{\realeps{}{}} \myprod{k'}{\bar{P}}
- \myprod{k}{q'} \myprod{\bar{P}}{\realeps{}{}} \myprod{\bar{P}}{\realeps{*}{}} \myprod{k'}{\Delta }\nonumber\\
&\hspace{80pt}+\myprod{k}{\bar{P}}
\left(
\myprod{\Delta }{\realeps{*}{}} \myprod{\bar{P}}{q'} \myprod{k'}{\realeps{}{}}
-\myprod{\Delta }{\realeps{*}{}} \myprod{\bar{P}}{\realeps{}{}} \myprod{k'}{q'}
+ \myprod{\bar{P}}{q'} \myprod{k'}{\Delta }
\right)\nonumber\\
&\hspace{80pt}-\myprod{k}{\Delta }
\left(
\myprod{\bar{P}}{q'} \myprod{\bar{P}}{\realeps{*}{}} \myprod{k'}{\realeps{}{}}
-\myprod{\bar{P}}{\realeps{}{}} 
\myprod{\bar{P}}{\realeps{*}{}} \myprod{k'}{q'}
+ \myprod{\bar{P}}{q'} \myprod{k'}{\bar{P}}
\right)
\Big)
\Big]
\Big\}
\label{sigmapol22}
\end{align}
\begin{align}
\sigmapol{2}{3}=&(2\,h) 4 M D_- D_+\nonumber\\
&\times
\Big\{
-G_M^2\Big[ 
\myprod{k}{S} \Big(\myprod{k}{\Delta } \myprod{k'}{q'}-\myprod{q'}{\Delta } \myprod{k'}{q'}+\myprod{k}{q'} \myprod{k'}{\Delta }\Big) \nonumber \\ 
&-\myprod{k'}{S} \left(\myprod{k}{\Delta } \myprod{k'}{q'}+\myprod{k}{q'} \left(\myprod{k'}{\Delta }+\myprod{q'}{\Delta }\right)\right) \nonumber \\ 
&+\myprod{P'}{S} \left(-\left(\myprod{k}{\Delta } \myprod{k'}{q'}+\myprod{k}{q'} \myprod{k'}{\Delta }\right)\right)
\Big] \nonumber \\ 
&-G_M F_2(1+\tau)
\Big[ 
- \myprod{k}{S} \left(\myprod{k}{\Delta } \myprod{k'}{q'}-\myprod{q'}{\Delta } \myprod{k'}{q'}+\myprod{k}{q'} \myprod{k'}{\Delta }\right) \nonumber \\ 
&+ \myprod{k'}{S} \left(\myprod{k}{\Delta } \myprod{k'}{q'}+\myprod{k}{q'} \left(\myprod{k'}{\Delta }+\myprod{q'}{\Delta }\right)\right) \nonumber \\ 
&+\frac{\myprod{P'}{S}}{2\,M^2(1+\tau)} \Big(\myprod{k}{\Delta } \myprod{k'}{q'} \left(\myprod{\bar{P}}{q'}+2 M^2 (\tau +1)\right)\nonumber\\
&\hspace{60pt}+\myprod{k}{q'} \left(\myprod{k'}{\Delta } \left(\myprod{\bar{P}}{q'}+2 M^2 (\tau +1)\right)-\myprod{q'}{\Delta } \myprod{k'}{\bar{P}}\right)\nonumber\\
&\hspace{60pt}-\myprod{q'}{\Delta } \myprod{k}{\bar{P}} \myprod{k'}{q'}\Big)
\Big]
\Big\}
\label{sigmapol23}
\\ 
\sigmapol{3}{3}=&(2\,h) 8 
M D_+^2   \nonumber \\ 
\times&
\,{\cal R}e\Big\{
-G_M^2
\Big[\myprod{q'}{\Delta } \myprod{S}{\realeps{*}{}} \left(\myprod{k}{\realeps{}{}} \myprod{k'}{q'}-\myprod{k}{q'} \myprod{k'}{\realeps{}{}}\right) \nonumber \\ 
&\hspace{50pt}+\myprod{k}{S} \left(
-\myprod{k}{\realeps{*}{}} \myprod{\Delta }{\realeps{}{}} \myprod{k'}{q'}
+\myprod{k}{q'} \myprod{\Delta }{\realeps{*}{}} \myprod{k'}{\realeps{}{}}\right) \nonumber \\ 
&\hspace{50pt}
+\myprod{k'}{S} 
\left(
\myprod{k}{\realeps{*}{}} \myprod{\Delta }{\realeps{}{}} \myprod{k'}{q'}
-\myprod{k}{q'} \myprod{\Delta }{\realeps{*}{}} \myprod{k'}{\realeps{}{}}\right) \nonumber \\ 
&\hspace{50pt}+\myprod{P'}{S} 
\left(
\myprod{k}{\realeps{*}{}} \myprod{\Delta }{\realeps{}{}} \myprod{k'}{q'}
-\myprod{k}{q'} \myprod{\Delta }{\realeps{*}{}} \myprod{k'}{\realeps{}{}}\right) \nonumber \\ 
&-G_M F_2 (1+\tau)
\Big[- \myprod{q'}{\Delta } \myprod{S}{\realeps{*}{}} \left(\myprod{k}{\realeps{}{}} \myprod{k'}{q'}-\myprod{k}{q'} \myprod{k'}{\realeps{}{}}\right) \nonumber \\ 
&\hspace{50pt}+  \myprod{k}{S} \left(
\myprod{k}{\realeps{*}{}} \myprod{\Delta }{\realeps{}{}} \myprod{k'}{q'}
-\myprod{k}{q'} \myprod{\Delta }{\realeps{*}{}} \myprod{k'}{\realeps{}{}}\right) \nonumber \\ 
&\hspace{50pt}- \myprod{k'}{S} \left(
\myprod{k}{\realeps{*}{}} \myprod{\Delta }{\realeps{}{}} \myprod{k'}{q'}
-\myprod{k}{q'} \myprod{\Delta }{\realeps{*}{}} \myprod{k'}{\realeps{}{}}\right) \nonumber \\ 
&\hspace{50pt}+\myprod{P'}{S} 
\Big(
\myprod{k}{\realeps{*}{}} \myprod{k'}{q'} \left(\myprod{q'}{\Delta } \myprod{\bar{P}}{\realeps{}{}}-\myprod{\Delta }{\realeps{}{}} \left(\myprod{\bar{P}}{q'}+2 M^2 (\tau +1)\right)\right)\nonumber\\
&\hspace{80pt}- \myprod{k'}{\realeps{*}{}}\myprod{k}{q'} \left(\myprod{q'}{\Delta } \myprod{\bar{P}}{\realeps{}{}}-\myprod{\Delta }{\realeps{}{}} \left(\myprod{\bar{P}}{q'}+2 M^2 (\tau +1)\right)\right)
\Big)
\Big]
\Big\}
\label{sigmapol33}\,\,,
\end{align}
where we have omitted terms that vanish after summing over the final photon polarization.
In the lab frame, for the longitudinal target polarization case, the azimuthal dependece of the BH amplitude squared comes only from the invariant $(k\,\Delta)$. For transverse  polarization, additional azimuthal dependence is introduced through $(P^\prime\,S)$, $(k\,S$ and $k^\prime\,S)$. 

\section{Details of the BH/DVCS Interference Coefficients}

Longitudinally Polarized Beam, Unpolarized Target

\begin{eqnarray}
A_{LU}^{\mathcal{I}} &=& \frac{1}{2(kq')(k'q')}\Bigg \{(Q^{2}+t)\Big((P^{+}q^{-}-P^-q^+)(kk)_{T}  + (k'^-k^+ - k'^+k^-)(Pk)_{T} + 2(P^+q'^- - P^-q'^+)(kk)_{T} \nonumber \\ && \hspace{2cm} - (P^+k^- - P^-k^+)(kq')_{T} + (q'^+k^- - q'^-k^+)(Pk)_{T} -  (P^+k'^- - P^-k'^+)(kq')_{T}  \nonumber \\ && \hspace{2cm} + (q'^+k'^- - q'^-k'^+)(Pk)_{T} + 2( q^{\prime -}P^{+}-q^{\prime +}P^{-})(kk')\Big) \nonumber \\
&-& (Q^{2}-t+4(k\Delta))\Big( 2(k^-k'^+ - k^+k'^-)(Pq')+(k'^-k^+ - k'^+k^-)(Pq')_{T} + (k'^+q'^- - k'^-q'^+ )(Pk)_{T}  \nonumber \\ && \hspace{2cm} - (k^+q'^- - k^-q'^+ )(Pk)_{T} \Big) \Bigg \} \sin{\phi}\\
B_{LU}^{\mathcal{I}} &=& \frac{\xi}{4(kq')(k'q')} \Bigg \{(Q^{2}+t)\Big((\Delta^+q^- - \Delta^-q^+)(kk)_{T}  + (k'^-k^+  - k'^+k^-) (k\Delta)_{T} + (\Delta^+q'^- - \Delta^-q'^+)(kk)_{T} \nonumber \\ && \hspace{2cm} -  (\Delta^+k^- - \Delta^-k^+)(kq')_{T}  + (q'^+k^- - q'^-k^+)(k\Delta)_{T} + (\Delta^+q'^- - \Delta^-q'^+)(kk)_{T} \nonumber \\
&& \hspace{2cm} -  (\Delta^+k'^- - \Delta^-k'^+)(kq')_{T}  + (q'^+k'^- - q'^-k'^+)(k\Delta)_{T} + 2( q^{\prime -}\Delta^{+}-q^{\prime +}\Delta^{-})(kk')\Big)\nonumber \\
&-&  (Q^{2}-t+4(k\Delta)) \Big( 2(k^-k'^+ - k^+k'^-)(\Delta q')+(k'^-k^+- k'^+k^-)(q'\Delta)_{T}+(k'^+q'^- - k'^-q'^+)(k\Delta)_{T} \nonumber \\ && \hspace{2cm}  +  (k^-q'^+  - k^+q'^-)(k\Delta)_{T}
 \Big) \Bigg \} \sin{\phi} \\
C_{LU}^{\mathcal{I}} &=& \frac{1}{4(kq')(k'q')}\Bigg \{(Q^{2}+t)\Big((k'^-k^+ - k'^+k^-)(k\Delta)_{T}+ (q'^-k^+ - q'^+k^- )(k\Delta)_{T}+ (q'^-k'^+ - q'^+k'^-)(k \Delta)_{T}\Big) \nonumber \\
&-&  (Q^{2}-t+4(k\Delta))\Big( (k^{\prime +}q^{\prime -}-k^{\prime -}q^{\prime +})(k\Delta)_{T}  + (k^{\prime -}k^{+}-k^{\prime +}k^{-})(q'\Delta)_{T}+(k^{-}q^{\prime +}-k^{+}q^{\prime -})(k\Delta)_{T} \Big) \Bigg \} \sin{\phi} \nonumber \\ 
\end{eqnarray}

Unpolarized Beam, Longitudinally Polarized Target

\begin{eqnarray}
A_{UL}^{\mathcal{I}} &=&\frac{1}{2(kq')(k'q')}\Bigg \{ (Q^{2}+t)\Big((k'P)(2(kk)_{T} -(kq')_{T} + 2(kq')) + (kP)(2(kk')_{T} -(k'q')_{T} + 2(k'q')) \nonumber \\
&&\hspace{2cm} +  2(kk')(kP)_T - (k'q')(kP)_T - (kq')(k'P)_T \Big)\nonumber \\
&-&  (Q^{2}-t+4(k\Delta))\Big((Pq')((kk')_{T}+(k'q')_{T} 
- 2(kk')) - (2(kk')(q'P)_T - (k'q')(kP)_T - (kq')(k'P)_T)\Big) \Bigg \}\sin{\phi} \nonumber \\
\\
B_{UL}^{\mathcal{I}} &=& \frac{\xi}{4(kq')(k'q')} \Bigg \{(Q^{2}+t)\Big((k'\Delta)(2(kk)_{T} -(kq')_{T} + 2(kq')) + (k\Delta)(2(kk')_{T} -(k'q')_{T} + 2(k'q'))\nonumber  \\
&& \hspace{2cm} + 2(kk')(k\Delta)_T - (k'q')(k\Delta)_T - (kq')(k'\Delta)_T \Big) \nonumber \\ &-& (Q^{2}-t+4(k\Delta))\Big((\Delta q')((kk')_{T}+(k'q')_{T}
- 2(kk')) - (2(kk')(q'\Delta)_T - (k'q')(k\Delta)_T - (kq')(k'\Delta)_T)\Big) \Bigg \} \sin{\phi} \nonumber \\
\\
C_{UL}^{\mathcal{I}} &=& \frac{1}{4(kq')(k'q')}\Bigg \{ (Q^{2}+t)\Big( 2(kk')(k\Delta)_{T} - (k'q')(k\Delta)_{T} - (kq')(k'\Delta)_{T} 
\nonumber \\ && \hspace{2cm} + 2\xi \big(2(kk')(kP)_{T} - (k'q')(kP)_{T} - (kq')(k'P)_{T}\big) \Big) \nonumber \\ &+& (Q^{2}-t+4(k\Delta))\Big( (kk')(q'\Delta)_{T} - (k'q')(k\Delta)_{T}- (kq')(k'\Delta)_{T} \nonumber \\ && \hspace{2cm}+ 2\xi \big( (kk')(q'P)_{T}- (k'q')(kP)_{T}- (kq')(k'P)_{T}\big) \Big) \Bigg \}\sin{\phi} \nonumber \\
\end{eqnarray}

Longitudinally Polarized Beam, Longitudinally Polarized Target

\begin{eqnarray}
A_{LL}^{\mathcal{I}} &=&  - \frac{1}{4(kq')(k'q')}\Bigg \{ (Q^{2}+t)\Big( (k^{\prime +}P^{-}- k^{\prime -}P^{+})(kk)_{T}  + (k^{-}P^{+} - k^{+}P^{-})(kk)_{T} + (k'^-k^+ - k'^+k^-)(Pk)_{T} \nonumber \\
&& \hspace{2cm} + 2(q^{\prime -}P^{+} - q^{\prime +}P^{-})(kk)_{T} + (k^{+}P^{-}- k^{-}P^{+})(kq')_{T}  + (q'^+k^- - q'^-k^+)(Pk)_{T} \nonumber\\ && \hspace{2cm} + (k'^{+}P^{-}-k'^{-}P^{+})(kq')_{T}
   + (q'^+k'^- - q'^-k'^+)(Pk)_{T} - 2(kk')(q'^+P^- -q'^-P^+)\Big) \nonumber\\
&-& (Q^{2}-t+4(k\Delta))\Big(2(Pq')(k'^+k^- - k'^-k^+  ) + (Pq')_{T}(k'^+k^- - k'^-k^+) + (Pk)_{T}(k^+q'^- k^-q'^+) 
\nonumber \nonumber \\ && \hspace{2cm} +  (Pk)_{T}(k'^-q'^+ - k'^+q'^-)    \Big) \Bigg \} \cos{\phi} \\
B_{LL}^{\mathcal{I}} &=&  -\frac{\xi}{8(kq')(k'q')}\Bigg \{ (Q^{2}+t) \Big((kk)_{T}(k^{-}\Delta^{+} - k^{+}\Delta^{-}) + (kk)_{T}(k^{\prime +}\Delta^{-} - k^{\prime -}\Delta^{+})  + (k'^-k^+ - k'^+k^-)(k\Delta)_{T} \nonumber \\
&& \hspace{2cm} +  2(kk)_{T}(q^{\prime -}\Delta^{+}- q^{\prime +}\Delta^{-}) + (kq')_{T}(k^{+}\Delta^{-} - k^{-}\Delta^{+})  + (q'^+k^- - q'^-k^+)(k\Delta)_{T} \nonumber \\
&& \hspace{2cm} +  (kq')_{T}(k'^{+}\Delta^{-}  - k'^{-}\Delta^{+} ) + (q'^+k'^- - q'^-k'^+)(k\Delta)_{T}  - 2(kk')(q'^+\Delta^- -q'^-\Delta^+)\Big)\nonumber \\
&-& (Q^{2}-t+4(k\Delta)) \Big(2(\Delta q')(k'^+k^- - k'^-k^+ + ) + (q'\Delta)_{T}(k'^+k^- - k'^-k^+) + (k\Delta)_{T}(k^+q'^- - k^-q'^+) \nonumber \\
&& \hspace{2cm} + (k\Delta)_{T}(k'^-q'^+  -  k'^+q'^- ) \Big) \Bigg \} \cos{\phi}\\
C_{LL}^{\mathcal{I}} &=& \frac{1}{4(kq')(k'q')}\Bigg \{ (Q^{2}+t))\Big((k^-k'^+ - k'^-k^+)(k\Delta)_{T} + (q'^+k^-  - q'^-k^+ )(k\Delta)_{T}  + (q'^+k'^- - q'^-k'^+ )(k\Delta)_{T} \Big) \nonumber\\
&-& (Q^{2}-t+4(k\Delta)) \Big((k^{\prime +}k^{-}-k^{\prime -}k^{+})(q'\Delta)_{T}+(k^{\prime -}q^{\prime +}-k^{\prime +}q^{\prime -})(k\Delta)_{T}+ (k^{+}q^{\prime -}-k^{-}q^{\prime +})(k\Delta)_{T} \Big) \Bigg \} \cos{\phi} \nonumber \\
\end{eqnarray}

\bibliography{DVCS_BH_bib}
\end{document}